\documentclass{article}
\usepackage{epubtk}    
\usepackage{booktabs}  
\usepackage{graphicx} 
\usepackage{upgreek}
\usepackage{mathrsfs}
\usepackage{yfonts}
\usepackage{amsmath}
\usepackage{amsfonts}
\usepackage{amssymb}
\usepackage{bm} 
\usepackage{epsfig} 
\usepackage{epstopdf}
\usepackage{eufrak,mathrsfs,yfonts}
\usepackage[breaklinks]{hyperref}
\usepackage{color}

\DeclareMathOperator\erf{erf} 

\newcommand{\RealNum}{\ensuremath{{\bf R}}}
\newcommand{\threegamma}{\gamma}
\newcommand{\threeGamma}{\Upgamma}
\newcommand{\threeK}{K}
\newcommand{\threeA}{A}
\newcommand{\threechi}{\chi}
\newcommand{\threeD}{D}

\def\be{\begin{equation}}
\def\ee{\end{equation}}
\def\beq{\begin{eqnarray}}
\def\eeq{\end{eqnarray}}
\def\tu{{\bar u}}
\def\tv{{\bar v}}

\def\trho{{\bar \rho}}

\newcommand{\A}{{\scriptscriptstyle{A}}}
\newcommand{\B}{{\scriptscriptstyle{B}}}
\newcommand{\I}{{\scriptscriptstyle{I}}}
\newcommand{\J}{{\scriptscriptstyle{J}}}
\newcommand{\M}{{\scriptscriptstyle{M}}}
\newcommand{\N}{{\scriptscriptstyle{N}}}
\newcommand{\K}{{\scriptscriptstyle{K}}}
\newcommand{\LL}{{\scriptscriptstyle{L}}}

\def\Lie{\mathcal{L}}

\showlistoftablesfalse

\begin{document}

\title{Exploring New Physics Frontiers Through Numerical Relativity}

\author{%
  \epubtkAuthorData{Vitor Cardoso}{%
    CENTRA, Departamento de F\'{\i}sica, Instituto Superior T\'ecnico,
    Universidade de Lisboa,\\
    Avenida~Rovisco Pais 1, 1049 Lisboa, Portugal\\
    and\\
    Perimeter Institute for Theoretical Physics, Waterloo, Ontario N2L
    2Y5, Canada}
  {vitor.cardoso@ist.utl.pt}{%
http://centra.ist.utl.pt/~vitor/}%
\and
\epubtkAuthorData{Leonardo Gualtieri}{%
  Dipartimento di Fisica, Universit\`a di Roma ``La Sapienza''\\ 
  \& Sezione INFN Roma1, P.A. Moro 5, 00185, Roma, Italy}{%
  Leonardo.Gualtieri@roma1.infn.it}{%
http://www.roma1.infn.it/teongrav/leonardo/homepage.html}
\and
\epubtkAuthorData{Carlos Herdeiro}{%
  Departamento de F\'\i sica da Universidade de Aveiro and CIDMA,\\
  Campus de Santiago, 3810-183 Aveiro, Portugal}{%
  herdeiro@ua.pt}{%
http://gravitation.web.ua.pt/herdeiro}
\and
\epubtkAuthorData{Ulrich Sperhake}{%
  DAMTP, Centre for Mathematical Sciences, University of Cambridge,\\
  Wilberforce Road, Cambridge, CB3 0WA, United Kingdom
}{%
  sperhake@tapir.caltech.edu}{%
  http://www.damtp.cam.ac.uk/user/us248/}
}

\date{}
\maketitle

\begin{abstract}
The demand to obtain answers to highly complex problems within
strong-field gravity has been met with significant progress in the
numerical solution of Einstein's equations -- along with some
spectacular results -- in various setups.

We review techniques for solving Einstein's equations in generic
spacetimes, focusing on fully nonlinear evolutions but also on how to
benchmark those results with perturbative approaches. The results
address problems in high-energy physics, holography, mathematical
physics, fundamental physics, astrophysics and cosmology.
\end{abstract}

\epubtkKeywords{Gravitation, Numerical methods, Black holes,
  Extensions of the standard model, Alternative theories of gravity,
  Extra dimensions, Trans-Planckian scattering}

\newpage
\tableofcontents

\newpage
\section*{Acronyms}

\begin{table}[h]
\begin{tabular}{ll}
ADM    & Arnowitt--Deser--Misner                              \\
(A)dS  & (Anti-)de Sitter                                     \\
AH     & Apparent horizon                                     \\
BH     & Black hole                                           \\
BSSN   & Baumgarte--Shapiro--Shibata--Nakamura                \\
CFT    & Conformal field theory                               \\
EOB    & Effective one body                                    \\
GHG    & Generalized Harmonic Gauge                           \\
EM     & Electromagnetism, Electromagnetic                    \\
GR     & General relativity                                   \\ 
GW     & Gravitational wave                                   \\
IBVP   & Initial-boundary-value problem                       \\
LHC    & Large Hadron Collider                                \\
LIGO   & Laser Interferometric Gravitational Wave Observatory \\
NR     & Numerical relativity                                 \\
NS     &  Neutron star                                        \\
PDE    & Partial differential equation                        \\
PN     & Post-Newtonian                                       \\
PPN    & Parametrized Post-Newtonian                          \\
QNM    & Quasi-normal mode                                    \\
RN     & Reissner--Nordstr\"om                                \\
SMT    & String/M theory                                      \\
ZFL    & Zero-frequency limit                                 
\end{tabular}
\end{table}

\bigskip

\clearpage

\section*{Notation and conventions}

Unless otherwise and explicitly stated, we use geometrized units where
$G=c=1$, so that energy and time have units of length. Geometric objects
are denoted with boldface symbols, whereas their components are not.
We also adopt the $(-+++\dots)$ convention for the metric. For reference, the following is a list of
symbols that are used often throughout the text.

\begin{table}[h]
\begin{tabular}{ll}
  $D$ & Total number of spacetime dimensions (we always consider one timelike\\
      & and $D-1$ spacelike dimensions).\\
  $L$ & Curvature radius of (A)dS spacetime, related to the (negative) positive \\
      & cosmological constant $\Lambda$ in the Einstein equations 
        ($G_{\mu\nu}+\Lambda g_{\mu\nu}=0$) \\
  & through $L^2=(D-2)(D-1)/(2|\Lambda|)$. \\
  $M$ & BH mass.\\
  $a$ & BH rotation parameter.\\
  $R_{\rm S}$ & Radius of the BH's event horizon in the chosen coordinates.\\
  $\omega$ & Fourier transform variable. The time dependence of any
  field is $\sim e^{-i\omega t}$.  \\
  & For stable spacetimes, ${\rm Im}(\omega)<0$. \\
  $s$ & Spin of the field.\\
  $l$ & Integer angular number, related to the eigenvalue $A_{lm}=l(l+D-3)$\\ 
  &of scalar spherical harmonics in $D$ dimensions.\\ 
  $a,\,b,\,\ldots,\,h$ & Index range referred to as ``early lower case
        Latin indices''\\ & (likewise for upper case indices).\\
  $i,\,j,\,\ldots,\,v$ & Index range referred to as ``late lower case
        Latin indices''\\ & (likewise for upper case indices).\\
  $g_{\alpha \beta}$ & Spacetime metric; greek indices run from 0 to $D-1$. \\
$\Gamma^{\alpha}_{\beta \gamma}$& $=\frac{1}{2} g^{\alpha \mu}
      \left( \partial_{\beta} g_{\gamma \mu} + \partial_{\gamma} g_{\mu \beta}
      - \partial_{\mu} g_{\beta \gamma} \right)$, Christoffel symbol
      associated with the \\
    & spacetime metric $g_{\alpha \beta}$.  \\
   $R^{\alpha}{}_{\beta \gamma \delta}$ & $= \partial_{\gamma}
      \Gamma^{\alpha}_{\delta \beta} - \partial_{\delta}
      \Gamma^{\alpha}_{\gamma \beta} + \Gamma^{\alpha}_{\gamma \rho}
      \Gamma^{\rho}_{\delta \beta} - \Gamma^{\alpha}_{\delta \rho}
      \Gamma^{\rho}_{\gamma \beta}$, Riemann curvature tensor of the \\
    & $D$-dimensional spacetime. \\
  $\nabla_{\alpha}$ & $D$-dimensional covariant derivative associated with
      $\Gamma^{\alpha}_{\beta \gamma}$. \\
  $\gamma_{ij}$ & Induced metric, also known as first fundamental form, on \\
    & $(D-1)$-dimensional spatial hypersurface;
      latin indices run from 1 to $D-1$. \\
  $K_{ij}$ & Extrinsic curvature, also known as second fundamental form,
      on \\
    & $(D-1)$-dimensional spatial hypersurface. \\
 $\Upgamma^i_{jj}$ & $(D-1)$-dimensional Christoffel symbol
      associated with $\gamma_{ij}$. \\
  $\mathcal{R}^i{}_{jkl}$ & $(D-1)$-dimensional Riemann curvature tensor of the
      spatial hypersurface. \\
  $D_i$ & $(D-1)$-dimensional covariant derivative associated with
      $\Upgamma^i_{jk}$.\\
   $S^n$& $n$-dimensional sphere.    
\end{tabular}
\end{table}
\clearpage
\newpage

\section{Prologue}
\label{sec:introduction}

\textit{``Wir m\"ussen wissen, wir werden wissen.''} (We must know, we will know) 

\noindent -- D.\ Hilbert, 
Address to the Society of German Scientists and Physicians, K\"onigsberg

\noindent (September 08, 1930).
\medskip

\noindent
One century of peering into Einstein's field equations has given us elegant and
simple solutions, and shown how they behave when slightly displaced from
equilibrium. We were rewarded with a beautiful mathematical theory of black
holes (BHs) and their perturbations, and a machinery which is able to handle all
weak-field phenomena. After all, one hundred years is not a very long time to
understand a theory with such conceptual richness. Left behind, as an annoying
nuisance, was the problem of dynamical strong-field effects such as the last
stages of BH mergers.

In the last few decades, it gradually became clear that analytical or
perturbative tools could only go so far: gravitational wave (GW) detectors were
promising to see the very last stages of BH-binary inspirals;
fascinating developments in String/M theory (SMT) were hinting
at a connection between gauge theories and strong gravity effects;
extensions of the standard model of particle physics were conjecturing
the existence of extra dimensions which only gravity had access to,
and were predicting BH formation at accelerators! This, and more,
required the ability to solve Einstein's equations (numerically) in
full generality in the nonlinear regime.  The small ``annoying
nuisance'' rapidly grew to become an elephant in the room that had to be tamed.

But necessity is the mother of inventions. In 2005, several groups achieved the
first long-term stable evolutions of BH-binaries in four-dimensional,
asymptotically flat spacetimes, starting a phase transition in the field.  It is
common to refer to such activity -- numerically solving Einstein's equations
\be
R_{\mu\nu}-\frac{1}{2}Rg_{\mu\nu}=\frac{8\pi G}{c^4}T_{\mu\nu}\,,
\ee
or extensions thereof -- as ``numerical relativity'' (NR). In practice, \emph{any}
numerical procedure is a means to an end, which is \emph{to know}.  In this
sense, NR is a gray area which could lie at the intersection between
numerical analysis, general relativity (GR) and high-energy physics. Many different
numerical techniques have been used to solve the field equations in a variety of
contexts. NR usually entails solving the full set of
nonlinear, time-dependent Einstein-type equations.

This is a review on NR. We will cover all aspects of the main
developments in the last decade, focusing for the most part on evolutions of
BH spacetimes. The numerical resolution of Einstein's equations in a
computer has a five-decade long history and many important ingredients. In fact,
NR is sufficiently complex that a number of outstanding review
works have already been dedicated to specific aspects, like construction of
initial data, finding horizons in numerical spacetimes, evolving the field
equations in the presence of matter, etc. We will not attempt to cover these in
any detail; we refer the reader to the relevant section of Living
Reviews%
\epubtkFootnote{\url{http://relativity.livingreviews.org/Articles/subject.html}}
for this and to textbooks on the subject at large
\cite{Alcubierre:2008,BaumgarteShapiro:2010,Palenzuela:2008,Gourgoulhon:2007ue}. The present work is mostly
intended to make the reader familiar with new developments, which have not
and could not have been covered in those works, given the pace at which the
field is evolving.

A few words about the range and applicability of NR methods are in order, as they
help clarify the content of this review work.  NR is but one,
albeit important and complex, tool which helps us to get through solving and
understanding certain processes.  Traditionally, the two-body problem in GR for
instance, was approached via a slow-motion, large separation post-Newtonian
expansion.  The PN expansion breaks down when the distances between the bodies
are small and the velocities are large. BH perturbation theory on the
other hand, can handle the two-body problem for any separation and velocity, but
as long as there is a decoupling of mass scales, i.e, one of the objects must be
much more massive than the other. The remaining is NR turf: large velocities,
small separations, strong field and similar masses.  This is depicted
in Figure~\ref{fig:NR_validity_diagram}, which we have extended to
allow for generic situations. NR methods typically break down (due to large
computational requirements) when there are extremely different scales in the problem, i.e., when
extremely large or small dimensionless quantities appear. 
For instance, the two-body problem in GR can be handled for a relatively short
timescale, \emph{and} as long as the two bodies do not have extreme mass
ratios. In spacetimes with other lengthscales, for instance AdS, NR
encounters difficulties when the binary lengthscale is much smaller than the AdS
lengthscale for example. While such simulations can in principle be
done, they may not capture the relevant physics associated with the AdS
boundary.

To conclude this discussion, neither NR nor perturbative techniques are paradisiac islands in isolation; 
input and interplay from and with other solutions is often required. As such, we will also discuss in some detail
some of the perturbative tools and benchmarks used in the field.

\epubtkImage{}{%
  \begin{figure}[htb]
    \centerline{\includegraphics[width=0.8\textwidth]{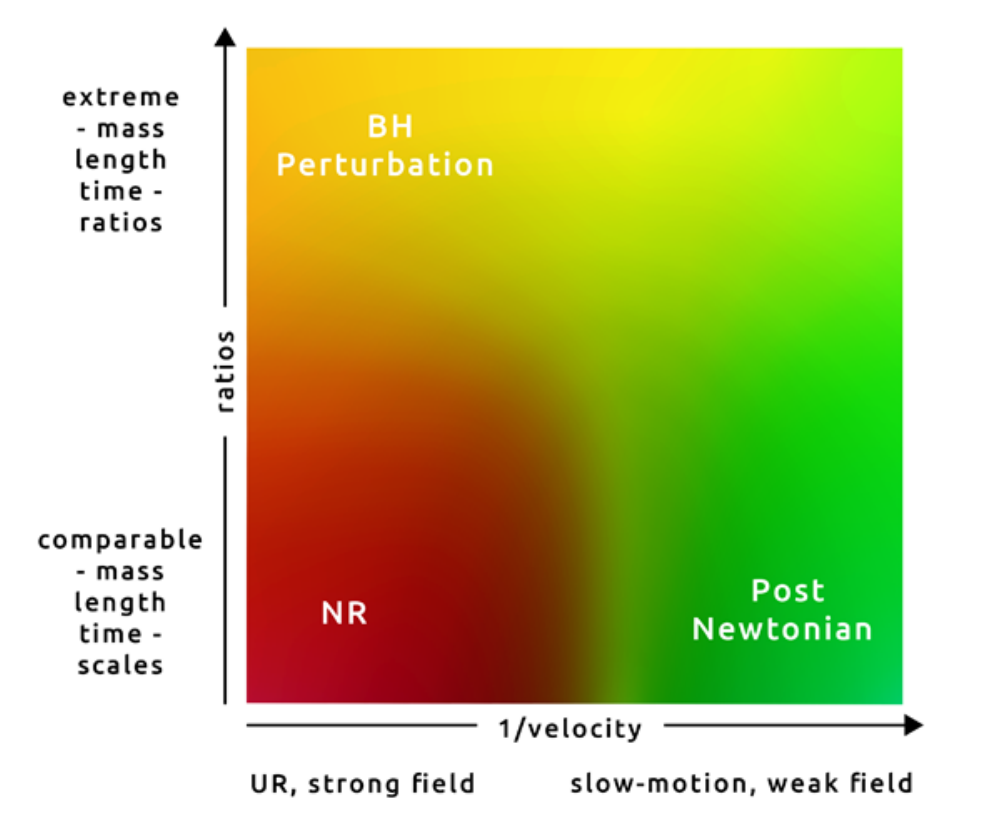}}
    \caption{Range of various approximation tools (``UR'' stands for
      ultra-relativistic). NR is mostly limited by resolution issues
      and therefore by possible different scales in the problem.}
    \label{fig:NR_validity_diagram}
\end{figure}}

NR has been crucial to answer important questions in astrophysics, GW physics,
high-energy physics and fundamental physics, and as such we thought it convenient -- and fun -- to start with a timeline
and main theoretical landmarks which have stimulated research in the last years. This will hopefully help the reader
getting started by understanding which are the main breakthroughs and where exactly do we stand.

\newpage
\section{Milestones}

Numerical solving is a thousand-year old art, which developed into modern
numerical analysis several decades ago with the advent of modern computers and
supercomputers. For a compelling account of the early history of numerical
analysis and computing we refer the reader to Goldstine
\cite{Goldstine:1972,Goldstine:1977}.

It is impossible to summarize all the important work on the subject in this
review, but we find it instructive to list a chronogram of several relevant
milestones taking us to 2014, in the context of GR. The following is a list --
necessarily incomplete and necessarily biased -- of works which, in our opinion,
have been instrumental to shape the evolution of the field. A more complete set
of references can be found in the rest of this review.

\noindent $\bullet$ 1910 -- The analysis of finite difference methods for PDEs
is initiated with Richardson~\cite{Richardson:1910}.

\noindent $\bullet$ 1915 -- Einstein develops GR~\cite{1915SPAW.......844E,Einstein:1916cc}. 

\noindent $\bullet$ 1916 -- Schwarzschild derives the first solution of
Einstein's equations, describing the gravitational field generated by a point
mass. Most of the subtleties and implications of this solution will only be
understood many years later~\cite{Schwarzschild:1916uq}.

\noindent $\bullet$ 1917 -- de Sitter derives a solution of Einstein's equations
describing a Universe with constant, positive curvature $\Lambda$. His solution
would later be generalized to the case $\Lambda<0$~\cite{1917KNAB...19.1217D}.

\noindent $\bullet$ 1921, 1926 -- In order to unify electromagnetism with GR, Kaluza
and Klein propose a model in which the spacetime has five dimensions, one of
which is compactified on a circle~\cite{Kaluza:1921tu,Klein:1926tv}.

\noindent $\bullet$ 1928 -- Courant, Friedrichs and Lewy use finite differences to
establish existence and uniqueness results for elliptic boundary-value and eigenvalue
problems, and for the initial-value problem for hyperbolic
and parabolic PDEs~\cite{Courant:1928}. 

\noindent $\bullet$ 1931 -- Chandrasekhar derives an upper limit for white dwarf masses,
above which electron degeneracy pressure cannot sustain the star~\cite{Chandrasekhar:1931ih}. 
The Chandrasekhar limit was subsequently extended to NSs by Oppenheimer and Volkoff~\cite{Oppenheimer:1939ne}.

\noindent $\bullet$ 1939 -- Oppenheimer and Snyder present the first dynamical collapse solution within GR~\cite{Oppenheimer:1939ue}.

\noindent $\bullet$ 1944 -- Lichnerowicz~\cite{Lichnerowicz1944} proposes
the conformal decomposition of the Hamiltonian constraint laying the
foundation for the solution of the initial data problem.

\noindent $\bullet$ 1947 -- Modern numerical analysis is considered by many to
have begun with the influential work of John von Neumann and Herman Goldstine
\cite{Neumann:1947}, which studies rounding error and includes a discussion of what one 
today calls scientific computing.

\noindent $\bullet$ 1952 -- Choquet-Bruhat~\cite{FouresBruhat:1952zz}
shows that the Cauchy problem obtained from the spacetime decomposition
of the Einstein equations has locally a unique solution.

\noindent $\bullet$ 1957 -- Regge and Wheeler~\cite{Regge:1957rw} analyze a
special class of gravitational perturbations of the Schwarzschild
geometry. This effectively marks the birth of BH perturbation theory, even 
before the birth of the BH concept itself. 

\noindent $\bullet$ 1958 -- Finkelstein understands that the $r=2M$ surface of
the Schwarzschild geometry is not a singularity but a horizon
\cite{Finkelstein:1958zz}. The so-called ``golden age of GR'' begins: in a few
years there would be enormous progress in the understanding of GR and of its solutions.

\noindent $\bullet$ 1961 -- Brans and Dicke propose an alternative theory of
gravitation, in which the metric tensor is non-minimally coupled with a scalar
field~\cite{Brans:1961sx}. 

\noindent $\bullet$ 1962 -- Newman and Penrose~\cite{Newman:1961qr} develop a
formalism to study gravitational radiation using spin coefficients.

\noindent $\bullet$ 1962 -- Bondi, Sachs and coworkers develop the characteristic formulation of the Einstein equations~\cite{Bondi:1962px,Sachs:1962wk}.

\noindent $\bullet$ 1962 -- Arnowitt, Deser and Misner
\cite{Arnowitt:1962hi} develop the canonical 3+1 formulation of the Einstein equations.

\noindent $\bullet$ 1963 -- Kerr~\cite{Kerr:1963ud} discovers the mathematical
solution of Einstein's field equations describing rotating BHs. In the same
year, Schmidt identifies the first quasar (quasi-stellar radio source)~\cite{Schmidt:1963}. Quasars
 are now believed to be supermassive BHs, described by the
Kerr solution.

\noindent $\bullet$ 1963 -- Tangherlini finds the higher-dimensional
generalization of  the Schwarzschild solution~\cite{Tangherlini:1963bw}.

\noindent $\bullet$ 1964 -- Chandrasekhar and Fock develop the post-Newtonian
theory~\cite{1965ApJ...142.1488C,fock1964theory}. 

\noindent $\bullet$ 1964 -- First documented attempt to solve Einstein's
equations numerically by Hahn \& Lindquist~\cite{Hahn1964}. Followed up
by Smarr \& Eppley about one decade later~\cite{Smarr1975, Eppley1975}.

\noindent $\bullet$ 1964 -- Seymour Cray designs the CDC 6600, generally
considered the first supercomputer. Speeds have increased by over one billion
times since.

\noindent $\bullet$ 1964 -- Using suborbital rockets carrying Geiger counters new sources of cosmic X-rays are discovered. One of these X-ray sources, Cygnus X-1, confirmed in 1971 with the UHURU orbiting X-ray observatory, is soon accepted as the first plausible
stellar-mass BH candidate (see, e.g., \cite{Bolton:1972}).
The UHURU orbiting X-ray observatory makes the
first surveys of the X-ray sky discovering over 300 X-ray ``stars''. 

\noindent $\bullet$ 1965 -- Penrose and Hawking prove that collapse of ordinary
matter leads, under generic conditions, to spacetime singularities (the
so-called ``singularity theorems'')~\cite{Penrose:1964wq,Hawking:1966vg}. A few
years later, Penrose conjectures that these singularities, where quantum
gravitational effects become important, are generically contained within BHs --
The \emph{cosmic censorship conjecture}~\cite{Penrose:1969,Wald:1997wa}.

\noindent $\bullet$ 1965 -- Weber builds the first GW detector, a resonant alluminium cylinder~\cite{PhysRevLett.17.1228,1968PhT....21d..34W}.

\noindent $\bullet$ 1966 -- May and White perform a full nonlinear numerical 
collapse simulation for some realistic equations of state~\cite{May:1966zz}.

\noindent $\bullet$ 1967 -- Wheeler~\cite{Ruffini:1971,wheeler} coins the term
\emph{black hole} (see the April 2009 issue of \textit{Physics Today}, and
Ref.~\cite{Wheeler:1998vs} for a fascinating, first-person historical
account). 

\noindent $\bullet$ 1967, 1971 -- Israel, Carter and Hawking prove that any
stationary, vacuum BH is described by the Kerr solution~\cite{Israel:1967wq,Carter:1971zc,Hawking:1971vc,Hawking:1973uf}. This result
motivates Wheeler's statement that ``a BH has no hair''~\cite{Ruffini:1971}.

\noindent $\bullet$ 1968 -- Veneziano proposes his dual resonance model, which
will later be understood to be equivalent to an oscillating string
\cite{Veneziano:1968yb}. This date is considered the dawn of SMT.

\noindent $\bullet$ 1969 -- Penrose shows that the existence of an ergoregion allows to extract energy and
angular momentum from a Kerr BH~\cite{Penrose:1969}. The wave analogue of the Penrose process is subsequently shown to occur by Zeldovich, who proves that dissipative rotating bodies (such as Kerr BHs, for which the dissipation is provided by the horizon) amplify incident waves in a process now called superradiance~\cite{zeldovich1,zeldovich2}.

\noindent $\bullet$ 1970 -- Zerilli~\cite{Zerilli:1970se,Zerilli:1971wd}
extends the Regge--Wheeler analysis to general perturbations of a Schwarzschild
BH. He shows that the problem can be reduced to the study of a pair of
Schr\"odinger-like equations, and applies the formalism to the problem of
gravitational radiation emitted by infalling test particles.

\noindent $\bullet$ 1970 -- Vishveshwara~\cite{vish} studies numerically the scattering of GWs by BHs: 
at late times the waveform consists of damped sinusoids (now called ringdown waves, or quasi-normal modes).

\noindent $\bullet$ 1971 -- Davis et~al.~\cite{Davis:1971gg} carry out the
first quantitative calculation of gravitational radiation emission within BH
perturbation theory, considering a particle falling radially into a
Schwarzschild BH. Quasi-normal mode (QNM) ringing is excited when the
particle crosses the maximum of the potential barrier of the Zerilli
equation, located close to the unstable circular orbit for photons.

\noindent $\bullet$ 1973 -- Bardeen, Carter and Hawking derive the four laws of BH mechanics~\cite{Bardeen:1973gs}.

\noindent $\bullet$ 1973 -- Teukolsky~\cite{Teukolsky:1972my} decouples and
separates the equations for perturbations in the Kerr geometry using the
Newman--Penrose formalism~\cite{Newman:1961qr}.

\noindent $\bullet$ 1973 -- York~\cite{York1973, York1974} introduces a split of the extrinsic
curvature leading to the \emph{Lichnerowicz--York conformal decomposition} which underlies most of the initial data calculations in NR.

\noindent $\bullet$ 1973 -- Thorne provides a criterium for BH formation,
the \emph{hoop conjecture}~\cite{Thorne:1972}; it predicts collapse to BHs
in a variety of situations including very high-energy particle collisions, 
which were to become important in TeV-scale gravity scenarios.

\noindent $\bullet$ 1974 -- Hulse and Taylor find the first pulsar, i.e., a
radiating neutron star (NS), in a binary star system~\cite{1975ApJ}. The continued
study of this system over time has produced the first solid observational
evidence, albeit indirect, for GWs.  This, in turn, has further
motivated the study of dynamical compact binaries and thus the development of
NR and resulted in the 1993 Nobel Prize for Hulse and Taylor.

\noindent $\bullet$ 1975 -- Using quantum field theory in curved space, Hawking
finds that BHs have a thermal emission~\cite{Hawking:1974sw}. This result is one
of the most important links between GR and quantum mechanics.

\noindent $\bullet$ 1977 -- NR is born with coordinated efforts to evolve BH spacetimes
\cite{1977NYASA.302..569S,Eardley:1978tr,1979sgrp.book.....S}.

\noindent $\bullet$ 1978 -- Cunningham, Price and Moncrief
\cite{Cunningham1,Cunningham2,Cunningham3} study radiation from relativistic
stars collapsing to BHs using perturbative methods. QNM ringing is excited.

\noindent $\bullet$ 1979 -- York~\cite{York1979} reformulates the
canonical decomposition by ADM, casting the Einstein equations in a form now commonly (and somewhat misleadingly)
referred to as the ADM equations.

\noindent $\bullet$ 1980 -- Bowen \& York
develop the \emph{conformal imaging} approach resulting in analytic
solutions to the momentum constraints under the assumption of
maximal slicing as well as conformal and asymptotic flatness~\cite{Bowen:1980yu}.

\noindent $\bullet$ 1983 -- Chandrasekhar's monograph~\cite{MTB} summarizes the
state of the art in BH perturbation theory, elucidating connections between
different formalisms.

\noindent $\bullet$ 1985 -- Stark and Piran~\cite{Stark:1985da} extract
GWs from a simulation of rotating collapse to a BH in NR.

\noindent $\bullet$ 1985 -- Leaver~\cite{Leaver:1985ax,leJMP,Leaver:1986gd}
provides the most accurate method to date to compute BH QNMs 
using continued fraction representations of the relevant wavefunctions.

\noindent $\bullet$ 1986 -- McClintock and Remillard~\cite{McClintock:1986}
show that the X-ray nova A0620-00 contains a compact object of mass almost
certainly larger than $3M_\odot$, paving the way for the identification of
many more stellar-mass BH candidates. 

\noindent $\bullet$ 1986 -- Myers and Perry construct higher-dimensional
rotating, topologically spherical, BH solutions~\cite{Myers:1986un}.

\noindent $\bullet$ 1987 -- 't Hooft~\cite{tHooft:1987rb} argues that the
scattering process of two point-like particles above the fundamental Planck scale
is well described and calculable using classical gravity. This idea is behind
the application of GR for modeling trans-Planckian particle collisions.

\noindent $\bullet$ 1989 -- Echeverria~\cite{Echeverria:1989hg} estimates the
accuracy with which one can estimate the mass and angular momentum of a BH from
QNM observations. The formalism is substantially refined in
Refs.~\cite{Berti:2005ys,Berti:2009kk}.

\noindent $\bullet$ 1992 -- The LIGO detector project is funded by the National Science Foundation. It reaches design sensitivity
in 2005~\cite{Abbott:2007kv}. A few years later, in 2009, the Virgo detector also reaches its
design sensitivity~\cite{Accadia:2010zza}.

\noindent $\bullet$ 1992 -- Bona and Mass\'o show that harmonic slicing has a 
singularity-avoidance property, setting the stage for the development of the ``1+log'' slicing~\cite{Bona:1992zz}. 

\noindent $\bullet$ 1992 -- D'Eath and Payne
\cite{DEath:1990de,DEath:1992hb,DEath:1992hd,DEath:1992qu} develop a perturbative method
to compute the gravitational radiation emitted in the head-on collision of two
BHs at the speed of light. Their second order result will be in good agreement
with later numerical simulations of high-energy collisions. 

\noindent $\bullet$ 1993 -- Christodoulou and Klainerman show that Minkowski spacetime is nonlinearly stable~\cite{1993gnsm.book.....C}.
 
\noindent $\bullet$ 1993 -- Anninos et~al.~\cite{Anninos:1993zj} first
succeed in simulating the head-on collision of two BHs, and observe QNM ringing
of the final BH.

\noindent $\bullet$ 1993 -- Gregory and Laflamme show that black strings, one of
the simplest higher-dimensional solutions with horizons, are unstable against
axisymmetric perturbations~\cite{Gregory:1993vy}. The instability is similar to
the Rayleigh--Plateau instability seen in fluids~\cite{Cardoso:2006ks,Camps:2010br}; the 
end-state was unclear.

\noindent $\bullet$ 1993 -- Choptuik finds evidence of universality and scaling
in gravitational collapse of a massless scalar field. ``Small'' initial data
scatter, while ``large'' initial data collapse to BHs~\cite{Choptuik:1992jv};
first use of mesh refinement in NR.

\noindent $\bullet$ 1994 -- The ``Binary Black Hole Grand Challenge Project'',
the first large collaboration with the aim of solving a specific NR problem
(modeling a binary BH coalescence), is launched~\cite{1995Sci...270..941M,1997ASPC..123..305C}.

\noindent $\bullet$ 1995, 1998 -- Through a conformal decomposition,
split of the extrinsic curvature and use of additional variables,
Baumgarte, Shapiro, Shibata and Nakamura~\cite{Shibata:1995we,
Baumgarte:1998te} recast the ADM equations as the so-called
BSSN system, partly building on earlier work by Nakamura, Oohara and
Kojima~\cite{1987PThPS..90....1N}.

\noindent $\bullet$ 1996 -- Br{\"u}gmann~\cite{Bruegmann:1996kz} uses mesh
refinement for simulations of BH spacetimes in 3+1 dimensions.

\noindent $\bullet$ 1997 -- \textsc{Cactus}~1.0 is released in April
1997. \textsc{Cactus}~\cite{cactus} is a freely available environment for collaboratively developing
parallel, scalable, high-performance multidimensional component-based
simulations. Many NR codes are based on this framework.
Recently, \textsc{Cactus} also became available in the form of the \textsc{Einstein Toolkit}~\cite{Loffler:2011ay,EinsteinToolkit}.

\noindent $\bullet$ 1997 -- Brandt \& Br{\"u}gmann~\cite{Brandt:1997tf}
present \emph{puncture} initial data as a generalization of
Brill-Lindquist data to the case of generic Bowen--York extrinsic curvature.

\noindent $\bullet$ 1997 -- Maldacena~\cite{Maldacena:1997re} formulates the
AdS/CFT duality conjecture. Shortly afterward, the papers by Gubser, Klebanov,
Polyakov~\cite{Gubser:1998bc} and Witten~\cite{Witten:1998qj} establish a
concrete quantitative recipe for the duality. The AdS/CFT era begins. In the
same year, the correspondence is generalized to non-conformal theories in a
variety of approaches (see~\cite{Aharony:1999ti} for a review). The terms
``gauge/string duality'', ``gauge/gravity duality'' and ``holography'' appear
(the latter had been previously introduced in the context of quantum gravity
\cite{tHooft:1993gx,Susskind:1994vu}), referring to these generalized settings.

\noindent $\bullet$ 1998 -- The hierarchy problem in physics -- the huge
discrepancy between the electroweak and the Planck scale -- is addressed
in the so--called \emph{braneworld scenarios}, in which we live on a four-dimensional 
subspace of a higher-dimensional spacetime, and the Planck scale can be lowered to the TeV
\cite{ArkaniHamed:1998rs,Antoniadis:1998ig,Randall:1999vf,Randall:1999ee}.

\noindent $\bullet$ 1998 -- First stable simulations of a single BH spacetime in
fully $D=4$ dimensional NR within a ``characteristic formulation''~\cite{Lehner:1998ti,Gomez:1998uj}, 
and two years later within a Cauchy formulation~\cite{Alcubierre:2000yz}.

\noindent $\bullet$ 1998 -- The possibility of BH formation in braneworld scenarios is first discussed~\cite{Argyres:1998qn,Banks:1999gd}. 
Later work suggests BH formation could occur at the LHC~\cite{Dimopoulos:2001hw,Giddings:2001bu} or in ultra-high energy cosmic ray collisions
\cite{Feng:2001ib,Anchordoqui:2001ei,Emparan:2001kf}.

\noindent $\bullet$ 1999 -- Friedrich \& Nagy~\cite{Friedrich:1998xt}
present the first well-posed formulation of the initial-boundary-value problem
 (IBVP) for the Einstein equations.

\noindent $\bullet$ 2000 -- Brandt et~al.~\cite{Brandt:2000yp}
simulate the first grazing collisions of BHs using a revised version
of the Grand Challenge Alliance code~\cite{Cook:1997na}.

\noindent $\bullet$ 2000 -- Shibata and Ury{\={u}}~\cite{Shibata:1999wm} perform the
first general relativistic simulation of the merger of two NSs. More
recent simulations~\cite{Baiotti:2008ra}, using a technique developed by Baiotti and Rezzolla that 
circumvents singularity excision~\cite{Baiotti:2006wm}, confirm that ringdown is excited when the merger leads to
BH formation. 
In 2006, Shibata and Ury{\={u}} perform NR simulations of BH-NS binaries~\cite{Shibata:2006ks}.

\noindent $\bullet$ 2001 -- Emparan and Reall provide the first example of a
stationary asymptotically flat vacuum solution with an event horizon of
non-spherical topology -- the ``black ring''~\cite{Emparan:2001wn}.

\noindent $\bullet$ 2001 -- Horowitz and Maeda suggest that black strings do not fragment and that the end-state of the Gregory--Laflamme instability may be an inhomogeneous string~\cite{Horowitz:2001cz}, driving the development of the field. Non-uniform strings are constructed perturbatively by Gubser~\cite{Gubser:2001ac}
and numerically by Wiseman, who, however, shows that these cannot be the end-state of the Gregory--Laflamme instability~\cite{Wiseman:2002zc}.

\noindent $\bullet$ 2003 -- In a series of papers
\cite{Kodama:2003jz,Ishibashi:2003ap,Kodama:2003kk}, Kodama and Ishibashi
extend the Regge--Wheeler--Zerilli formalism to higher dimensions.

\noindent $\bullet$ 2003 -- Schnetter et~al.~\cite{Schnetter:2003rb}
present the publically available \textsc{Carpet} mesh refinement package
which has constantly been updated since and is being used by many
NR groups.

\noindent $\bullet$ 2005 -- Pretorius~\cite{Pretorius:2005gq} achieves the
first long-term stable numerical evolution of a BH binary. Soon afterwards,
other groups independently succeed in evolving merging BH binaries using
different techniques~\cite{Campanelli:2005dd,Baker:2005vv}. The waveforms
indicate that ringdown contributes a substantial amount to the radiated
energy.

\noindent $\bullet$ 2007 -- First results from NR simulations show that
spinning BH binaries can coalesce to produce BHs with very large
recoil velocities~\cite{Gonzalez:2007hi,Campanelli:2007cga}.

\noindent $\bullet$ 2007 -- Boyle et~al.~\cite{Boyle:2007ft}
achieve unprecedented accuracy and number of orbits in
simulating a BH binary through inspiral and merger
with a spectral code that later becomes known as ``SpEC'' and
uses multi-domain decomposition~\cite{Pfeiffer:2002wt} and a dual
coordinate frame~\cite{Scheel:2006gg}.

\noindent $\bullet$ 2008 -- The first simulations of high-energy collisions of two BHs are performed~\cite{Sperhake:2008ga}. 
These were later generalized to include spin and finite impact parameter collisions, yielding zoom-whirl
behavior and the largest known luminosities~\cite{Shibata:2008rq,Sperhake:2009jz,Sperhake:2010uv,Sperhake:2012me}.

\noindent $\bullet$ 2008 -- First NR simulations in AdS for studying the isotropization of a strongly
coupled $\mathcal{N}=4$ supersymmetric Yang--Mills plasma through the gauge/gravity duality~\cite{Chesler:2008hg}.

\noindent $\bullet$ 2009 -- Dias et~al. show that rapidly spinning Myers--Perry BHs present zero-modes, signalling linear instability against axially symmetric perturbations~\cite{Dias:2009iu}, as previously argued by Emparan and Myers~\cite{Emparan:2003sy}. Linearly unstable modes were subsequently explored in Refs.~\cite{Dias:2010maa,Dias:2010eu}.

\noindent $\bullet$ 2009 -- Shibata and Yoshino evolve Myers--Perry BHs nonlinearly and show that a non-axisymmetric instability is present~\cite{Shibata:2009ad}. 

\noindent $\bullet$ 2009 -- Collisions of boson stars show that at large enough energies a BH forms, in agreement with
the hoop conjecture~\cite{Choptuik:2009ww}. Subsequent investigations extend these results to fluid stars~\cite{East:2012mb,Rezzolla:2013kt}.

\noindent $\bullet$ 2010 -- Building on previous work~\cite{Choptuik:2003qd}, Lehner and Pretorius study the nonlinear development of the Gregory--Laflamme instability in five dimensions, which shows hints of pinch-off and cosmic censorship violation~\cite{Lehner:2010pn}. 

\noindent $\bullet$ 2010, 2011 -- First nonlinear simulations describing collisions of higher-dimensional BHs, by Zilh\~ao et~al.,
Witek et~al. and Okawa et~al.~\cite{Zilhao:2010sr,Witek:2010xi,Okawa:2011fv}.

\noindent $\bullet$ 2011 -- Bizo\'n and Rostworowski extend Choptuik's collapse simulations to asymptotically AdS spacetimes~\cite{Bizon:2011gg}, 
finding evidence that generic initial data collapse to BHs, thereby conjecturing a nonlinear instability of AdS.

\noindent $\bullet$ 2013 -- Collisions of spinning BHs provide evidence that multipolar structure
of colliding objects is not important at very large energies~\cite{Sperhake:2012me}.

\newpage
\section{Strong Need for Strong Gravity}
\label{sec:motivation}
The need for NR is almost as old as GR itself, but the real push to develop these tools 
came primarily from the necessity to understand conceptual issues such as the end-state of collapse
and the two-body problem in GR as well as from astrophysics and GW astronomy.
The breakthroughs in the last years have prompted a serious reflexion and examination of the 
multitude of problems and fields which stand to gain from NR tools and results, if extended to encompass general spacetimes.
The following is a brief description of each of these topics. The range
of fundamental issues for which accurate strong gravity simulations are required will hopefully become clear.

\subsection{Astrophysics}

\subsubsection{Gravitational wave astronomy}
\label{gwastro}
GWs are one of the most fascinating predictions of GR. First conceived by
Einstein~\cite{Einstein:1916cc,Einstein:1918cc}, it was unclear for a long time
whether they were truly physical. Only in the 1960s were their existence and
properties founded on a sound mathematical basis (see
\cite{Isaacson:1967zz,Isaacson:1968zza} and references therein). In the same
period, after the seminal work of Weber~\cite{Weber:1960zz}, the scientific
community was starting a growing experimental effort to directly detect GWs. The
first detectors were resonant antennas; their sensitivity was far too low to
detect any signal (unless a nearby galactic supernova exploded when the detector
was taking data), and they were eventually replaced by interferometric
detectors. The first generation of such detectors (LIGO, Virgo, GEO600, TAMA) did
not reveal any gravitational signal, but the second generation (Advanced
LIGO/Virgo~\cite{LIGO,VIRGO}) should be operative by 2015 and is expected
to make the first detection of GWs. In parallel, Pulsar Timing Arrays are
promising to detect ultra-low frequency GWs~\cite{Lee:2011et} whereas the polarization of the cosmic microwave background
can be used as a detector of GWs from an inflationary epoch in the very early universe~\cite{Seljak:1996gy,Grishchuk:1974ny,Starobinsky:1982ee,Rubakov:1982df,Abbott:1984fp}. In the
subsequent years more sensitive detectors, such as the underground cryogenic
interferometer KAGRA~\cite{KAGRA} (and, possibly, ET~\cite{ET}) and possibly a
space-based detector such as LISA/eLISA~\cite{ELISA}, will allow us to know the
features of the signal in more detail, and then to use this information to learn
about the physics of the emitting sources, and the nature of the gravitational
interaction.

Soon after the beginning of the experimental efforts to build a GW detector, it
became clear that the detection of GWs emitted by astrophysical
sources would open a new window of observational astronomy, in addition to the
electromagnetic spectrum, neutrinos, cosmic rays, etc.  The impact of such a
detection would be similar to that of $X$-rays from astrophysical sources, i.e.,
the birth of a new branch of astronomy: ``GW astronomy''
\cite{Press:1972am,Grishchuk:1988kq,Schutz:1999xj}. In this new field, source modelling 
is crucial, since a theoretical understanding of the expected GW sources is needed
to enhance the chances of detection and to extract the relevant physics. 
Indeed, template-matching techniques -- frequently used in data analysis -- can be helpful to extract the signal from
the detector noise, but they require an a-priori knowledge of the
waveforms~\cite{thorne-87}.

A wide scientific community formed, with the aim to model the physical processes
which are expected to produce a detectable GW signal, and to compute the emitted
gravitational waveform (which depends on the unknown parameters of the source
and of the emitting process). 
Together with the understanding of the two-body problem in GR,
this effort was one of the main driving forces leading to the development
of NR. Indeed, many promising GW sources can only be modeled by solving the fully
non-linear Einstein equations numerically. 

Ground based interferometers are (and are expected to be in the next decades)
sensitive to signals with frequencies ranging from some tens of Hz to about one
kHz. Space-based interferometers would be sensitive at much lower frequencies:
from some mHz to about one tenth of Hz. GW astronomy, of course,
is presently concerned with sources emitting GWs in these frequency bands. 

Many astrophysical processes are potential sources for GW detectors.  In the
following, we shall briefly discuss only some of them, i.e., those that require
NR simulations to be modeled: compact binary inspirals, and instabilities of
rotating NSs. We shall not discuss supernova core collapse -- one of the first GW
sources which have been studied with NR, and one of the most problematic to
model -- since it will be discussed in Section~\ref{sec:collapse}.

Compact binary inspirals, i.e., the inspiral and merger of binary systems formed
by BHs and/or NSs, are the most promising GW sources to be detected.
Advanced LIGO/Virgo are expected to detect some tens of these sources per year
\cite{Abadie:2010cf}.  While the inspiral phase of a compact binary system can
be accurately modeled through PN approaches, and the final (``ringdown'') phase, when the BH resulting from the coalescence oscillates in its
characteristic proper modes, can be accurately described through perturbative approaches, the
intermediate merger phase can only be modeled by NR. This task has posed
formidable theoretical and computational challenges to the scientific community.

The numerical simulation of the merger phase of a BH-BH binary coalescence, and
the determination of the emitted gravitational waveform, had been an open
problem for decades, until it was solved in 2005~\cite{Pretorius:2005gq,Campanelli:2005dd,Baker:2005vv}. This challenge forced
the gravitational community to reflect on deep issues and problems arising
within
Einstein's theory, such as the role of singularities and horizons, and the
possible ways to locally define energy and momentum.

BH-NS and NS-NS binary coalescences pose a different sort of problems than those
posed by BH-BH coalescences.  They are not a ``clean'' system such as purely
vacuum BH spacetimes, characterized by the gravitational interaction
only. An accurate numerical modeling involves various branches of physics
(nuclear physics, neutrinos, electromagnetic fields), and requires the
understanding of many different processes. Typically, NR simulations of BH-NS
and NS-NS mergers make simplifying assumptions, both because taking into account all
aspects at the same time would be too complicated, and because some of them are not fully understood.  
Currently, the behaviour of matter in the inner core of a NS is one of the challenges to be tackled. 
Indeed, nuclear physicists still do
not understand which is the equation of state of matter at such extreme conditions of density
and temperature (see, e.g., \cite{Lattimer:2006xb} and references therein). This
uncertainty reflects our ignorance on the behaviour of the hadronic interactions
in the non-perturbative regime. On the other hand, understanding the NS equation of state is
considered one of the main outcomes expected from the detection of a GW signal
emitted by NSs, for instance in compact binary coalescences
\cite{Oechslin:2007gn,Read:2009yp,Bauswein:2011tp,Takami:2014zpa}.

Neutron star oscillations are also a candidate GW source for ground based
interferometers. When perturbed by an external or internal event, a NS can be
set into non-radial damped oscillations, which are associated to the emission of
GWs. The characteristic frequencies of oscillation, the QNMs, are characterized by their complex frequency $\omega=\sigma+{\rm
  i}/\tau$, where $\sigma$ is the pulsation frequency, and $\tau$ is the damping
time of the oscillation (for detailed discussions on the QNMs of NSs and BHs see
\cite{Kokkotas:1999bd,Nollert:1999ji,Ferrari:2007dd,Berti:2009kk} and references
therein).

If a NS rotates, its oscillations can become unstable. In this case, the
oscillation grows until the instability is suppressed by some damping mechanism or by non-linear effects; this process can
be associated to a large GW emission (see, e.g., \cite{Andersson:2002ch} and references therein). These instabilities may explain the observed values of the NS rotation rates~\cite{Bildsten:1998ey}. Their numerical modeling, however, is not an easy task. Perturbative approaches, which easily allow one to compute the QNMs of non-rotating NSs, become very involved in the presence of rotation.
Therefore, the perturbation equations can only be solved with simplifying assumptions, which make the model less accurate. Presently, NR is the only way to model stationary, rapidly rotating NSs (see, e.g., \cite{Stergioulas:2003yp} and
references therein), and it has recently been applied to model their oscillations~\cite{Zink:2010bq}.

\subsubsection{Collapse in general relativity}
\label{sec:collapse}
Decades before any observation of supermassive compact objects, and long before
BHs were understood, Chandrasekhar showed that the electron degeneracy pressure
in very massive white dwarfs is not enough to prevent them from imploding
\cite{Chandrasekhar:1931ih}. Similar conclusions were reached later by
Oppenheimer and Volkoff, for neutron degeneracy pressure in NSs
\cite{Oppenheimer:1939ne}. We can use Landau's original argument to understand
these results~\cite{Landau:1932,Landau:1938,Shapiro:1983du}: consider a star of
radius $R$ composed of $N$ fermions, each of mass $m_F$. The momentum of each
fermion is $p_F\sim \hbar n^{1/3}$, with $n=N/R^3$ the number density of
fermions. In the relativistic regime, the Fermi energy per particle then reads
$E_F=p_Fc=\hbar cN^{1/3}/R$. The gravitational energy per fermion is approximately $E_G \sim -Gm_F^2/R$, and the star's total energy is thus,
\be
E \equiv E_F+E_G= \frac{\hbar cN^{1/3}-GNm_F^2}{R}\,.
\ee
For small $N$, the total energy is positive, and we can decrease it by
increasing $R$. At some point the fermion becomes non-relativistic and $E_F\sim
p_F^2\sim 1/R^2$. In this regime, the gravitational binding energy $E_G$
dominates over $E_F$, the total energy is negative and tends to zero as $R\to
\infty$. Thus there is a local minimum and the star is stable. However, for large $N$ in the relativistic regime the total energy is
negative, and can be made even more negative by \emph{decreasing} $R$: it is
energetically favoured for the star to continually collapse!  The threshold for
stability occurs at a zero of the total energy, when
\beq
N_{\rm max}&>&\left(\frac{\hbar c}{Gm_F^2}\right)^{3/2}\,,\\
M_{\rm max}&\sim& N_{\rm max} m_F\sim \left(\frac{\hbar c}{Gm_F^{4/3}}\right)^{3/2}\,.
\eeq
For neutrons, stars with masses above $\sim 3M_{\odot}$ cannot attain equilibrium.

What is the fate of massive stars whose pressure cannot counter-balance gravity?
Does the star's material continually collapse to a single point, or is it
possible that pressure or angular momentum become so important that the material
bounces back? The answer to these questions would take several decades more, and
was one of the main driving forces to develop solid numerical schemes to handle
Einstein's equations.

Other developments highlighted the importance of understanding
gravitational collapse in GR. One was the advent of GW detectors. The strongest
sources of GWs are compact and moving relativistically, and supernovae are
seemingly ideal: they occur frequently and are extremely violent.
Unfortunately, Birkhoff's theorem implies that spherically symmetric sources do
not radiate. Thus a careful, and much more complex analysis of collapse is
required to understand these sources.

In parallel, BH physics was blooming. In the 1970's one key result was
established: the uniqueness theorem, stating that -- under general regularity assumptions -- the only stationary, asymptotically flat, vacuum
solution of Einstein's field equations is the Kerr BH. Thus, \emph{if} a horizon forms, the final stationary configuration
is expected to be of the Kerr family. This important corollary
of Einstein's field equations calls for a dynamical picture of BH formation
through collapse and an understanding of how the spacetime multipolar structure
dynamically changes to adapt to the final Kerr solution as a BH forms.

\subsubsection{Kicks}
It has been known since the early 1960s that GWs
emitted by accelerated particles do not only carry energy but also
momentum away from the system on which thus is imparted a
\emph{kick} or \emph{recoil}. This effect was first studied
by Bonnor \& Rotenberg~\cite{Bonnor1961} for the case of a system of
oscillating particles, and has been identified by Peres~\cite{Peres1962}
to be at leading order due to the interference of the mass
quadrupole radiation with the mass octupole or flow quadrupole.

From an astrophysical point of view, the most important processes
generating such gravitational recoil are the collapse of a stellar
core to a compact object and the inspiral and merger of compact
binaries. Supermassive BHs with masses in the range
of $10^5~M_{\odot}$ to $10^{10}~M_{\odot}$ in particular are
known to reside at the centre of many galaxies and are likely
to form inspiralling binary systems as a consequence of galaxy mergers.
Depending on the magnitude of the resulting velocities, kicks can in principle displace or eject BHs from their
hosts and therefore play an important role in the formation history
of these supermassive BHs.

The first calculations of recoil velocities based on perturbative techniques
have been applied to gravitational collapse scenarios by
Bekenstein~\cite{Bekenstein:1973zz} and Moncrief~\cite{Moncrief1979}.
The first analysis of GW momentum flux generated by
binary systems was performed by Fitchett~\cite{Fitchett1983} in 1983
for two masses in Keplerian orbit. The following two decades saw various
(semi-)analytic calculations for inspiraling compact binary systems
using the particle approximation, post-Newtonian techniques and the
close-limit approach (see Section~\ref{sec:exact_approximation} for a description of these techniques and main results). 
In conclusion of these studies,
it appeared likely that the gravitational recoil from non-spinning
binaries was unlikely to exceed a few hundred km/s. Precise estimates,
however, are dependent on an accurate modeling of the highly non-linear
late inspiral and merger phase and therefore required NR simulations. Furthermore, the impact of spins on
the resulting velocities remained essentially uncharted territory
until the 2005 breakthroughs of NR made possible
the numerical simulations of these systems. As it turned out,
some of the most surprising and astrophysically influencial
results obtained from NR
concern precisely the question of the gravitational recoil of spinning
BH binaries.

\subsubsection{Astrophysics beyond Einstein gravity}
\label{sec:beyondastro}
Although GR is widely accepted as the standard theory of gravity and has
survived all experimental and observational (weak field) scrutiny, there is
convincing evidence that it is not the ultimate theory of gravity: since GR is
incompatible with quantum field theory, it should be considered as the low
energy limit of some, still elusive, more fundamental theory. In addition, GR
itself breaks down at small length scales, since it predicts singularities. For
large scales, on the other hand, cosmological observations show that our
Universe is filled with dark matter and dark energy, of as yet unknown nature.

This suggests that the strong-field regime of gravity -- which has barely been
tested so far -- could be described by some modification or extension of GR. In
the next few years both GW detectors~\cite{Will:2004xi,Yunes:2013dva} and
astrophysical observations~\cite{Psaltis:2008bb} will provide an unprecedented
opportunity to probe the strong-field regime of the gravitational interaction,
characterized by large values of the gravitational field $\sim\frac{GM}{rc^2}$
or of the spacetime curvature $\sim\frac{GM}{r^3c^2}$ (it is a matter of debate
which of the two parameters is the most appropriate for characterizing the
strong-field gravity regime~\cite{Psaltis:2008bb,Yunes:2013dva}). However, our
present theoretical knowledge of strong-field astrophysical processes is based,
in most cases, on the a-priori assumption that GR \emph{is} the correct theory of
gravity.  This sort of \emph{theoretical bias}~\cite{Yunes:2009ke} would strongly
limit our possibility of testing GR.

It is then of utmost importance to understand the behaviour of astrophysical
processes in the strong gravity regime beyond the assumption that GR is the
correct theory of gravity. The most powerful tool for this purpose is probably
NR; indeed, although NR has been developed to solve Einstein's
equations (possibly coupled to other field equations), it can in principle be
extended and modified, to model physical processes in alternative theories of
gravity. In summary, NR can be applied to specific, well motivated theories of
gravity. These theories should derive from -- or at least be inspired by -- some
more fundamental theories or frameworks, such as for instance SMT
\cite{green1987superstring,polchinski1998string} (and, to some extent, Loop
Quantum Gravity~\cite{rovelli1998loop}).  In addition, such theories should
allow a well-posed initial-value formulation of the field equations.  Various
arguments suggest that the modifications to GR could involve
\cite{Yunes:2013dva} (i) additional degrees of freedom (scalar fields, vector
fields); (ii) corrections to the action at higher order in the spacetime
curvature; (iii) additional dimensions.

Scalar-tensor theories for example (see, e.g., \cite{fujii2003scalar,Will:2014va}
and references therein), are the most natural and simple generalizations of GR
including additional degrees of freedom.
In these theories, which include for instance Brans--Dicke gravity
\cite{Brans:1961sx}, the metric tensor is non-minimally coupled with one or
more scalar fields. In the case of a single scalar field (which can be
generalized to multi-scalar-tensor theories~\cite{Damour:1992we}), the action
can be written as
\begin{equation} 
S=\frac{1}{16\pi G}\int d^4x \sqrt{-g}\left[F(\phi)R-8\pi GZ(\phi)g^{\mu\nu}\partial_\mu
\phi\partial_\nu\phi-U(\phi)\right]+S_m(\psi_m,g_{\mu\nu})\label{SJordan}
\end{equation} 
where $R$ is the Ricci scalar associated to the metric $g_{\mu\nu}$, $F,Z,U$ are
arbitrary functions of the scalar field $\phi$, and $S_m$ is the action
describing the dynamics of the other fields (which we call ``matter fields'',
$\psi_m$). A more general formulation of scalar-tensor theories yielding second order equations of motion has been
proposed by Horndeski~\cite{Horndeski:1974wa} (see also Ref.~\cite{Deffayet:2011gz}).

Scalar-tensor theories can be obtained as low-energy limits of SMT~\cite{Garay:1992ej}; this provides motivation for studying these theories on
the grounds of fundamental physics. An additional motivation comes from the
recently proposed ``axiverse''
scenario~\cite{Arvanitaki:2009fg,Arvanitaki:2010sy}, in which ultra-light axion
fields (pseudo-scalar fields, behaving under many respects as scalar fields)
arise from the dimensional reduction of SMT, and play a role in cosmological
models.

Scalar-tensor theories are also appealing alternatives to GR because they
predict new phenomena, which are not allowed in GR. In these theories, the GW
emission in compact binary coalescences has a dipolar ($\ell=1$) component,
which is absent in GR; if the scalar field has a (even if extremely small) mass,
superradiant instabilities occur~\cite{Cardoso:2005vk,Pani:2012vp,Witek:2012tr},
which can determine the formation of floating orbits in extreme mass ratio
inspirals~\cite{Cardoso:2011xi,Yunes:2011aa}, and these orbits affect the
emitted GW signal; last but not least, under certain conditions isolated NSs can
undergo a phase transition, acquiring a nontrivial scalar field profile ({\it
  spontaneous scalarization}~\cite{Damour:1992we,Damour:1993hw}) while
dynamically evolving NSs -- requiring full NR simulations to understand -- may
display a similar effect (\emph{dynamical scalarization}
\cite{Barausse:2012da,Palenzuela:2013hsa}). A detection of one of these phenomena
would be a smoking gun of scalar-tensor gravity.

These theories, whose well posedness has been proved
\cite{Salgado:2005hx,Salgado:2008xh}, are a perfect arena for NR.  Recovering
some of the above smoking-gun effects is extremely challenging, as the required
timescales are typically very large when compared to any other timescales in the
problem.

Other examples for which NR can be instrumental include theories in which the
Einstein-Hilbert action is modified by including terms quadratic in the
curvature (such as $R^2$, $R_{\mu\nu}R^{\mu\nu}$,
$R_{\mu\nu\alpha\beta}R^{\mu\nu\alpha\beta}$,
$\epsilon_{\mu\nu\alpha\beta}R^{\mu\nu\rho\sigma}R^{\alpha\beta}_{~~\rho\sigma}$),
possibly coupled with scalar fields, or theories which explicitly break Lorentz
invariance.  In particular, Einstein-Dilaton-Gauss-Bonnet gravity and Dynamical
Chern-Simons gravity~\cite{Pani:2009wy,Alexander:2009tp} can arise from SMT
compactifications, and Dynamical Chern-Simons gravity also arises in Loop
Quantum Gravity; theories such as Einstein-Aether~\cite{Jacobson:2008aj} and
``Horava-Lifshitz'' gravity~\cite{Horava:2009uw}, which break Lorentz invariance
(while improving, for instance, renormalizability properties of GR), allow the
basic tenets of GR to be challenged and studied in depth.

\subsection{Fundamental and mathematical issues}

\subsubsection{Cosmic censorship}
\label{sec_cc}
Spacetime singularities signal the breakdown of the geometric description of the
spacetime, and can be diagnosed by either the blow-up of observer-invariant
quantities or by the impossibility to continue timelike or null geodesics past
the singular point.  For example, the Schwarzschild geometry has a curvature
invariant $R^{abcd}R_{abcd}=48G^2M^2/(c^4r^6)$ in Schwarzschild coordinates,
which diverges at $r=0$, where tidal forces are also infinite. Every timelike or
null curve crossing the horizon necessarily hits the origin in finite proper
time or affine parameter and therefore the theory breaks down at these points:
it fails to predict the future development of an object which reaches the
singular point.  Thus, the classical theory of GR, from which spacetimes with
singularities are obtained, is unable to describe these singular points and
contains its own demise.  Adding to this classical breakdown, it is likely that
quantum effects take over in regions where the curvature radius becomes comparable to the scale of quantum processes,
much in the same way as quantum electrodynamics is necessary in regions where EM
fields are large enough (as characterized by the invariant $E^2-B^2$) that
pair creation occurs.  Thus, a quantum theory of gravity might be needed close
to singularities.

It seems therefore like a happy coincidence that the Schwarzschild singularity
is cloaked by an event horizon, which effectively causally disconnects the
region close to the singularity from outside observers. This coincidence
introduces a miraculous cure to GR's apparently fatal disease: one can continue
using classical GR for all practical purposes, while being blissfully ignorant
of the presumably complete theory that smoothens the singularity, as all those
extra-GR effects do not disturb processes taking place outside the horizon.

Unfortunately, singularities are expected to be quite generic: in a remarkable
set of works, Hawking and Penrose have proved that, under generic conditions and
symmetries, collapse leads to singularities
\cite{Penrose:1964wq,Hawking:1967ju,Hawking:1969sw,Natario:2006gt}. Does this
always occur, i.e., are such singularities \emph{always} hidden to
outside observers by event horizons?  This is the content of Penrose's ``cosmic
censorship conjecture'', one of the outstanding unsolved questions in
gravity. Loosely speaking, the conjecture states that physically reasonable
matter under generic initial conditions only forms singularities hidden behind
horizons~\cite{Wald:1997wa}.

The cosmic censorship conjecture and the possible existence of naked
singularities in our Universe has triggered interest in complex problems which
can only be addressed by NR. This is a very active line of research, with
problems ranging from the collapse of matter to the nonlinear stability of
``black'' objects. 

\subsubsection{Stability of black hole interiors}
\label{section_interior}
As discussed in Section~\ref{sec:collapse}, the known fermionic degeneracies are
unable to prevent the gravitational collapse of a sufficiently massive
object. Thus, if no other (presently unkown) physical effect can prevent it,
according to GR, a BH forms. From the uniqueness theorems (cf. Section~\ref{section_4dexact}), this BH is described
by the Kerr metric. Outside the event horizon, the Kerr family -- a 2-parameter
family described by mass $M$ and angular momentum $J$ -- varies smoothly with its
parameters. But inside the event horizon a puzzling feature occurs. The interior
of the $J=0$ solution -- the Schwarzschild geometry -- is qualitatively different
from the $J\neq 0$ case.  Indeed, inside the Schwarzschild event horizon a
point-like, spacelike singularity creates a boundary for spacetime. Inside the
$0<J\le M^2$ Kerr event horizon, by contrast, there is a ring-like, timelike
singularity, beyond which another asymptotically flat spacetime region, with
$r<0$ in Boyer-Lindquist coordinates, may be reached by causal trajectories. The
puzzling feature is then the following: according to these exact solutions, the
interior of a Schwarzschild BH, when it absorbes an infinitesimal particle with
angular momentum, must drastically change, in particular by creating another
asymptotically flat region of spacetime.

This latter conclusion is quite unreasonable. It is more reasonable to expect
that the internal structure of an eternal Kerr BH must be very different from
that of a Kerr BH originating from gravitational collapse. Indeed, there are
arguments, of both physical~\cite{Penrose68} and mathematical nature
\cite{Chandrasekhar:1982}, indicating that the Cauchy horizon (i.e., inner
horizon) of the eternal charged or rotating hole is unstable against small
(linear) perturbations, and therefore against the accretion of any material. The
natural question is then, what is the endpoint of the instability?

As a toy model for the more challenging Kerr case, the aforementioned question
was considered in the context of spherical perturbations of the
RN BH by Poisson and Israel. In their seminal work the
phenomenon of \emph{mass inflation} was unveiled
\cite{Poisson:1989zz,Poisson:1990eh}: if ingoing and outgoing streams of matter are
simultaneously present near the inner horizon, then relativistic
counter-streaming\epubtkFootnote{That is, the existence of mass currents in opposite spatial directions (in between the horizons) and at relativistic velocities in the centre of energy frame.} between those streams leads to exponential growth of
gauge-invariant quantities such as the interior
(Misner--Sharp~\cite{Misner:1964je}) mass, the center-of-mass energy
density, or curvature scalar invariants. Since this effect is causally
disconnected from any external observers, the mass of the BH measured
by an outside observer remains unchanged by the mass inflation going
on in the interior. But this inflation phenomenon causes the spacetime
curvature to grow to Planckian values in the neighbourhood of the
Cauchy horizon. The precise nature of this evolution for the Kerr case
is still under study. For the simpler RN case, it has been argued by
Dafermos, using analytical methods, that the singularity that forms is
not of space-like nature~\cite{Dafermos:2012np}. Fully non-linear
numerical simulations will certainly be important for understanding
this process.

\subsubsection{Most luminous events in the Universe}
\label{sec_most_lum}
The most advanced laser units on the planet can output luminosities as high as
$\sim 10^{18}\,{\rm W}$~\cite{ELI}, while at $\sim 10^{26}\,{\rm W}$ the Tsar
Bomba remains the most powerful artificial explosion ever~\cite{TsarBomba}.
These numbers pale in comparison with strongly dynamical astrophysical events: a
$\gamma$-ray burst, for instance, reaches luminosities of approximately $\sim
10^{45} \,{\rm W}$.  A simple order of magnitude estimate can be done to
estimate the total luminosity of the Universe in the EM spectrum, by counting
the total number of stars, roughly $10^{23}$~\cite{Herschel}.  If all of them
have a luminosity equal to our Sun, we get a total luminosity of approximately
$\sim 10^{49}\,{\rm W}$, a number which can also be arrived at through more
careful considerations~\cite{Wijers:2005pn}.  Can one possibly surpass this
astronomical number?

In four spacetime dimensions, there is only one constant with dimensions of
energy per second that can be built out of the classical universal
constants. This is the Planck luminosity ${\cal L}_G$,
\be
{\cal L}_G\equiv\frac{c^5}{G}=3.7\times 10^{52}\,{\rm W}\,.\label{eq:dyson}
\ee
The quantity ${\cal L}_G$ should control gravity-dominated dynamical processes; 
as such it is no wonder that these events release huge luminosities. Take the
gravitational collapse of a compact star with mass $M$ and radius $R\sim
GM/c^2$. During a collapse time of the order of the infall time, $\tau \sim
R/\sqrt{GM/R}\sim GM/c^3$, the star can release an energy of up to $Mc^2$. The
process can therefore yield a power as large as $c^5/G={\cal L}_G$. It was
conjectured by Thorne~\cite{Thorne:1983} that the Planck luminosity is in fact
an upper limit for the luminosity of \emph{any} process in the Universe.\epubtkFootnote{This bound has an interesting story. Kip Thorne, and others
  after him, attribute the conjecture to Freeman Dyson; Freeman Dyson denies he
  ever made such a conjecture, and instead attributes such notion
  to his 1962 paper~\cite{Dyson}, where he works out the power emitted by a
  binary of compact objects. (We thank Gary Gibbons and Christoph Schiller for correspondence on this matter.)} The conjecture was put on a somewhat firmer
footing by Gibbons who has shown that there is an upper limit to the
\emph{tension} of $c^4/(4G)$, implying a limit in the luminosity of ${\cal
  L}_G/4$~\cite{Gibbons:2002iv}.

\emph{Are} such luminosities ever attained in practice, is there any process
which can reach the Planck luminosity and outshine the entire Universe?  The
answer to this issue requires once again a peek at gravity in strongly dynamical
collisions with full control of strong-field regions.  It turns out that high
energy collisions of BHs do come close to saturating the bound \eqref{eq:dyson}
and that in general colliding BH binaries are more luminous than the entire Universe in
the EM spectrum~\cite{Sperhake:2008ga,Sperhake:2009jz,Sperhake:2010uv,Sperhake:2012me}.

\subsubsection{Higher dimensions}
\label{sec_hi_dim}
Higher-dimensional spacetimes are a natural framework for mathematicians and
have been of general interest in physics, most notably as a tool to unify
gravity with the other fundamental interactions.  The quest for a unified theory
of all known fundamental interactions is old, and seems hopeless in
four-dimensional arenas.  In a daring proposal however, Kaluza and Klein, 
already in 1921 and 1926 showed that such a programme might be attainable if one is
willing to accept higher-dimensional theories as part of the fundamental picture 
\cite{Kaluza:1921tu,Klein:1926tv} (for a historical view, see~\cite{Duff:1994tn}).

Consider first for simplicity the $D$-dimensional Klein--Gordon equation $\Box
\phi(x^{\mu},z^i)=0\,\,(\mu=0,\ldots ,3,\,\,\,i=4,\ldots ,D-1)$, where the $(D-4)$ extra
dimensions are compact of size $L$. Fourier decompose in $z^j$, i.e,
$\phi(x^\mu,z^j)=\sum_n \psi(x^\mu)e^{in z^j/L}$, to get $\Box\psi-\frac{n^2}{L^2}\psi=0$, where here $\Box$ is the four-dimensional d'Alembertian operator.  As a consequence,

\noindent i) the fundamental, homogenous mode $n=0$ is a massless
\emph{four-dimensional} field obeying the same Klein--Gordon equation, whereas

\noindent ii) even though we started with a higher-dimensional massless theory, we
end up with a tower of massive modes described by the four-dimensional massive
Klein--Gordon equation, with mass terms proportional to $n/L$. One important,
generic conclusion is that the higher-dimensional (fundamental) theory imparts
mass terms as imprints of the extra dimensions. As such, the effects of
extra dimensions are in principle testable. However, for very small $L$ these
modes have a very high-energy and are very difficult to excite (to ``see'' an
object of length $L$ one needs wavelengths of the same order or smaller),
thereby providing a plausible explanation for the non-observation of
extra dimensions in everyday laboratory experiments.

The attempts by Kaluza and Klein to unify gravity and electromagnetism
considered five-dimensional Einstein field equations with the metric
appropriately decomposed as,
\be
d{\hat s}^2=e^{\alpha \phi} \, ds^2+e^{-2\alpha\phi}\left(dz+{\cal A}\right)^2\,.
\ee
Here, $ds^2=g_{\mu\nu}dx^{\mu}dx^\nu$ is a four-dimensional geometry, ${\cal
  A}=A_{\mu}dx^{\mu}$ is a gauge field and $\phi$ is a scalar; the constant $\alpha$ can be chosen to yield the four-dimensional theory in the Einstein frame. \emph{Assuming} all
the fields are independent of the extra dimension $z$, one finds a set of
\emph{four-dimensional} Einstein-Maxwell-scalar equations, thereby \emph{almost}
recovering both GR and EM~\cite{Duff:1994tn}.  This is the basic idea behind the
Kaluza--Klein procedure, which unfortunately failed due to the presence of the
(undetected) scalar field.

The idea of using higher dimensions was to be revived decades later in a more
sophisticated model, eventually leading to SMT. The development of the gauge/gravity duality (see Section~\ref{sec:gauge-gravity} below)
and TeV-scale scenarios in high-energy physics (see Section~\ref{sec:tev}) highlighted the importance of understanding Einstein's
equations in a generic number of dimensions. Eventually, the study of Einstein's
field equations in $D$-dimensional backgrounds branched off as a subject of its
own, where $D$ is viewed as just another parameter in the theory.  This area has
been extremely active and productive and provides very important information on
the content of the field equations and the type of solutions it admits. Recently, GR in the large $D$ limit has been suggested as a new tool to gain insight into the $D$ dependence of physical processes~\cite{Emparan:2013moa}. 

The uniqueness theorems, for example, are known to break down in higher dimensions,
at least in the sense that solutions are uniquely characterized by asymptotic
charges.  BHs of spherical topology -- the extension of the Kerr
solution to higher dimensions -- can co-exist with black rings~\cite{Emparan:2001wn}. In
fact, a zoo of black objects are known to exist in higher dimensions, but the
dynamical behavior of this zoo (of interest to understand stability of the
solutions and for collisions at very high energies) is poorly known, and
requires NR methods to understand.

One other example requiring NR tools is the instability of black strings. Black
strings are one of the simplest vacuum solutions one can construct, by extending
trivially a four-dimensional Schwarzschild BH along an extra, flat
direction. Such solutions are unstable against long wavelength perturbations in
the fifth dimension which act to fragment the string. This instability is known
as the Gregory--Laflamme instability~\cite{Gregory:1993vy}. The instability is
similar in many aspects to the Rayleigh--Plateau instability seen in fluids,
which does fragment long fluid cylinders~\cite{Cardoso:2006ks}. However, the
same scenario in the black string case would seem to lead to cosmic censorship
violation, since the pinch-off would be accompanied by (naked) regions of
unbounded curvature.%
\epubtkFootnote{Such fragmentation may however not be a counterexample
  to the spirit of the cosmic censorhip conjecture, if black strings
  do not form in generic collapse situations. One hint that this may
  indeed be the case comes from the Dyson--Chandrasekhar--Fermi
  instability of higher-dimensional cylindrical matter
  configurations~\cite{Cardoso:2006sj}: if cylindrical matter
  configurations are themselves unstable it is unlikely that their
  collapse leads to black strings.}
Evidence that the Gregory--Laflamme does lead to disruption of strings
was recently put forward~\cite{Lehner:2010pn}.

\subsection{High-energy physics}

\subsubsection{The gauge/gravity duality}
\label{sec:gauge-gravity}
The gauge/gravity duality, or AdS/CFT
correspondence, is the conjecture, first proposed by Maldacena in 1998
\cite{Maldacena:1997re}, and further developed in
\cite{Witten:1998qj,Gubser:1998bc}, that string theory on an AdS
spacetime (times a compact space) is \emph{dual} (i.e., equivalent under an appropriate mapping) to a CFT defined on the boundary
of the AdS space. Since its proposal, this conjecture has been supported by
impressive and compelling evidence, it has branched off to, e.g., the
AdS/Condensed Matter correspondence~\cite{Hartnoll:2009sz}, 
and it has inspired other proposals of duality relations with a similar spirit, such as the dS/CFT correspondence
\cite{Strominger:2001pn} and the Kerr/CFT correspondence~\cite{Guica:2008mu}. All these dualities
are examples of the \emph{holographic principle} which has been proposed in the context of quantum gravity
\cite{tHooft:1993gx,Susskind:1994vu}, stating that the information contained in a $D$-dimensional gravitational 
system is encoded in some boundary of the system. The paradigmatic example of this idea is a BH spacetime, 
whose entropy is proportional to the horizon area.

These dualities -- in which strong gravity systems play a crucial role -- offer
tools to probe strongly coupled gauge theories (in $D-1$ dimensions) by studying
classical gravity (in $D$ dimensions). For instance, the
confinement/deconfinement phase transition in quantum chromodynamics-like
theories has been identified with the Hawking-Page phase transition for AdS
BHs~\cite{Witten:1998zw}. Away from thermal equilibrium, the quasi-normal
frequencies of AdS BHs have been identified with the poles of retarded
correlators describing the relaxation back to equilibrium of a perturbed dual
field theory~\cite{Horowitz:1999jd,Birmingham:2001pj}.  The strongly coupled
regime of gauge theories is inaccessible to perturbation theory and therefore
this new tool has created expectations for understanding properties of the
plasma phase of non-Abelian quantum field theories at non-zero temperature,
including the transport properties of the plasma and the propagation and
relaxation of plasma perturbations, experimentally studied at the Relativistic Heavy Ion Collider 
and now also at the LHC~\cite{CasalderreySolana:2011us}. Strong coupling can be tackled by
lattice-regularized thermodynamical calculations of quantum chromodynamics, but the generalization
of these methods beyond static observables to characterizing transport
properties has limitations, due to computational costs. An example of an experimentally accessible transport property 
is the dimensionless ratio of the shear viscosity to the entropy
density. Applying the gauge/gravity duality, this property can be computed by determining the absorption cross section of low-energy gravitons in
the dual geometry (a BH/black brane)~\cite{Kovtun:2004de}, obtaining a
result compatible with the experimental data. This has offered the holographic
description of heavy ion collisions phenomenological credibility.  An
outstanding theoretical challenge in the physics of heavy ion collisions is the
understanding of the `early thermalization problem': the mechanism driving the short -
less than 1 fm/c~\cite{Heinz:2004pj} -- time after which experimental data
agrees with a hydrodynamic description of the quark-gluon plasma. Holography uses
$\mathcal{N}$=4 Super Yang Mills theory as a learning ground for the real
quark-gluon plasma. Then, the formation of such plasma in the collision of
high-energy ions has been modeled, in its gravity dual, by colliding
gravitational shock waves in five-dimensional AdS space
\cite{Chesler:2008hg}. These strong gravity computations have already offered
some insight into the early thermalization problem, by analyzing the formation
and settling down of an AdS BH in the collision process. But the use of
shock waves is still a caricature of the process, which could be rendered more
realistic, for instance, by colliding other highly boosted lumps of energy or
BHs in AdS.

Another example of gauge/gravity duality is the AdS/Condensed Matter
correspondence, between field theories that may describe superconductors and
strong gravity~\cite{Hartnoll:2009sz,Horowitz:2010nh,Hartnoll:2011fn}. The
simplest gravity theory in this context is Einstein-Maxwell-charged scalar
theory with negative cosmological constant. The RN-AdS BH 
solution of this theory, for which the scalar field vanishes, is unstable
for temperatures $T$ below a critical temperature $T_c$. If triggered, the
instability leads the scalar field to condense into a non-vanishing profile
creating a scalar hair for the BH and breaking the $U(1)$-gauge symmetry
spontaneously. The end point of the instability is a static solution that has
been constructed numerically and has properties similar to those of a superconductor
\cite{Hartnoll:2008kx}. Thus, this instability of the RN-AdS
BH at low temperature was identified with a superconducting phase
transition, and the RN-AdS and hairy BHs in the
gravitational theory, respectively, were identified with the normal and
superconducting phases of a holographic superconductor, realized within the dual
field theory. Holographic superconductors are a promising approach to
understanding strongly correlated electron systems. In particular,
non-equilibrium processes of strongly correlated systems, such as
superconductors, are notoriously difficult and this holographic method offers a
novel tool to tackle this longstanding problem. In the gauge/gravity approach,
the technical problem is to solve the classical dynamics of strong gravitational
systems in the dual five-dimensional spacetime. Using the AdS/CFT dictionary,
one then extracts the dynamics of the phase transition for the boundary theory
and obtains the time dependence of the superconducting order parameter and the
relaxation time scale of the boundary theory.

\subsubsection{Theories with lower fundamental Planck scale}
\label{sec:tev}
As discussed in Section~\ref{sec_hi_dim}, higher-dimensional theories have been
suggested since the early days of GR as a means to achieve unification of
fundamental interactions. The extra dimensions have traditionally been envisaged
as compact and very small ($\sim$ Planck length), in order to be compatible with
high energy experiments. Around the turn of the millennium, however, a new set of
scenarios emerged wherein the extra dimensions are only probed by the
gravitational interaction, because a confining mechanism ties the standard model
interactions to a 3+1-dimensional subspace (which is called the ``brane'', while
the higher-dimensional spacetime is called the ``bulk''). These models -- also called
``braneworld scenarios'' -- can be considered SMT inspired. The main ideas
behind them are provided by SMT, including the existence of extra dimensions and
also the existence of subspaces, namely \emph{Dirichlet-branes}, on which a well
defined mechanism exists to confine the standard model fields.

Our poor knowledge of the gravitational interaction at very short scales (below
the millimeter at the time of these proposals, below $\lesssim 10^{-4}$ meters
at the time of writing ~\cite{PhysRevLett.108.081101,Will:2014va}), allows
large~\cite{Antoniadis:1998ig,ArkaniHamed:1998rs,Dimopoulos:2001hw} or
infinitely large extra dimensions~\cite{Randall:1999vf,Randall:1999ee}. The
former are often called \emph{ADD models}, whereas the latter are known as
\emph{Randall--Sundrum scenarios}. Indeed these types of extra dimensions are
compatible with high energy phenomenology. Besides being viable, these models
(or at least some of them) have the conceptual appeal of providing an
explanation for the ``hierarchy problem'' of particle physics: the large
hierarchy between the electroweak scale ($\sim$ 250 GeV) and the Planck scale
($\sim$ $10^{19}$ GeV), or in other words, why the gravitational interaction
seems so feeble as compared to the other fundamental interactions. The reason
would be that whereas nuclear and electromagnetic interactions propagate in
3+1 dimensions, gravity propagates in $D$ dimensions. A 3+1 dimensional
application of Gauss's law then yields an incomplete account of the total
gravitational flux. Thus, the apparent (3+1 dimensional) gravitational
coupling appears smaller than the real ($D$ dimensional) one.
Or, equivalently, the real fundamental Planck energy scale becomes much smaller than the apparent one.  An estimate is obtained considering
the $D$-dimensional gravitational action and integrating the compact dimensions
by assuming the metric is independent of them:
\begin{equation}
\mathcal{S}\propto \frac{1}{G_D}\int d^Dx \sqrt{{}^{D}g} \ {}^{D}R= \frac{V_{D-4}}{G_D}\int d^4x \sqrt{{}^{4}g} \  {}^{4}R \, ,
\label{Daction}
\end{equation}
thus the four-dimensional Newton's constant is related to the $D$-dimensional
one by the volume of the compact dimensions $G_4=G_D/V_{D-4}$. 

In units such that $c=\hbar=1$ (different from the units $G=c=1$ used in
the rest of this paper), the mass-energy Planck scale in four dimensions
$E^{(4)}_{\rm Planck}$ is related to Newton's constant by 
$G_4=(E^{(4)}_{\rm Planck})^{-2}$, since $\int d^4x \sqrt{{}^{4}g} \ {}^{4}R$ has the dimension
of length squared; similar dimensional arguments in Eq.~(\ref{Daction})
show that in $D$ dimensions $G_D=(E^{(D)}_{\rm Planck})^{-(D-2)}$.  Therefore,
the $D$-dimensional Planck energy $E^{(D)}_{\rm Planck}$ is related to the four-dimensional one by
\begin{equation}
\frac{E^{(D)}_{\rm Planck}}{E^{(4)}_{\rm Planck}}=\left(\frac{1}{(E^{(4)}_{\rm Planck})^{D-4}V_{D-4}}\right)^{\frac{1}{D-2}}=
\left(\frac{(L^{(4)}_{\rm Planck})^{D-4}}{V_{D-4}}\right)^{\frac{1}{D-2}} \ ,\label{energyratio}
\end{equation}
where we have defined the four-dimensional Planck length as $L^{(4)}_{\rm Planck}=1/E^{(4)}_{\rm Planck}$. For instance,
for $D=10$ and taking the six extra dimensions of the order of the Fermi,
Eq.~(\ref{energyratio}) shows that the fundamental Planck scale would be of the order of
a TeV. For a more detailed account of the braneworld scenario, we
refer the reader to the reviews~\cite{Rubakov:2001kp,Maartens:2003tw}.

The real fundamental Planck scale sets the regime in particle physics beyond
which gravitational phenomena cannot be ignored and actually become dominant
\cite{tHooft:1987rb}; this is the trans-Planckian regime in which particle
collisions lead to BH formation and sizeable GW emission. A Planck scale at the
order of TeV (\emph{TeV gravity scenario}) could then imply BH formation in
particle accelerators, such as the LHC, or in ultra high-energy cosmic rays
\cite{Banks:1999gd,Dimopoulos:2001hw,Giddings:2001bu}.  Well into the
trans-Planckian regime, i.e., for energies significantly larger than the Planck
scale, classical gravity described by GR in $D$-dimensions is the appropriate
description for these events, since the formed BHs are large enough so that
quantum corrections may be ignored on and outside the horizon.

In this scenario, phenomenological signatures for BH formation would be obtained
from the Hawking evaporation of the micro BHs, and include a large multiplicity
of jets and large transverse momentum as compared to standard model backgrounds
\cite{Aad:2009wy}. Preliminary searches of BH formation events in the LHC data
have already been carried out, considering $pp$ collisions with center-of-mass energies up to 8~TeV; 
up to now, no evidence of BH creation has been found~\cite{Chatrchyan:2012taa,Aad:2012ic,Chatrchyan:2013xva,Aad:2013gma}.  
To filter experimental data from particle colliders, Monte Carlo event generators have
been coded, e.g., \cite{Frost:2009cf}, which need as input the cross section for
BH formation and the inelasticity in the collisions (gravitationally radiated
energy). The presently used values come from apparent horizon (AH)
estimates, which in $D=4$ are known to be off by a factor of 2 (at
least). In $D$-dimensions, these values must be obtained from
numerical simulations colliding highly boosted lumps of energy, BHs or
shock waves, since it is expected that in this regime `matter does not
matter'; all that matters is the amount of gravitational charge, i.e.,
energy, carried by the colliding objects.

\newpage
\section{Exact Analytic and Numerical Stationary Solutions}
\label{sec:elliptic} 

Any numerical or analytic analysis of dynamical processes must start
with a careful analysis of the static or stationary solutions
underlying those dynamics. In GR this is particularly relevant, as
stationary solutions are known and have been studied for many decades,
and important catalogs have been built. Furthermore, stationary
solutions are also relevant in a NR context: they can be used as
powerful benchmarks, initial data for nonlinear evolutions, and as a
final state reference to interpret results. We now briefly review some
of the most important, and recent, work on the subject directly
relevant to ongoing NR efforts. This Section does not dispense with
the reading of other reviews on the subject, for instance
Refs.~\cite{Stephani:2003tm,Emparan:2008eg,Horowitz:2012nnc,Wiseman:2011by}.

\subsection{Exact solutions} 

\subsubsection{Four-dimensional, electrovacuum general relativity with $\Lambda$} 
\label{section_4dexact}

Exact solutions of a non-linear theory, such as GR, provide invaluable insights
into the physical properties of the theory. Finding such solutions analytically
and through a direct attack, that is by inserting an educated ansatz into the
field equations, can be a \emph{tour de force}, and, in general, only leads to
success if a large isometry group is assumed from the beginning for the
spacetime geometry. For instance, assuming spherical symmetry, in vacuum, leads
to a fairly simple problem, whose general solution is the Schwarzschild metric
\cite{Schwarzschild:1916uq}. This simplicity is intimately connected with the
inexistence of a spherically symmetric mode for gravitational radiation in
Einstein gravity, which means that, in vacuum, a spherically symmetric
solution must be static, as recognized by Birkhoff~\cite{birkhoff1923}. On the
other hand, assuming only axial symmetry leads to a considerably more difficult
problem, even under the additional assumption of staticity. This problem was
first considered by Weyl~\cite{Weyl:1917} who unveiled a curious and helpful
mapping from these solutions to axially symmetric solutions of Newtonian gravity
in an auxiliary $3$-dimensional flat space; under this mapping, a solution to the
latter problem yields a solution to the vacuum Einstein equations: \emph{a Weyl solution}. For instance, the Schwarzschild solution of mass $M$ can be
recovered as a Weyl solution from the Newtonian gravitational field of an
infinitely thin rod of linear density $1/2$ and length $2M$. As we shall discuss
in Section~\ref{exact_beyond}, the generalization of Weyl solutions plays an important
role in the construction of qualitatively new solutions to the higher-dimensional Einstein equations.

Within the axially symmetric family of solutions, the most interesting case from
the astrophysical viewpoint is the solution for a rotating source, which could
describe the gravitational field exterior to a rotating star or the one of a
rotating BH. An exact solution of Einstein's equations describing the exterior
of a rotating star has not been found (rotating stars are described using
perturbative and numerical approaches~\cite{Stergioulas:2003yp}),%
\epubtkFootnote{In the context of quantum gravity, it has ben shown
  that including a fundamental minimal length, a solution exists in
  which an interior regular solution is matched to the exterior Kerr
  metric. Such configuration, however, is a ``regularized'' BH rather
  than a description of stars~\cite{Smailagic:2010nv}.}
but in the case of a rotating BH, such a solution does exist.
 To obtain this stationary, rather than static, geometry, the Weyl
approach by itself is unhelpful and new methods had to be developed. These new
methods started with Petrov's work on the classification of the Weyl tensor
types~\cite{Petrov:1954}. The Weyl tensor determines four null complex `eigenvectors' at
each point, and the spacetime is called `algebraically special' if at least two
of these coincide. Imposing the algebraically special condition has the
potential to reduce the complicated nonlinear PDEs in two variables, obtained
for a vacuum axially symmetric stationary metric, to ordinary differential equations. Using the (then)
recently shown Goldberg-Sachs theorem~\cite{GS:1962}, Kerr eventually
succeeded in obtaining the celebrated Kerr metric in 1963
\cite{Kerr:1963ud}. This family of solutions was generalized to include charge
by Newman et~al. -- the Kerr--Newman solution~\cite{Newman:1965my} -- and to
include a cosmological constant by Carter~\cite{Carter:1968ks}. In
Boyer-Lindquist coordinates, the Kerr--Newman-(A)dS metric reads:
\begin{equation}
ds^2=-\frac{\Delta_r}{\rho^2}\left[dt-\frac{a\sin^2\theta}{\Sigma}d\phi\right]^2
+\frac{\rho^2}{\Delta_r}dr^2+\frac{\rho^2}{\Delta_\theta}d\theta^2+\frac{\Delta_\theta \sin^2\theta}{\rho^2}\left[adt-\frac{r^2+a^2}{\Sigma}d\phi\right]^2 \ ,
\end{equation}
where
\begin{equation}
\rho^2=r^2+a^2\cos^2\theta \ , \qquad \Sigma=1+\frac{a^2\Lambda}{3} \ , 
\end{equation}
\begin{equation}
\Delta_r=(r^2+a^2)\left(1-\frac{r^2\Lambda}{3}\right)-2Mr+Q^2+P^2 \ , \qquad \Delta_\theta=1+\frac{a^2\Lambda}{3}\cos^2\theta \ .
\end{equation} 
Here, $M,aM,Q,P,\Lambda$ are respectively, the BH mass, angular momentum,
electric charge, magnetic charge and cosmological constant.

At the time of its discovery, the Kerr metric was presented as an example of a
stationary, axisymmetric (BH) solution. The outstanding importance of the Kerr
metric was only realized some time later with the establishment of the
\emph{uniqueness theorems}~\cite{Carter:1971zc,Robinson:1974nf}: the only
asymptotically flat, stationary and axisymmetric, electrovacuum solution to
the Einstein equations, which is non-singular on and outside an event horizon is
the Kerr--Newman geometry. Moreover, Hawking's rigidity theorem
\cite{Hawking:1973uf} made the axisymmetric assumption unnecessary: a
stationary BH must indeed be axisymmetric. Although the stability of the Kerr
metric is not a closed subject, the bottom line is that it is widely believed
that the final equilibrium state of the gravitational collapse of an enormous
variety of different stars is described by the Kerr geometry, since the electric
charge should be astrophysically negligible. If true, this is indeed a truly
remarkable fact (see, however, Section~\ref{sec_elliptic} for ``hairy'' BHs).

Even if we are blessed to know precisely the metric that describes the final
state of the gravitational collapse of massive stars or of the merger of two BHs, the
geometry of the time-dependent stages of these processes seems desperately out
of reach as an exact, analytic solution. To understand these processes we must
then resort to approximate or numerical techniques.

\subsubsection{Beyond four-dimensional, electrovacuum general
  relativity with $\Lambda$}
\label{exact_beyond}
As discussed in Section~\ref{sec:motivation} there are various motivations to
consider generalizations of (or alternative theories to) four-dimensional
electrovacuum GR with $\Lambda$. A natural task is then to
address the exact solutions of such theories. Here we shall briefly address the
exact solutions in two different classes of modifications of Einstein
electrovacuum gravity: i) changing the dimension, $D\neq 4$; ii) changing the
equations of motion, either by changing the right-hand side -- i.e., theories with
different matter fields, including non-minimally coupled ones --, or by changing
the left-hand side -- i.e., higher curvature gravity. We shall focus on relevant
solutions for the topic of this review article, referring to the specialized
literature where appropriate.

\begin{description}

\item[$\bullet$]{Changing the number of dimensions: GR in $D\neq 4$.} Exact solutions in
  \emph{higher-dimensional} GR, $D>4$, have been explored intensively for
  decades and an excellent review on the subject is
  Ref.~\cite{Emparan:2008eg}. In the following we shall focus on the vacuum
  case.

  The first classical result is the $D>4$ generalization of the Schwarzschild
  BH, i.e the vacuum, spherically -- that is $SO(D-1)$ -- symmetric solution to
  the $D$-dimensional Einstein equations (with or without cosmological
  constant), obtained by Tangherlini~\cite{Tangherlini:1963bw} in the same year
  the Kerr solution was found. Based on his solution, Tangherlini suggested an
  argument to justify the (apparent) dimensionality of spacetime. But apart from
  this insight, the solution is qualitatively similar to its four-dimensional
  counterpart: an analog of Birkhoff's theorem holds and it is perturbatively
  stable.

  On the other hand, the existence of extra dimensions accommodates a variety of
  extended objects with \emph{reduced} spherical symmetry -- that is
  $SO(D-1-p)$ -- surrounded by an event horizon, generically dubbed as
  \emph{$p$-branes}, where $p$ stands for the spatial dimensionality of the
  object~\cite{Horowitz:1991cd,Duff:1993ye}. Thus a point-like BH is a
  $0$-brane, a string is a $1$-brane and so on. The charged counterparts of
  these objects have played a central role in SMT, especially when
  charged under a type of gauge field called `Ramond-Ramond' fields, in which
  case they are called $Dp$-branes or simply $D$-branes~\cite{Duff:1994an}. Here
  we wish to emphasize that the Gregory--Laflamme instability discussed in
  Section~\ref{sec_hi_dim} was unveiled in the context of $p$-branes, in particular
  black strings~\cite{Gregory:1993vy,Gregory:1994bj}. The understanding of the
  non-linear development of such instability is a key question requiring
  numerical techniques.

  The second classical result was the generalization of the Kerr solution to
  higher dimensions, i.e., a vacuum, stationary, axially -- that is%
\epubtkFootnote{Here, [$\frac{D-1}{2}]$ denotes the integer part of
  $\frac{D-1}{2}$.}
  $SO(2)^{[\frac{D-1}{2}]}$ -- symmetric solution to the $D$-dimensional
  Einstein equations, obtained in 1986 by Myers and Perry~\cite{Myers:1986un}
  (and later generalized to include a cosmological constant~\cite{Gibbons:2004uw,Gibbons:2004js}). The derivation of this solution was
  quite a technical achievement, made possible by using a Kerr-Schild type
  ansatz. The solution exhibits a number of new qualitative features, in
  particular in what concerns its stability. It has $[\frac{D-1}{2}]$
  independent angular momentum parameters, due to the nature of the rotation
  group in $D$ dimensions. If only one of these rotation parameters is
  non-vanishing, i.e., for the singly spinning Myers--Perry solution, in
  dimensions $D\ge 6$ there is no bound on the angular momentum $J$ in terms of
  the BH mass $M$.  
  \emph{Ultra-spinning} Myers--Perry BHs are then possible and their
  horizon appears highly deformed, becoming locally analogous to that of a
  $p$-brane. This similarity suggests that ultra-spinning BHs should suffer from
  the Gregory--Laflamme instability. Entropic arguments also support the
  instability of these BHs~\cite{Emparan:2003sy} (see Section~\ref{sec_spacetime_stability} for recent developments).

  The third classical result was the recent discovery of the black ring in $D=5$~\cite{Emparan:2001wn}, a black object
  with a non-simply connected horizon, having spatial sections which are
  topologically $S^2\times S^1$. Its discovery raised questions about how the
  $D=4$ results on uniqueness and stability of vacuum solutions generalized to
  higher-dimensional gravity. Moreover, using the generalization to higher
  dimensions of Weyl solutions~\cite{Emparan:2001wk} and of the inverse
  scattering technique~\cite{Harmark:2004rm}, geometries with a non-connected
  event horizon -- i.e., multi-object solutions -- which are asymptotically flat,
  regular on and outside an event horizon have been found, most notably the
  black Saturn~\cite{Elvang:2007rd}. Such solutions rely on the existence of
  black objects with non-spherical topology; regular multi-object solutions with
  only Myers--Perry BHs do not seem to exist~\cite{Herdeiro:2008en}, just as
  regular multi-object solutions with only Kerr BHs in $D=4$ are inexistent
 ~\cite{Neugebauer:2009su,Herdeiro:2008kq}.

  Let us briefly mention that BH solutions in \emph{lower dimensional} GR
  have also been explored, albeit new ingredients are necessary for such
  solutions to exist. $D=3$ vacuum GR has no BH solutions, a fact related to the
  lack of physical dimensionality of the would be Schwarzschild radius
  $MG^{(3)}$, where $G^{(3)}$ is the $3$-dimensional Newton's constant. The
  necessary extra ingredient is a negative cosmological constant; considering it
  leads to the celebrated Ba\~nados-Teitelboim-Zanelli (BTZ) BH
 ~\cite{Banados:1992wn}. In $D=2$ a BH spacetime was obtained by Callan,
  Giddings, Harvey and Strominger (the CGHS BH), by considering GR non-minimally
  coupled to a scalar field theory~\cite{Callan:1992rs}. This solution provides
  a simple, tractable toy model for numerical investigations of dynamical
  properties; for instance see~\cite{Ashtekar:2010hx,Ashtekar:2010qz} for a numerical study of the evaporation of these BHs.

\item[$\bullet$]{Changing the equations: Different matter fields and higher
    curvature gravity.}

  The uniqueness theorems of four-dimensional electrovacuum GR make clear that
  BHs are selective objects. Their equilibrium state only accommodates a
  specific gravitational field, as is clear, for instance, from its
  constrained multipolar structure. In enlarged frameworks where other matter
  fields are present, this selectiveness may still hold, and various
  ``no-hair theorems'' have been demonstrated in the literature,
  i.e., proofs that under a set of assumptions no stationary regular BH solutions
  exist, supporting (nontrivial) specific types of fields. A prototypical case
  is the set of no-hair theorems for asymptotically flat, static, spherically
  symmetric BHs with scalar fields~\cite{Mayo:1996mv}. Note, however, that hairy BHs, do exist in various contexts, cf. Section~\ref{sec_elliptic}.

  The inexistence of an exact stationary BH solution, i.e., of an equilibrium
  state, supporting (say) a specific type of scalar field does not mean, however,
  that a scalar field could not exist long enough around a BH so that its effect
  becomes relevant for the observed dynamics. To analyse such possibilities
  dynamical studies must be performed, typically involving numerical techniques,
  both in linear and non-linear analysis. A similar discussion applies equally to the study of scalar-tensor theories of
  gravity, where the scalar field may be regarded as part of the gravitational
  field, rather than a matter field. Technically, these two perspectives may be
  interachanged by considering, respectively, the Jordan or the Einstein
  frame. The emission of GWs in a binary system, for instance, may depend on the
  `halo' of other fields surrounding the BH and therefore provide smoking guns
  for testing this class of alternative theories of gravity.

  Finally, the change of the left-hand side of the Einstein equations may be achieved by considering higher
  curvature gravity, either motivated by ultraviolet corrections to GR,
  i.e., changing the theory at small distance scales, such as Gauss-Bonnet
 ~\cite{Zwiebach:1985uq} (in $D\ge 5$), Einstein-Dilaton-Gauss-Bonnet gravity
  and Dynamical Chern-Simons gravity~\cite{Pani:2009wy,Alexander:2009tp}; or infrared corrections, changing the theory at large distance scales,
  such as certain $f(R)$ models. This leads, generically, to modifications of the
  exact solutions. For instance, the spherically symmetric solution to Gauss-Bonnet theory has been discussed in Ref.~\cite{Boulware:1985wk} and differs from, but
  asymptotes to, the Tangherlini solution. In specific cases, the higher
  curvature model may share some GR solutions. For instance, Chern--Simons
  gravity shares the Schwarzschild solution but not the Kerr
  solution~\cite{Alexander:2009tp}. Dynamical processes in these
  theories are of interest but their numerical formulation, for fully
  non-linear processes, may prove challenging or even, apart from
  special cases (see, e.g., \cite{Deppe:2012wk} for a study of
  critical collapse in Gauss--Bonnet theory), ill-defined.

\end{description}

\subsubsection{State of the art} 

\begin{description}

\item[$\bullet$] $D\neq 4$: The essential results in
  higher-dimensional vacuum gravity are the
  Tangherlini~\cite{Tangherlini:1963bw} and
  Myers--Perry~\cite{Myers:1986un} BHs, the (vacuum) black
  $p$-branes~\cite{Horowitz:1991cd,Duff:1993ye} and the Emparan--Reall
  black ring~\cite{Emparan:2001wn}. Solutions with multi-objects can
  be obtained explicitly in $D=5$ with the inverse scattering
  technique. Their line element is typically quite involved and given
  in Weyl coordinates (see~\cite{Emparan:2008eg} for a list and
  references). The Myers--Perry geometry with a cosmological constant
  was obtained in $D=5$ in Ref.~\cite{Hawking:1998kw} and for general
  $D$ and cosmological constant
  in~\cite{Gibbons:2004uw,Gibbons:2004js}. Black rings have been
  generalized, as numerical solutions, to higher $D$ in
  Ref.~\cite{Kleihaus:2012xh}. Black $p$-branes have been discussed,
  for instance, in Ref.~\cite{Horowitz:1991cd,Duff:1993ye}. In $D=3,2$
  the best known examples of BH solutions are, respectively, the
  BTZ~\cite{Banados:1992wn} and the CGHS BHs~\cite{Callan:1992rs}.

\item[$\bullet$] Changing the equations of motion: Hawking
  showed~\cite{Hawking:1972qk} that in Brans--Dicke gravity the only
  stationary BH solutions are the same as in GR. This result was
  recently extended by Sotiriou and Faraoni to more general
  scalar-tensor theories~\cite{Sotiriou:2011dz}. Such type of no-hair
  statements have also been proved for spherically symmetric solutions
  in GR (non-)minimally coupled to scalar
  fields~\cite{Bekenstein:1995un} \emph{and} to the electromagnetic
  field~\cite{Mayo:1996mv}; but they are not universal: for instance,
  a harmonic time dependence for a (complex) scalar field or a generic
  potential (together with gauge fields) are ways to circumvent these
  results (see Section~\ref{sec_elliptic} and e.g. the BH solutions
  in~\cite{Gibbons:1987ps}). BHs with scalar hair have also been
  recently argued to exist in generalized scalar-tensor
  gravity~\cite{Sotiriou:2013qea}.

\end{description}

\subsection{Numerical stationary solutions} 
\label{sec_elliptic}

Given the complexity of the Einstein equations, it is not surprising that, in
many circumstances, \emph{stationary} exact solutions cannot be found in
closed analytic form. In this subsection we shall very briefly mention numerical
solutions to such \emph{elliptic problems} for cases relevant to this review.

The study of the Einstein equations coupled to \emph{non-linear matter sources} must often be done numerically,
even if stationarity and spatial symmetries -- typically spherical or
axisymmetry -- are imposed.%
\epubtkFootnote{Numerical solutions of axially symmetric, rotating NSs
  in GR have been derived by several groups (see
  \cite{2013rrs..book.....F} and the Living Reviews article
  \cite{Stergioulas:2003yp}, and references therein), and in some
  cases their codes have been made publically available
  \cite{Stergioulas:1994ea,Bonazzola:1998qx}. These solutions are used
  to build initial data for NR simulations of NS-NS and BH-NS binary
  inspiral and merger.}
The study of numerical solutions of elliptic problems also
connects to research on soliton-like solutions in non-linear field theories
without gravity. Some of these solitons can be promoted to \emph{gravitating
  solitons} when gravity is included. Skyrmions are one such
case~\cite{Bizon:1992gb}. In other cases, the non-linear field theory does not
have solitons but, when coupled to gravity, gravitating solitons arise. This is
the case of the Bartnik--McKinnon particle-like solutions in Einstein--Yang--Mills
theory~\cite{Bartnik:1988am}. Moreover, for some of these gravitating solitons
it is possible to include a BH at their centre giving rise to ``hairy BHs''. For instance, in the case of Einstein--Yang--Mills theory, these
have been named ``colored BHs''~\cite{Bizon:1990sr}. We refer the
reader interested in such gravitating solitons connected to hairy BHs to
the review by Bizo\'n~\cite{Bizon:1994dh} and to the paper by Ashtekar
et~al.~\cite{Ashtekar:2000nx}.

A particularly interesting type of gravitating solitons are \emph{boson stars}
(see~\cite{Schunck:2003kk,Liebling:2012fv} for reviews) which
have been suggested as BH mimickers and dark matter candidates. These are
solutions to Einstein's gravity coupled to a complex massive scalar field, which
may, or may not, have self-interactions. Boson stars are horizonless gravitating
solitons kept in equilibrium by a balance between their self-generated gravity
and the dispersion effect of the scalar field's wave-like character. All known
boson star solutions were obtained numerically; and both static and rotating
configurations are known. The former ones have been used in numerical high
energy collisions to model particles and test the hoop
conjecture~\cite{Choptuik:2009ww} (see Section~\ref{sec:hoop_review} and also
Ref.~\cite{Palenzuela:2006wp} for earlier boson star collisions and
\cite{Mundim:2010hi} for a detailed description of numerical studies of boson
star binaries). The latter ones have been shown to connect to rotating BHs, both
for $D=5$ Myers--Perry BHs in AdS~\cite{Dias:2011at} and for $D=4$ Kerr
BHs~\cite{Herdeiro:2014goa}, originating families of rotating BHs with scalar
hair. Crucial to these connections is the phenomenon of superradiance (see
Section~\ref{sec:fundamental_fields}), which also afflicts rotating boson
stars~\cite{Cardoso:2007az}. The BHs with scalar hair branch off from the Kerr
or Myers--Perry-AdS BHs precisely at the threshold of the superradiant
instability for a given scalar field mode~\cite{Herdeiro:2014ima}, and display new physical properties, e.g. new shapes of ergo-regions~\cite{Herdeiro:2014jaa}.

The situation we have just described, i.e., the branching off of a solution to
Einstein's field equations into a new family at the onset of a classical
instability, is actually a recurrent situation. An earlier and paradigmatic
example -- occurring for the vacuum Einstein equations in higher dimensions --
is the branching off of black strings at the onset of the Gregory--Laflamme
instability~\cite{Gregory:1993vy} (see Section~\ref{sec_hi_dim} and
Section~\ref{sec:cosmic_censor_review}) into a family of non-uniform black
strings. The latter were found numerically by Wiseman~\cite{Wiseman:2002zc}
following a perturbative computation by Gubser~\cite{Gubser:2001ac}. We refer
the reader to Ref.~\cite{Kleihaus:2006ee} for more non-uniform string solutions,
to Refs.~\cite{Adam:2011dn,Wiseman:2011by} for a discussion of the techniques to
construct these numerical (vacuum) solutions and to~\cite{Horowitz:2011cq} for a
review of (related) Kaluza--Klein solutions. Also in higher dimensions, a number
of other numerical solutions have been reported in recent years, most notably
generalizations of the Emparan--Reall black
ring~\cite{Kleihaus:2010pr,Kleihaus:2012xh,Kleihaus:2013zpa} and BH solutions
with higher curvature corrections (see,
e.g.,~\cite{Brihaye:2010wx,Kleihaus:2012qz,Brihaye:2013vsa}). Finally, numerical
rotating BHs with higher curvature corrections but in $D=4$, within dilatonic
Einstein--Gauss--Bonnet theory, were reported in~\cite{Kleihaus:2011tg}.

In the context of holography (see Section~\ref{sec:gauge-gravity} and
Section~\ref{sec_holography7}), numerical solutions have been of paramount
importance. Of particular interest to this review are the hairy AdS BHs that
play a role in the AdS-Condensed matter duality, by describing the
superconducting phase of holographic superconductors. These were first
constructed (numerically) in~\cite{Hartnoll:2008kx}. See also the reviews
\cite{Hartnoll:2009sz,Horowitz:2010gk} for further developments.

In the context of Randall--Sundrum scenarios, large BHs were first shown to exist via a numerical calculation~\cite{Figueras:2011gd}, and later shown to agree with analytic expansions~\cite{Abdolrahimi:2012pb}.

Finally, let us mention, as one application to mathematical physics of numerical stationary solutions, the computation of Ricci-flat metrics on Calabi--Yau manifolds~\cite{Headrick:2005ch}. 

\newpage
\section{Approximation Schemes}
\label{sec:exact_approximation} 

The exact and numerically-constructed stationary solutions we outlined above are,
as a rule, objects which can also have interesting dynamics. A full understanding of these dynamics
is the subject of NR, but before attempting fully nonlinear evolutions of the field equations,
approximations are often useful. These work as benchmarks for numerical evolutions, as order-of-magnitude estimates and in 
some cases (for example extreme mass ratios) remain the only way to attack the problem, as it becomes prohibitively
costly to perform full nonlinear simulations, see Figure~\ref{fig:NR_validity_diagram}.
The following is a list of tools, techniques and results which have
been instrumental in the field. For an analysis of approximation schemes and their interface with
NR in four dimensional, asymptotically flat spacetimes, see Ref.~\cite{Tiec:2014lba}.
\subsection{Post-Newtonian schemes} 

\subsubsection{Astrophysical systems in general relativity}
For many physical phenomena involving gravity, GR predicts small deviations from
Newtonian gravity because for weak gravitational fields and low velocities
Einstein's equations reduce to the Newtonian laws
of physics. Soon after the formulation of GR, attempts were therefore made
(see, e.g., \cite{einstein1916sitzungsber,1916MNRAS..77..155D,lorentz1917collected,einstein1938gravitational,fock1939motion,papapetrou1951equations,plebanski1959general,1965ApJ...142.1488C,ehlers1980isolated}) to express
the dynamics of GR as deviations from the Newtonian limit
in terms of an expansion
parameter $\epsilon$. This parameter can be identified, for
instance, with the typical velocities of the matter composing the source, or
with the compactness of the source:
\begin{equation}
\label{eps1}
\epsilon\sim\frac{v}{c}\sim\sqrt{\frac{GM}{rc^2}}\,,
\end{equation}
which uses the fact that, for bound systems, the virial theorem
implies $v^2\sim GM/r$. In this approach, called ``post-Newtonian'', the
laws of GR are expressed in terms of the quantities and concepts of Newtonian
gravity (velocity, acceleration, etc.). A more rigorous definition of the
parameter $\epsilon$ can be found elsewhere~\cite{Blanchet:2014av}, but as a
book-keeping parameter it is customary to consider $\epsilon=v/c$. The
spacetime metric and the stress-energy tensor are expanded in powers of
$\epsilon$ and terms of order $\epsilon^n$ are commonly referred to as $(n/2)$-PN corrections. 
The spacetime metric and the motion of the source
are found by solving, order by order, Einstein's equations.

Strictly speaking, the PN expansion can only be defined in the near zone,
which is the region surrounding the source, with dimensions much smaller than
the wavelength $\lambda_{GW}$ of the emitted GWs. Outside this region, and in
particular in the wave zone (e.g., at a distance $\gg\lambda_{GW}$ from
the source), radiative processes make the PN expansion ill-defined, and
different approaches have to be employed, such as the post-Minkowskian
expansion, which assumes weak fields
but not slow motion. In the post-Minkowskian
expansion the gravitational field, described by the quantities
$h^{\alpha\beta}=\eta^{\alpha\beta}-\sqrt{-g}g^{\alpha\beta}$ (in harmonic
coordinates, such that $h^{\mu\nu}_{~,\nu}=0$) is formally expanded in powers of
Newton's constant $G$. Using a variety of different tools (PN expansion in
the near zone, post-Minkowskian expansion in the wave zone, multipolar
expansions, regularization of point-like sources, etc.), it is possible
to solve Einstein's equations, and to determine both the motion of the source
and its GW emission. Since each term of the post-Minkowskian expansion can
itself be PN-expanded, the final output of this computation has the form of a PN
expansion; therefore, these methods are commonly referred to as \emph{PN approximation schemes}.

PN schemes are generally used to study the motion of $N$-body systems in GR, and
to compute the GW signal emitted by these systems. More specifically, most of
the results obtained so far with PN schemes refer to
the \emph{relativistic two-body problem}, which can be applied to study
compact binary systems formed by BHs and/or NSs (see Section~\ref{gwastro}). In the
following we shall provide a brief summary of PN schemes, their main
features and results as
applied to the study of compact binary systems. For a more
detailed description, we refer the reader to one of the many reviews which have
been written on the subject; see
e.g.~\cite{Blanchet:2014av,KYpoisson-lecture-notes,Schafer:2009dq,iyer2011gravitational}.

Two different but equivalent approaches have been developed to solve the
relativistic two-body problem, finding the equations of motion of the source and
the emitted gravitational waveform: the multipolar post-Minkowskian approach of
Blanchet, Damour and Iyer~\cite{Blanchet:2014av}, and the direct integration of
the relaxed Einstein's equations, developed by Will and Wiseman
\cite{Will:1996zj}. In these approaches, Einstein's equations are solved
iteratively in the near zone, employing a PN expansion, and in the wave zone,
through a post-Minkowskian expansion. In both cases, multipolar expansions are
performed. The two solutions, in the near and in the wave zone, are then
matched. These approaches yield the equations of motion of the bodies, i.e.,
their accelerations as functions of their positions and velocities, and allow the energy balance equation of the system 
to be written as
\begin{equation}
\label{balance}
\frac{dE}{dt}=-{\cal L}\,.
\end{equation}
Here $E$ (which depends on terms of integer PN orders) can be considered as the
energy of the system), and ${\cal L}$ (depending on terms of half-integer PN
orders) is the emitted GW flux. The lowest PN order in the GW flux is given by
the quadrupole formula~\cite{Einstein:1918} (see also~\cite{Misner:1973cw}),
${\cal L}=G/(5c^5)(\dddot Q_{ab}\dddot Q_{ab}+O(1/c^3))$ where $Q_{ab}$ is the
(traceless) quadrupole moment of the source.  The leading term in ${\cal L}$ is
then of 2.5-PN order (i.e., $\sim1/c^5$), but since $Q_{ab}$ is computed in
the Newtonian limit, it is often considered as a ``Newtonian'' term.  A
remarkable result of the multipolar post-Minkowskian approach and of the direct
integration of relaxed Einstein's equations, is that once the equations are
solved at $n$-th PN order both in the near zone and in the wave zone, $E$ is
known at $n$-PN order, and ${\cal L}$ is known at $n$-PN order with respect to
its leading term, i.e., at $(n+2.5)$-PN order. Once the energy and the GW flux
are known with this accuracy, the gravitational waveform can be determined, in
terms of them, at $n$-PN order.

Presently, PN schemes determine the motion of a compact binary, and the
emitted gravitational waveform, up to 3.5-PN order for non-spinning binaries
in circular orbits~\cite{Blanchet:2014av}, but up to lower PN-orders for
eccentric orbits and for spinning binaries
\cite{Arun:2009mc,Buonanno:2012rv}. It is estimated that Advanced
LIGO/Virgo data analysis requires 3.5-PN templates~\cite{Boyle:2009dg},
and therefore
some effort still has to go into the modeling of
eccentric orbits and spinning binaries. It
should also be remarked that the state-of-the-art PN waveforms have been
compared with those obtained with NR simulations, showing a remarkable agreement
in the inspiral phase (i.e., up to the late inspiral stage)
\cite{Boyle:2007ft,Hannam:2007wf}.

An alternative to the schemes discussed above is the ADM-Hamiltonian approach
\cite{Schafer:2009dq}, in which using the ADM formulation of GR, the source is
described as a canonical system in terms of its Hamiltonian. The
ADM-Hamiltonian approach is equivalent to the multipolar post-Minkowskian
approach and to the direct integration of relaxed Einstein's equations, as long
as the evolution of the source is concerned~\cite{Damour:2000ni}, but since
Einstein's equations are not solved in the wave zone, the radiative effects are only
known with the same precision as the motion of the source. This framework has
been extended to spinning binaries (see~\cite{Steinhoff:2010zz} and references
therein). Recently, an alternative way to compute
the Hamiltonian of a post-Newtonian source has been developed, the effective
field theory approach~\cite{Goldberger:2007hy,Burgess:2007pt,Porto:2008tb,Galley:2008ih}, in which techniques originally
derived in the framework of quantum field theory are employed. This approach was also extended to spinning binaries 
\cite{Porto:2006bt,Porto:2010zg}. ADM-Hamiltonian and effective field theory are probably the most promising 
approaches to extend the accuracy of PN computations for spinning binaries.

The effective one body (EOB) approach developed at the end of the last century
\cite{Buonanno:1998gg} and recently improved~\cite{Damour:2009kr,Pan:2011gk}
(see, e.g., \cite{Damour:2012mv,Damour:2009ic} for a more detailed account) is an
extension of PN schemes, in which the PN Taylor series is suitably resummed, in
order to extend its validity up to the merger of the binary system. This 
approach maps the dynamics of the two compact objects into the dynamics of a
single test particle in a deformed Kerr spacetime. It is a canonical approach,
so the Hamiltonian of the system is computed, but the radiative part of the
dynamics is also described. Since the mapping between the two-body system and
the ``dual'' one-body system is not unique, the EOB Hamiltonian depends on a
number of free parameters, which are fixed using results of PN schemes, of
gravitational self-force computations, and of NR simulations. After this
calibration, the waveforms reproduce with good accuracy those obtained in
NR simulations (see,
e.g.,~\cite{Damour:2012mv,Damour:2009ic,Pan:2011gk,Baiotti:2011am}). In the same
period, a different approach has been proposed to extend PN templates to the
merger phase, matching PN waveforms describing the inspiral phase, with NR
waveforms describing the merger~\cite{Ajith:2007qp,Santamaria:2010yb}. Both this
``phenomenological waveform'' approach and the EOB approach use results from
approximation schemes and from NR simulations in order to describe the entire
waveform of coalescing binaries, and are instrumental for data analysis~\cite{Ohme:2011rm}.

To conclude this Section, we mention that PN schemes originally treated compact
objects as point-like, described by delta functions in the stress-energy tensor,
and employing suitable regularization procedures. This is appropriate for BHs,
and, as a first approximation, for NSs, too. Indeed, finite size effects are
formally of 5-PN order (see, e.g., \cite{damour-87,Blanchet:2014av}). However,
their contribution can be larger than what a naive counting of PN orders may
suggest~\cite{Mora:2003wt}. Therefore, the PN schemes and the EOB approach have
been extended to include the effects of tidal deformation of NSs in compact
binary systems and
on the emitted gravitational waveform using a set of parameters
(the ``Love numbers'') encoding the tidal deformability of the
star~\cite{Flanagan:2007ix,Damour:2009wj,Vines:2011ud,Bini:2012gu}.

\subsubsection{Beyond general relativity}

PN schemes are also powerful tools to study the nature of the gravitational
interaction, i.e., to describe and design observational tests of GR. They have
been applied either to build general parametrizations, or to determine observable
signatures of specific theories (two kinds of approaches which have been dubbed
\emph{top-down} and \emph{bottom-up}, respectively~\cite{Psaltis:2009xf}).

Let us discuss \emph{top-down} approaches first. Nearly fifty years ago Will and
Nordtvedt developed the PPN formalism
\cite{Will:1972zz,Nordtvedt:1972zz}, in which the PN metric of an $N$-body system is extended to a
more general form, depending on a set of parameters describing possible
deviations from GR. This approach (which is an extension of a similar approach
by Eddington~\cite{EddingtonBook}) facilitates tests of the weak-field regime of
GR. It is particularly well suited to perform tests in the Solar System. All
Solar System tests can be expressed in terms of constraints on the PPN
parameters, which translates into constraints on
alternative theories of gravity. For instance, the measurement of the Shapiro
time-delay from the Cassini spacecraft~\cite{Bertotti:2003rm} yields the
strongest bound on one of the PPN parameters; this bound determines the
strongest constraint to date on many modifications of GR, such as
Brans--Dicke theory.

More recently a different parametrized extension of the PN formalism has been proposed which, instead of the PN metric,
expands the gravitational waveform emitted by a compact binary inspiral in a set of parameters describing deviations from GR
\cite{Yunes:2009ke,Chatziioannou:2012rf}. The advantage of this so-called ``parametrized post-Einsteinian'' approach -
which is different in spirit from the PPN expansion, since it does not try to
describe the spacetime metric -- is its specific design to study the GW output of
compact binary inspirals which are the most promising sources for GW detectors
(see Section~\ref{gwastro}).

As mentioned above, PN approaches have also been applied \emph{bottom-up}, i.e., in a manner that directly calculates
the observational consequences of specific theories. For instance, the motion of binary pulsars has been studied, using PN schemes, in specific
alternative theories of gravity, such as scalar-tensor theories~\cite{Damour:1996ke}. The most promising observational quantity to look for
evidence of GR deviations is probably the gravitational waveform emitted in compact binary inspirals, as computed using PN approaches.
In the case of theories with additional fundamental fields, the leading effect is the
increase in the emitted gravitational flux arising from the additional degrees of freedom. This increase typically induces a
faster inspiral, which affects the phase of the gravitational waveform (see, e.g., \cite{Berti:2004bd}). For instance, in the case of scalar-tensor theories a dipolar component of the radiation can appear~\cite{Will:1989sk}. In other cases, as in massive graviton theories, the radiation has $\ell\ge2$ as in GR, but the flux is different. For further details, we refer the interested reader to~\cite{Will:1993ns} and references therein.

\subsubsection{State of the art}
The post-Newtonian approach has mainly been
used to study the relativistic two-body problem, i.e., to study the motion of
compact binaries and the corresponding GW emission. The first computation of
this kind, at leading order, was done by Peters and Mathews for generic
eccentric orbits~\cite{Peters:1963ux,Peters:1964zz}. It took about thirty
years to understand how to extend this computation at higher PN orders,
consistently modeling the motion and the gravitational emission of a compact
binary~\cite{Blanchet:2014av,Will:1996zj}.  The state-of-the-art computations
give the gravitational waveform emitted by a compact binary system, up to
3.5-PN order for non-spinning binaries in circular orbits
\cite{Blanchet:2014av}, up to 3-PN order for eccentric orbits
\cite{Arun:2009mc}, and up to 2-PN order for spinning binaries
\cite{Buonanno:2012rv}. An alternative approach, based on the computation of
the Hamiltonian~\cite{Schafer:2009dq}, is currently being extended to higher
PN orders~\cite{Steinhoff:2010zz,Jaranowski:2012eb,Hartung:2013dza}; however,
in this approach the gravitational waveform is computed with less accuracy
than the motion of the binary.

Recently, different approaches have been proposed to extend the validity of PN
schemes up to the merger, using results from NR to fix some of the parameters
of the model (as in the EOB approach
\cite{Damour:2009ic,Pan:2011gk,Baiotti:2011am,Damour:2012mv}), or matching NR
with PN waveforms (as in the ``phenomenological waveform'' approach
\cite{Ajith:2007qp,Santamaria:2010yb}). PN and EOB approaches have also been
extended to include the effects of tidal deformation of NSs
\cite{Flanagan:2007ix,Damour:2009wj,Vines:2011ud,Bini:2012gu}. 

PN approaches have been extended to test GR against alternative theories of
gravity.  Some of these extensions are based on a parametrization of specific
quantities, describing possible deviations from GR. This is the case in the PPN
approach~\cite{Will:1972zz,Nordtvedt:1972zz}, most suitable for Solar System
tests (see~\cite{Will:1993ns,Will:2014va} for extensive reviews on the subject),
and in the parametrized post-Einsteinian approach~\cite{Yunes:2009ke,Chatziioannou:2012rf}, most suitable
for the analysis of data from GW detectors.  Other extensions, instead, start from
specific alternative theories and compute -- using PN schemes -- their
observational consequences. In particular, the motion of compact binaries and
the corresponding gravitational radiation have been extensively studied in
scalar-tensor theories~\cite{Damour:1996ke,Will:1989sk,Alsing:2011er}.

\subsection{Spacetime perturbation approach}
\label{sec:st_perturbation} 
\subsubsection{Astrophysical systems in general relativity}
\label{classicpert}
The PN expansion is less successful at describing strong-field, relativistic
phenomena. Different tools have been devised to include this regime and
one of the most successful schemes consists of describing the spacetime as a small
deviation from a known exact solution. Systems well described by such
a perturbative approach include, for instance, the inspiral of a NS or
a stellar-mass BH of mass $\mu$ into a supermassive BH of mass
$M\gg \mu$~\cite{Glampedakis:2005hs,AmaroSeoane:2007aw}, or a BH undergoing small
oscillations around a stationary configuration~\cite{Kokkotas:1999bd,Ferrari:2007dd,Berti:2009kk}.

In this approach, the spacetime is assumed to be, at any instant, a small deviation from the background
geometry, which, in the cases mentioned above, is described by the Schwarzschild
or the Kerr solution here denoted by $g_{\mu\nu}^{(0)}$. The deformed spacetime
metric $g_{\mu\nu}$ can then be decomposed as
\begin{equation}
g_{\mu\nu}=g_{\mu\nu}^{(0)}+h_{\mu\nu}\,,\label{defpert}
\end{equation}
where $h_{\mu\nu}\ll1$ describes a small perturbation induced by a small object
or by any perturbing event.%
\epubtkFootnote{If matter or energy is present, there is a
  stress-energy tensor which is also perturbed,
  $T_{\mu\nu}=T_{\mu\nu}^{(0)}+\delta T_{\mu\nu}$. If $T_{\mu\nu}$
  describes a fluid, its perturbation can be described in terms of the
  perturbations of the thermodynamic quantities characterizing the
  fluid and of the matter velocity. We will only consider vacuum
  spacetimes here.}
Einstein's equations are linearized around the background solution, by
keeping only first-order terms in $h_{\mu\nu}$ (and in the other
perturbation quantities, if present).

The simple expansion (\ref{defpert}) implies a deeper geometrical
construction (see, e.g., \cite{stewart}), in which one considers a
family of spacetime manifolds ${\cal M}_\lambda$, parametrized by a parameter
$\lambda$; their metrics ${\mathbf g}(\lambda)$  
satisfy Einstein's equations, for each $\lambda$. The
$\lambda=0$ element of this family is the background spacetime, and the first
term in the Taylor expansion in $\lambda$ is the perturbation. Therefore, in the
spacetime perturbation approach it is the spacetime manifold itself to be
perturbed and expanded.  However, once the perturbations are defined (and the
gauge choice, i.e., the mapping between quantities in different manifolds, is
fixed), perturbations can be treated as genuine fields
living on the background spacetime ${\cal M}_0$. In particular, the linearized
Einstein equations can be considered as linear equations on the background
spacetime, and all the tools to solve linear differential equations on a
curved manifold can be applied.

The real power of this procedure comes into play once one knows how to separate
the angular dependence of the perturbations $h_{\mu\nu}$. This was first
addressed by Regge and Wheeler in their seminal paper~\cite{Regge:1957rw}, where
they showed that in the case of a Schwarzschild background, the metric
perturbations can be expanded in tensor spherical harmonics
\cite{mathewsharmonics}, in terms of a set of perturbation functions which only
depend on the coordinates $t$ and $r$. They also noted that the terms of this
expansion belong to two classes (even and odd perturbations, sometimes also
called polar and axial), with different behaviour under parity transformations
(i.e., $\theta\rightarrow\pi\!-\!\theta$, $\phi\rightarrow\phi\!+\!\pi$). The
linearized Einstein equations, expanded in tensor harmonics, yield the
dynamical equations for the perturbation functions. Furthermore, perturbations
corresponding to different
harmonic components or different parities decouple due to the fact
that the background is spherically symmetric. After a Fourier transformation in time,
the dynamical equations reduce to ordinary differential equations in $r$.

Regge and Wheeler worked out the equations for axial perturbations of
Schwarzschild BHs; later on, Zerilli derived the equations for polar
perturbations~\cite{Zerilli:1971wd}. With their gauge choice (the
``Regge--Wheeler gauge'', which allows us to set to zero some of the perturbation
functions), the harmonic expansion of the metric perturbation is
\begin{equation}
\label{decom}
h_{\mu\nu}(t,r,\theta,\phi)=\sum_{l,m}\int_{-\infty}^{+\infty}e^{-i\omega t}\left[
h^{{\rm ax},lm}_{\mu\nu}(\omega,r,\theta,\phi)+h^{{\rm pol},lm}_{\mu\nu}(
\omega,r,\theta,\phi)\right]d\omega
\end{equation}
with
\begin{eqnarray}
h^{{\rm ax},lm}_{\mu\nu} \, dx^\mu \, dx^\nu&=&2\left[h^{lm}_0(\omega,r)dt\!+\!h^{lm}_1(\omega,r)dr\right]
\left[\csc\theta\partial_{\phi}Y_{lm}(\theta,\phi)d\theta\!-\!\sin\theta\partial_{\theta}Y_{lm}(\theta,\phi)d\phi\right]
\label{oddpart}\\
h^{{\rm pol},lm}_{\mu\nu} \, dx^\mu \, dx^\nu&=&\left[f(r)H_0^{lm}(\omega,r)dt^2+2H_1 (\omega,r)^{lm}dtdr+H_2^{lm}(\omega,r)
dr^2\right.\nonumber\\
&&\left.+r^2K^{lm}(\omega,r)(d\theta^2+\sin^2\theta d\phi^2)\right]Y_{lm}(\theta,\phi)\,,\label{evenpart}
\end{eqnarray}
where $f(r)=1-2M/r$, and $Y_{lm}(\theta,\phi)$ are the scalar spherical harmonics.

It turns out to be possible to define a specific combination $Z^{lm}_{\rm RW}(\omega,r)$
of the axial perturbation functions $h^{lm}_0\,,~h^{lm}_1$, and a combination $Z^{lm}_{\rm Zer}(\omega,r)$
of the polar perturbation functions $H_{0,1,2}^{lm}\,,~K^{lm}$ which
describe completely the propagation of GWs. These functions, called
the Regge--Wheeler and the Zerilli function, satisfy Schroedinger-like wave
equations of the form
\be
\frac{d^2\Psi_{\rm RW,\,Zer}}{dr_*^2}+\left(\omega^2-V_{\rm RW,\,Zer}\right)\Psi_{\rm RW,\,Zer}={\cal S}_{\rm RW,\,Zer}\,.\label{eq:RW_Zer}
\ee
Here, $r_*$ is the tortoise coordinate~\cite{Misner:1973cw} and
$\cal S$ represents nontrivial source terms. The energy flux emitted
in GWs can be calculated straightforwardly from the solutions
$\Psi_{\rm RW,\,Zer}$.

This approach was soon extended to general spherically symmetric BH backgrounds
and a gauge-invariant formulation in terms of specific combinations of the
perturbation functions that remain unchanged under perturbative
coordinate transformations
\cite{Moncrief:1974am,Gerlach:1979rw}. In the same period, an alternative
spacetime perturbation approach was developed by Bardeen, Press and Teukolsky
\cite{Bardeen:1973xb,Teukolsky:1973ap}, based on the Newman-Penrose formalism
\cite{Newman:1961qr},
in which the spacetime perturbation is not described by the metric perturbation
$h_{\mu\nu}$, but by a set of gauge-invariant complex scalars, the Weyl scalars,
obtained by projecting the Weyl tensor $C_{\alpha\beta\gamma\delta}$ onto a
complex null tetrad $\boldsymbol{\ell},~\boldsymbol{k},~
\boldsymbol{m},~\boldsymbol{\bar{m}}$ defined such that all
their inner products vanish except $-\boldsymbol{k}\cdot
\boldsymbol{\ell} = 1 = \boldsymbol{m}\cdot
\boldsymbol{\bar{m}}$. One of these scalars, $\Psi_4$, describes the
(outgoing) gravitational radiation; it is defined as
\begin{equation}
  \Psi_4 \equiv -C_{\alpha \beta \gamma \delta}
        \ell^{\alpha}
        \bar{m}^{\beta}
        \ell^{\gamma}
        \bar{m}^{\delta}\,.
        \label{eq:Psi4}
\end{equation}
In the literature one may also find $\Psi_4$ defined without the minus sign,
but all physical results derived from $\Psi_4$ are invariant under this
ambiguity. We further note that the Weyl and Riemann tensors are identical
in vacuum. Most BH studies in NR consider vacuum
spacetimes, so that we can replace $C_{\alpha \beta \gamma \delta}$ in
Eq.~(\ref{eq:Psi4}) with $R_{\alpha \beta \gamma \delta}$.

In this framework, the perturbation equations reduce to a wave equation for (the
perturbation of) $\Psi_4$, which is called the Teukolsky equation
\cite{Teukolsky:1972my}. For a general account on the theory of BH perturbations
(with both approaches) see Chandrasekhar's book~\cite{MTB}.

The main advantage of the Bardeen-Press-Teukolsky approach is that it is
possible to separate the angular dependence of perturbations of the Kerr
background, even though such background is not spherically symmetric. Its main
drawback is that it is very difficult to extend it beyond its original setup,
i.e., perturbations of Kerr BHs. The tensor harmonic approach is much more
flexible. In particular, spacetime perturbation theory (with tensor harmonic
decomposition) has been extended to spherically symmetric stars
\cite{1967ApJ...149..591T,1983ApJS...53...73L,1985ApJ...292...12D,Chandrasekhar:1991fi}
(the extension to rotating stars is much more problematic
\cite{2013rrs..book.....F}).  As we discuss in Section~\ref{pertdgt4}, spacetime
perturbation theory with tensor harmonic decomposition can be extended to
higher-dimensional spacetimes. It is not clear whether such generalizations are
possible with the Bardeen-Press-Teukolsky approach.

The sources ${\cal S}_{\rm RW,\,Zer}$ describe the objects that excite the
spacetime perturbations, and can arise either directly from a non-vanishing
stress-energy tensor or by imposing suitable initial conditions on the
spacetime. These two alternative forms of exciting BH spacetimes have branched
into two distinct tools, which can perhaps be best classified as the ``point
particle''~\cite{Davis:1971gg,Cardoso:2002ay,Nakamura:1987zz,Berti:2010ce} and
the ``close limit'' approximations~\cite{Price:1994pm,Pullin:1999rg}.

In the point particle limit the source term is a nontrivial perturbing
stress-tensor, which describes for instance the infall of a small object along
generic geodesics.  The ``small'' object can be another BH, or a star, or even
matter accreting into the BH. While the framework is restricted to objects of
mass $\mu \ll M$, it is generically expected that the extrapolation to $\mu\sim
M$ yields at least a correct order of magnitude.  Thus, the
spacetime perturbation
approach is in principle able to describe qualitatively, if not quantitatively,
highly dynamic BHs under general conditions. The original approach treats the
small test particle moving along a geodesic of the background
spacetime. Gravitational back-reaction can be included by taking into account
the energy and angular momentum loss of the particle due to GW emission
\cite{Cutler:1994pb,Hughes:1999bq,Mino:2003yg}.  More sophisticated computations
are required to take into account the conservative part of the
``self-force''. For a general account on the self-force problem, we refer the
interested reader to the Living Reviews article on the subject~\cite{Poisson:2011nh}. In this
approach $\mu$ is restricted to be a very small quantity. It has been observed
by many authors~\cite{Anninos:1993zj,Sperhake:2011ik} that promoting $\mu/M$ to
the symmetric mass ratio $M_1M_2/(M_1+M_2)$ describes surprisingly well the dynamics of generic BHs with masses $M_1,M_2$.

In the close limit approximation the source term can be traced back to
nontrivial initial conditions. In particular, the original approach tackles the
problem of two colliding, equal-mass BHs, from an initial separation small
enough that they are initially surrounded by a common horizon. Thus, this
problem can be looked at as a single perturbed BH, for which some initial
conditions are known~\cite{Price:1994pm,Pullin:1999rg}.

A universal feature of the dynamics of BH spacetimes as given by either the
point particle or the close limit approximation is that the waveform $\Psi$
decays at late times as a universal, exponentially damped sinusoid called
ringdown or QNM decay. Because at late times the forcing caused by the source
term ${\cal S}$ has died away, it is natural to describe this phase as the free
oscillations of a BH, or in other words as solutions of the homogeneous version
of Eq.\eqref{eq:RW_Zer}. Together with the corresponding boundary conditions,
the Regge--Wheeler and Zerilli equations then describe a freely oscillating
BH. In vacuum, such boundary conditions lead to an eigenvalue equation for the
possible frequencies $\omega$.  Due to GW emission, these oscillations are
damped, i.e., they have discrete, complex frequencies called \emph{quasi-normal
  mode} frequencies of the BH~\cite{Kokkotas:1999bd,Ferrari:2007dd,Berti:2009kk}.  Such intuitive picture of
BH ringdown can be given a formally rigorous meaning through contour integration
techniques~\cite{Leaver:1986gd,Berti:2009kk}.

The extension of the Regge--Wheeler--Zerilli approach to asymptotically dS or AdS
spacetimes follows with the procedure outlined above and decomposition
\eqref{decom}; see also Ref.~\cite{Cardoso:2001bb}.  It turns out that the Teukolsky procedure
can also be generalized to these spacetimes~\cite{Chambers:1994ap,Dias:2012pp,Dias:2013sdc}.

\subsubsection{Beyond electrovacuum GR}
\label{pertaltern}
The Regge--Wheeler--Zerilli approach has proved fruitful also in other contexts
including alternative theories of gravity. Generically, the decomposition works
by using the same metric ansatz as in Eq.~\eqref{decom}, but now augmented to
include perturbations in matter fields, such as scalar or vector fields, or
further polarizations for the gravitational field.  Important examples where
this formalism has been applied include scalar-tensor theories
\cite{Saijo:1996iz,Cardoso:2011xi,Yunes:2011aa}, Dynamical Chern--Simons
theory~\cite{Cardoso:2009pk,Molina:2010fb,Pani:2011xj},
Einstein-Dilaton--Gauss--Bonnet~\cite{Pani:2009wy}, Horndeski gravity
\cite{Kobayashi:2012kh,Kobayashi:2014wsa}, and massive theories of
gravity~\cite{Brito:2013wya}.

\subsubsection{Beyond four dimensions}
\label{pertdgt4}
Spacetime perturbation theory is a powerful tool to study BHs in higher-dimensional spacetimes.
The tensor harmonic approach has been successfully extended by Kodama and
Ishibashi~\cite{Kodama:2003jz,Ishibashi:2003ap} to GR in higher-dimensional
spacetimes, with or without cosmological constant. Their approach generalizes the
gauge-invariant formulation of the Regge--Wheeler-Zerilli construction to
perturbations of Tangherlini's solution describing
spherically symmetric BHs.

Since many dynamical processes involving higher-dimensional BHs (in particular,
the collisions of BHs starting from finite distance) can be described in the far
field limit by a perturbed spherically symmetric BH spacetime, the Kodama and Ishibashi
approach can be useful to study the GW emission in these
processes. The relevance of this approach therefore extends well beyond the
study of spherically symmetric solutions. For applications
of this tool to the wave extraction of NR simulations see
for instance~\cite{Witek:2010xi}.

In the Kodama and Ishibashi approach, the $D$-dimensional spacetime metric is assumed to have the
form $g_{\mu\nu}=g_{\mu\nu}^{(0)}+h_{\mu\nu}$ where $g_{\mu\nu}^{(0)}$ is the
Tangherlini solution and $h_{\mu\nu}$ represents a small perturbation.
Decomposing the
$D$-dimensional spherical coordinates into $x^\mu=(t,r,\vec\phi)$ with
$D-2$ angular coordinates $\vec\phi=\{\phi^a\}_{a=1,\dots D-2}$, the perturbation $h_{\mu\nu}$ can be expanded
in spherical harmonics, as in the four-dimensional case (see
Section~\ref{classicpert}).  However the expansion in $D>4$ is more complex
than its four-dimensional counterpart: there are three classes of
perturbations called the ``scalar'',
``vector'' and ``tensor'' perturbations. The former two classes correspond, in
$D=4$, to polar and axial perturbations, respectively. These perturbations are
decomposed into scalar (${\cal S}^{ll'\dots}$), vector (${\cal V}^{ll'\dots}_a$)
and tensor (${\cal T}^{ll'\dots}_{ab}$) harmonics on the $(D-2)$-sphere $S^{D-2}$ and their
gradients, as follows:
\begin{equation}
\label{decomki}
h_{\mu\nu}(t,r,\vec\phi)=\sum_{ll'\dots}\int_{-\infty}^{+\infty}e^{-i\omega t}\left[
h^{{\rm S},ll'\dots}_{\mu\nu}(\omega,r,\vec\phi)+h^{{\rm V},ll'\dots}_{\mu\nu}(
\omega,r,\vec\phi) +h^{{\rm T},ll'\dots}_{\mu\nu}(
\omega,r,\vec\phi)\right]d\omega\,,
\end{equation}
where $ll'\dots$ denote harmonic indices on $S^{D-2}$ and the
superscripts S,V,T refer to scalar, vector and tensor perturbations,
respectively. Introducing early upper case Latin indices $A,\,B,\,\ldots =
0,\,1$ and $x^{\A}=(t,\,r)$, the metric perturbations can be written as
\begin{eqnarray}
&&h^{{\rm S},ll'\dots}_{\mu\nu}(\omega,r,\vec\phi) \, dx^\mu \, dx^\nu=\nonumber\\
&&\left[f_{AB}^{{\rm S}\,ll'\dots}(\omega,r)dx^Adx^B+H_{L}^{{\rm S}\,ll'\dots}(\omega,r)
\Omega_{ab}d\phi^ad\phi^b\right]{\cal S}^{ll'\dots}(\vec\phi)\nonumber\\
&&+f_A^{{\rm S}\,ll'\dots}(\omega,r)dx^A{\cal S}^{ll'\dots}_a(\vec\phi)d\phi^{a}+H_T^{{\rm S}
\,ll'\dots}(\omega,r){\cal S}_{ab}^{ll'\dots}(\vec\phi)d\phi^a d\phi^b\nonumber\\ 
&&h^{{\rm V},ll'\dots}_{\mu\nu}(\omega,r,\vec\phi) \, dx^\mu \, dx^\nu=\nonumber\\
&&\left[f_A^{{\rm V}\,ll'\dots}(\omega,r)dx^A\right]{\cal V}_a^{ll'\dots}(\vec\phi)d\phi^a
+H_T^{{\rm V}\,ll'\dots}(\omega,r){\cal V}_{ab}^{ll'\dots}(\vec\phi)d\phi^a d\phi^b\nonumber\\
&&h^{{\rm T},ll'\dots}_{\mu\nu}(\omega,r,\vec\phi) \, dx^\mu \, dx^\nu=H_T^{{\rm T}\,ll'\dots}(
\omega,r){\cal T}_{ab}^{ll'\dots}(\vec\phi)d\phi^a d\phi^b\,,
\end{eqnarray}
where $f_{AB}^{{\rm S}ll'\dots}(\omega,r),f_{A}^{{\rm
    S}ll'\dots}(\omega,r),\dots$ are the spacetime perturbation functions.  In
the above expressions, $\Omega_{ab}$ is the metric on $S^{D-2}$, ${\cal
  S}_a=-{\cal S}_{,a}/k$, ${\cal S}_{ab}= {\cal S}_{:ab}/k^2$ minus trace terms,
where $k^2=l(l+D-3)$ is the eigenvalue of the scalar harmonics, and the ``:'' denotes the covariant derivative on $S^{D-2}$;
the traceless ${\cal V}_{ab}$ is defined in a similar way.

A set of gauge-invariant variables and
the so-called ``master functions'', generalizations of the
Regge--Wheeler and Zerilli functions, can be constructed out of the
metric perturbation functions and satisfy wave-like differential equations
analogous to Eq.~(\ref{eq:RW_Zer}). The GW amplitude and its energy and
momentum fluxes can be expressed in terms of these master functions.

For illustration of this procedure, we consider here the special case
of scalar perturbations. We define the gauge-invariant quantities
\begin{equation}
 F = H_L + \frac{1}{D-2} H_T + \frac{1}{r} X_{\A} \hat{D}^{\A} r\,,
 ~~~~~
 F_{\A \B} = f_{\A \B} + \hat{D}_{\B} X_{\A} + \hat{D}_{\A} X_{\B}\,,
\end{equation}
where we have dropped harmonic indices,
\begin{equation}
 X_{\A} \equiv \frac{r}{k}\left( f_{\A} + \frac{r}{k}
       \hat{D}_{\A} H_T \right)\,,
\end{equation}
and $\hat{D}_{\A}$ denotes the covariant derivative associated with
$(t,\,r)$ sub-sector of the background metric.
A master function $\Phi$ can be conveniently defined in terms
of its time derivative according to
\begin{equation}
 \partial_t \Phi = (D-2) r^{\frac{D-4}{2}}
       \frac{-F^r{}_t + 2r\partial_t F}{k^2-D+2
       +\frac{(D-2)(D-1)}{2} \frac{R_{\rm S}^{D-3}}{r^{D-3}}}\,.
\end{equation}
From the master function, we can calculate the GW energy flux
\begin{equation}
 \frac{dE_{\ell m}}{dt} =
       \frac{1}{32\pi} \frac{D-3}{D-2} k^2 (k^2-D+2)
       (\partial_t \Phi_{\ell m})^2\,.
\end{equation}
The total radiated energy is obtained from integration in time and
summation over all multipoles
\begin{equation}
E=\sum_{\ell=2}^{\infty} \sum_{m=-\ell}^{\ell}
  \int_{-\infty}^{\infty} \frac{dE_{\ell m}}{dt} \, dt\,.
\end{equation}

In summary, this approach can be used, in analogy with the Regge--Wheeler--Zerilli
formalism in four dimensions, to determine the quasi-normal mode
spectrum (see, e.g., the review~\cite{Berti:2009kk} and references
therein), to determine the gravitational-wave
emission due to a test source~\cite{Berti:2003si,Berti:2010gx}, or to evaluate
the flux of GWs emitted by a dynamical spacetime which tends
asymptotically to a perturbed Tangherlini solution~\cite{Witek:2010xi}.

The generalization of this setup to higher-dimensional rotating
(Myers--Perry~\cite{Myers:1986un}) BHs is still an open issue, since
the decoupling of the perturbation equations has so far
only been obtained in specific cases and for a
subset of the perturbations~\cite{Murata:2007gv,Kunduri:2006qa,Kodama:2009bf}.

Spacetime perturbation theory has also been used to study other types of
higher-dimensional objects as for example black strings.
Gregory and Laflamme~\cite{Gregory:1993vy,Gregory:1994bj} considered a very
specific sector of the possible gravitational perturbations of these objects,
whereas Kudoh~\cite{Kudoh:2006bp} performed a complete analysis that builds on
the Kodama--Ishibashi approach.

\subsubsection{State-of-the-art} 

\begin{description}

\item[$\bullet$]{Astrophysical systems.} 
  Perturbation theory has been applied extensively to the modelling of
  BHs and compact stars, either without source terms, including in particular
  quasi-normal modes~\cite{Kokkotas:1999bd,Ferrari:2007dd,Berti:2009kk},
  or with point particle sources. Note that
  wave emission
  from extended matter distributions can be understood as interference
  of waves from point
  particles~\cite{Haugan:1982fb,1982ApJ...260..838S,1985ApJS...58..297P}. Equations
  for BH perturbations have been derived for
  Schwarzschild~\cite{Regge:1957rw,Zerilli:1971wd},
 RN~\cite{Zerilli:1974ai}, Kerr~\cite{Teukolsky:1973ap} and
  slowly rotating Kerr--Newman BHs~\cite{Pani:2013ija}. Equations for
  perturbations of stars have been derived for spherically
  symmetric~\cite{1967ApJ...149..591T,1983ApJS...53...73L,Chandrasekhar:1991fi}
  and slowly rotating stars~\cite{ChandraFerrari91,Kojima:1992ie}.

  Equations of BH perturbations with a point particle source have been studied
  as a tool to understand BH dynamics. This is a decades old topic, historically
  divided into investigations of circular and quasi-circular motion, and head-ons
  or scatters.

\noindent \textit{Circular and quasi-circular motion.}
Gravitational radiation from point particles in circular geodesics was studied
in Refs.~\cite{Misner:1972jf,Davis:1972dm,Breuer:1973kp} for non-rotating BHs
and in Ref.~\cite{Detweiler:1978ge} for rotating BHs. This problem was
reconsidered and thoroughly analyzed by Poisson, Cutler and collaborators, and
by Tagoshi, Sasaki and Nakamura in a series of elegant works, where contact
was also made with the PN expansion (see the Living Reviews article~\cite{Sasaki:2003xr} and
references therein). The emission of radiation, together with the self-gravity
of the objects implies that particles do not follow geodesics of the
background spacetime. Inclusion of dissipative effects is usually done by
balance-type arguments
\cite{Hughes:1999bq,Hughes:2001jr,Sundararajan:2008zm,Fujita:2009us} but it can
also be properly accounted for by computing the self-force effects of the
particle motion (see the Living Reviews article~\cite{Poisson:2011nh} and references therein).
EM waves from particles
in circular motion around BHs were studied in
Refs.~\cite{Davis:1972dm,Breuer:1973kp,Breuer:1973kt}.

\noindent \textit{Head-on or finite impact parameter collisions: non-rotating BHs.}
Seminal work by Davis et~al.~\cite{Davis:1971gg,Davis:1972pa} models the
gravitational radiation from BH collisions by a point particle
falling from rest at infinity into a Schwarzschild BH. This work has been
generalized to include head-on collisions at non-relativistic velocities~\cite{Ruffini:1973ky,Ferrari:1981dh,Lousto:1996sx,Berti:2010ce}, at exactly the
speed of light~\cite{Cardoso:2002ay,Berti:2010ce}, and to non-head-on collisions
at non-relativistic velocities~\cite{Detweiler:1979xr,Berti:2010ce}.

The infall of multiple point particles has been explored in
Ref.~\cite{Berti:2006hb} with particular emphasis on
resonant excitation of QNMs. Shapiro and collaborators have investigated the infall or collapse of extended
matter distributions through superpositions of point particle waveforms
\cite{Haugan:1982fb,1982ApJ...260..838S,1985ApJS...58..297P}.

Electromagnetic radiation from high-energy collisions of charged
particles with uncharged BHs was studied in Ref.~\cite{Cardoso:2003cn}
including a comparison with zero-frequency limit (ZFL)
predictions. Gravitational and EM radiation generated in collisions of
charged BHs has been considered in
Refs.~\cite{Johnston:1973cd,Johnston:1974vf}.

\noindent \textit{Head-on or finite impact parameter collisions: rotating BHs.}
Gravitational radiation from point particle collisions with Kerr BHs has
been studied in Refs.~\cite{Kojima:1983ua,Kojima:1984pz,Kojima:1984cc,Kojima:1984cj}.
Suggestions that cosmic censorship might fail in high-energy collisions
with near-extremal Kerr BHs, have recently
inspired further scrutiny of these scenarios
\cite{Barausse:2010ka,Barausse:2011vx} as well as the investigation
of enhanced absorption effects
in the ultra-relativistic regime~\cite{Gundlach:2012aj}.

\noindent \textit{Close Limit approximation.} The close limit approximation was
first compared against nonlinear simulations of equal-mass, non-rotating BHs
starting from rest~\cite{Price:1994pm}. It has since been generalized to
unequal-mass~\cite{Andrade:1996pc} or even the point particle limit
\cite{Lousto:1996sx}, rotating BHs~\cite{Krivan:1998er} and boosted BHs at
second-order in perturbation theory~\cite{Nicasio:1998aj}. Recently the close
limit approximation has also been applied to initial configurations
constructed with PN methods~\cite{LeTiec:2009yf}.

\item[$\bullet$]{Beyond electrovacuum GR.} The resurgence of scalar-tensor
  theories as a viable and important prototype of alternative theories of
  gravity, as well as the conjectured existence of a multitude of fundamental
  bosonic degrees of freedom, has revived interest in BH dynamics in the
  presence of fundamental fields. Radiation from collisions of scalar-charged particles with BHs was studied in Ref.~\cite{Brito:2012gj}.
Radiation from massive scalar fields around rotating BHs was studied in
Ref.~\cite{Cardoso:2011xi} and shown to lead to floating orbits. Similar effects
do \emph{not} occur for massless gravitons~\cite{Kapadia:2013kf}.

\item[$\bullet$]{Beyond four-dimensions and asymptotic flatness.} The
  gauge/gravity duality and related frameworks highlight the importance of
  (A)dS and higher-dimensional background spacetimes. The formalism to handle
  gravitational perturbations of four-dimensional, spherically symmetric asymptotically (A)dS BHs has
  been developed in Ref.~\cite{Cardoso:2001bb}, whereas perturbations of rotating
  AdS BHs were recently tackled~\cite{Chambers:1994ap,Dias:2012pp,Dias:2013sdc}. Gravitational perturbations
  of higher-dimensional BHs can be handled through the elegant approach by
  Kodama and Ishibashi~\cite{Kodama:2003jz,Kodama:2003kk}, generalized in
  Ref.~\cite{Kudoh:2006bp} to include perturbations of black
  strings. Perturbations of higher-dimensional, rotating BHs can be expressed in
  terms of a single master variable only in few special cases
 ~\cite{Kunduri:2006qa}. The generic case has been handled by numerical methods in the linear regime
 ~\cite{Dias:2010eu,Hartnett:2013fba}.

  Scalar radiation by particles around Schwarzschild-AdS BHs has been studied in
  Refs.~\cite{Cardoso:2002up,Cardoso:2001tw,Cardoso:2002cf}. We are not aware of
  any studies on gravitational or electromagnetic radiation 
emitted by particles in orbit about BHs in spacetimes with a cosmological constant.

  The quadrupole formula was generalized to higher-dimensional spacetimes in
  Ref.~\cite{Cardoso:2002pa}.  The first fully relativistic calculation of GWs
  generated by point particles falling from rest into a higher-dimensional
  asymptotically flat non-rotating BH was done in Ref.~\cite{Berti:2003si}, and
  later generalized to arbitrary velocity in Ref.~\cite{Berti:2010gx}. The mass multipoles 
  induced by an external gravitational field  (i.e., the ``Love numbers'') to a higher-dimensional BH, 
  have been determined in Ref.~\cite{Kol:2011vg}.

  The close limit approximation was extended to higher-dimensional,
  asymptotically flat, spacetimes in Refs.~\cite{Yoshino:2005ps,Yoshino:2006kc}.

\end{description}

\subsection{The zero-frequency limit}
\label{sec:ZFLsection}

\subsubsection{Astrophysical systems in general relativity} 
While conceptually simple, the spacetime perturbation approach does involve
solving one or more second-order, non-homogeneous differential equations.  A
very simple and useful estimate of the energy spectrum and total radiated
gravitational energy can be obtained by using what is known as the ZFL or instantaneous collision approach.

The technique was derived by Weinberg in
1964~\cite{Weinberg:1964ew,Weinberg:1965nx} from quantum arguments,
but it is equivalent to a purely classical calculation~\cite{Smarr:1977fy}. The approach
is a consequence of the identity
\be
\left.\overline{(\dot{h})}\, \right|_{\omega=0}=\lim_{\omega\to 0}\int_{-\infty}^{+\infty}
\dot{h}e^{-i\omega t} \, dt=h(t=+\infty)-h(t=-\infty)\,,
\ee
for the Fourier transform $\overline{(\dot{h})}(\omega)$ of the time derivative of any
metric perturbation $h(t)$ (we omitted unimportant constant overall factors in the
definition of the transform). Thus, the low-frequency spectrum depends
exclusively on the asymptotic state of the colliding particles which can be
readily computed from their Coulomb gravitational fields. Because the energy
spectrum is related to $\bar{\dot{h}}(\omega)$ via
\be
\frac{d^2E}{d\Omega \, d\omega}\propto r^2\left(\overline{(\dot{h})}\right)^2\,,
\ee
we immediately conclude that the energy spectrum at low-frequencies depends only
on the asymptotic states
\cite{Weinberg:1965nx,Adler:1975dj,Smarr:1977fy,Berti:2010ce,Kovacs:1977uw,Lemos:thesis}.
If furthermore the asymptotic states are an accurate description of the
collision at all times, as for instance if the colliding particles are
point-like, then one expects the ZFL to be an accurate description of the
problem.

For the head-on collision of two equal-mass objects each with mass $M\gamma/2$,
Lorentz factor $\gamma$ and velocity $v$ in the center-of-mass frame, one finds
the ZFL prediction~\cite{Smarr:1977fy,Lemos:thesis}
\be
\frac{d^2E}{d\omega \, d\Omega}=\frac{M^2\gamma^2v^4}{4\pi^2}\frac{\sin^4
\theta}{\left(1-v^2\cos^2\theta\right)^2}\,.
\ee
The particles collide head-on along the $z$-axis and we use standard spherical
coordinates. The spectrum is flat, i.e., $\omega$-independent, thus the total
radiated energy is formally divergent. The approach neglects the details of the
interaction and the internal structure of the colliding and final objects, and
the price to pay is the absence of a lengthscale, and therefore the appearance
of this divergence.  The divergence can be cured by introducing a
phenomenological cutoff in frequency. If the final object has typical size $R$,
we expect a cutoff $\omega_{\rm cutoff}\sim 1/R$ to be a reasonable
assumption. BHs have a more reasonable cutoff in frequency given by their
lowest QNMs; because QNMs are defined within a multipole decomposition, one
needs first to decompose the ZFL spectrum into multipoles (see Appendix B of
Ref.~\cite{Berti:2010ce} and Appendix B2 of Ref.~\cite{Lemos:thesis}). Finally,
one observes that the high-energy limit $v\to 1$ yields isotropic emission; when
translated to a multipole dependence, it means that the energy in each multipole
scales as $1/l^2$ in this limit.

The ZFL has been applied in a variety of contexts, including electromagnetism
where it can be used to compute the electromagnetic radiation given away in
$\beta$-decay~\cite{Cardoso:2003cn,jackson}; Wheeler used the ZFL to estimate
the emission of gravitational and electromagnetic radiation from impulsive
events~\cite{Wheeler:1962}; the original treatment by Smarr considered only
head-on collisions and computed only the
spectrum and total emitted energy. These
results have been generalized to include collisions with finite impact parameter
and to a computation of the radiated momentum as well
\cite{Lemos:thesis,Berti:2010ce}.  Finally, recent nonlinear simulations of
high-energy BH or star collisions yield impressive agreement with ZFL
predictions~\cite{Sperhake:2008ga,Berti:2010ce,East:2012mb,Brito:2012gj}.

\subsubsection{State-of-the-art} 

\begin{description}

\item[$\bullet$]{Astrophysical systems.}
  The zero-frequency limit for head-on
  collisions of particles was used by Smarr~\cite{Smarr:1977fy} to understand
  gravitational radiation from BH collisions and in Ref.~\cite{Adler:1975dj} to understand radiation from supernovae-like
  phenomena. It was later generalized to the nontrivial finite impact parameter case~\cite{Lemos:thesis}, and compared extensively with fully nonlinear numerical simulations~\cite{Berti:2010ce}. Ref.~\cite{Cardoso:2003cn} reports on collisions of an
  electromagnetic charge with a non-rotating BH in a spacetime perturbation
  approach and compares the results with a ZFL calculation.

\item[$\bullet$]{Beyond four-dimensional, electrovacuum GR.}
  Recent work has started applying the ZFL to other spacetimes and theories. Brito~\cite{Brito:2012gj} used the ZFL to understand head-on
  collisions of \emph{scalar} charges with four-dimensional BHs. The ZFL
  has been extended to higher dimensions in Refs.~\cite{Cardoso:2002pa,Lemos:thesis} and
  recently to specific AdS soliton spacetimes in Ref.~\cite{Cardoso:2013vpa}.

\end{description}

\subsection{Shock wave collisions}
\label{sec:shock_waves} 
An alternative
technique to model the dynamics of collisons of two
particles (or two BHs) at high energies describes the particles
as gravitational shock waves. This method yields a bound on the emitted
gravitational radiation using an exact solution, and provides an estimate of the
radiation using a perturbative method. In the following we shall review both.

In $D=4$ vacuum GR, a point-like particle is described by the Schwarzschild
metric of mass $M$. The gravitational field of a particle moving with velocity
$v$ is then obtained by boosting the Schwarzschild metric. Of particular
interest is the limiting case where the velocity approaches the speed of
light $v\rightarrow c$. Taking simultaneously the limit $M\rightarrow 0$ so that
the zeroth component of the $4$-momentum, $E$, is held fixed,
$E=M/\sqrt{1-v^2/c^2}={\rm constant}$, one observes an infinite Lorentz
contraction of the curvature in the spatial direction of the motion. In this
limit, the geometry becomes that of an \emph{impulsive or shock} gravitational
$pp$-wave, i.e., a plane-fronted gravitational wave with parallel rays, sourced by a null particle. This is the Aichelburg-Sexl geometry
\cite{Aichelburg:1970dh} for which the curvature has support only on a null
plane. In Brinkmann coordinates, the line element is:
\be
ds^2 = -du \, dv + \kappa\Phi(\rho)\delta(u) \, du^2 + d\rho^2 + \rho^2 \, d\phi^2  \ , \qquad 
-\Delta \left[\kappa\Phi(\rho)\right]=4\pi \kappa \delta (\rho) \ .
\label{shockwave}
\ee 
Here the shock wave is moving in the positive $z$-direction, where
$(u=t-z,v=t+z)$. This geometry solves the Einstein equations with energy
momentum tensor $T_{uu}=E\delta(u)\delta(\rho)$ -- corresponding to a null
particle of energy $E=\kappa/4G$, traveling along $u=0=\rho$ -- provided the
equation on the right-hand side of \eqref{shockwave} is satisfied, where the
Laplacian is in the flat 2-dimensional transverse space. Such a solution
is given in closed analytic form by $\Phi(\rho)=-2\ln(\rho)$.

The usefulness of shock waves in modelling collisions of particles or BHs at very
high energies relies on the following fact. Since the geometry of a single shock
wave is flat outside a null plane, one can superimpose two shock wave solutions
traveling in opposite directions and still obtain an \emph{exact} solution of
the Einstein equations, valid up to the moment when the two shock waves
collide. The explicit metric is obtained by superimposing two copies of
\eqref{shockwave}, one with support at $u=0$ and another one with support at
$v=0$. But it is more convenient to write down the geometry in coordinates for
which test particle trajectories vary continuously as they cross the
shock. These are called Rosen coordinates, $(\bar{u},\bar{v},\bar{\rho},\phi)$; their relation with Brinkmann coordinates can be found in~\cite{Herdeiro:2011ck} and the line element for the superposition becomes
\begin{eqnarray}
ds^2 = -d\tu \, d\tv + &&\left[\Big(1+\dfrac{\kappa \tu
      \theta(\tu)}{2}
    \Phi''\Big)^2+ \Big(1+\dfrac{\kappa \tv \theta(\tv)}{2}\Phi''\Big)^2-1\right]d\trho^{2} \\
  + \trho^2 &&\left[\Big(1+ \frac{\kappa \tu \, \theta(\tu)}{2 \bar
      \rho}\Phi'\Big)^2 + \Big(1+ \frac{\kappa \tv \, \theta(\tv)}{2 \bar
      \rho}\Phi'\Big)^2-1\right] d\phi^2 \ .\label{collision}
\end{eqnarray}
This metric is a valid description of the spacetime with the two shock waves
except in the future light-cone of the collision, which occurs at
$\bar{u}=0=\bar{v}$. Remarkably, and despite not knowing anything about the
future development of the collision, an AH can be found for this
geometry within its region of validity, as first pointed out by Penrose. Its
existence indicates that a BH forms and moreover its area provides a lower bound
for the mass of the BH~\cite{waldbook}. This AH is the union of
two surfaces,
\[ \{ \mathcal{S}_1, \ {\rm on} \ \bar{u}=0 \ {\rm and} \
\bar{v}=-\psi_1(\bar{\rho})\le 0\} \ , \qquad {\rm and} \qquad \{ \mathcal{S}_2,
\ {\rm on} \ \bar{v}=0 \ {\rm and} \ \bar{u}=-\psi_2(\bar{\rho})\le 0\} \ ,
\]
for some functions $\psi_1,\psi_2$ to be determined. The relevant null normals
to $\mathcal{S}_1$ and $\mathcal{S}_2$ are, respectively,
\be
l_1=\partial_{\bar{u}}-\frac{1}{2}\psi'_1g^{\bar{\rho}\bar{\rho}}\partial_{\bar{\rho}}+
\frac{1}{4}\left(\psi_1'\right)^2g^{\bar{\rho}\bar{\rho}}\partial_{\bar{v}} \ , \qquad l_2
=\partial_{\bar{v}}-\frac{1}{2}\psi'_2g^{\bar{\rho}\bar{\rho}}\partial_{\bar{\rho}}+
\frac{1}{4}\left(\psi_2'\right)^2g^{\bar{\rho}\bar{\rho}}\partial_{\bar{u}} \ .
\ee
One must then guarantee that these normals have zero expansion and are
continuous at the intersection $\bar{u}=0=\bar{v}$. This yields the solution
$\psi_1(\bar{\rho})=\kappa\Phi(\bar{\rho}/\kappa)=\psi_2(\bar{\rho})$. In
particular, at the intersection, the AH has a polar radius
$\bar{\rho}=\kappa$. The area of the AH is straightforwardly
computed to be
$2\pi^2\kappa^2$, and provides a lower bound on the area of a section of the
event horizon, and hence a lower bound on the mass of the BH:
$M/\kappa>1/\sqrt{8}$. By energy conservation, we then obtain an upper bound on
the inelasticity $\epsilon$, i.e the fraction of the initial centre of
mass energy which can be emitted in gravitational radiation: 
\be
\epsilon_{\rm AH}\le 1-\frac{1}{\sqrt{2}}\simeq 0.29 \ .
\label{ahbound}
\ee

Instead of providing a bound on the inelasticity, a more ambitious program is to
determine the exact inelasticity by solving the Einstein equations in the future
of the collision. Whereas an analytic exact solution seems out of reach, a
numerical solution of the fully non-linear field equations might be achievable,
but none has been reported. The approach that has produced the
most interesting results, so far, is to solve the Einstein equations
perturbatively in the future of the collision.

To justify the use of a perturbative technique and introduce a perturbation
expansion parameter, D'Eath and Payne
\cite{DEath:1992hb,DEath:1992hd,DEath:1992qu} made the following argument. In
a boosted frame, say in the negative $z$ direction, one of the shock waves will
become blueshifted whereas the other will become redshifted. These are,
respectively, the waves with support on $u=0$ and $v=0$. The geometry is still
given by \eqref{collision}, but with the energy parameter $\kappa$ multiplying
$\bar{u}$ terms ($\bar{v}$ terms) replaced by a new energy parameter $\nu$
(parameter $\lambda$).
For a large boost, $\lambda/\nu\ll 1$, or in other words, in the
boosted frame there are a strong shock (at $u=0$) and a weak
shock (at $v=0$). The weak shock is regarded as a perturbation of the
spacetime of the strong shock, and $\lambda/\nu$ provides the expansion
parameter to study this perturbation. Moreover, to set up initial conditions for
the post-collision perturbative expansion, one recasts the exact solution on the
immediate future of the strong shock, $u=0^+$, in a perturbative form,
even though it is an \emph{exact} solution. It so happens that expressing the exact solution
in such perturbative fashion only has terms up to second order:
\be
g_{\mu\nu}|_{u=0^+}=\nu^2\left[\eta_{\mu\nu}+\frac{\lambda}{\nu}h^{(1)}_{\mu\nu}
+\left(\frac{\lambda}{\nu}\right)^2h^{(2)}_{\mu\nu}\right] \ .
\label{bcsw}
\ee 
This perturbative expansion is performed in dimensionless
coordinates of Brinkmann type, as in Eq.\eqref{shockwave},
since the latter are more intuitive than Rosen
coordinates.
The geometry to the future of the strong shock, on the other hand, will be of the form
\be
g_{\mu\nu}|_{u>0}=\nu^2\left[\eta_{\mu\nu}+\sum_{i=1}^\infty\left(\frac{\lambda}{\nu}\right)^ih^{(i)}_{\mu\nu}\right] \ ,
\ee
where each of the $h_{\mu\nu}^{(i)}$ will be obtained by solving the Einstein
equations to the necessary order. For instance, to obtain $h^{(1)}_{\mu\nu}$ one
solves the linearized Einstein equations. In the de Donder gauge these yield a
set of decoupled wave equations of the form $\Box \bar{h}^{(1)}_{\mu\nu}=0$,
where the $\bar{h}_{\mu\nu}^{(1)}$ is the trace reversed metric
perturbation. The wave equation must then be subjected to the boundary
conditions \eqref{bcsw}. At higher orders, the problem can also be reduced to
solving wave equations for $h_{\mu\nu}^{(i)}$, but now with sources provided by
the perturbations of lower order~\cite{Coelho:2012sy}.

After obtaining the metric perturbations to a given order, one must still
compute the emitted gravitational radiation, in order to obtain the
inelasticity. In the original work
\cite{DEath:1990de,DEath:1992hb,DEath:1992hd,DEath:1992qu}, the metric perturbations were
computed to second order and the gravitational radiation was extracted using
Bondi's formalism and the Bondi mass loss formula. The first-order results can
equivalently be obtained using the Landau--Lifshitz pseudo-tensor for GW
extraction~\cite{Herdeiro:2011ck}. The results in first and second order are,
respectively: 
\be \epsilon^{(1)}=0.25 \ , \qquad \epsilon^{(2)}=0.164 \ .\label{eq:shock}  
\ee
Let us close this subsection with three remarks on these results. Firstly, the results \eqref{eq:shock}
are below the AH bound \eqref{ahbound}, as they should. Secondly,
and as we shall see in Section~\ref{sec:HEcollisions}, the second-order result is in excellent
agreement with results from NR simulations. Finally, as we
comment in the next subsection, the generalisation to higher dimensions of the
first-order result reveals a remarkably simple pattern.

\subsubsection{State-of-the-art}
\label{sec:StateOfArt}

The technique of superimposing two Aichelburg--Sexl shock
waves~\cite{Aichelburg:1970dh} was first used by Penrose in
unpublished work but quoted, for instance, in
Ref.~\cite{DEath:1992hb}. Penrose showed the existence of an AH for
the case of a head-on collision, thus suggesting BH
formation. Computing the area of the AH yields an upper bound on the
fraction of the overall energy radiated away in GWs, i.e., the
inelasticity. In the early 2000s, the method of superimposing shock
waves and finding an AH was generalized to $D\ge 5$ and non-zero
impact parameter in Refs.~\cite{Eardley:2002re,Yoshino:2002tx} and
refined in Ref.~\cite{Yoshino:2005hi} providing, in addition to a measure of the
inelasticity, an estimate of the cross section for BH formation in a high-energy
particle collision. A potential improvement to the AH based
estimates was carried out in a series of papers by D'Eath and Payne
\cite{DEath:1990de,DEath:1992hb,DEath:1992hd,DEath:1992qu}. They computed the metric to
the future of the collision perturbatively to second order
in the head-on case. This method was generalized to $D\ge 5$ in first-order
perturbation theory~\cite{Herdeiro:2011ck, Coelho:2012sya} yielding a very
simple result: $\epsilon^{(1)}=1/2-1/D$. A formalism for higher order and the
caveats of the method in the presence of electric charge were exhibited in
\cite{Coelho:2012sy}. AH formation in shock wave collisions with
generalized profiles and asymptotics has been studied in
\cite{Albacete:2009ji,Taliotis:2012sx,AlvarezGaume:2008fx,DuenasVidal:2010vi}.

\newpage
\section{Numerical Relativity}
\label{sec:NR}

Generating time-dependent solutions to the Einstein equations
using numerical methods involves an extended list of ingredients
which can be loosely summarized as follows.
\begin{itemize}
  \item Cast the field equations as an IBVP.
  \item Choose a specific formulation that admits a
        \emph{well-posed} IBVP,
        i.e., there exist suitable choices for the following ingredients
        that ensure well posedness.
  \item Choose numerically suitable coordinate or \emph{gauge} conditions.
  \item Discretize the resulting set of equations.
  \item Handle singularities such that they do not result in the generation of
    \emph{non-assigned numbers} which rapidly swamp the computational domain.
  \item Construct initial data that solve the Einstein
        constraint equations and represent a realistic snapshot of
        the physical system under consideration.
  \item Specify suitable outer boundary conditions.
  \item Fix technical aspects: mesh refinement and/or multi-domains
        as well as use of multiple computer processors through parallelization.
  \item Apply diagnostic tools that measure GWs,
        BH horizons, momenta and masses, and other fields.
\end{itemize}
In this section we will discuss
state-of-the-art choices for these ingredients.

\subsection{Formulations of the Einstein equations}
\label{sec:NRFormulations}

\subsubsection{The ADM equations}
\label{sec:ADM}
The Einstein equations in $D$ dimensions describing a spacetime with
cosmological constant $\Lambda$ and energy-matter content $T_{\alpha \beta}$
are given by
\begin{equation}
  R_{\alpha \beta} - \frac{1}{2}R g_{\alpha \beta}
        + \Lambda g_{\alpha \beta}
      = 8\pi T_{\alpha \beta}
  ~~~\Leftrightarrow~~~
  R_{\alpha \beta} = 8\pi \left( T_{\alpha \beta} - \frac{1}{D-2}
        g_{\alpha \beta} T \right) + \frac{2}{D-2} \Lambda
        g_{\alpha \beta}.
  \label{eq:EinsteinEqs_D}
\end{equation}
Elegant though this tensorial form of the equations is from a mathematical
point of view, it is not immediately suitable for a numerical implementation.
For one thing, the character of the equations as a
\emph{hyperbolic}, \emph{parabolic} or \emph{elliptic} system is not evident.
In other words, are we dealing with an \emph{initial-value} or a
\emph{boundary-value problem}? In fact, the Einstein equations
are of mixed character in this regard and represent an
IBVP. Well posedness of the IBVP then requires a suitable formulation
of the evolution equations, boundary conditions and initial data.
We shall discuss this particular aspect in more detail further below,
but first consider the general structure of the equations. The
multitude of possible ways of writing the Einstein equations are
commonly referred to as \emph{formulations} of the equations and
a good starting point for their discussion is the canonical ``3+1''
or ``$(D-1)+1$'' split originally developed by
Arnowitt, Deser \& Misner~\cite{Arnowitt:1962hi} and later reformulated
by York~\cite{York1979,York1983}. \\[5pt]

The tensorial form of the Einstein equations (\ref{eq:EinsteinEqs_D})
fully reflects the unified viewpoint of space and time; it is only
through the Lorentzian signature $(-,\,+,\,\ldots,\,+)$ of the metric that
the timelike character of one of the coordinates manifests
itself.\epubtkFootnote{Strictly speaking, the signature represents the signs
of the eigenvalues of the metric: $g_{\alpha \beta}$ has
$1$ negative and $D-1$ positive eigenvalues even when
the timelike coordinate is replaced in terms of one or two null coordinates.}
It turns out crucial for understanding the character of Einstein's
equations to make the distinction between spacelike and timelike
coordinates more explicit.

Let us consider for this purpose a spacetime described by
a manifold $\mathcal{M}$ equipped with a metric $g_{\alpha \beta}$
of Lorentzian signature. We shall further assume
that there exists a \emph{foliation} of the spacetime in the sense that
there exists a scalar function $t:\mathcal{M} \rightarrow \mathbb{R}$
with the following properties. (i) The 1-form $\boldsymbol{\mathsf{d}}t$
associated with the function $t$ is timelike everywhere; (ii) The
hypersurfaces $\Sigma_t$ defined by $t=\mathrm{const}$ are
non-intersecting and
$\displaystyle \cup_{t \in \mathbb{R}} \Sigma_t = \mathcal{M}$.
Points inside each hypersurface $\Sigma_t$ are labelled by
spatial coordinates $x^{\I},~\,I=1,\ldots,D-1$, and we refer to the
coordinate system $(t,~x^{\I})$ as \emph{adapted} to the spacetime
split.

Next, we define the \emph{lapse function} $\alpha$ and \emph{shift
vector} $\boldsymbol{\beta}$ through
\begin{equation}
  \alpha \equiv \frac{1}{||\boldsymbol{\mathsf{d}}t||},~~~~~~~~~~
  \beta^{\mu} \equiv (\partial_t)^{\mu}-\alpha n^{\mu},
\end{equation}
where $\boldsymbol{n}\equiv -\alpha \boldsymbol{\mathsf{d}}t$
is the timelike unit normal field.
The geometrical interpretation of these quantities in terms of the
timelike unit normal field $n^{\alpha}$ and the coordinate basis
vector $\partial_t$ is illustrated in Figure~\ref{fig:foliation}.
Using the relation $\langle \boldsymbol{\mathsf{d}}t,{\boldsymbol{\partial}}_t \rangle = 1$
\epubtkImage{}{%
  \begin{figure}
    \centerline{\includegraphics[clip=true,height=250pt]{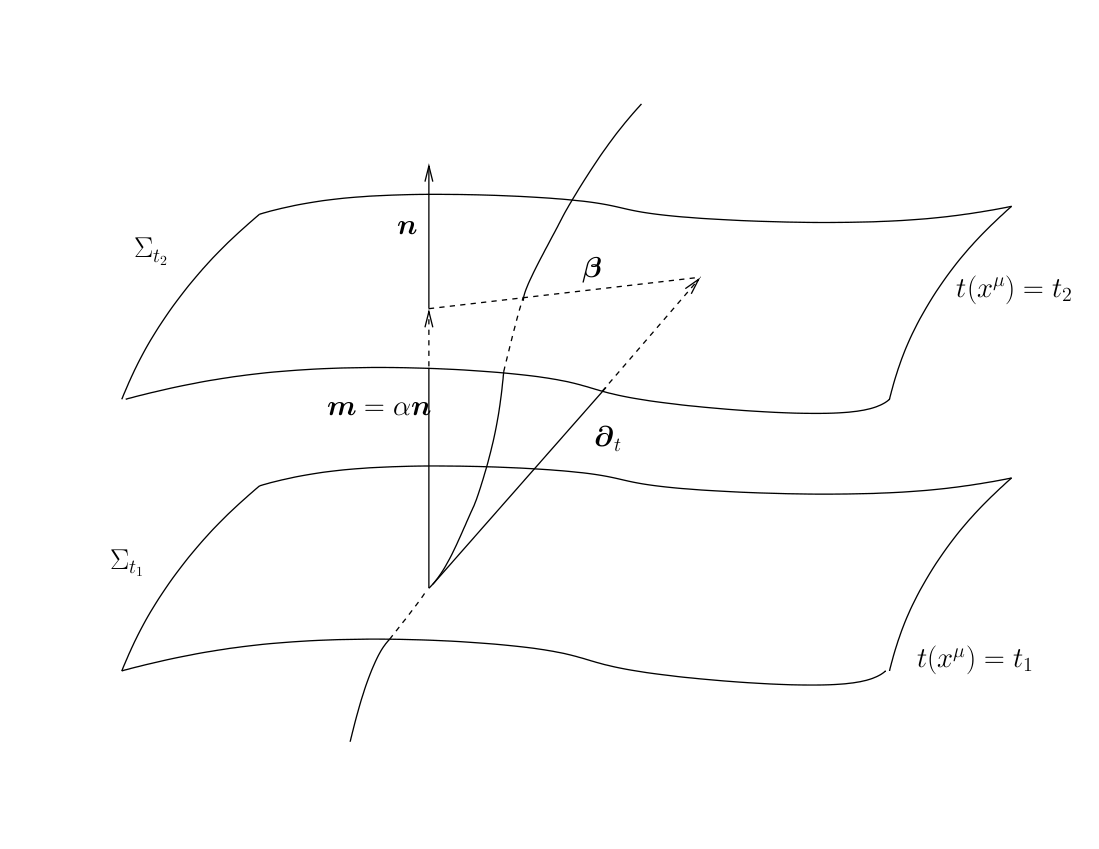}}
    \caption{Illustration of two hypersurfaces of a foliation
    $\Sigma_t$. Lapse $\alpha$ and shift $\beta^{\mu}$ are defined by
    the relation of the timelike unit normal field $n^{\mu}$ and the
    basis vector ${\boldsymbol{\partial}}_t$ associated with the
    coordinate $t$. Note that $\langle
    \boldsymbol{\mathsf{d}}t,\,\alpha \boldsymbol{n}\rangle = 1$
    and, hence, the shift vector $\boldsymbol{\beta}$ is tangent to
    $\Sigma_t$.}
  \label{fig:foliation}
\end{figure}}
and the definition of $\alpha$ and $\boldsymbol{\beta}$, one directly finds
$\langle \boldsymbol{\mathsf{d}} t, \boldsymbol{\beta} \rangle = 0$,
so that the shift $\boldsymbol{\beta}$ is
tangent to the hypersurfaces $\Sigma_t$. It measures the deviation of
the coordinate vector ${\boldsymbol{\partial}}_t$ from the normal direction $\boldsymbol{n}$.
The lapse function relates the proper time
measured by an observer moving with four velocity $n^{\alpha}$
to the coordinate time $t$: $\Delta \tau = \alpha \Delta t$.

A key ingredient for the spacetime split of the equations is the projection of
tensors onto time and space directions. For this purpose, the space
projection operator is defined as $\bot^{\alpha}{}_{\mu} \equiv
\delta^{\alpha}{}_{\mu} + n^{\alpha}n_{\mu}$. For a generic tensor
$T^{\alpha_1 \alpha_2\ldots}{}_{\beta_1 \beta_2\ldots}$, its spatial
projection is given by projecting each index speparately
\begin{equation}
  (\bot T)^{\alpha_1 \alpha_2\ldots}{}_{\beta_1 \beta_2\ldots}
      \equiv \bot^{\alpha_1}{}_{\mu_1}
             \bot^{\alpha_2}{}_{\mu_2}
             \ldots
             \bot^{\nu_1}{}_{\beta_1}
             \bot^{\nu_2}{}_{\beta_2}
             \ldots
             T^{\mu_1 \mu_2 \ldots}{}_{\nu_1 \nu_2 \ldots}\,.
\end{equation}
A tensor $\boldsymbol{S}$ is called \emph{tangent} to $\Sigma_t$ if it is invariant
under projection, i.~e.~$\boldsymbol{\bot} \boldsymbol{S}= \boldsymbol{S}$. In adapted coordinates, we can ignore
the time components of such spatial tensors and it is common practice to
denote their components with Latin indices $I,\,J,\,\ldots = 1,\,...,\,(D-1)$.
We similarly obtain time projections of a tensor by contracting its
indices with $n^{\alpha}$. Mixed projections are obtained by contracting
any combination of tensor indices with $n^{\alpha}$ and projecting the
remaining ones with $\bot^{\alpha}{}_{\mu}$.
A particularly important tensor is obtained from the spatial projection
of the spacetime metric
\begin{equation}
  \gamma_{\alpha \beta} \equiv
        (\bot g)_{\alpha \beta} = \bot^{\mu}{}_{\alpha} \bot^{\nu}{}_{\beta}
        g_{\mu \nu} = (\delta^{\mu}{}_{\alpha} + n^{\mu}n_{\alpha})
        (\delta^{\nu}{}_{\beta} + n^{\nu}n_{\beta}) g_{\mu \nu}
        = g_{\alpha \beta} + n_{\alpha} n_{\beta}
        = \bot_{\alpha \beta}\,.
  \label{eq:ff1}
\end{equation}
$\gamma_{\alpha \beta}$ is known as the \emph{first fundamental form} or
\emph{spatial metric} and describes the intrinsic geometry of the
spatial hypersurfaces $\Sigma_t$. As we see from Eq.~(\ref{eq:ff1}),
it is identical to the projection operator. In the remainder,
we will use both the $\boldsymbol{\bot}$ and $\boldsymbol{\gamma}$ symbols
to denote this tensor depending on whether the emphasis is on the
projection or the hypersurface geometry.

With our definitions, it is straightforward to show that the
spacetime metric in adapted coordinates $(t,x^{\I})$ can be written
as $ds^2 = -\alpha^2 \, dt^2 +\gamma_{\I \J} (dx^{\I} + \beta^{\I} \, dt)
(dx^{\J} + \beta^{\J} \, dt)$ or, equivalently,
\begin{equation}
  g_{\alpha \beta} = \left(
        \begin{array}{c|c}
          -\alpha^2 + \beta_{\M} \beta^{\M} & \beta_{\J} \\
          \hline
          \beta_{\I} & \gamma_{\I \J}
        \end{array}
        \right)
        ~~~\Leftrightarrow~~~
  g^{\alpha \beta} = \left(
        \begin{array}{c|c}
          -\alpha^{-2} & \alpha^{-2} \beta^{\J} \\
          \hline
          \alpha^{-2} \beta^{\I} & \gamma^{\I \J}
          - \alpha^{-2}\beta^{\I} \beta^{\J}
        \end{array}
        \right)\,.
  \label{eq:3+1metric}
\end{equation}
It can be shown~\cite{Gourgoulhon:2007ue} that the spatial metric
$\gamma_{\I \J}$ defines a unique, torsion-free and metric-compatible
connection $\Upgamma^{\I}_{\J \K} = \frac{1}{2} \gamma^{\I \M}
(\partial_{\J} \gamma_{\K \M} + \partial_{\K} \gamma_{\M \J}
- \partial_{\M} \gamma_{\J \K})$
whose covariant derivative for an arbitrary spatial tensor is given by
\begin{equation}
  D_{\gamma} S^{\alpha_1 \alpha_2 \ldots}{}_{\beta_1 \beta_2 \ldots}
  = \bot^{\lambda}{}_{\gamma} \bot^{\alpha_1}{}_{\mu_1}
    \bot^{\alpha_2}{}_{\mu_2} \ldots
    \bot^{\nu_1}{}_{\beta_1} \bot^{\nu_2}{}_{\beta_2} \ldots
    \nabla_{\lambda} S^{\mu_1 \mu_2 \ldots}{}_{\nu_1 \nu_2 \ldots}\,.
\end{equation}

The final ingredient required for the spacetime split of the Einstein
equations is the \emph{extrinsic curvature} or \emph{second fundamental
form} defined as
\begin{equation}
  K_{\alpha \beta} \equiv -\bot \nabla_{\beta} n_{\alpha}\,.
  \label{eq:ff2}
\end{equation}
The sign convention employed here is common in NR but the ``$-$'' is sometimes
omitted in other studies of GR. The definition (\ref{eq:ff2}) provides an
intuitive geometric interpretation of the extrinsic curvature as the change in
direction of the timelike unit normal field $\boldsymbol{n}$ as we move across
the hypersurface $\Sigma_t$. As indicated by its name, the extrinsic curvature
thus describes the embedding of $\Sigma_t$ inside the higher-dimensional
spacetime manifold. The projection
$\bot \nabla_{\beta} n_{\alpha}$ is symmetric under exchange of its indices in
contrast to its non-projected counterpart $\nabla_{\beta} n_{\alpha}$. For the
formulation of the Einstein equations in the spacetime split, it is helpful to
introduce the vector field $m^{\mu} \equiv \alpha n^{\mu} = (\partial_t)^{\mu} -
\beta^{\mu}$.  A straightforward calculation shows that the extrinsic curvature
can be expressed in terms of the Lie derivative of the spatial metric along
either $\boldsymbol{n}$ or $\boldsymbol{m}$ according to
\begin{equation}
  K_{\alpha \beta} = -\frac{1}{2} \mathcal{L}_{\boldsymbol{n}}
        \gamma_{\alpha \beta}
        = -\frac{1}{2\alpha} \mathcal{L}_{\boldsymbol{m}}
        \gamma_{\alpha \beta}\,.
\end{equation}

We have now assembled all tools to calculate the spacetime
projections of the Riemann tensor. In the following order,
these are known as the Gauss, the contracted Gauss, the scalar
Gauss, the Codazzi, the contracted Codazzi equation,
as well as the final projection
of the Riemann tensor and its contractions:
\begin{eqnarray}
  \bot^{\mu}{}_{\alpha} \bot^{\nu}{}_{\beta} \bot^{\gamma}{}_{\rho}
  \bot^{\sigma}{}_{\delta} R^{\rho}{}_{\sigma \mu \nu}
  &=& \mathcal{R}^{\gamma}{}_{\delta \alpha \beta}
      + K^{\gamma}{}_{\alpha} K_{\delta \beta}
      - K^{\gamma}{}_{\beta} K_{\delta \alpha}, \nonumber \\
  \bot^{\mu}{}_{\alpha} \bot^{\nu}{}_{\beta} R_{\mu \nu}
  + \bot_{\mu \alpha} \bot^{\nu}{}_{\beta} n^{\rho} n^{\sigma}
  R^{\mu}{}_{\rho \nu \sigma}
  &=& \mathcal{R}_{\alpha \beta} + KK_{\alpha \beta}
      - K^{\mu}{}_{\beta} K_{\alpha \mu},  \nonumber\\
  R+ 2R_{\mu \nu} n^{\mu} n^{\nu}
  &=& \mathcal{R} +K^2 - K^{\mu \nu} K_{\mu \nu},  \nonumber\\
  \bot^{\gamma}{}_{\rho} n^{\sigma} \bot^{\mu}{}_{\alpha}
  \bot^{\nu}{}_{\beta} R^{\rho}{}_{\sigma \mu \nu}
  &=& D_{\beta} K^{\gamma}{}_{\alpha} - D_{\alpha} K^{\gamma}{}_{\beta},
      \nonumber \\
  n^{\sigma}\bot^{\nu}{}_{\beta} R_{\sigma \nu}
  &=& D_{\beta} K - D_{\mu} K^{\mu}{}_{\beta}, \nonumber \\
  \bot_{\alpha \mu} \bot^{\nu}{}_{\beta} n^{\sigma} n^{\rho}
  R^{\mu}{}_{\rho \nu \sigma}
  &=& \frac{1}{\alpha} \mathcal{L}_{\boldsymbol{m}} K_{\alpha \beta}
      + K_{\alpha \mu} K^{\mu}{}_{\beta}
      + \frac{1}{\alpha} D_{\alpha} D_{\beta} \alpha, \nonumber \\
  \bot^{\mu}{}_{\alpha} \bot^{\nu}{}_{\beta} R_{\mu \nu}
  &=& -\frac{1}{\alpha} \mathcal{L}_{\boldsymbol{m}} K_{\alpha \beta}
      - 2K_{\alpha \mu} K^{\mu}{}_{\beta}
      - \frac{1}{\alpha}D_{\alpha} D_{\beta} \alpha
      + \mathcal{R}_{\alpha \beta} + KK_{\alpha \beta}, \nonumber \\
  R &=& -\frac{2}{\alpha} \mathcal{L}_{\boldsymbol{m}} K
      - \frac{2}{\alpha} \gamma^{\mu \nu} D_{\mu} D_{\nu} \alpha
      + \mathcal{R} + K^2 + K^{\mu \nu} K_{\mu \nu}\,.
  \label{eq:projections_Riemann}
\end{eqnarray}
Here,
$\mathcal{R}$ denotes the Riemann tensor and its contractions as
defined in standard fashion from the spatial metric $\gamma_{\I \J}$.
For simplicity, we have kept all spacetime indices here even for
spatial tensors. As mentioned above, the time components can and will
be discarded eventually.

By using Eq.~(\ref{eq:projections_Riemann}), we can express the
space and time projections of the Einstein equations
(\ref{eq:EinsteinEqs_D}) exclusively in terms of the first and
second fundamental forms and their derivatives. It turns out
helpful for this purpose to introduce the corresponding
projections of the energy-momentum tensor which are given by
\begin{eqnarray}
  &\rho = T_{\mu \nu}n^{\mu} n^{\nu},~~~~~&j_{\alpha} = -\bot^{\nu}{}_{\alpha}
          T_{\mu \nu}n^{\mu}, \\
  &S_{\alpha \beta} = \bot^{\mu}{}_{\alpha} \bot^{\nu}{}_{\beta} T_{\mu \nu},
  &S = \gamma^{\mu \nu} S_{\mu \nu}\,.
\end{eqnarray}
Then, the energy-momentum tensor is reconstructed according to
$T_{\alpha \beta} = S_{\alpha \beta} + n_{\alpha} j_{\beta}
+ n_{\beta} j_{\alpha} + \rho n_{\alpha} n_{\beta}$.
Using the explicit expressions for the Lie derivatives
\begin{eqnarray}
  \mathcal{L}_{\boldsymbol{m}} K_{\I \J} =
        \mathcal{L}_{{\boldsymbol{\partial}}_t - {\boldsymbol{\beta}}} K_{\I \J}
        = \partial_t K_{\I \J} -\beta^{\M} \partial_{\M} K_{\I \J}
          -K_{\M \J} \partial_{\I} \beta^{\M}
          - K_{\I \M} \partial_{\J} \beta^{\M}\,,\\
  \mathcal{L}_{\boldsymbol{m}} \gamma_{\I \J}
        = \mathcal{L}_{{\boldsymbol{\partial}}_t - {\boldsymbol{\beta}}} \gamma_{\I \J}
        = \partial_t \gamma_{\I \J} - \beta^{\M} \partial_{\M} \gamma_{\I \J}
          - \gamma_{\M \J} \partial_{\I} \beta^{\M}
          - \gamma_{\I \M} \partial_{\J} \beta^{\M}\,,
\end{eqnarray}
we obtain the spacetime split of the Einstein equations
\begin{eqnarray}
  \partial_t \gamma_{\I \J} &=& \beta^M \partial_M \gamma_{\I \J}
      + \gamma_{\M \J} \partial_{\I} \beta^{\M} + \gamma_{\I \M}
      \partial_{\J} \beta^{\M} - 2\alpha K_{\I \J}\,,
  \label{eq:dgammaADM} \\
  \partial_t K_{\I \J} &=& \beta^{\M} \partial_{\M} K_{\I \J}
        + K_{\M \J} \partial_{\I} \beta^{\M}
        + K_{\I \M} \partial_{\J} \beta^{\M}
        - D_{\I} D_{\J} \alpha
        + \alpha \left( \mathcal{R}_{\I \J} +KK_{\I \J}
        -2K_{\I \M}K^{\M}{}_{\J} \right)
      \nonumber
  \label{eq:dKADM} \\
     && + 8\pi \alpha \left( \frac{S-\rho}{D-2} \gamma_{\I \J} - S_{\I \J}
        \right)
        - \frac{2}{D-2} \alpha \Lambda \gamma_{\I \J}, \\
   0 &=& \mathcal{R} + K^2 - K^{\M \N}K_{\M \N} - 2\Lambda - 16\pi \rho,
   \label{eq:HamADM} \\
   0 &=& D_{\I} K -D_{\M} K^{\M}{}_{\I} +8\pi j_{\I}\,.
   \label{eq:momADM}
\end{eqnarray}
By virtue of the Bianchi identities, the constraints
(\ref{eq:HamADM}) and (\ref{eq:momADM}) are preserved under the evolution
equations.
Furthermore, we can see that $D(D-1)/2$ second-order-in-time
evolution equations for the $\gamma_{\I\J}$ are
written as a first-order-in-time system through introduction of the
extrinsic curvature. Additionally, we have obtained
$D$ constraint equations, the
Hamiltonian and momentum constraints, which relate
data within a hypersurface $\Sigma_t$. We note that the Einstein equations
do not determine the lapse $\alpha$ and shift $\beta^I$.
For the case of $D=4$, these equations are often referred to as the
ADM equations, although we note that Arnowitt, Deser \& Misner used
the canonical momentum in place of the extrinsic curvature
in their original work~\cite{Arnowitt:1962hi}. Counting the degrees
of freedom, we start with $D(D+1)/2$
components of the spacetime metric.
The Hamiltonian and momentum constraints determine $D$ of these while
$D$ gauge functions represent the gauge freedom, leaving $D(D-3)/2$ physical
degrees of freedom as expected.

\subsubsection{Well posedness}
The suitability of a given system of differential equations for a numerical
time evolution critically depends on a continuous dependency of
the solution on the initial data. This aspect is referred to as
\emph{well posedness} of the IBVP and is discussed in
great detail in Living Reviews articles and other works~\cite{Reula:1998ty,Sarbach:2012pr,Gustafsson1995,Hilditch:2013sba}. Here, we merely list the basic concepts and
refer the interested reader to these articles.

Consider for simplicity an initial-value problem in one space and
one time dimension for a single variable $u(t,x)$
on an unbounded domain. Well posedness
requires a norm $||\cdot ||$,
i.e., a map from the space of functions $f(x)$ to the real numbers
$\mathbb{R}$, and a function $F(t)$ independent of the initial
data such that
\begin{equation}
  || \delta u(t,\cdot) || \le F(t) || \delta u(0,\cdot)||\,,
\end{equation}
where $\delta u$ denotes a linear perturbation relative to a solution
$u_0(t,x)$~\cite{Gundlach:2006tw}.
We note that $F(t)$ may be a rapidly growing function,
for example an exponential, so that well posedness represents
a necessary but not sufficient criterion for suitability of a numerical
scheme.

Well posedness of formulations of the Einstein equations is typically
studied in terms of the \emph{hyperbolicity} properties of the system
in question.
Hyperbolicity of a system of PDEs is often defined in terms of the
\emph{principal part}, that is, the terms of the PDE which contain the
highest-order derivatives. We consider for simplicity a quasilinear
first-order system for a set of variables $\boldsymbol{u}(t,x)$
\begin{equation}
  \partial_t \boldsymbol{u} = P(t,x,\boldsymbol{u},\partial_x) \boldsymbol{u}\,.
\end{equation}
The system is called strongly hyperbolic if $P$ is a smooth differential
operator and its associated principal symbol is symmetrizeable
\cite{Nagy:2004td}. For the special case of constant coefficient systems
this definition simplifies to the requirement that the principal symbol
has only imaginary eigenvalues and a complete set of linearly
independent eigenvectors. 
If linear independence of the eigenvectors is not satisfied,
the system is called weakly hyperbolic.
For more complex systems
of equations, strong and weak hyperbolicity can be defined in a more
general fashion~\cite{Reula:1998ty,Nagy:2004td,Reula:2004xd,Sarbach:2012pr}.

In our context, it is of particular importance that strong hyperbolicity
is a necessary condition for a well posed IBVP
\cite{Taylor1981,Taylor1991}. The ADM equations
(\ref{eq:dgammaADM})\,--\,(\ref{eq:dKADM}), in contrast, have been shown to
be weakly but not strongly hyperbolic for fixed gauge~\cite{Nagy:2004td};
likewise, a first-order reduction of the ADM equations has been
shown to be weakly hyperbolic~\cite{Kidder:2001tz}. These results
strongly indicate that the ADM formulation is not suitable for numerical
evolutions of generic spacetimes.

A modification of the ADM equations which has been used with
great success in NR is the BSSN system
\cite{Baumgarte:1998te,Shibata:1995we} which is the subject of the next
section.
\subsubsection{The BSSN equations}
It is interesting to note that the BSSN formulation had been developed
in the 1990s before a comprehensive understanding
of the hyperbolicity properties of the Einstein equations had been
obtained; it was only about a decade after its first numerical application that
strong hyperbolicity of the BSSN system~\cite{Gundlach:2006tw} was demonstrated.
We see here an example of how powerful a largely empirical
approach can be in the derivation of successful numerical methods.
And yet, our understanding of the mathematical properties is of more
than academic interest as we shall see in Section~\ref{sec:BeyondBSSN}
below when we discuss recent investigations of potential improvements
of the BSSN system.

The modification of the ADM equations which results in the BSSN formulation
consists of a trace split of the extrinsic curvature, a conformal decomposition
of the spatial metric
and of the traceless part of the extrinsic curvature and the
introduction of the contracted Christoffel symbols as independent variables. For
generality we will again write the definitions of the variables and the
equations for the case of an arbitrary number $D$ of spacetime dimensions. We
define
\begin{eqnarray}
  &\chi = \gamma^{-1/(D-1)}\,,
  ~~~~~
  &K = \gamma^{\M \N} K_{\M \N}, \nonumber \\
  &\tilde{\gamma}_{\I \J} = \chi \gamma_{\I \J}
  ~~~
  &\Leftrightarrow
  ~~~
  \tilde{\gamma}^{\I \J} = \frac{1}{\chi} \gamma^{\I \J}, \nonumber \\
  &\tilde{A}_{\I \J} = \chi \left( K_{\I \J} - \frac{1}{D-1}
      \gamma_{\I \J} K \right)
  ~~~
  &\Leftrightarrow
  ~~~
  K_{\I \J} = \frac{1}{\chi} \left( \tilde{A}_{\I \J} + \frac{1}{D-1}
      \tilde{\gamma}_{\I \J} K \right)\,, \nonumber \\
  &\tilde{\Upgamma}^{\I} = \tilde{\gamma}^{\M \N} \tilde{\Upgamma}^{\I}_{\M \N}\,,&
  \label{eq:BSSNvars}
\end{eqnarray}
where $\gamma \equiv \det \gamma_{\I \J}$ and $\tilde{\Upgamma}^{\I}_{\M \N}$
is the Christoffel symbol defined in the usual manner in terms of
the conformal metric $\tilde{\gamma}_{\I \J}$.
Note that the definition (\ref{eq:BSSNvars}) implies two algebraic
and one differential constraints
\begin{equation}
  \tilde{\gamma} = 1,~~~~~\tilde{\gamma}^{\M \N} \tilde{A}_{\M \N} = 0,~~~~~
  \mathcal{G}^{\I} = \tilde{\Upgamma}^{\I} - \tilde{\gamma}^{\M \N}
        \tilde{\Upgamma}^{\I}_{\M \N}=0\,.
\end{equation}

Inserting the definition (\ref{eq:BSSNvars}) into the ADM equations
(\ref{eq:dgammaADM})\,--\,(\ref{eq:dKADM}) and using the Hamiltonian
and momentum constraints respectively in the evolution equations
for $K$ and $\tilde{\Upgamma}^I$ results in the BSSN evolution system
\begin{eqnarray}
  \partial_t \chi &=& \beta^{\M} \partial_{\M} \chi
        + \frac{2}{D-1} \chi (\alpha K - \partial_{\M} \beta^{\M})\,,
        \label{eq:BSSNdtphi} \\
  \partial_t \tilde{\gamma}_{\I \J} &=& \beta^{\M} \partial_{\M}
        \tilde{\gamma}_{\I \J}
        + 2\tilde{\gamma}_{\M (\I} \partial_{\J )} \beta^{\M}
        - \frac{2}{D-1}\tilde{\gamma}_{\I \J} \partial_{\M} \beta^{\M}
        - 2\alpha \tilde{A}_{\I \J}\,,\\
  \partial_t K &=& \beta^{\M} \partial_{\M} K
        - \chi \tilde{\gamma}^{\M \N} D_{\M} D_{\N} \alpha
        + \alpha \tilde{A}^{\M \N}\tilde{A}_{\M \N}
        + \frac{1}{D-1}\alpha K^2 \nonumber \\
     && + \frac{8\pi}{D-2} \alpha [S+(D-3)\rho]
        - \frac{2}{D-2}\alpha \Lambda\,, \\
  \partial_t \tilde{A}_{\I \J} &=& \beta^{\M} \partial_{\M} \tilde{A}_{\I \J}
        + 2\tilde{A}_{\M (\I} \partial_{\J )} \beta^{\M}
        - \frac{2}{D-1} \tilde{A}_{\I \J} \partial_{\M} \beta^{\M}
        + \alpha K\tilde{A}_{\I \J}
        - 2\alpha \tilde{A}_{\I \M} \tilde{A}^{\M}{}_{\J} \nonumber \\
     && + \chi \left(
          \alpha \mathcal{R}_{\I \J} - D_{\I} D_{\J} \alpha
          - 8\pi \alpha S_{\I \J} \right)^{\rm TF}\,,\\
  \partial_t \tilde{\Upgamma}^{\I} &=& \beta^{\M} \partial_{\M}
          \tilde{\Upgamma}^{\I}
        + \frac{2}{D-1} \tilde{\Upgamma}^{\I} \partial_{\M} \beta^{\M}
        - \tilde{\Upgamma}^{\M}\partial_{\M} \beta^{\I}
        + \tilde{\gamma}^{\M \N} \partial_{\M} \partial_{\N} \beta^{\I}
        + \frac{D-3}{D-1}\tilde{\gamma}^{\I \M} \partial_{\M}
          \partial_{\N} \beta^{\N}
          \nonumber \\
     && - \tilde{A}^{\I \M} \left[
          (D-1) \alpha \frac{\partial_{\M} \chi}{\chi}
          + 2\partial_{\M} \alpha \right]
        + 2\alpha \tilde{\Upgamma}^{\I}_{\M \N} \tilde{A}^{\M \N}
        - 2\frac{D-2}{D-1} \alpha \tilde{\gamma}^{\I \M} \partial_{\M} K
        - 16\pi \frac{\alpha}{{\chi}} j^{\I}.
        \label{eq:BSSNdGamma}
\end{eqnarray}
Here the superscript ``TF'' denotes the trace-free part and we further
use the following expressions which relate physical to conformal
variables:
\begin{eqnarray}
  \Upgamma^{\I}_{\J \K} &=& \tilde{\Upgamma}^{\I}_{\J \K}
        - \frac{1}{2\chi}\left( \delta^{\I}{}_{\K} \partial_{\J} \chi +
          \delta^{\I}{}_{\J} \partial_{\K} \chi
          - \tilde{\gamma}_{\J \K} \tilde{\gamma}^{\I \M} \partial_{\M} \chi
          \right)\,, \\
  \mathcal{R}_{\I \J} &=& \tilde{\mathcal{R}}_{\I \J}
        + \mathcal{R}^{\chi}_{\I \J}\,, \\
  \mathcal{R}^\chi_{\I \J} &=&
        \frac{\tilde{\gamma}_{\I \J}}{2\chi} \left[
        \tilde{\gamma}^{\M \N} \tilde{D}_\M \tilde{D}_\N \chi
        - \frac{D-1}{2\chi} \tilde{\gamma}^{\M \N}
          \partial_\M \chi \,\, \partial_\N \chi \right]
        + \frac{D-3}{2\chi} \left( \tilde{D}_\I \tilde{D}_\J \chi
          - \frac{1}{2\chi} \partial\I \chi\,\,\partial_\J \chi \right)\,,
        \\
  \tilde{\mathcal{R}}_{\I \J} &=&
        - \frac{1}{2} \tilde{\gamma}^{\M \N} \partial_{\N} \partial_{\N}
          \tilde{\gamma}_{\I \J}
        + \tilde{\gamma}_{\M(\I} \partial_{\J)} \tilde{\Upgamma}^{\M}
        + \tilde{\Upgamma}^{\M} \tilde{\Upgamma}_{(\I \J)\M}
        + \tilde{\gamma}^{\M \N} \left[
          2\tilde{\Upgamma}^{\K}_{\M(\I} \tilde{\Upgamma}_{\J)\K \N}
          + \tilde{\Upgamma}^{\K}_{\I \M} \tilde{\Upgamma}_{\K \J \N} \right], \\
  D_{\I} D_{\J} \alpha &=&
        \tilde{D}_{\I} \tilde{D}_{\J} \alpha
        + \frac{1}{\chi}\partial_{(\I} \chi\,\partial_{\J )} \alpha
        - \frac{1}{2\chi}\tilde{\gamma}_{\I \J} \tilde{\gamma}^{\M \N}
          \partial_{\M} \chi\,\partial_{\N} \alpha\,.
        \label{eq:BSSNDDalpha}
\end{eqnarray}
In practical applications, it turns out necessary for numerical stability
to enforce the algebraic constraint $\tilde{\gamma}^{\M \N} \tilde{A}_{\M \N}=0$
whereas enforcement of the unit determinant $\tilde{\gamma}=1$ appears
to be optional. A further subtlety is concerned with the presence
of the conformal connection functions $\tilde{\Upgamma}^{\I}$
on the right-hand side of
the BSSN equations. Two recipes have been identified that provide
long-term stable numerical evolutions. (i) The independently
evolved $\tilde{\Upgamma}^{\I}$ are only used when they appear
in differentiated form but are replaced by their definition in terms
of the conformal metric $\tilde{\gamma}_{\I \J}$ everywhere else
\cite{Alcubierre:2000yz}. (ii) Alternatively, one can add
to the right-hand side of Eq.~(\ref{eq:BSSNdGamma})
a term $-\sigma \mathcal{G}^{\I} \partial_{\M} \beta^{\M}$, where $\sigma$
is a positive constant~\cite{Yo:2002bm}.

We finally note that in place of the variable $\chi$, alternative
choices for evolving the conformal factor are in use in some
NR codes, namely $\phi \equiv -(\ln\,\chi)/4$~\cite{Baker:2005vv}
or $W\equiv \sqrt{\chi}$~\cite{Marronetti:2007wz}.
An overview of the specific choices of variables and treatment
of the BSSN constraints for the present generation of codes
is given in Section~4 of~\cite{Hinder:2013oqa}.

\subsubsection{The generalized harmonic gauge formulation}
\label{sec:GHG}
It has been realized a long time ago that the Einstein equations
have a mathematically appealing form if one imposes the
\emph{harmonic gauge} condition $\Box x^{\alpha}=-g^{\mu \nu}
\Gamma^{\alpha}_{\mu \nu}=0$~\cite{Einstein:1916cc}.
Taking the derivative
of this condition eliminates a specific combination of second derivatives
from the Ricci tensor such that its principal part is that of
the scalar
wave operator
\begin{equation}
  R_{\alpha \beta} = -\frac{1}{2} g^{\mu \nu} \partial_{\mu} \partial_{\nu}
        g_{\alpha \beta} + \ldots\,,
\end{equation}
where the dots denote terms involving at most the first derivative of
the metric. In consequence of this simplification of the principal part,
the Einstein equations in harmonic gauge can straightforwardly be
written as a strongly hyperbolic system.
This formulation even satisfies
the stronger condition of \emph{symmetric hyperbolicity} which is defined
in terms of the existence of a conserved, positive energy
\cite{Sarbach:2012pr},
and harmonic coordinates have played a key part in establishing
local uniqueness of the solution to the Cauchy problem in GR~\cite{FouresBruhat:1952zz, Bruhat1962, Fischer1972}.

This particularly appealing property of the Ricci tensor
can be maintained for arbitrary coordinates by introducing the functions
\cite{Friedrich1985,Garfinkle:2001ni}
\begin{equation}
  H^{\alpha} \equiv
      \Box x^{\alpha} = -g^{\mu \nu} \Gamma^{\alpha}_{\mu \nu}\,,
  \label{eq:GHG_H}
\end{equation}
and promoting them to the status of independently evolved
variables; see also~\cite{Pretorius:2004jg, Lindblom:2005qh}. This
is called the \emph{Generalized Harmonic Gauge formulation}. 

With this definition, it turns out convenient to consider the
generalized class of equations
\begin{equation}
  R_{\alpha \beta} -\nabla_{(\alpha} \mathcal{C}_{\beta)}
        = 8\pi \left( T_{\alpha \beta} - \frac{1}{D-2}Tg_{\alpha \beta}
          \right) + \frac{2}{D-2} \Lambda g_{\alpha \beta}\,,
  \label{eq:GHG_modEinstein}
\end{equation}
where $\mathcal{C}^{\alpha} \equiv H^{\alpha} - \Box x^{\alpha}$.
The addition of the term $\nabla_{(\alpha} \mathcal{C}_{\beta)}$
replaces the contribution of $\nabla_{(\alpha} \Box x_{\beta)}$ to
the Ricci tensor in terms of $\nabla_{(\alpha} H_{\beta)}$ and
thus changes the principal part to that of the scalar wave operator.
A solution to the Einstein equations is now obtained by solving
Eq.~(\ref{eq:GHG_modEinstein}) subject to the constraint
$\mathcal{C}_{\alpha}=0$.

The starting point for a Cauchy evolution are initial data
$g_{\alpha \beta}$ and $\partial_t g_{\alpha \beta}$ which
satisfy the constraints
$\mathcal{C}^{\alpha}=0=\partial_t \mathcal{C}^{\alpha}$.
A convenient manner to construct such initial data is to
compute the initial $H^{\alpha}$ directly from Eq.~(\ref{eq:GHG_H})
so that $\mathcal{C}^{\alpha}=0$ by construction. It can then be
shown~\cite{Lindblom:2005qh}
that the ADM constraints (\ref{eq:HamADM}), (\ref{eq:momADM})
imply $\partial_t \mathcal{C}^{\mu}=0$.
By virtue of the contracted Bianchi identities, the evolution
of the constraint system obeys the equation
\begin{equation}
  \Box \mathcal{C}_{\alpha} = -\mathcal{C}^{\mu} \nabla_{(\mu}
        \mathcal{C}_{\alpha)} - \mathcal{C}^{\mu}
        \left[ 8\pi \left( T_{\mu \alpha} - \frac{1}{D-2}Tg_{\mu \alpha}
        \right) + \frac{2}{D-2}\Lambda g_{\mu \alpha} \right]\,,
\end{equation}
and the constraint $\mathcal{C}^{\alpha}=0$ is preserved
under time evolution in the continuum limit.

A key addition to the GHG formalism has been devised by
Gundlach et~al.~\cite{Gundlach:2005eh} in the form of
damping terms which prevent growth of numerical violations
of the constraints $\mathcal{C}^{\alpha}=0$ due to
discretization or roundoff errors.

Including these damping terms and
using the definition (\ref{eq:GHG_H}) to substitute higher
derivatives in the Ricci tensor, the generalized Einstein equations
(\ref{eq:GHG_modEinstein}) can be written as
\begin{eqnarray}
  g^{\mu \nu} \partial_{\mu} \partial_{\nu} g_{\alpha \beta} &=&
        - 2\partial_{\nu} g_{\mu (\alpha}\,\partial_{\beta)} g^{\mu \nu}
        - 2\partial_{(\alpha} H_{\beta)}
        + 2H_{\mu} \Gamma^{\mu}_{\alpha \beta}
        - 2\Gamma^{\mu}_{\nu \alpha} \Gamma^{\nu}_{\mu \beta} \nonumber \\
     && - 8\pi T_{\alpha \beta}
        + \frac{8\pi T -2\Lambda}{D-2} g_{\alpha \beta}
        - 2\kappa \left[2n_{(\alpha}\mathcal{C}_{\beta)}
          - \lambda g_{\alpha \beta} n^{\mu} \mathcal{C}_{\mu} \right]\,,
\end{eqnarray}
where $\kappa$, $\lambda$ are user-specified constraint-damping parameters.
An alternative first-order system of the GHG formulation has been
presented in Ref.~\cite{Lindblom:2005qh}.

\subsubsection{Beyond BSSN: Improvements for future applications}
\label{sec:BeyondBSSN}
The vast majority of BH evolutions in generic $4$-dimensional
spacetimes have been performed with the GHG and the BSSN formulations.
It is interesting to note in this context the complementary nature of
the two formulations' respective strengths and weaknesses. In particular,
the constraint subsystem of the BSSN equations contains a zero-speed mode
\cite{Beyer:2004sv,Gundlach:2004jp,Gundlach:2004ri} which may lead to
large Hamiltonian constraint violations. The GHG system does not contain
such modes and furthermore admits a simple way of
controlling constraint violations in the form
of damping terms~\cite{Gundlach:2005eh}. Finally, the wave-equation-type
principal part of the GHG system allows for the straightforward
construction of constraint-preserving boundary conditions
\cite{Rinne2005a,Kreiss:2006mi,Ruiz:2007hg}. On the other hand,
the BSSN formulation is remarkably robust and allows for
the simulation of BH binaries over a wide
range of the parameter space with little if any modifications of the
gauge conditions; cf.~Section~\ref{sec:gauge}.
Combination of these advantages in a single system has motivated
the exploration of improvements to the BSSN system and in recent years
resulted in the identification of a conformal version of the $Z4$ system,
originally developed in
Refs.~\cite{Bona:2003fj,Bona:2003qn,Bona:2004ky}, as a highly promising
candidate
\cite{Alic:2011gg,Cao:2011fu,Weyhausen:2011cg,Hilditch:2012fp}.

The key idea behind the $Z4$ system is to replace the Einstein equations
with a generalized class of equations given by
\begin{equation}
  G_{\alpha \beta} = 8\pi T_{\alpha \beta} - \nabla_{\alpha} Z_{\beta}
      - \nabla_{\beta} Z_{\alpha} + g_{\alpha \beta} \nabla_{\mu}
      Z^{\mu}
      + \kappa_1 [n_{\alpha} Z_{\beta} + n_{\beta} Z_{\alpha}
        +\kappa_2 g_{\alpha \beta} n_{\M} Z^{\M}]\,,
\end{equation}
where $Z_{\alpha}$ is a vector field of constraints which is decomposed
into space and time components according to $\Theta\equiv
-n^{\mu} Z_{\mu}$ and $Z_{\I}=\bot^{\mu}{}_{\I} Z_{\mu}$. Clearly, a
solution to the Einstein equations is recovered provided the constraint
$Z_{\mu}=0$ is satisfied. The conformal version of the $Z4$ system
is obtained in the same manner as for the BSSN system and leads to time
evolution equations for a set of variables nearly identical to the
BSSN variables but augmented by the constraint variable $\Theta$.
The resulting evolution equations given in the literature vary in
details, but clearly represent relatively minor modifications for
existing BSSN codes
\cite{Alic:2011gg, Cao:2011fu, Hilditch:2012fp}. Investigations have
shown that the conformal $Z4$ system is indeed suitable for implementation
of constraint preserving boundary conditions~\cite{Ruiz:2010qj}
and that
constraint violations in simulations of gauge waves and BH
and NS spacetimes
are indeed smaller than those obtained for the BSSN
system, in particular when constraint damping is actively enforced
\cite{Alic:2011gg, Hilditch:2012fp}. This behaviour also manifests
itself in more accurate results for the gravitational radiation
in binary inspirals~\cite{Hilditch:2012fp}. In summary, the conformal
$Z4$ formulation is a very promising candidate for future numerical
studies of BH spacetimes, including in particular
the asymptotically AdS case where a rigorous control of the
outer boundary is of utmost importance;
cf.~Section~\ref{sec:boundaries} below.

Another modification of the BSSN equations is based on the use of
densitized versions of the trace of the extrinsic curvature and
the lapse function as well as the traceless part of the extrinsic
curvature with mixed indices~\cite{Laguna:2002zc, Witek:2010es}.
Some improvements in simulations of colliding BHs in higher-dimensional
spacetimes have been found by careful exploration of the densitization
parameter space~\cite{Witek:2013koa}.

\subsubsection{Alternative formulations}
The formulations discussed in the previous subsections are based on a spacetime
split of the Einstein equations.
A natural alternative to such a split is given
by the characteristic approach pioneered by Bondi et~al.
and Sachs~\cite{Bondi:1962px,Sachs:1962wk}. Here, at least one coordinate is null and thus adapted to the
characteristics of the vacuum Einstein equations.  For generic four-dimensional
spacetimes with no symmetry assumptions, the characteristic formalism results in
a natural hierarchy of 2 evolution equations, 4 hypersurface equations relating
variables on hypersurfaces of constant retarded (or advanced) time, as well as 3
supplementary and 1 trivial equations. A comprehensive overview of
characteristic methods in NR is given in the Living Reviews article~\cite{Winicour2012}.  Although characteristic codes have been
developed with great success in spacetimes with additional symmetry assumptions,
evolutions of generic BH spacetimes face the problem of formation of caustics,
resulting in a breakdown of the coordinate system; see~\cite{Babiuc:2013rra} for
a recent investigation. One possibility to avoid the problem of caustic
formation is \emph{Cauchy-characteristic matching}, the combination of a
$(D-1)+1$ or Cauchy-type numerical scheme in the interior strong-field region
with a characteristic scheme in the outer parts. In the form of
Cauchy-characteristic extraction, i.e., ignoring the injection of information
from the characteristic evolution into the inner Cauchy region, this approach
has been used to extract GWs with high accuracy from numerical simulations of
compact objects~\cite{Reisswig:2009us, Babiuc:2010ze}.

All the Cauchy and characteristic or combined approaches we have discussed
so far, evolve the physical spacetime, i.e., a manifold with metric
$(\mathcal{M},g_{\alpha \beta})$.
An alternative approach for asymptotically flat spacetimes
dating back to H{\"u}bner~\cite{Hubner:1994pd} instead considers the
numerical construction of a conformal spacetime
$(\tilde{\mathcal{M}},\tilde{g}_{\alpha \beta})$ where
$\tilde{g}_{\alpha \beta} = \Omega^2 g_{\alpha \beta}$ subject to the
condition that $g_{\alpha \beta}$ satisfies the Einstein equations
on $\mathcal{M}$. The conformal factor $\Omega$ vanishes
at null infinity $\mathscr{I} = \mathscr{I}^+ \cup \mathscr{I}^-$
of the physical spacetime which is thus conformally related to an
interior of the unphysical manifold $\tilde{\mathcal{M}},
\tilde{g}_{\alpha \beta}$ which extends beyond the physical
manifold. A version of these \emph{conformal field equations}
that overcomes the singular nature of the transformed Einstein
equations at $\mathscr{I}$ has been developed by
Friedrich~\cite{Friedrich:1981at,Friedrich:1981wx}. This formulation
is suitable for a 3+1 decomposition into a symmetric hyperbolic
system%
\epubtkFootnote{Decompositions in terms of null foliations have to our
  knowledge not been studied yet, although there is no evident reason
  that speaks against such an approach.}
of evolution equations for an enhanced (relative to the
ADM decomposition) set of variables. The additional cost
resulting from the larger set of variables, however, is mitigated
by the fact that these include projections of the Weyl tensor
that directly encode the GW content. Even though the conformal
field equations have as yet not resulted in simulations of
BH systems analogous to those achieved in BSSN or GHG, their
elegance in handling the entire spacetime without truncation
merits further investigation.
For more details about the
formulation and numerical applications,
we refer the reader to the above articles,
Lehner's review~\cite{Lehner:2001wq}, Frauendiener's Living Reviews
article~\cite{Frauendiener:2004} as well as
\cite{Frauendiener:2002ix,Alcubierre:2003pc} and references therein.
A brief historic overview of many formulations of the Einstein
equations (including systems not discussed in this work)
is given in Ref.~\cite{Shinkai:2008yb}; see in particular Figures~3
and 4 therein.

We finally note that
for simulations of spacetimes with high degrees of symmetry,
it often turns out convenient to directly impose the symmetries
on the shape of the line element rather than use one of the general
formalisms discussed so far. As an example, we consider the classic
study by May and White~\cite{May1966, May1967} of the dynamics
of spherically symmetric perfect fluid stars. A four-dimensional spherically symmetric
spacetime can be described in terms of the simple line element
\begin{equation}
ds^2 = -a^2(x,t) \, dt^2 + b^2(x,t) \, dx^2 + R^2(x,t) \, d\Omega_2^2 \,,
\label{eq:MayWhiteds}
\end{equation}
where $d\Omega_2^2$ is the line element of the $2$-sphere. May and White employ
Lagrangian coordinates co-moving with the fluid shells which is imposed through
the form of the energy-momentum tensor $T^0{}_0=-\rho(1+\epsilon)$,
$T^1{}_1=T^2{}_2=T^3{}_3=P$. Here, the rest-mass density $\rho$, internal energy
$\epsilon$, and pressure $P$ are functions of the radial and time
coordinates. Plugging the line element (\ref{eq:MayWhiteds}) into the Einstein
equations~(\ref{eq:EinsteinEqs_D}) with $D=4$, $\Lambda=0$ and the equations of
conservation of energy-momentum $\nabla_{\mu}T^{\mu}{}_{\alpha}=0$, result in a
set of equations for the spatial and time derivatives of the metric and matter
functions amenable for a numerical treatment; cf.~Section~II in Ref.~\cite{May1966}
for details.

\subsubsection{Einstein's equations extended to include fundamental
  fields}
\label{sec:matter}
The addition of matter to the spacetime can, in principle, be done using
the formalism just laid down\epubtkFootnote{\dots but beware! For many realistic
types of matter, novel effects -- such as shocks -- can hamper an efficient
evolution. These have to be handled with care and would require a review
of its own.}. The simplest extension of the field equations to include
matter is described by the Einstein--Hilbert action (in $4$-dimensional
asymptotically flat spacetimes) minimally coupled to a complex, massive
scalar field $\Phi$ with mass parameter $\mu_S=m_{S}/\hbar$,
\begin{equation}
  \label{eq:action}
  S =\int {\rm d}^{4}x \sqrt{-g}
        \left( \frac{R}{16\pi}  -\tfrac{1}{2} g^{\mu\nu}\partial_{\mu}
        \Phi^{\ast}{}\partial_{\nu}\Phi
        -\tfrac{1}{2}\mu_{S}^{2} \Phi^{\ast}{}\Phi  \right)\,.
\end{equation}
If we introduce a time reduction variable defined as
\begin{equation}
  \label{eq:defKijKPhi}
  \Pi =  - \tfrac{1}{\alpha} \left(\partial_{t} - \Lie_{\boldsymbol{\beta}}
        \right) \Phi\,,
\end{equation}
we recover the equations of motion and constraints
\eqref{eq:dgammaADM}\,--\,\eqref{eq:momADM} with $D=4,\Lambda=0$ and with
energy density $\rho$, energy-momentum flux $j_{i}$ and spatial components
$S_{ij}$ of the energy-momentum tensor given by
\begin{eqnarray}
  \label{eq:rho}
  \rho &=& \tfrac{1}{2} \Pi^{\ast}{} \Pi + \tfrac{1}{2} \mu_{S}^{2}
        \Phi^{\ast}{}\Phi + \tfrac{1}{2} D^{i}\Phi^{\ast}{} D_{i}\Phi\,,\\
  \label{eq:jj}
  j_i &=& \tfrac{1}{2}\left( \Pi^{\ast}{} D_{i}\Phi
        + \Pi D_{i} \Phi^{\ast}{}\right)\,,\\
  \label{eq:Sij}
  S_{ij} &=& \tfrac{1}{2}\left( D_{i}\Phi^{\ast}{} D_{j}\Phi
        + D_{i}\Phi D_{j}\Phi^{\ast}\right) + \tfrac{1}{2}\gamma_{ij}\left(
        \Pi^{\ast}{} \Pi - \mu_S^2\Phi^{\ast}{}\Phi
        - D^{k}\Phi^{\ast}{}D_{k}\Phi \right)\,.
\end{eqnarray}
Vector fields can be handled in similar fashion, we refer the
reader to Ref.~\cite{Witek:2012tr} for linear studies and to
Refs.~\cite{Palenzuela:2009yr,Palenzuela:2010nf,Zilhao:2012gp,Zilhao:2013nda}
for full nonlinear evolutions.

In summary, a great deal of progress has been made in recent years
concerning the well-posedness of the numerical methods used for the
construction of spacetimes. We note, however, that the well-posedness
of many problems beyond electrovacuum GR remains unknown at
present. This includes, in particular, a wide class of alternative
theories of gravity where it is not clear whether they admit
well-posed IBVPs.

\subsection{Higher-dimensional NR in effective ``3+1'' form}
\label{sec:NRD}
Performing numerical simulations in generic higher-dimensional spacetimes
represents a major challenge for simple computational reasons.  Contemporary
simulations of compact objects in four spacetime dimensions require
$\mathcal{O}(100)$ cores and $\mathcal{O}(100)~{\rm Gb}$ of memory for storage
of the fields on the computational domain. In the absence of spacetime
symmetries, any extra spatial dimension needs to be resolved by
$\mathcal{O}(100)$ grid points resulting in an increase by about two orders of
magnitude in both memory requirement and computation time.  In spite of the
rapid advance in computer technology, present computational power is pushed to
its limits with $D=5$ or, at best, $D=6$ spacetime dimensions. For these
reasons, as well as the fact that the community already has robust codes
available in $D=4$ dimensions, NR applications to higher-dimensional spacetimes
have so far focussed on symmetric spacetimes that allow for a reduction to an
effectively four-dimensional formalism. Even though this implies a reduced class
of spacetimes available for numerical study, many of the most important
questions in higher-dimensional gravity actually fall into this class of
spacetimes. In the following two subsections we will describe two different
approaches to achieve such a dimensional reduction, for the cases of spacetimes
with $SO(D-2)$ or $SO(D-3)$ isometry, i.e., the rotational symmetry leaving
invariant ${S}^{D-3}$ or ${S}^{D-4}$, respectively (we denote with ${S}^n$ the $n$-dimensional
sphere). The group $SO(D-2)$ is the isometry of, for instance, head-on
collisions of non-rotating BHs, while the group $SO(D-3)$ is the isometry of
non-head-on collisions of non-rotating BHs; $SO(D-3)$ is also the isometry of
non-head-on collisions of rotating BHs with one nonvanishing angular momentum, 
generating rotations on the orbital plane (see Figure~\ref{fig:isocoll}). Furthermore, the $SO(D-3)$ group is the isometry of a
single rotating BH, with one non-vanishing angular momentum. We remark that, in
order to implement the higher-dimensional system in (modified) four-dimensional
evolution codes, it is necessary to perform a $4+(D-4)$ splitting of the
spacetime dimensions. With such splitting, the equations have a manifest
$SO(D-3)$ symmetry, even when the actual isometry is larger.

\epubtkImage{}{%
  \begin{figure}[htb]
    \centerline{\includegraphics[width=0.8\textwidth]{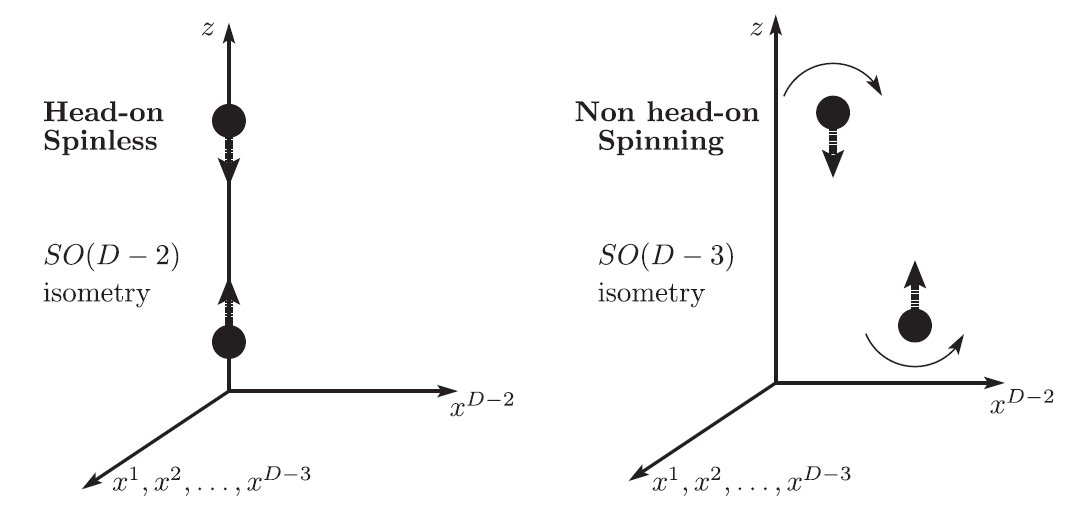}}
    \caption{$D$-dimensional representation of head-on collisons for
      spinless BHs, with isometry group $SO(D-2)$ (left), and
      non-head-on collisons for BHs spinning in the orbital plane,
      with isometry group $SO(D-3)$ (right) \cite{Zilhao:2010sr}.}
    \label{fig:isocoll}
\end{figure}}

We shall use the following conventions for indices. As before, Greek indices
$\alpha,\,\beta,\,\ldots$ cover all spacetime dimensions and late upper case capital
Latin indices $I,\,J,\,\ldots=1,\,\ldots\,D-1$ cover the $D-1$ spatial
dimensions, whereas late lower case Latin indices $i,\,j,\,\ldots = 1,\,2,\,3$
cover the three spatial dimensions of the eventual computational domain. In
addition, we introduce barred Greek indices
$\bar{\alpha},\,\bar{\beta},\,\ldots=0,\, \ldots,\,3$ which also include time,
and early lower case Latin indices $a,\,b,\,\ldots=4,\,\ldots,\,D-1$ describing
the $D-4$ spatial directions associated with the rotational symmetry. Under the
$4+(D-4)$ splitting of spacetime dimensions, then, the coordinates $x^\mu$
decompose as $x^\mu\rightarrow(x^{\bar\mu},x^a)$.  When explicitly stated, we
shall consider instead a $3+(D-3)$ splitting, e.g., with barred Greek indices
running from $0$ to $2$, and early lower case Latin indices running from $3$ to
$D-1$.

\subsubsection{Dimensional reduction by isometry}
\label{sec:isometry}
The idea of dimensional reduction had originally been developed by Geroch
\cite{Geroch:1970nt} for four-dimensional spacetimes possessing one Killing
field as for example in the case of axisymmetry; for numerical applications see
for example Refs.~\cite{1980PThPh..63..719M,Sjodin:2000zd, Sperhake:2000fe, Choptuik:2003as}. The case of
arbitrary spacetime dimensions and number of Killing vectors has been discussed
in Refs.~\cite{Cho:1986wk,Cho:1987jf}.\epubtkFootnote{We remark that the dimensional reduction here discussed is different from Kaluza--Klein dimensional reduction~\cite{Kaluza:1921tu,Duff:1994tn}, an idea first proposed about one century ago, which in recent decades has attracted a lot of interest in the context of SMT. Indeed, a crucial feature of Kaluza--Klein dimensional reduction is spacetime compactification, which does not occur in our case.} More recently, this idea has been used to develop
a convenient formalism to perform NR simulations of BH dynamical systems in
higher dimensions, with $SO(D-2)$ or $SO(D-3)$ isometry~\cite{Zilhao:2010sr,Witek:2010xi}. Comprehensive summaries of this approach are
given in Refs.~\cite{Zilhao:2013gu,Witek:2013koa,Witek:2013ora}.

The starting point is the general $D$-dimensional spacetime metric written in
coordinates adapted to the symmetry
\begin{equation}
  ds^2 = g_{\alpha \beta} \, dx^\alpha \, dx^\beta =
        \left( g_{\bar{\mu} \bar{\nu}} + e^2 \kappa^2 g_{ab}B^a{}_{\bar{\mu}}
        B^b{}_{\bar{\nu}} \right) \, dx^{\bar{\mu}} \, dx^{\bar{\nu}}
        + 2e \kappa B^a{}_{\bar{\mu}} g_{ab} \, dx^{\bar{\mu}} \, dx^b
        + g_{ab} \, dx^a \, dx^b\,.
  \label{eq:metricCho}
\end{equation}
Here, $\kappa$ and $e$ represent a scale parameter and a coupling constant
that will soon drop out and play no role in the eventual spacetime reduction.
We note that 
the metric (\ref{eq:metricCho}) is fully general in the same sense
as the spacetime metric in the ADM split discussed in Section~\ref{sec:ADM}.

The special case of a $SO(D-2)$ ($SO(D-3)$) isometry 
admits $(n+1)n/2$
Killing fields $\boldsymbol{\xi}_{(\boldsymbol{i})}$ where $n \equiv D-3$ ($n \equiv D-4$)
stands for the number of extra dimensions. For $n=2$, for instance, there exist
three Killing fields given in spherical coordinates by
$\boldsymbol{\xi}_{(\boldsymbol{1})}=
\boldsymbol{\partial}_{\phi}$,
$\boldsymbol{\xi}_{(\boldsymbol{2})} =
\sin\phi\,\boldsymbol{\partial}_{\theta}+\cot\theta\,\cos\phi\,
\boldsymbol{\partial}_{\phi}$, $\boldsymbol{\xi}_{(\boldsymbol{3})} =
\cos\phi\,\boldsymbol{\partial}_{\theta}
- \cot \theta\,\sin\phi\, \boldsymbol{\partial}_{\phi}$.

Killing's equation $\mathcal{L}_{\boldsymbol{\xi}_{(\boldsymbol{i})}} g_{AB}=0$
implies that
\begin{equation}
  \mathcal{L}_{\boldsymbol{\xi}_{(\boldsymbol{i})}} g_{ab}=0\,,~~~
  \mathcal{L}_{\boldsymbol{\xi}_{(\boldsymbol{i})}} B^a{}_{\bar{\mu}}=0\,,~~~
  \mathcal{L}_{\boldsymbol{\xi}_{(\boldsymbol{i})}} g_{\bar{\mu} \bar{\nu}}=0\,,
\end{equation}
where, as discussed above, the decomposition
$x^\mu\rightarrow(x^{\bar\mu},x^a)$ describes a $4+(D-4)$ splitting in the case
of $SO(D-3)$ isometry, and a $3+(D-3)$ splitting in the case of $SO(D-2)$
isometry.

From these conditions, we draw the following conclusions:  (i) $g_{ab} =
e^{2\psi(x^{\bar{\mu}})} \Omega_{ab}$, where $\Omega_{ab}$ is the metric on the
$S^n$ sphere with unit radius and $\psi$ is a free field; (ii) $g_{\bar\mu\bar\nu} =
g_{\bar\mu\bar\nu}(x^{\bar\sigma})$ in adapted coordinates; (iii)
$[\boldsymbol{\xi}_{(\boldsymbol{i})},\boldsymbol{B}_{\bar\mu}]=0$.
We here remark an interesting consequence
of the last property.  Since, for $n\ge 2$, there exist no nontrivial vector
fields on $S^n$ that commute with all Killing fields, all vector fields
$B^a{}_{\bar{\mu}}$ vanish; when, instead, $n=0,1$ (i.e., when $D=4$, or $D=5$
for $SO(D-3)$ isometry), this conclusion can not be made. In this approach, as
it has been developed up to now~\cite{Zilhao:2010sr,Witek:2010xi,Witek:2014mha}, one restricts
to the $n\ge2$ case, and it is then possible to assume
$B^a{}_{\bar{\mu}}\equiv0$.  Eq.~(\ref{eq:metricCho}) then reduces to
the form%
\epubtkFootnote{There is a length-squared factor multiplying the
  exponential which we set to unity.}
\begin{equation}
  ds^2 = g_{\bar\mu\bar\nu} \, dx^{\bar\mu} \, dx^{\bar\nu} +
  e^{2\psi(x^{\bar{\mu}})} \Omega_{ab} \, dx^a \, dx^b\,.
\label{eq:metricus}
\end{equation}
For this reason, this approach can only be applied when $D\ge5$ in the case of
$SO(D-2)$ isometry, and $D\ge6$ in the case of $SO(D-3)$ isometry. 

As mentioned above, since the Einstein equations have to be implemented in a
four-dimensional NR code, we eventually have to perform a $4+(D-4)$ splitting,
even when the spacetime isometry is $SO(D-2)$. This means that the line element
is (\ref{eq:metricus}), with $\bar{\alpha},\,\bar{\beta},\,\ldots=0,\,
\ldots,\,3$ and $a,\,b,\,\ldots=4,\,\ldots,\,D-1$.  In this case, only a subset
$SO(D-3)\subset SO(D-2)$ of the isometry is manifest in the line element; the
residual symmetry yields an extra relation among the components
$g_{\bar\mu\bar\nu}$. If the isometry group is $SO(D-3)$, the line element is
the same, but there is no extra relation.

A tedious but straightforward calculation~\cite{Zilhao:2013gu} shows that the
components of the $D$-dimensional Ricci tensor can then be written as 
\begin{eqnarray}
  R_{ab} &=& \left\{ (D-5) - e^{2\psi}\left[ (D-4) \partial^{\bar{\mu}} \psi
        \partial_{\bar{\mu}} \psi + \bar{\nabla}^{\bar{\mu}}
        \partial_{\bar{\mu}} \psi\right] \right\} \Omega_{ab}\,,\nonumber \\
  R_{\bar{\mu} a} &=& 0\,, \nonumber \\
  R_{\bar{\mu} \bar{\nu}} &=& \bar{R}_{\bar{\mu} \bar{\nu}}
        - (D-4) (\bar{\nabla}_{\bar{\nu}} \partial_{\bar{\mu}}
        \psi +
        \partial_{\bar{\mu}} \psi\,\partial_{\bar{\nu}} \psi)\,,
        \nonumber \\
  R &=& \bar{R} + (D-4) \left[ (D-5) e^{-2\psi}
        -2\bar{\nabla}^{\bar{\mu}} \partial_{\bar{\mu}}\psi
        - (D-3)\partial^{\bar{\mu}} \psi\,\partial_{\bar{\mu}} \psi \right]\,,
\end{eqnarray}
where $\bar{R}_{\bar\mu\bar\nu}$, $\bar{R}$ and $\bar{\nabla}$
respectively denote the
3+1-dimensional Ricci tensor, Ricci scalar and covariant derivative
associated with the 3+1 metric
$\bar{g}_{\bar{\mu}\bar{\nu}} \equiv g_{\bar{\mu} \bar{\nu}}$.
The $D$-dimensional vacuum Einstein equations with cosmological
constant $\Lambda$ can then be formulated in terms of fields on a
3+1-dimensional manifold
\begin{eqnarray}
  && \bar{R}_{\bar{\mu} \bar{\nu}} = (D-4) (\bar{\nabla}_{\bar{\nu}}
        \partial_{\bar{\mu}} \psi
        - \partial_{\bar{\mu}}\psi \,\partial_{\bar{\nu}} \psi)
        - \Lambda \bar{g}_{\bar{\mu} \bar{\nu}}\,,
        \label{eq:EinsteinIso1} \\
  &&e^{2\psi}\left[ (D-4)\partial^{\bar{\mu}}\psi\,\partial_{\bar{\mu}}\psi
        + \bar{\nabla}^{\bar{\mu}} \partial_{\bar{\mu}} \psi - \Lambda \right]
        = (D-5)\,.
        \label{eq:EinsteinIso2}
\end{eqnarray}
One important comment is in order at this stage. If we describe the
three spatial dimensions in terms of Cartesian coordinates $(x,\,y,\,z)$,
one of these is now a quasi-radial coordinate. Without loss of generality,
we choose $y$ and the computational domain is given by
$x,\,z \in \mathbb{R}$, $y\ge 0$. In consequence of the radial nature
of the $y$ direction, $e^{2\psi}=0$ at $y=0$. Numerical problems arising
from this coordinate singularity can be avoided by working instead
with a rescaled version of the variable $e^{2\psi}$. More specifically,
we also include the BSSN conformal factor $e^{-4\phi}$ in the redefinition and
write
\begin{equation}
  \zeta \equiv \frac{e^{-4\phi}}{y^2} e^{2\psi}\,.
\end{equation}
The BSSN version of the $D$-dimensional vacuum Einstein equations
(\ref{eq:EinsteinIso1}), (\ref{eq:EinsteinIso2}) with $\Lambda=0$ in its dimensionally reduced form
on a 3+1 manifold
is then given by Eqs.~(\ref{eq:BSSNdtphi})\,--\,(\ref{eq:BSSNdGamma}) with the
following modifications. (i) Upper-case capital indices $I,\,J,\,\ldots$
are replaced with their lower case counterparts $i\,\,j,\,\ldots=1,\,2,\,3$.
(ii) The $(D-1)$ dimensional metric $\gamma_{\I \J}$,
Christoffel symbols $\Upgamma^{\I}_{\J \K}$, covariant derivative $D$,
conformal factor $\chi$ and extrinsic curvature variables
$K$ and $\tilde{A}_{\I \J}$ are replaced by the $3$ dimensional metric
$\threegamma_{ij}$, the $3$ dimensional Christoffel symbols $\threeGamma^i_{jk}$, the covariant derivative
$\threeD$, as well as the conformal factor $\threechi$, $\threeK$ and
$\threeA_{ij}$ defined in analogy to Eq.~(\ref{eq:BSSNvars}) with $D=4$, i.e.
\begin{eqnarray}
  &\threechi = \threegamma^{-1/3}\,,
  ~~~~~
  &\threeK = \threegamma^{nm} \threeK_{mn}, \nonumber \\
  &\tilde{\threegamma}_{i j} = \threechi \threegamma_{i j}
  ~~~
  &\Leftrightarrow
  ~~~
  \tilde{\threegamma}^{ij} = \frac{1}{\threechi} \threegamma^{ij}, \nonumber \\
  &\tilde{\threeA}_{ij} = \threechi \left( \threeK_{ij} - \frac{1}{3}
      \threegamma_{ij} \threeK \right)
  ~~~
  &\Leftrightarrow
  ~~~
  \threeK_{ij} = \frac{1}{\threechi} \left( \tilde{\threeA}_{ij} + \frac{1}{3}
      \tilde{\threegamma}_{ij} \threeK \right)\,, \nonumber \\
  &\tilde{\threeGamma}^i = \tilde{\threegamma}^{mn}
      \tilde{\threeGamma}^i_{mn}\,.&
  \label{eq:BSSN3vars}
\end{eqnarray}
(iii) The extra dimensions manifest themselves as quasi-matter terms
given by
\begin{eqnarray}
  \frac{4\pi (\rho + S)}{D-4} &=& (D-5) \frac{\threechi}{\zeta}
     \frac{\tilde{\threegamma}^{yy} \zeta - 1}{y^2}
     -\frac{2D-7}{4\zeta} \tilde{\threegamma}^{mn} \partial_m
      \eta~\partial_n \threechi
     - \threechi \frac{\tilde{\threeGamma}^y}{y}
     + \frac{D-6}{4} \frac{\threechi}{\zeta^2} \tilde{\threegamma}^{mn}
     \partial_m \zeta~\partial_n \zeta \nonumber \\
  && + \frac{1}{2\zeta} \tilde{\threegamma}^{mn} (\threechi \tilde{\threeD}_m
       \partial_n \zeta
     - \zeta \tilde{\threeD}_m \partial_n \threechi)
     + (D-4) \frac{\tilde{\threegamma}^{ym}}{y} \left( \frac{\threechi}{\zeta}
     \partial_m \zeta - \partial_m \threechi \right)
     - \frac{\threeK K_{\zeta}}{\zeta} - \frac{\threeK^2}{3}
     \nonumber \\
  &&
     - \frac{1}{2}\frac{\tilde{\threegamma}^{ym}}{y} \partial_m \threechi
     + \frac{D-1}{4} \tilde{\threegamma}^{mn}
       \frac{\partial_m \threechi~\partial_n \threechi}
       {\threechi}
     - (D-5) \left( \frac{K_{\zeta}}{\zeta} + \frac{\threeK}{3} \right)^2\,, \\
  \frac{8\pi \threechi S_{ij}^{\rm TF}}{D-4} &=&
     - \left( \frac{K_{\zeta}}{\zeta} + \frac{\threeK}{3} \right)
       \tilde{\threeA}_{ij}
     +\frac{1}{2} \left[
     \frac{2\threechi}{y\zeta}(\delta^y{}_{(j} \partial_{i)} \zeta
     - \zeta \tilde{\threeGamma}^y_{ij})
     + \frac{1}{2\threechi}\partial_i \threechi~\partial_j \threechi
     - \tilde{\threeD}_i \partial_j \threechi
     + \frac{\threechi}{\zeta} \tilde{\threeD}_i \partial_j \zeta
     \right. \nonumber \\
  && \left.
     + \frac{1}{2\threechi} \tilde{\threegamma}_{ij}
       \tilde{\threegamma}^{mn} \partial_n \threechi
       \left( \partial_m \threechi - \frac{\threechi}{\zeta}
       \partial_m \zeta \right)
     - \tilde{\threegamma}_{ij} \frac{\tilde{\threegamma}^{ym}}{y}
       \partial_m \threechi
     - \frac{\threechi}{2\zeta^2}\partial_i \zeta~\partial_j \zeta
       \right]^{\rm TF}
     \\
  \frac{16\pi j_i}{D-4} &=&
     \frac{2}{y} \left( \delta^y{}_i \frac{K_{\zeta}}{\zeta}
     -\tilde{\threegamma}^{ym} \tilde{\threeA}_{mi} \right)
     + \frac{2}{\zeta} \partial_i K_{\zeta}
     - \frac{K_{\zeta}}{\zeta} \left( \frac{1}{\threechi}\partial_i \threechi
       + \frac{1}{\zeta} \partial_i \zeta \right)
     + \frac{2}{3} \partial_i \threeK \nonumber \\
  && - \tilde{\threegamma}^{nm} \tilde{\threeA}_{mi}
     \left( \frac{1}{\zeta} \partial_n
     \zeta - \frac{1}{\threechi} \partial_n \threechi \right)\,.
\end{eqnarray}
Here, $K_{\zeta}\equiv -(2\alpha y^2)^{-1}(\partial_t-\mathcal{L}_{\beta})(\zeta y^2)$.
The evolution of the field $\zeta$ is determined by Eq.~(\ref{eq:EinsteinIso2})
which in terms of the BSSN variables becomes
\begin{eqnarray}
  \partial_t \zeta &=&
        \beta^m \partial_m \zeta -2\alpha K_{\zeta}
        - \frac{2}{3}\zeta \partial_m \beta^m
        + 2\zeta \frac{\beta^y}{y}\,, \\
  \partial_t K_{\zeta} &=& \beta^m \partial_m K_{\zeta}
        - \frac{2}{3}K_{\zeta} \partial_m \beta^m
        + 2\frac{\beta^y}{y} K_{\zeta}
        - \frac{1}{3} \zeta (\partial_t - \mathcal{L}_{\beta} ) \threeK
        - \threechi \zeta \frac{\tilde{\threegamma}^{ym}}{y} \partial_m \alpha
        \nonumber \\
     && - \frac{1}{2} \tilde{\threegamma}^{mn} \partial_m \alpha
          ~(\threechi \partial_n \zeta - \zeta \partial_n \threechi)
        + \alpha \left[
          (5-D) \threechi \frac{\zeta \tilde{\threegamma}^{yy}-1}{y^2}
          + (4-D) \threechi \frac{\tilde{\threegamma}^{ym}}{y} \partial_m \zeta
          \right.
        \nonumber \\
     && + \frac{2D-7}{2} \zeta \frac{\tilde{\threegamma}^{ym}}{y}
        \partial_m \threechi
        + \frac{6-D}{4} \frac{\threechi}{\zeta} \tilde{\threegamma}^{mn}
          \partial_m \zeta~\partial_n \zeta
        + \frac{2D-7}{4} \tilde{\threegamma}^{mn} \partial_m
        \zeta~\partial_n \threechi
        \nonumber \\
     && + \frac{1-D}{4} \frac{\zeta}{\threechi} \tilde{\threegamma}^{mn}
        \partial_m \threechi ~\partial_n \threechi
        + (D-6) \frac{K_{\zeta}^2}{\zeta}
        + \frac{2D-5}{3} \threeK K_{\zeta}
        + \frac{D-1}{9}\zeta \threeK^2
        \nonumber \\
     && \left.
        + \frac{1}{2} \tilde{\threegamma}^{mn}
        (\zeta \tilde{\threeD}_m \partial_n \threechi
          - \threechi \tilde{\threeD}_m \partial_n \zeta)
        + \threechi \zeta \frac{\tilde{\threeGamma}^y}{y}
        \right] \,.
\end{eqnarray}
It has been demonstrated in Ref.~\cite{Zilhao:2010sr} how all terms containing factors of $y$ in the denominator can be
regularized using the symmetry properties of tensors and their
derivatives across $y=0$ and assuming that the spacetime does not
contain a conical singularity.

\subsubsection{The cartoon method}
\label{sec:Cartoon}
The \emph{cartoon method} has originally been
developed in Ref.~\cite{Alcubierre:1999ab} for evolving axisymmetric four-dimensional
spacetimes using an effectively two-dimensional spatial grid which
employs ghostzones, i.e., a small number of extra gridpoints off the
computational plane required for evaluating finite
differences in the third spatial direction. 
Integration in time, however, is performed exclusively on the two-dimensional plane whereas the
ghostzones are filled in after each timestep by appropriate interpolation
of the fields in the plane and subsequent rotation of the solution
using the axial spacetime symmetry. A version of this method
has been applied to 5-dimensional spacetimes in Ref.~\cite{Yoshino:2009xp}. For arbitrary spacetime dimensions, however, even
the relatively small number of ghostzones required in every extra dimension
leads to a substantial increase in the computational resources; for
fourth-order finite differencing, for example, four ghostzones
are required in each extra dimension resulting in an increase
of the computational domain by an overall factor $5^{D-4}$. An elegant
scheme to avoid this difficulty while preserving all advantages
of the cartoon method has been developed
in Ref.~\cite{Pretorius:2004jg}
and is sometimes referred to as the \emph{modified cartoon method}.
This method has been applied to $D >5$
dimensions in Refs.~\cite{Shibata:2010wz,Lehner:2011wc,Yoshino:2011zz}
and we will discuss it now in more detail.

Let us consider for illustrating this method a $D$-dimensional
spacetime with $SO(D-3)$ symmetry and Cartesian coordinates
$x^{\mu}=(t,\,x,\,y,\,z,\,w^a)$, where $a=4,\,\ldots,\,D-1$.
Without loss of generality, the coordinates are chosen such
that the $SO(D-3)$ symmetry implies rotational symmetry in the
planes spanned by each choice of two coordinates from\epubtkFootnote{Note
that Ref.~\cite{Shibata:2010wz} chooses $z$ instead of $y$.}
$(y,\,w^a)$. The goal is to obtain a formulation of the
$D$-dimensional Einstein equations (\ref{eq:BSSNdtphi})\,--\,(\ref{eq:BSSNDDalpha})
with $SO(D-3)$ symmetry that can be evolved exclusively on the
$xyz$ hyperplane. The tool employed for this purpose is to use
the spacetime symmetries in order to trade derivatives off the
hyperplane, i.e., in the $w^a$ directions, for derivatives inside
the hyperplane. Furthermore, the symmetry implies relations between
the $D$-dimensional components of the BSSN variables.

These relations are obtained by applying a coordinate transformation
from Cartesian to polar coordinates in any of the two-dimensional
planes spanned by $y$ and $w$, where $w\equiv w^a$ for any particular
choice of $a \in \{4,\,\ldots,\,D-1\}$
\begin{eqnarray}
  &\rho = \sqrt{y^2+w^2}\,,& y = \rho \cos \varphi\,, \nonumber \\
  &\varphi = \arctan \frac{w}{y}\,,& w = \rho \sin \varphi\,.
  \label{eq:coordtrafo1}
\end{eqnarray}
Spherical symmetry in $n\equiv D-4$ dimensions implies the existence of
$n(n+1)/2$ Killing vectors, one for each plane with rotational symmetry.
For each Killing vector $\boldsymbol{\xi}$, the Lie derivative of the spacetime
metric vanishes. For the $yw$ plane,
in particular, the Killing vector field is
$\boldsymbol{\xi}=\boldsymbol{\partial}_{\varphi}$
and the Killing condition is given by the simple relation
\begin{equation}
  \partial_{\varphi} g_{\mu \nu} = 0\,.
\end{equation}
All ADM and BSSN variables are constructed from the spacetime metric
and a straightforward calculation demonstrates that the
Lie derivatives along $\boldsymbol{\partial}_{\varphi}$
of all these variables vanish.
For $D\ge 6$, we can always choose the coordinates such that for
$\mu \ne \varphi$, $g_{\mu \varphi}=0$ which implies the vanishing
of the BSSN variables $\beta^{\varphi}= \tilde{\gamma}^{\mu \varphi}
= \tilde{\Upgamma}^{\varphi} = 0$. The case of $SO(D-3)$ symmetry in $D=5$
dimensions is special in the same sense as already discussed in 
Section~\ref{sec:isometry} and the vanishing of $\tilde{\Upgamma}^{\varphi}$
does not in general hold. As before, we therefore consider in $D=5$
the more restricted class of $SO(D-2)$ isometry which implies
$\tilde{\Upgamma}^{\varphi}=0$. Finally, the Cartesian coordinates
$w^a$ can always be chosen such that the diagonal metric components
are equal, 
\begin{equation}
  \gamma_{w^1 w^1} = \gamma_{w^2 w^2} = \ldots \equiv \gamma_{ww}\,.
\end{equation}
We can now
exploit these properties in order to trade derivatives in the desired
manner. We shall illustrate this for the second $w$ derivative
of the $ww$ component of a symmetric $\tbinom{0}{2}$ tensor density
$\boldsymbol{S}$ of
weight $\mathcal{W}$ which transforms under change of coordinates
$x^{\mu} \leftrightarrow x^{\hat{\alpha}}$
according to
\begin{equation}
  S_{\hat{\alpha} \hat{\beta}} = \mathcal{J}^{\mathcal{W}}
        \frac{\partial x^{\mu}}{\partial x^{\hat{\alpha}}}
        \frac{\partial x^{\nu}}{\partial x^{\hat{\beta}}}
        S_{\mu \nu}\,,~~~~~~~~~~
  \mathcal{J} \equiv \det \left(
          \frac{\partial x^{\mu}}{\partial x^{\hat{\alpha}}}
        \right) \,.
\end{equation}
Specifically, we consider the coordinate transformation
(\ref{eq:coordtrafo1}) where $\mathcal{J}=\rho$.
In particular, this transformation
implies
\begin{equation}
  \partial_w S_{ww} = \frac{\partial \rho}{\partial w}
        \partial_{\rho} S_{ww}
        + \frac{\partial \varphi}{\partial w}
        \partial_\varphi S_{ww}\,,
  \label{eq:mydTwwdw}
\end{equation}
and we can substitute
\begin{equation}
  S_{ww} = \mathcal{J}^{-\mathcal{W}} \left(
        \frac{\partial \rho}{\partial w} \frac{\partial \rho}{\partial w}
        S_{\rho \rho}
        + 2\frac{\partial \rho}{\partial w} \frac{\partial \varphi}{\partial w}
        S_{\rho \varphi}
        + \frac{\partial \varphi}{\partial w}\frac{\partial \varphi}{\partial w}
        S_{\varphi \varphi}
        \right)\,.
  \label{eq:myTww}
\end{equation}
Inserting (\ref{eq:myTww}) into (\ref{eq:mydTwwdw}) and setting
$S_{\rho \varphi}=0$ yields a lengthy expression involving derivatives of
$S_{\rho \rho}$ and $S_{\varphi \varphi}$ with respect to $\rho$ and
$\varphi$. The latter vanish due to symmetry and we substitute for
the $\rho$ derivatives using
\begin{eqnarray}
  \partial_{\rho} S_{\rho \rho} &=&
        \left( \frac{\partial y}{\partial \rho} \partial_y
               + \frac{\partial w}{\partial \rho} \partial_w \right)
        \left[ \mathcal{J}^{\mathcal{W}} \left(
               \frac{\partial y}{\partial \rho}
               \frac{\partial y}{\partial \rho} S_{yy}
               + 2 \frac{\partial y}{\partial \rho}
               \frac{\partial w}{\partial \rho} S_{yw}
               + \frac{\partial w}{\partial \rho}
               \frac{\partial w}{\partial \rho} S_{ww} \right) \right]\,,
        \nonumber \\
  \partial_{\rho} S_{\varphi \varphi} &=&
        \left( \frac{\partial y}{\partial \rho} \partial_y
               + \frac{\partial w}{\partial \rho} \partial_w \right)
        \left[ \mathcal{J}^{\mathcal{W}} \left(
               \frac{\partial y}{\partial \varphi}
               \frac{\partial y}{\partial \varphi} S_{yy}
               + 2 \frac{\partial y}{\partial \varphi}
               \frac{\partial w}{\partial \varphi} S_{yw}
               + \frac{\partial w}{\partial \varphi}
               \frac{\partial w}{\partial \varphi} S_{ww} \right) \right]\,.
\end{eqnarray}
This gives a lengthy expression relating the $y$ and $w$ derivatives of
$S_{ww}$. Finally, we recall that we need these derivatives
in the $xyz$ hyperplane and therefore set $w=0$. In order to obtain an
expression for the second $w$ derivative of $S_{ww}$, we first differentiate
the expression with respect to $w$ and then set $w=0$. The final result
is given by
\begin{equation}
  \partial_w S_{ww} = 0\,,~~~~~\partial_w \partial_w S_{ww}
        = \frac{\partial_y S_{ww}}{y}
        + 2\frac{S_{yy} - S_{ww}}{y^2}\,.
  \label{eq:dwTww12}
\end{equation}
Note that the density weight dropped out of this calculation, so that
Eq.~(\ref{eq:dwTww12}) is valid for the BSSN variables $\tilde{A}_{\mu \nu}$
and $\tilde{\gamma}_{\mu \nu}$ as well.

Applying a similar procedure to all components of scalar, vector
and symmetric tensor densities gives all expressions necessary to
trade derivatives off the $xyz$ hyperplane for those inside it.
We summarize the expressions recalling our notation:
a late Latin index, $i=1,\ldots,3$
stands for either $x$, $y$ or $z$ whereas an early
Latin index, $a=4,\ldots,D-1$ represents any of the $w^a$ directions.
For scalar, vector and tensor fields $\Psi$, $V$ and $T$
we obtain
\begin{eqnarray}
  0 &=& \partial_a \Psi =
  \partial_i \partial_a \Psi =
  V^a =
  \partial_i V^a =
  \partial_a \partial_b V^c =
  \partial_a V^i =
  \partial_a S_{bc} =
  \partial_i \partial_a S_{bc} =
  S_{ia} \nonumber \\
  &=& \partial_a \partial_b S_{ic} =
  \partial_a S_{ij} =
  \partial_i \partial_a S_{jk}\,, \nonumber \\
  \partial_a \partial_b \Psi &=& \delta_{ab} \frac{\partial_y \Psi}{y} \,,
        \nonumber \\
  \partial_a V^b &=& \delta^b{}_a \frac{V^y}{y}\,, \nonumber \\
  \partial_i \partial_a V^b &=& \delta^b{}_{a} \left(
        \frac{\partial_i V^y}{y} - \delta_{iy} \frac{V^y}{y^2} \right)\,,
        \nonumber \\
  \partial_a \partial_b V^i &=& \delta_{ab} \left(
        \frac{\partial_y V^i}{y} - \delta^{i}_{y} \frac{V^y}{y^2} \right) \,,
        \nonumber \\
  S_{ab} &=& \delta_{ab} S_{ww}\,, \nonumber \\
  \partial_a \partial_b S_{cd} &=& \left( \delta_{ac} \delta_{bd}
        + \delta_{ad} \delta_{bc} \right) \frac{S_{yy}-S_{ww}}{y^2}
        + \delta_{ab} \delta_{cd} \frac{\partial_y S_{ww}}{y}\,,
        \nonumber \\
  \partial_a S_{ib} &=& \delta_{ab} \frac{S_{iy} - \delta_{iy} S_{ww}}{y}\,,
        \nonumber \\
  \partial_i \partial_a S_{jb} &=& \delta_{ab} \left(
        \frac{\partial_i S_{jy} - \delta_{jy} \partial_i S_{ww}}{y}
        - \delta_{iy} \frac{S_{jy}-\delta_{jy} S_{ww}}{y^2} \right)\,,
        \nonumber \\
  \partial_a \partial_b S_{ij} &=& \delta_{ab} \left(
        \frac{\partial_y S_{ij}}{y}
        - \frac{\delta_{iy} S_{jy} + \delta_{jy} S_{iy}
          - 2\delta_{iy} \delta_{jy} S_{ww}}{y^2} \right)\,.
\end{eqnarray}
By trading or eliminating derivatives using these relations, a numerical
code can be written to evolve $D$-dimensional spacetimes with $SO(D-3)$
symmetry on a strictly three-dimensional computational grid. We finally
note that $y$ is a quasi-radial variable so that $y\ge0$.

\subsection{Initial data}
\label{sec:initdata}
In Section~\ref{sec:NRFormulations} we have discussed different ways
of casting the Einstein equations into a form suitable for numerical
simulations. At the start of Section~\ref{sec:NR}, we have
listed a number of additional ingredients
that need to be included for a complete numerical study and physical
analysis of BH spacetimes. We will now discuss the main choices
used in practical computations to address these remaining items, starting with the initial conditions.

As we have seen in Section~\ref{sec:NRFormulations}, initial data to be
used in time evolutions of the Einstein equations need to satisfy the
Hamiltonian and momentum constraints~(\ref{eq:HamADM}), (\ref{eq:momADM}).
A comprehensive
overview of the approach to generate BH initial data is given
by Cook's Living Reviews article~\cite{Cook:2000vr}. Here we merely summarize
the key concepts used in the construction of vacuum initial data,
but discuss in some more detail how solutions to the constraint
equations can be generated in the presence of specific matter fields
that play an important role in the applications discussed in
Section~\ref{sec:NRapplications}.

One obvious way to obtain constraint-satisfying initial data is
to directly use analytical solutions to the Einstein equations as
for example the Schwarzschild solution in $D=4$ in isotropic
coordinates
\begin{equation}
  ds^2 = -\left(\frac{M-2r}{M+2r}\right)^2 \, dt^2
        + \left( 1+ \frac{M}{2r} \right)^4
          \left[ dr^2 + r^2 (d\theta^2 + \sin^2 \theta\,d\phi^2)\right]\,.
\end{equation}
Naturally, the numerical evolution of an analytically known spacetime
solution does not generate new physical insight. It still serves as
an important way to test numerical codes and, more importantly,
analytically known solutions often form the starting point to
construct generalized classes of initial data whose time evolution
is not known without numerical study. Classic examples of such
analytic initial data are the Misner~\cite{Misner:1960zz}
and Brill--Lindquist~\cite{Brill:1963yv} solutions
describing $n$ non-spinning
BHs at the moment of time symmetry. In Cartesian coordinates,
the Brill--Lindquist data generalized to arbitrary $D$ are given by
\begin{equation}
  K_{\I \J} = 0\,,~~~
  \gamma_{\I \J} = \psi^{4/(D-3)} \delta_{\I \J}\,,~~~
  \psi = 1 + \sum_A \frac{\mu_A}{4 \left[\sum_{\K=1}^{D-1}
         (x^{\K} - x_0^{\K})^2\right]^{(D-3)/2}} \,,\label{BLid}
\end{equation}
%
where the summations over $A$ and $\K$ run over the number of BHs and the
spatial coordinates, respectively, and $\mu_A$ are parameters related to the
mass of the $A$-th BH through the surface area $\Omega_{D-2}$ of the
$(D-2)$-dimensional sphere by $\mu_A = 16\pi M / [(D-2)\Omega_{D-2}]$.
We remark that in the case of a
single BH, the Brill--Lindquist initial data (\ref{BLid}) reduce to the
Schwarzschild spacetime in Cartesian, isotropic coordinates (see
Eq.~(\ref{gciso}) in Section~\ref{global}).

A systematic way to generate solutions to the constraints describing BHs in $D=4$ dimensions
is based on the York--Lichnerowicz split~\cite{Lichnerowicz1944,
York:1971hw, York:1972sj}.
This split
employs a conformal spatial metric defined by $\gamma_{ij} = \psi^4
\bar{\gamma}_{ij}$; note that in contrast to the BSSN variable
$\tilde{\gamma}_{ij}$, in general $\det \bar{\gamma}_{ij}\ne 1$.
Applying a \emph{conformal traceless split}
to the extrinsic curvature according to
\begin{equation}
  K_{ij} = A_{ij} + \frac{1}{3} \gamma_{ij} K\,,~~~
  A^{ij} = \psi^{-10} \bar{A}^{ij}~~~\Leftrightarrow~~~
  A_{ij} = \psi^{-2} \bar{A}_{ij}\,,
  \label{eq:CTsplit}
\end{equation}
and further decomposing $\bar{A}_{ij}$ into a longitudinal and a
transverse traceless part, the momentum constraints simplify significantly;
see~\cite{Cook:2000vr} for details as well as a discussion of the
alternative \emph{physical transverse-traceless split} and
\emph{conformal thin-sandwich decomposition}~\cite{York1999}.
The conformal thin-sandwich
approach, in particular, provides a method to generate initial
data for the lapse and shift which minimize the initial rate of
change of the spatial metric, i.e., data in a quasi-equilibrium
configuration~\cite{Cook2004, Caudill2006}.

Under the further assumption of vanishing trace of the extrinsic curvature
$K=0$, a flat conformal metric $\bar{\gamma}_{ij} = f_{ij}$, where
$f_{ij}$ describes a flat Euclidean space, and asymptotic flatness
$\lim_{r\rightarrow \infty}\psi =1$, the momentum constraint admits an
analytic solution known as Bowen--York data~\cite{Bowen:1980yu}
\begin{equation}
  \bar{A}_{ij} = \frac{3}{2r^2} \left[ P_i n_j + P_j n_i
        - (f_{ij} - n_i n_j) P^k n_k \right]
        + \frac{3}{r^3} \left( \epsilon_{kil} S^l n^k n_j
          + \epsilon_{kjl} S^l n^k n_i \right)\,,
  \label{eq:BYAij}
\end{equation}
with $r=\sqrt{x^2+y^2+z^2}$, $n^i = x^i/r$ the unit radial vector and
user-specified parameters $P^i$, $S^i$. By calculating the momentum associated with the
asymptotic translational and rotational Killing vectors $\xi^i_{(k)}$
\cite{York1980}, one can show that $P^i$ and $S^i$ represent the components of
the total linear and angular momentum of the initial hypersurface.  The
linearity of the momentum constraint further allows us to superpose solutions
$\bar{A}^{(a)}_{ij}$ of the type (\ref{eq:BYAij}) and the total linear momentum
is merely obtained by summing the individual $P_{(a)}^i$.  The total angular
momentum is given by the sum of the individual spins $S_{(a)}^i$ plus an
additional contribution representing the orbital angular momentum. For the
generalization of Misner data, it is necessary to construct inversion-symmetric
solutions of the type (\ref{eq:BYAij}) using the method of images
\cite{Bowen:1980yu, Cook:2000vr}. Such a procedure is not required for
generalizing Brill-Lindquist data where a superposition of solutions
$\bar{A}^{(a)}_{ij}$ of the type (\ref{eq:BYAij}) can be used directly to
calculate the extrinsic curvature from Eq.~(\ref{eq:CTsplit}) and insert the
resulting expressions into the vacuum Hamiltonian constraint given with the
above listed simplifications by
\begin{equation}
  \bar{\nabla}^2 \psi + \frac{1}{8} K^{mn}K_{mn} \psi^{-7} = 0\,,
  \label{eq:PunctureHam}
\end{equation}
where $\bar{\nabla}^2$ is the Laplace operator associated with the
flat metric $f_{ij}$. This elliptic equation is commonly solved
by decomposing $\psi$ into a Brill-Lindquist piece
$\psi_{\rm BL}=\sum_{a=1}^N m_a/|\vec{r}-\vec{r}_a|$ plus
a regular piece $u=\psi-\psi_{\rm BL}$, where $\vec{r}_a$ denotes
the location of the $a$-th BH and $m_a$ a parameter that determines
the BH mass and is sometimes referred to as the \emph{bare mass}.
Brandt \& Br{\"u}gmann~\cite{Brandt:1997tf} have proven existence
and uniqueness of $C^2$ regular solutions $u$ to
Eq.~(\ref{eq:PunctureHam}) and the resulting \emph{puncture} data
are the starting point of the majority of numerical BH evolutions
using the BSSN moving puncture technique.
The simplest example of this type of initial data is given by
Schwarzschild's solution in isotropic coordinates where
\begin{equation}
  K_{mn}=0\,,~~~~~~~~\psi=1+\frac{m}{2r}\,.
  \label{eq:isotropicpsi}
\end{equation}
In particular, this solution admits the isometry $r\rightarrow
m^2/(4r)$ which leaves the coordinate sphere $r=m/2$ invariant, but
maps the entire asymptotically flat spacetime $r>m/2$ into the interior
and vice versa. The solution therefore consists of 2 asymptotically
flat regions connected by a ``throat'' and spatial infinity of the
far region is compactified into the single point $r=0$ which
is commonly referred to as the \emph{puncture}. Originally, time evolutions
of puncture initial data split the conformal factor,
in analogy to the initial-data construction, into a singular
Brill-Lindquist contribution given by the $\psi$ in
Eq.~(\ref{eq:isotropicpsi}) plus a deviation $u$ that is regular
everywhere; cf.~Section~IV B in~\cite{Alcubierre:2002kk}. In this
approach, the puncture locations remain fixed on the computational
domain. The simulations through inspiral and
merger by~\cite{Campanelli:2005dd,Baker:2005vv}, in contrast,
evolve the entire conformal factor using gauge conditions
that allow for the puncture to move across the domain
and are therefore often referred to as ``moving puncture evolutions''.

In spite of its popularity, there remain a few caveats with puncture
data that have inspired explorations of alternative initial data.
In particular, it has been shown that there exist
no maximal, conformally flat spatial slices of the Kerr
spacetime~\cite{Garat2000,ValienteKroon:2003ux}.
Constructing
puncture data of a single BH with non-zero Bowen--York parameter
$S^i$ will therefore inevitably result in a hypersurface
containing a BH plus some additional content which typically manifests
itself in numerical evolutions as spurious GWs,
colloquially referred to as ``junk radiation''. For rotation parameters
close to the limit of extremal Kerr BHs, the amount of spurious
radiation rapidly increases leading to an upper limit of the
dimensionless spin parameter $J/M^2 \approx 0.93$ for conformally
flat Bowen--York-type data~\cite{Cook1989, Dain:2002ee, Dain:2008ck,
Lovelace:2008tw}; BH initial data of Bowen--York type with a spin parameter
above this value rapidly relax to rotating BHs with spin
$\chi \approx 0.93$, probably through absorption of some fraction
of the spurious radiation. This limit has been overcome~\cite{Lovelace:2008tw,Lovelace:2010ne}
by instead constructing initial data with an
extended version of the conformal thin-sandwich method using
superposed Kerr--Schild BHs~\cite{Kerr1965}.
In an alternative approach, most of the above outlined puncture method
is applied but using a non-flat conformal metric; see for instance
\cite{Krivan:1998td,Hannam:2006zt}.

In practice, puncture data are the method-of-choice for most
evolutions performed with the BSSN-moving-puncture technique%
\epubtkFootnote{Generalizations to higher dimensions have been studied
  for Einstein gravity~\cite{Zilhao:2011yc}
    and for five-dimensional
  Gauss--Bonnet gravity~\cite{Yoshino:2011qp}.}
whereas GHG evolution schemes commonly start from conformal thin-sandwich
data using either conformally flat or Kerr--Schild background data.
Alternatively to both these approaches, initial data containing scalar fields which rapidly collapse to one
or more BHs has also been employed~\cite{Pretorius:2005gq}. \\[5pt]

The constraint equations in the presence of matter become more complex. A
simple procedure can however be used to yield analytic solutions to
the initial data problem in the presence of minimally coupled scalar
fields~\cite{Okawa:2014nda,Okawa:2013jba}.  Although in general the
constraints~\eqref{eq:HamADM}\,--\,\eqref{eq:momADM} have to be solved
numerically, there is a large class of analytic or semi-analytic initial
data for the Einstein equations extended to include scalar fields. The
construction of constraint-satisfying initial data starts from a conformal
transformation of the ADM variables~\cite{Cook:2000vr}
\begin{eqnarray}
  \label{eq:conformaltrafoID}
  \gamma_{ij}&=&\psi^4\bar{\gamma}_{ij} \,,\quad
       \bar{\gamma}=\det\bar{\gamma}_{ij} = 1 \,,\\
  K_{ij} &=& A_{ij} +\tfrac{1}{3}\gamma_{ij}K\,,\quad
        A_{ij}=\psi^{-2}\bar{A}_{ij}\,,
\end{eqnarray}
which can be used to re-write the constraints as 
\begin{eqnarray}
  \label{eq:ConstraintsID}
  \mathcal{H}&=& \bar{\bigtriangleup}\psi -\tfrac{1}{8}\bar{R}\psi
        -\tfrac{1}{12}K^2\psi^{5} +\tfrac{1}{8}\bar{A}^{ij}
         \bar{A}_{ij}\psi^{-7}+\pi\psi\left[ \bar{D}^{i}
         \Phi^{\ast}\bar{D}_{i}\Phi + \psi^{4}\left(\Pi^{\ast}\Pi
         + \mu_{S}^{2}\Phi^{\ast}\Phi\right) \right],\\
  \label{eq:mom_const_t}
  \mathcal{M}_i &=& \bar{D}_j\bar{A}^j_i -\tfrac{2}{3}\psi^{6}\bar{D}_i
        K -4\pi \psi^{6}\left(\Pi^{\ast}\bar{D}_i\Phi+\Pi\bar{D}_i
        \Phi^{\ast}\right)\,.
\end{eqnarray}
Here $\bar{\bigtriangleup}=\bar{\gamma}^{ij}\bar{D}_i\bar{D}_j$,
$\bar{D}$ and $\bar{R}$ denote the conformal covariant derivative and
Ricci scalar and $\Pi$ is a time reduction variable defined
in~\eqref{eq:defKijKPhi}.

Take for simplicity a single, non-rotating BH surrounded
by a scalar field (more general cases are studied in
Ref.~\cite{Okawa:2014nda,Okawa:2013jba}).
If we adopt the maximal slicing condition $K=0$ and set
$\bar{A}_{ij}=0,\Phi=0$, then the momentum constraint is immediately
satisfied, and one is left with the the Hamiltonian constraint, which
for conformal flatness, i.e., $\bar{\gamma}_{ij}=f_{ij}$ reads
\begin{eqnarray}
  \label{eq:ham_const_k0}
  \bigtriangleup_{\rm flat} \psi =\left[
        \frac{1}{r^2}\frac{\partial}{\partial r}r^2\frac{\partial}{\partial r}
       +\frac{1}{r^2\sin\theta}\frac{\partial}{\partial \theta}
        \sin\theta\frac{\partial}{\partial \theta}
       +\frac{1}{r^2\sin^2\theta}\frac{\partial^2}{\partial \Phi^2}
        \right] \psi= -\pi\psi^{5} \Pi\, \Pi^*
        \,.
\end{eqnarray}
The ansatz
\begin{eqnarray}
  \label{eq:ansatz_Pi_SBH}
  \Pi  &=& \frac{\psi^{-5/2}}{\sqrt{r\pi}} F(r)Z(\theta,\phi)\,,\\
  \label{eq:ansatz_conf_SBH}
  \psi &=&1 +\frac{M}{2r} +\sum_{lm} \frac{u_{lm}(r)}{r} Y_{lm}(\theta,\phi)
  \,,
\end{eqnarray}
reduces the Hamiltonian constraint to
\begin{equation}
  \label{eq:usol}
  \sum_{lm} \left(u_{lm}''-\frac{l(l+1)}{r^2}u_{lm}\right)
        Y_{lm}=-F(r)^2Z(\theta, \phi)^2\,.
\end{equation}
By a judicious choice of the angular function $Z(\theta,\phi)$,
or in other words, by projecting $Z(\theta,\phi)$ onto spherical
harmonics $Y_{lm}$, the above equation reduces to a single second-order,
ordinary differential equation.  Thus, the complex problem of finding
appropriate initial data for massive scalar fields was reduced to
an almost trivial problem, which admits some interesting analytical
solutions~\cite{Okawa:2014nda,Okawa:2013jba}.  Let us focus for
definiteness on spherically symmetric solutions (we refer the reader to
Ref.~\cite{Okawa:2014nda,Okawa:2013jba} for the general case), by taking
a Gaussian-type solution ansatz,
\begin{equation}
  \label{eq:ansatz_l0m0}
  Z(\theta,\phi)=\frac{1}{\sqrt{4\pi}}\,,\quad F(r)= A_{00}\times
        \sqrt{r}e^{-\tfrac{(r-r_0)^2}{w^2}}\,,
\end{equation}
where $A_{00}$ is the scalar field amplitude and $r_0$ and $w$ are the
location of the center of the Gaussian and its width.
By solving Eq.~\eqref{eq:usol}, we obtain the only non-vanishing component
of $u_{lm}(r)$
\begin{equation}
  \label{eq:u00}
  u_{00}=A_{00}^2\frac{w[w^2-4r_0(r-r_0)]}{16\sqrt{2}}
        \left[\erf{\left(\frac{\sqrt{2}(r-r_0)}{w}\right)}-1\right]
        -A_{00}^2\frac{r_0w^2}{8\sqrt{\pi}}e^{-2(r-r_0)^2/w^2}\,,
\end{equation}
where we have imposed that $u_{lm}\to 0$ at infinity. Other solutions
can be obtained by adding a constant to \eqref{eq:u00}.
%
\subsection{Gauge conditions}
\label{sec:gauge}

We have seen in Section~\ref{sec:NRFormulations}, that the Einstein equations
do not make any predictions about the gauge functions; the ADM equations
leave lapse $\alpha$ and shift $\beta^i$ unspecified and the GHG
equations make no predictions about the source functions $H^{\alpha}$.
Instead, these functions can be freely specified by the user and
represent the coordinate or gauge-invariance of the
theory of GR. Whereas the physical properties of a
spacetime remain unchanged under gauge transformations,
the performance of numerical evolution schemes depends
sensitively on the gauge choice. It is well-known, for example,
that evolutions of the Schwarzschild spacetime employing geodesic slicing
$\alpha=1$ and vanishing shift $\beta^i=0$ inevitably reach a
hypersurface containing the
BH singularity after a coordinate time interval $t=\pi M$
\cite{Smarr:1977uf}; computers respond to singular functions with
non-assigned numbers which rapidly swamp the entire
computational domain and render further evolution in time practically
useless. This problem can be avoided by controlling the lapse
function such that the evolution in proper time slows down
in the vicinity of singular points in the spacetime
\cite{Estabrook:1973ue}.
Such slicing conditions are called \emph{singularity avoiding}
and have been studied systematically in the form of the
Bona-Mass{\'o} family of slicing conditions~\cite{Bona:1994dr};
see also~\cite{Garfinkle:2001ni, Alcubierre:2002iq}.
A potential problem arising
from the use of singularity avoiding slicing is the different progress
in proper time in different regions of the computational domain
resulting in a phenomenon often referred to as ``grid stretching''
or ``slice stretching'' which can be compensated
with suitable non-zero choices for the shift vector
\cite{Alcubierre:2002kk}.

The particular coordinate conditions used with great success
in the BSSN-based moving puncture approach~\cite{Campanelli:2005dd,
Baker:2005vv} in $D=4$ dimensions
are variants of the ``1+log'' slicing and ``$\Gamma$-driver''
shift condition~\cite{Alcubierre:2002kk}
\begin{eqnarray}
  \partial_t \alpha &=& \beta^m \partial_m \alpha - 2 \alpha K\,,
  \label{eq:BSSNdtalpha1} \\
  \partial_t \beta^i &=& \beta^m \partial_m \beta^i + \frac{3}{4} B^i\,,
  \label{eq:BSSNdtbeta1} \\
  \partial_t B^i &=& \beta^m \partial_m B^i + \partial_t
        \tilde{\Upgamma}^i - \eta B^i\,.
  \label{eq:BSSNdBt1}
\end{eqnarray}
We note that the variable $B^i$ introduced here is an auxiliary variable
to write the second-order-in-time equation for the shift vector
as a first-order system and has no relation with the variable of the same
name introduced in Eq.~(\ref{eq:metricCho}).
The ``damping'' factor $\eta$ in Eq.~(\ref{eq:BSSNdBt1})
is specified either as a constant, a function depending
on the coordinates $x^i$ and BH parameters~\cite{Schnetter:2010cz},
a function of the BSSN variables~\cite{Mueller:2009jx,
Mueller:2010zze}, or evolved as an independent variable
\cite{Alic:2010wu}. A first-order-in-time evolution equation
for $\beta^i$ has been suggested in
\cite{vanMeter:2006vi} which results from integration
of Eqs.~(\ref{eq:BSSNdtbeta1}), (\ref{eq:BSSNdBt1})
\begin{equation}
  \partial_t \beta^i = \beta^m \partial_m \beta^i
        + \frac{3}{4} \tilde{\Upgamma}^i - \eta \beta^i\,.
  \label{eq:BSSNdtbeta2}
\end{equation}
Some NR codes omit the advection derivatives
of the form $\beta^m \partial_m$ in
Eqs.~(\ref{eq:BSSNdtalpha1})\,--\,(\ref{eq:BSSNdtbeta2}).
Long-term stable numerical simulations of BHs in higher dimensions require modifications
in the coefficients in Eqs.~(\ref{eq:BSSNdtalpha1})\,--\,(\ref{eq:BSSNdtbeta2})
\cite{Shibata:2010wz} and/or the addition of extra terms~\cite{Zilhao:2010sr}.
Reference~\cite{Etienne:2014tia} recently suggested a modification of
Eq.~(\ref{eq:BSSNdtalpha1}) for the lapse function $\alpha$ that
significantly reduces noise generated by a sharp initial gauge wave pulse
as it crosses mesh refinement boundaries.

BH simulations with the GHG formulation employ a wider range of
coordinate conditions. For example, Pretorius' breakthrough
evolutions~\cite{Pretorius:2005gq} set $H_i=0$ and
\begin{equation}
  \Box H_t = -\xi_1 \frac{\alpha - 1}{\alpha^\eta}
        + \xi_2 n^{\mu} \partial_{\mu} H_t\,,
\end{equation}
with parameters $\xi_1=19/m$, $\xi_2=2.5/m$, $\eta=5$ where $m$
denotes the mass of a single BH.
An alternative choice used with great success in
long binary BH inspiral simulations~\cite{Szilagyi:2009qz}
sets $H_{\alpha}$
such that the dynamics are minimized at early stages of the
evolution, gradually changes to harmonic gauge $H_{\alpha}=0$
during the binary inspiral and uses a damped harmonic gauge
near merger
\begin{equation}
  H_{\alpha} = \mu_0 \left[ \ln \left(\frac{\sqrt{\gamma}}{\alpha}
        \right) \right]^2
        \left[ \ln \left( \frac{\sqrt{\gamma}}{\alpha} \right) n_{\alpha}
        - \alpha^{-1} g_{\alpha m} \beta^m \right]\,,
\end{equation}
where $\mu_0$ is a free parameter. We note in this context that for $D=4$,
the GHG source functions $H^{\alpha}$ are related to the ADM lapse
and shift functions through~\cite{Pretorius:2004jg}
\begin{eqnarray}
  n^{\mu} H_{\mu} &=& -K -n^{\mu} \partial_{\mu} \ln \alpha\,, \\
  \gamma^{\mu i} H_{\mu} &=& -\gamma^{mn} \Upgamma^i_{mn}
        + \gamma^{im} \partial_m \ln \alpha
        + \frac{1}{\alpha} n^{\mu} \partial_{\mu} \beta^i\,.
\end{eqnarray}
%

\subsection{Discretization of the equations}
In the previous sections we have derived formulations of the Einstein equations
in the form of an IBVP. Given an initial snapshot of the physical system under
consideration, the evolution equations, as for example in the form of the BSSN
equations~(\ref{eq:BSSNdtphi})\,--\,(\ref{eq:BSSNdGamma}), then predict the evolution
of the system in time. These
evolution equations take the form of a set of non-linear partial differential
equations which relate a number of grid variables and their time and spatial
derivatives.  Computers, on the other hand, exclusively operate with (large sets
of) numbers and for a numerical simulation we need to translate the differential
equations into expressions relating arrays of numbers.

The common methods to implement this \emph{discretization} of
the equations are \emph{finite differencing},
the \emph{finite element}, \emph{finite volume} and \emph{spectral} methods.
Finite element and volume methods are popular choices in various computational
applications, but have as yet not been applied to time evolutions of BH
spacetimes. Spectral methods provide a particularly efficient and
accurate approach for numerical modelling provided the functions do not
develop discontinuities. Even though BH spacetimes contain singularities,
the use of singularity excision provides a tool to remove these from the
computational domain. 
This approach has been used with great success
in the SpEC code to evolve inspiralling and merging
BH binaries with very high accuracy; see, e.g., \cite{Boyle:2007ft,
Chu:2009md,Lovelace:2011nu}. Spectral methods have also been used successfully
for the modelling of spacetimes with high degrees of symmetry
\cite{Chesler:2008hg,Chesler:2009cy,Chesler:2010bi} and
play an important role in the construction of initial data
\cite{Ansorg:2004ds,Ansorg:2006gd,Zilhao:2011yc}. An indepth
discussion of spectral methods is given in the
Living Reviews article~\cite{Grandclement2009}.
The main advantage of finite differencing methods
is their comparative simplicity. Furthermore, they have proved
very robust in the modelling of rather extreme BH configurations
as for example BHs colliding near the
speed of light~\cite{Sperhake:2008ga,Okawa:2011fv,Sperhake:2012me}
or binaries with mass ratios up to $1:100$~\cite{Lousto:2010ut,Lousto:2010qx,
Sperhake:2011ik}.

\paragraph*{Mesh refinement and domain decomposition:}
BH spacetimes often involve lengthscales that differ by
orders of magnitude. The BH horizon extends over lengths of the
order $\mathcal{O}(1)~M$ where $M$ is the mass of the BH. Inspiralling
BH binaries, on the other hand, emit GWs with wavelengths
of $\mathcal{O}(10^2)~M$. Furthermore, GWs are rigorously
defined only at infinity. In practice, wave extraction is often performed at
finite radii but these need to be large enough to ensure that systematic
errors are small. In order to accomodate accurate wave extraction,
computational domains used for the modelling of asymptotically
flat BH spacetimes typically have a size of $\mathcal{O}(10^3)~M$.
With present computational infrastructure it is not possible to evolve
such large domains with a uniform, high resolution that is sufficient to
accurately model the steep profiles arising near the BH horizon.
The solution to this difficulty is the use of mesh refinement,
i.e., a grid
resolution that depends on the location in space and may also vary in time.
The use of mesh refinement in BH modelling is simplified by the remarkably
rigid nature of BHs which rarely exhibit complicated structure beyond
some mild deformation of a sphere. The requirements of increased
\epubtkImage{}{%
\begin{figure}[htb]
  \centerline{\includegraphics[height=180pt]{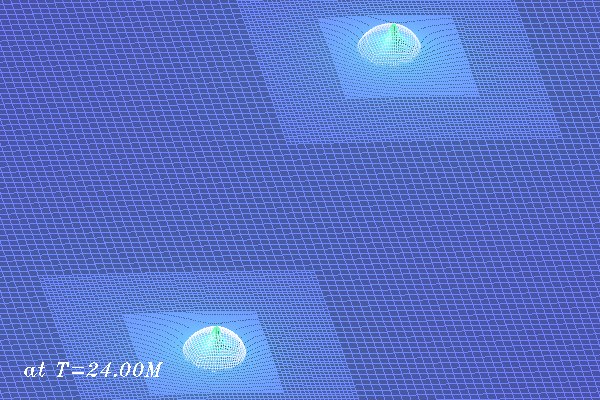}}
  \caption{Illustration of mesh refinement for a BH binary with one
        spatial dimension suppressed. Around each
        BH (marked by the spherical AH),
        two nested boxes are visible. These are immersed within one
        large, common grid or \emph{refinement level}.}
  \label{fig:moving_boxes}
\end{figure}}
resolution are therefore simpler to implement than, say, in the modelling of
airplanes or helicopters. In BH spacetimes the grid resolution must be highest
near the BH horizon and it decreases gradually at larger and larger distances
from the BH. In terms of the internal book-keeping, this allows for a
particularly efficient manner to arrange regions of refinement which is often
referred to as \emph{moving boxes}.  A set of nested boxes with outwardly
decreasing resolution is centered on each BH of the spacetime and follows the BH
motion.  These sets of boxes are immersed in one or more common boxes which are
large enough to accomodate those centered on the BHs. As the BHs approach each
other, boxes originally centered on the BHs merge into one and become part of
the common-box hierarchy.  A snapshot of such moving boxes is displayed in
Figure~\ref{fig:moving_boxes}.

Mesh refinement in NR has been pioneered by Choptuik in his seminal study on
critical phenomena in the collapse of scalar fields~\cite{Choptuik:1992jv}. The
first application of mesh refinement to the evolution of BH binaries was
performed by Br{\"u}gmann~\cite{Bruegmann:1996kz}. There exists a variety of mesh
refinement packages available for use in NR
including
\textsc{Bam}~\cite{Bruegmann:1996kz},
\textsc{Had}~\cite{HAD},
\textsc{Pamr/Amrd}~\cite{PAMRAMRD},
\textsc{Paramesh}~\cite{MacNeice2000},
\textsc{Samrai}~\cite{Samraiweb} and
the \textsc{Carpet}~\cite{Schnetter:2003rb, Carpetweb}
package integrated into the
\textsc{Cactus Computational Toolkit}~\cite{Cactusweb}.
For additional information on \textsc{Cactus} see also the
\textsc{Einstein Toolkit} webpage
\cite{EinsteinToolkit} and the lecture notes~\cite{Zilhao:2013hia}.
A particular mesh-refinement algorithm used for many BH applications
is the Berger--Oliger~\cite{Berger:1984zza} scheme where coarse
and fine levels communicate through interpolation in the form of the
\emph{prolongation} and \emph{restriction} operation;
see~\cite{Schnetter:2003rb} for details.
Alternatively, the different lengthscales
can be handled efficiently through the use of multiple domains of different
shapes. Communication between the individual subdomains is performed
either through overlaps or directly at the boundary for touching
domains. Details of this domain decomposition can be found in
\cite{Pfeiffer:2002wt,Buchman:2012dw} and references therein.

\subsection{Boundary conditions}
\label{sec:boundaries}
In NR, we typically encounter two types of physical boundaries, (i)
inner boundaries due to the treatment of spacetime singularities in BH
solutions and (ii) the outer boundary either at infinite distance from
the strong-field sources or, in the form of an approximation to this
scenario, at the outer edge of the computational domain at large but
finite distances.

\paragraph*{Singularity excision:}
BH spacetimes generically contain singularities, either physical
singularities with a divergent Ricci scalar or coordinate
singularities where the spacetime curvature is well behaved but some
tensor components approach zero or inifinite values. In the case of
the Schwarzschild solution in Schwarzschild coordinates, for example,
$r=0$ corresponds to a physical singularity whereas the singular
behaviour of the metric components $g_{tt}$ and $g_{rr}$ at $r=2M$
merely reflects the unsuitable nature of the coordinates as
$r\rightarrow 2M$ and can be cured, for example, by transforming to
Kruskal--Szekeres coordinates; cf.~for example Chapter~7 in
Ref.~\cite{Carroll:1997ar}. Both types of singularities give rise to
trouble in the numerical modelling of spacetimes because computers
only handle finite numbers. Some control is available in the form of
gauge conditions as discussed in Section~\ref{sec:gauge}; the
evolution of proper time is slowed down when the evolution gets close
to a singularity. In general, however, BH singularities require some
special numerical treatment.

\epubtkImage{}{%
\begin{figure}[htb]
  \centerline{\includegraphics[height=200pt]{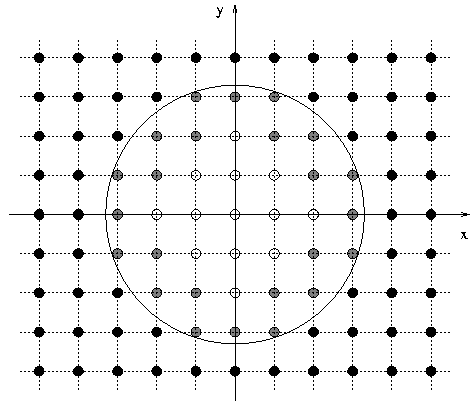}}
  \caption{Illustration of singularity excision. The small circles
           represent vertices of a numerical grid on a two-dimensional
           cross section of the computational domain containing the
           spacetime singularity, in this case at the origin.
           A finite region around the singularity but within the
           event horizon (large circle) is excluded from
           the numerical evolution (white circles). Gray circles
           represent the excision boundary where function values can
           be obtained through regular evolution in time using
           sideways derivative
           operators as appropriate (e.g.,~\cite{Pretorius:2004jg}) or
           regular update with spectral methods
           (e.g.,~\cite{Scheel:2002yj,Scheel:2006gg}), or
           through extrapolation (e.g.,~\cite{Shoemaker:2003td,Sperhake:2003fc}).
           The regular evolution of
           exterior grid points (black circles) is performed with
           standard techniques using information also on the excision
           boundary.}
  \label{fig:excision}
\end{figure}}

Such a treatment is most commonly achieved in the form of \emph{singularity}
or \emph{BH excision} originally suggested by Unruh as quoted in
\cite{Thornburg1987}. According to Penrose's cosmic
censorship conjecture, a spacetime singularity should be cloaked inside
an event horizon and the spacetime region outside the event horizon
is causally disconnected from the dynamics inside (see Section~\ref{sec_cc}).
The excision technique is based on the corresponding assumption that the numerical treatment
of the spacetime inside the horizon has no causal effect on the exterior.
In particular, excising a finite region around the singularity but within
the horizon should leave the exterior spacetime unaffected.
This is illustrated in Figure~\ref{fig:excision} where the excision region
is represented by small white circles which are excluded from the numerical
evolution. Regular grid points, represented in the figure by
black circles, on the other hand are evolved normally. As we have seen in the
previous section, the numerical evolution in time of functions at a particular
grid point typically requires information from neighbouring grid points.
The updating of variables at regular points therefore requires data
on the excision boundary represented in Figure~\ref{fig:excision} by gray
circles.
Inside the BH horizon, represented by the large circle in the
figure, however, information can only propagate inwards, so that
the variables on the excision boundary can be obtained through
use of sideways derivative operators (e.g.,~\cite{Pretorius:2004jg}),
extrapolation (e.g.,~\cite{Shoemaker:2003td,Sperhake:2003fc})
or regular update with spectral methods
(e.g.,~\cite{Scheel:2002yj,Scheel:2006gg}).
Singularity excision has been used with great success in many numerical
BH evolutions~\cite{Alcubierre:2000yz,Sperhake:2003fc,
Sperhake:2005uf,Pretorius:2005gq,Scheel:2006gg,Bantilan:2012vu,
Hemberger:2012jz}.

Quite remarkably, the moving puncture method for evolving BHs
does not employ any such specific numerical treatment near BH singularities,
but instead applies the same evolution procedure for points arbitrarily close
to singularities as for points far away and appears ``to get away with it''.
In view of the remarkable success of the moving puncture method,
various authors have explored the behaviour of the puncture singularity
in the case of a single Schwarzschild BH
\cite{Hannam:2006vv, Brown:2007nt, Brown:2007tb, Hannam:2008sg, Brown:2009ki,
Dennison:2010wd}. Initially, the puncture represents spatial infinity on the
\emph{other} side of the wormhole geometry compactified into a single point.
Under numerical evolution using moving puncture gauge conditions, however,
the region immediately around this singularity rapidly evolves into
a so-called trumpet geometry which is partially covered by the
numerical grid to an extent that depends on the numerical resolution;
cf.\ Figure~1 in~\cite{Brown:2009ki}. In practice, the singularity
falls through the inevitably finite resolution of the computational
grid which thus facilitates a \emph{natural excision}
of the spacetime singularity without the need of any special numerical
treatment.

\paragraph*{Outer boundary:}
Most physical scenarios of interest for NR involve
spatial domains of infinite extent and there arises the question how
these may be accomodated inside the finite memory of a computer system.
Probably the most elegant and rigorous method is to apply a spatial
compactification, i.e., a coordinate transformation that maps the entire
domain including spatial infinity to a finite coordinate range. Such
compactification is best achieved in characteristic formulations
of the Einstein equations where the spacetime foliation in terms
of ingoing and/or outgoing light cones may ensure adequate
resolution of in- or outgoing radiation
throughout the entire domain. In principle, such a compactification
can also be implemented in Cauchy-type formulations,
but here it typically leads to an increasing blueshift of
radiative signals as they propagate towards spatial infinity.
As a consequence, any discretization method applied will eventually
fail to resolve the propagating features.
This approach has been used in Pretorius' breakthrough
\cite{Pretorius:2005gq} and the effective damping of radiative signals at large
distances through underresolving them approximates a
no-ingoing-radiation boundary condition. An intriguing alternative
consists in using instead a space-time slicing of \emph{asymptotically} null
hypersurfaces which play a key role in the conformal field equations
\cite{Frauendiener:2004}. To our knowledge, this method has not
yet been applied successfully to BH simulations in either astrophysical
problems or simulations of the type reviewed here, but may well merit
more study in the future.

The vast majority of Cauchy-based NR applications
instead resort to an approximative treatment where the infinite spatial
domain is truncated and modeled as a compact domain with ``suitable''
outer boundary conditions. Ideally, the boundary conditions would
satisfy the following requirements~\cite{Rinne:2006vv}: They
(i) ensure well posedness of the IBVP,
(ii) are compatible with the constraint equations,
and (iii) correctly represent the physical conditions, which in almost
all practical applications means that they control or minimize
the ingoing gravitational radiation.

Boundary conditions meeting these requirements at least partially have been
studied most extensively for the harmonic or generalized harmonic
formulation of the Einstein equations
\cite{Kreiss:2006mi,Rinne:2006vv,Babiuc:2006ik,Rinne:2007ui,Ruiz:2007hg}.

For the BSSN system, such boundary conditions have as yet
not been identified and practical applications commonly apply
\emph{outgoing radiation} or
\emph{Sommerfeld} boundary conditions, which are an approximation in this context, where they
are applied at large but finite distances from the strong-field region.  Let us
assume, for this purpose, that a given grid variable $f$ asymptotes to a
constant background value $f_0$ in the limit of large $r$ and contains a leading
order deviation $u(t-r)/r^n$ from this value, where $u$ remains finite as
$r\rightarrow \infty$,  and $n$ is a constant positive
integer number. For $r\rightarrow \infty$, we therefore have
\begin{equation}
  f(t,r) = f_0 + \frac{u(t-r)}{r^n}\,,
\end{equation}
where the dependence on retarded time represents the outgoing nature of
the radiative deviations. In consequence, $\partial_t u + \partial_r u=0$,
which translates into the following conditon for the grid variable $f$
\begin{equation}
  \partial_t f + n \frac{f-f_0}{r} + \frac{x^{\I}}{r} \partial_{\I} f = 0\,,
\end{equation}
where $x^{\I}$ denote Cartesian coordinates. Because information only propagates outwards, the
spatial derivative $\partial_{\I} f$ is evaluated using a one-sided
stencil. This method is straightforwardly generalized to asymptotically expanding
cosmological spacetimes of dS type containing BHs; cf.~Eq.~(9)
in Ref.~\cite{Zilhao:2012bb}. Even though this approximation appears to work rather (one might be tempted
to say surprisingly) well in practice
\cite{Rinne:2007ui}, it is important to bear in mind the
following caveats. (i) The number of conditions imposed in this way
exceeds the number of ingoing characteristics calling into question the
well-posedness of the resulting system. (ii) Sommerfeld conditions are
not constraint satisfying which leads to systematic errors that do not
converge away as resolution is increased. (iii) Some spurious reflections
of gravitational waves may occur, especially when applied at too small
radii. These potential difficulties of BSSN evolutions have motivated
studies of generalizing the BSSN system, in particular the conformal
Z4 formulations discussed in Section~\ref{sec:BeyondBSSN} which
accomodate constraint preserving boundary conditions which facilitate
control of the ingoing gravitational radiation~\cite{Hilditch:2012fp}.

In asymptotically AdS spacetimes, the outer boundary represents a more
challenging problem and the difficulties just discussed are likely to impact
numerical simulations more severely if not handled appropriately.  This is
largely a consequence of the singular behaviour of the AdS metric even in the
absence of a BH or any matter sources. The AdS metric (see Section
\ref{sec:gauge-gravity}) is the maximally symmetric
solution to the Einstein equations (\ref{eq:EinsteinEqs_D}) with $T_{\alpha
  \beta}=0$ and $\Lambda < 0$.  This solution can be represented by the
hyperboloid $X_0^2 + X_D^2 - \sum_{i=1}^{D-1} X_i^2$ embedded in a flat
$D+1$-dimensional spacetime with metric
$ ds^2= -dX_0^2 - dX_D^2 + \sum_{i=1}^{D-1} dX_i^2\,.$
It can be represented in global coordinates, as
\begin{equation}
  ds^2 = \frac{L^2}{\cos^2\rho} (-d\tau^2 + d\rho^2 + \sin^2\rho \,
        d\Omega_{D-2}^2)\,, \label{eq:AdSglobal}
\end{equation}
where $0\le \rho <\pi/2,~-\pi < \tau \le \pi$ and
$\Lambda = -(D-1)(D-2)/(2L^2)$ (by unwrapping the cylindrical time
direction, the range of the time coordinate is often extended
to $\tau \in \mathbb{R}$), or in the Poincar\`e coordinates:
\begin{equation}
  ds^2 = \frac{L^2}{z^2} \left[ -dt^2 + dz^2 + \sum_{i=1}^{D-2}
        (dx^i)^2 \right]\,,
        \label{eq:AdSPoincare}
\end{equation}
with $z>0,~t\in \mathbb{R}$.  It can be shown that Poincar{\'e} coordinates only
cover half the hyperboloid and that the other half corresponds to $z<0$
\cite{Bayona:2005nq}.

Clearly, both the global (\ref{eq:AdSglobal}) and the Poincar{\'e}
(\ref{eq:AdSPoincare}) versions of the AdS metric become singular at
their respective outer boundaries $\rho \rightarrow \pi/2$ or
$z\rightarrow 0$. The induced metric at infinity is therefore
only defined up to a conformal rescaling. This remaining freedom
manifests itself in the boundary topology of the global and
Poincar{\'e} metrics which, respecitvely, become in the limit
$\rho \rightarrow \pi/2$ and $z\rightarrow 0$
\begin{equation}
  ds_{\rm gl}^2 \sim -d\tau^2 + d\Omega_{D-2}^2, \qquad
  ds_{\rm P}^2 \sim -dt^2 + \sum_{i=1}^{D-2} (dx^i)^2\,.
\end{equation}
In the context of the gauge/gravity duality, gravity in global or Poincar{\'e} AdS is related to CFTs on spacetimes of different topology:
$\mathbb{R}\times S_{D-2}$ in the former and $\mathbb{R}^{D-1}$
in the latter case.

The boundary treatment inside a numerical modelling of asymptotically
AdS spacetimes needs to take care of the singular nature of the
metric. In practice, this is achieved through some form of
regularization which makes use of the fact that the singular piece
of an asymptotically AdS spacetime is known in analytic form,
e.g., through Eqs.~(\ref{eq:AdSglobal}) or (\ref{eq:AdSPoincare}).
In Ref.~\cite{Bantilan:2012vu} the spacetime metric
is decomposed into an analytically known AdS part plus a deviation
which is regular at infinity.
In this approach, particular care needs to be taken of
the gauge conditions to ensure that the coordinates remain
compatible with this decomposition throughout the simulation.
An alternative approach consists in factoring out appropriate factors
involving the bulk coordinate as for example the term $\cos \rho$
in the denominator on the right-hand side of Eq.~(\ref{eq:AdSglobal}).
This method is employed in several recent works~\cite{Chesler:2010bi,Heller:2011ju,Bizon:2011gg}.

We finally note that the boundary plays an active role in AdS spacetimes.
The visualization of the AdS spacetime in the form of a
Penrose diagram demonstrates that it is not globally hyperbolic,
i.e., there exists no Cauchy surface on which initial
data can be specified in such a way that the entire future of the spacetime
is uniquely determined. This is in marked contrast to the
Minkowski spacetime. Put in other words, the outer boundary of
asymptotically flat spacetimes is represented in a Penrose diagram by a
null surface such that information cannot propagate from infinity
into the interior spacetime. In contrast, the outer boundary of asymptotically
AdS spacetimes is timelike and, hence, the outer boundary actively influences
the evolution of the interior. The specification of boundary conditions
in NR applications to the gauge/gravity duality or AdS/CFT
correspondence therefore reflects part of the description of the
physical system under study; cf.~Section~\ref{sec_holography7}.

\subsection{Diagnostics}
\label{sec:diagnostics}
Once we have numerically generated a spacetime, there still remains
the question of how to extract physical information from the large
chunk of numbers the computer has written to the hard drive. This
analysis of the data faces two main problems in NR
applications, (i) the gauge or coordinate dependence of the results
and (ii) the fact that many quantities we are familiar with from
Newtonian physics are hard or not even possible to define in a rigorous
fashion in GR. In spite of these difficulties,
a number of valuable diagnostic tools have been developed and the purpose
of this section is to review how these are extracted.

The physical information is often most conveniently calculated from
the ADM variables and we assume for this discussion that a numerical
solution is available in the form of the ADM variables
$\gamma_{\I \J}$, $K_{\I \J}$, $\alpha$ and $\beta^{\I}$. Even if the
time evolution has been performed using other variables as for example
the BSSN or GHG variables the conversion between these and their
ADM counterparts according to Eq.~(\ref{eq:BSSNvars}) or
(\ref{eq:3+1metric}) is straightforward.

One evident diagnostic directly arises from the structure of the Einstein
equations where the number of equations exceeds the number of free
variables; cf.\ the discussion following Eq.~(\ref{eq:momADM}).
Most numerical applications employ ``free evolutions''
where the evolution equations are used for updating the grid variables.
The constraints are thus not directly used in the numerical evolution
but need to be satisfied by any solution to the Einstein equations.
A convergence analysis of the constraints (see for example Figure~3
in Ref.~\cite{Sperhake:2006cy}) then provides an important consistency
check of the simulations.

Before reviewing the extraction of physical information from a numerical
simulation, we note a potential subtlety arising from the convention
used for Newton's constant in higher-dimensional spacetimes. We wrote
the Einstein equations in the form (\ref{eq:EinsteinEqs_D}) and chose units
where $G=1$ and $c=1$. 
This implies that the Einstein equations have the form $R_{\alpha\beta}-1/2Rg_{\alpha\beta}+\Lambda g_{\alpha\beta}=8\pi
GT_{\alpha\beta}$ for all spacetime dimensionalities
(here and in Section~\ref{global} we explicitly keep $G$ in the equations).
As we shall see below, with this convention the
Schwarzschild radius of a
static BH in $D$ dimensions is given by
\begin{equation}
  R_{\rm S}^{D-3} = \frac{16\pi G M}{(D-2)\Omega_{D-2}}\,,~~~~~
  \Omega_{D-2}= \frac{2\pi^{(D-1)/2}}{\Gamma\left( \frac{D-1}{2} \right)}\,,
  \label{eq:TangherliniRS}
\end{equation}
where $\Omega_{D-2}$ denotes the area of the unit $S^{D-2}$ sphere.

\subsubsection{Global quantities and horizons}
\label{global}
For spacetimes described by a metric that is asymptotically flat and
time independent, the total mass-energy and linear momentum are
given by the ADM mass and ADM momentum, respectively. These quantities arise
from boundary terms in the Hamiltonian of GR and
were derived by Arnowitt, Deser \& Misner~\cite{Arnowitt:1962hi}
in their canonical analysis of the theory. They are given in terms
of the ADM variables by
\begin{eqnarray}
  M_{\rm ADM} &=& \frac{1}{16\pi G} \lim_{r\rightarrow \infty}
        \oint_{S_r} \delta^{\M \N}
        (\partial_{\N} \gamma_{\M \K} - \partial_{\K}
        \gamma_{\M \N}) \hat{r}^{\K} dS\,,
        \label{eq:ADMmass} \\
  P_{\I} &=& \frac{1}{8\pi G} \lim_{r\rightarrow \infty}
        \oint_{S_r} (K_{\M \I} - \delta_{\M \I} K)
        \hat{r}^{\M} dS\,.
\end{eqnarray}
Here, the spatial tensor components $\gamma_{\I \J}$ and
$K_{\I \J}$ are assumed to be given in Cartesian coordinates,
$\hat{r}^{\M}=x^\M/r$ is the outgoing unit vector normal to the area
element $dS$ of the $S^{D-2}$ sphere and $dS=r^{D-2} \, d\Omega_{D-2}$.
The above integral is defined only for a restricted
class of coordinate systems, known as asymptotic Euclidian
coordinates for which the metric components are required to be of
the form $g_{\mu\nu}=\eta_{\mu\nu}+{\cal O}(1/r)$.
Under a more restrictive set of assumptions
about the fall-off behaviour of the metric and extrinsic curvature
components (see Secs.~7.5.1 and 7.5.2 in \cite{Gourgoulhon:2007ue}
and references therein for a detailed discussion),
one can also derive an expression for the global angular momentum
\begin{equation}
  J_{\I} = \frac{1}{8\pi} \lim_{r\rightarrow \infty}
        \oint_{S_r} (K_{\J \K} - K \gamma_{\J \K}) \xi_{(\boldsymbol{\I})}^{\J}
        \hat{r}^{\K} dS\,,
\end{equation}
where $\xi_{(\boldsymbol{\I})}$ are the Killing vector fields associated
with the asymptotic rotational symmetry given, in $D=4$, by
$\boldsymbol{\xi}_{(\boldsymbol{x})} =
         -z\boldsymbol{\partial}_y + y \boldsymbol{\partial}_z$,
$\boldsymbol{\xi}_{(\boldsymbol{y})} =
         -x\boldsymbol{\partial}_z + z \boldsymbol{\partial}_x$ and
$\boldsymbol{\xi}_{(\boldsymbol{z})} =
         -y\boldsymbol{\partial}_x + x \boldsymbol{\partial}_y$.
For a more-in-depth discussion of the ADM mass and
momentum as well as the conditions required for the definition
of the angular momentum the reader is referred to Section~7 of
\cite{Gourgoulhon:2007ue}.
Expressions for $M_{\rm ADM}$, $P_{\I}$ and $J_{\I}$
can also be derived in more
general (curvilinear) coordinate systems as long as the metric approaches the
flat-space form in those curvilinear coordinates at an appropriate rate;
see, e.g., Section~7 in~\cite{Gourgoulhon:2007ue} for a detailed review.

As an example, we calculate the ADM mass of
the $D$-dimensional Schwarzschild BH in
Cartesian, isotropic coordinates $(t,x^{\I})$
described by the spatial metric
\begin{equation}
  \gamma_{\I \J} = \psi^{\frac{4}{D-3}} \delta_{\I \J}\,,~~~~~
  \psi = 1+ \frac{\mu}{4r^{D-3}}\,,\label{gciso}
\end{equation}
and vanishing extrinsic curvature $K_{\I \J}=0$. A straightforward
calculation shows that
\begin{equation}
  \partial_{\K} \gamma_{\I \J} = -\psi^{\frac{4}{D-3}-1}\frac{\mu}{r^{D-1}}
        x_{\K} \delta_{\I \J}\,,
\end{equation}
%
so that (since $x^\K=r\hat{r}^\K$)
\begin{equation}
  \delta^{\M \N} (\partial_{\N} \gamma_{\M \K} - \partial_{\K} \gamma_{\M \N})
        \frac{x^{\K}}{r}
        = (D-2) \psi^{\frac{4}{D-3}-1} \frac{\mu x_{\K}}{r^{D-1}}
          \frac{x^{\K}}{r} = (D-2) \frac{\mu}{r^{D-2}}\,,
\end{equation}
where we have used the fact that in the limit $r\rightarrow \infty$
we can raise and lower indices with the Euclidean metric $\delta_{IJ}$
and $\psi \rightarrow 1$. From Eq.~(\ref{eq:ADMmass}) we thus obtain
\begin{eqnarray}
  M_{\rm ADM} &=& \frac{1}{16\pi G} \lim_{r\rightarrow \infty}
        \oint_{S_r} (D-2) \frac{\mu}{r^{D-2}} r^{D-2} \, d\Omega_{D-2}
        = \frac{D-2}{16\pi G} \mu \oint d\Omega_{D-2}
        \nonumber \\
    &=& \frac{D-2}{16\pi G}\mu \Omega_{D-2}
        = \frac{D-2}{16\pi G}\mu
          \frac{2\pi^{\frac{D-1}{2}}}{\Gamma\left( \frac{D-1}{2} \right)}\,.
  \label{eq:TangherliniMADM}
\end{eqnarray}
The Schwarzschild radius in areal coordinates is given by
$R_{\rm S}^{D-3} = \mu$ and we have recovered
Eq.~(\ref{eq:TangherliniRS}).

The event horizon is defined as the boundary between points in the spacetime
from which null geodesics can escape to infinity and points from which they
cannot. The event horizon is therefore by definition a concept that depends on
the entire spacetime. In the context of numerical simulations, this implies that
an event horizon can only be computed if information about the entire spacetime
is stored which results in large data sets even by contemporary
standards. Nevertheless, event horizon finders have been developed in
Refs.~\cite{Diener:2003jc,Cohen:2008wa}. For many purposes, however, it is more
convenient to determine the existence of a horizon using data from a spatial
hypersurface $\Sigma_t$ only. Such a tool is available in the form of an AH. AHs
are one of the most important diagnostic tools in NR and are reviewed in detail
in the Living Reviews article~\cite{Thornburg2007a}. It can be shown
under the assumption of cosmic censorship and reasonable energy conditions, that
the existence of an AH implies an event horizon whose cross section with
$\Sigma_t$ either lies outside the AH or coincides with it;
see~\cite{Hawking:1973uf,waldbook} for details and proofs.

The key concept underlying the AH
is that of a trapped surface defined as a surface where the expansion
$\Theta \equiv \nabla_{\mu} k^{\mu}$ of a congruence of outgoing null geodesics
with tangent vector $k^{\mu}$ satisfies $\Theta \le 0$. A marginally trapped surface is defined as
a surface where $\Theta=0$ and an AH is
defined as
the outermost marginally trapped surface on a spatial hypersurface $\Sigma_t$.
Translated into the ADM variables, the condition $\Theta=0$ can be
shown to lead to an elliptic equation for the unit normal direction
$s^{\I}$ to the $D-2$-dimensional horizon surface
\begin{equation}
  q^{\M \N} D_{\M} s_{\N} - K + K_{\M \N} s^{\M} s^{\N} = 0\,.
\end{equation}
Here, $q_{\M \N}$ denotes the $(D-2)$-dimensional metric induced
on the horizon surface. Numerical
algorithms to solve this equation have been developed by several
authors~\cite{Gundlach:1997us,Alcubierre:1998rq,
Thornburg:1995cp,Schnetter:2003pv,Thornburg:2003sf}.

In the case of a static, spherically symmetric BH,
it is possible to use 
the formula $A_{\rm hor} = \Omega_{D-2} R_{\rm S}^{D-2}$ for the area
of a $D-2$ sphere to eliminate $R_{\rm S}$ in Eq.~(\ref{eq:TangherliniRS}). 
We thus obtain an expression that relates the horizon area to a mass commonly referred
to as the \emph{irreducible mass}
\begin{equation}
  M_{\rm irr} = \frac{(D-2)\Omega_{D-2}}{16\pi G}
        \left( \frac{A_{\rm hor}}{\Omega_{D-2}} \right)^{\frac{D-3}{D-2}}\,.\label{irr_def}
\end{equation}
It is possible to derive the same expression in the more general case of a
stationary BH, such as the Kerr BH in $D=4$, or the Myers--Perry BH in $D>4$.

The irreducible mass, as defined by Eq.~\ref{irr_def}, is identical to the ADM mass for a static BH.
This equation can be used to define the irreducible mass for stationary BHs as well~\cite{Christodoulou:1970wf}.
In $D=4$ dimensions this becomes $16\pi\,G M_{\rm irr}^2=A_{\rm hor}$.
Furthermore, a rotating BH in $D=4$ is described by a single spin parameter
$S$ and the BH mass consisting of rest mass and rotational energy has
been shown by Christodoulou~\cite{Christodoulou:1970wf} to be given by
\begin{equation}
  M^2 = M_{\rm irr}^2 + \frac{S^2}{4G^2M_{\rm irr}^2}\,.
  \label{eq:MChr}
\end{equation}
By adding the square of the linear momentum $P^2$ to the right-hand side
of this equation we obtain the total energy of a spacetime containing
a single BH with spin $S$ and linear momentum $P$. In $D=4$,
Christodoulou's
formula (\ref{eq:MChr}) can be used to calculate the spin from the
equatorial circumference $C_e$ and the horizon area according to
\cite{Sperhake:2009jz}
\begin{equation}
  \frac{2\pi A_{\rm hor}}{C_e^2} = 1+\sqrt{1-j^2}\,,
\end{equation}
where $j=S/(GM^2)$ is the dimensionless spin parameter of the BH.
Even though this relation is strictly valid only for the case of single
stationary BHs, it provides a useful approximation in binary spacetimes
as long as the BHs are sufficiently far apart.

It is a remarkable feature of BHs that their local properties such as mass
and angular momentum can be determined in the way summarized here. In
general it is not possible to assign in such a well-defined manner a
local energy or momentum content to compact subsets of spacetimes
due to the non-linear nature of GR. For BHs, however, it is
possible to derive expressions analogous to the ADM integrals
discussed above, but now applied to the apparent horizon. Ultimately,
this feature rests on the \emph{dynamic} and \emph{isolated horizon}
framework; for more details see~\cite{Dreyer:2002mx,Ashtekar:2003hk}
and the Living Reviews article by Ashtekar \& Krishnan~\cite{Ashtekar:2004cn}.

\subsubsection{Gravitational wave extraction}
\label{sec:GWextraction}
Probably the most important physical quantity to be extracted from dynamical BH
spacetimes is the gravitational radiation. It is commonly extracted from
numerical simulations in the form of either the Newman--Penrose scalar or a
master function obtained through BH perturbation theory (see
Section~\ref{classicpert}). Simulations using a characteristic formulation also
facilitate wave extraction in the form of the Bondi mass loss formula. The
Landau--Lifshitz pseudo-tensor~\cite{landau1975classical}, which has been
generalized to $D>4$ in~\cite{Yoshino:2009xp}, has been used for gravitational
radiation extraction in Ref.~\cite{Shibata:2010wz} for studies of BH
stability in higher dimensions; for applications in $D=4$ see,
e.g.,~\cite{Lovelace:2009dg}. Here we will focus on the former two
methods; wave extraction using the Bondi formalism is discussed in
detail in Ref.~\cite{Winicour2012}.

\paragraph*{Newman--Penrose scalar:}
The formalism to extract GWs in the form of the Newman--Penrose scalar is
currently fully understood only in $D=4$ dimensions. Extension of this method is
likely to require an improved understanding of the Goldberg--Sachs theorem in
$D>4$ which is subject to ongoing research~\cite{Ortaggio:2012hc}.
The following discussion is therefore limited to $D=4$ and we shall further focus on the case of asymptotically flat
spacetimes. The Newman--Penrose formalism~\cite{Newman:1961qr} (see Section~\ref{classicpert}) 
is based on a tetrad of null vectors, two of them real and referred to as
$\boldsymbol{\ell}$, $\boldsymbol{k}$ in this work,
and two complex conjugate vectors referred to as $\boldsymbol{m}$
and $\boldsymbol{\bar{m}}$; cf.~Eq.~(\ref{eq:Psi4}) and
the surrounding discussion. Under certain
conditions the projections of the Weyl tensor onto these
tetrad directions may allow for a particularly convenient way to
identify the physical properties of the spacetime. More specifically,
the 10 independent components of the Weyl tensor are rearranged in the
form of 5 complex scalars defined as (see, e.g., \cite{Nerozzi:2008ng})
\begin{eqnarray}
  \Psi_0 &=& -C_{\alpha \beta \gamma \delta}
           k^{\alpha}
           m^{\beta}
           k^{\gamma}
           m^{\delta}\,, \nonumber \\
  \Psi_1 &=& -C_{\alpha \beta \gamma \delta}
           k^{\alpha}
           \ell^{\beta}
           k^{\gamma}
           m^{\delta}\,, \nonumber \\
  \Psi_2 &=& -C_{\alpha \beta \gamma \delta}
           k^{\alpha}
           m^{\beta}
           \bar{m}^{\gamma}
           \ell^{\delta}\,, \nonumber \\
  \Psi_3 &=& -C_{\alpha \beta \gamma \delta}
           k^{\alpha}
           \ell^{\beta}
           \bar{m}^{\gamma}
           \ell^{\delta}\,, \nonumber \\
  \Psi_4 &=& -C_{\alpha \beta \gamma \delta}
           \ell^{\alpha}
           \bar{m}^{\beta}
           \ell^{\gamma}
           \bar{m}^{\delta}\,.
  \label{eq:allPsi4}
\end{eqnarray}
The identification of these projections with gravitational radiation
is based on the work of Bondi et~al. and Sachs
\cite{Bondi:1962px,Sachs:1962wk} and the geometrical construction
of Penrose~\cite{Penrose:1962ij} but crucially relies on a correct
choice of the null tetrad in Eq.~(\ref{eq:allPsi4})
which needs to correspond to a Bondi frame.
One example of this type, frequently considered in numerical
applications, is the Kinnersley tetrad~\cite{Kinnersley:1969zza,Teukolsky:1973ap}.
More specifically, one employs a tetrad that converges to the Kinnersley tetrad
as the spacetime approaches Petrov type D\epubtkFootnote{``Petrov type D'' is a class of algebraically
special spacetimes, which includes in particular the Schwarzschild and Kerr
solutions.}.
Tetrads with this property are often referred to as
\emph{quasi-Kinnersley tetrads} and belong to a class of tetrads which
are related to each other by spin/boost transformations;
see~\cite{Beetle:2004wu,Nerozzi:2004wv,Zhang:2012ky} and references therein.
A particularly convenient choice consists in the transverse
frame where $\Psi_1=\Psi_3=0$ and the remaining scalars encode the
ingoing gravitational radiation ($\Psi_0$), the outgoing radiation
($\Psi_4$) and the static or \emph{Coulomb} part of the gravitational
field ($\Psi_2$). The construction of suitable tetrads in dynamic,
numerically generated spacetimes represents a non-trivial task
and is the subject of ongoing research
(see for example
\cite{Campanelli:2005ia,Lehner:2007ip,Nerozzi:2011pn,Zhang:2012ky}).

For reasons already discussed in Section~\ref{sec:boundaries},
extraction of gravitational waves is often performed
at finite distance from the sources; but see
Refs.~\cite{Reisswig:2009us, Babiuc:2010ze} for Cauchy-characteristic
extraction that facilitates GW calculation at future null infinity.
GW extraction at finite distances requires further ingredients
which are discussed in more detail in~\cite{Lehner:2007ip}.
These include a specific asymptotic behaviour of the Riemann tensor,
the so-called \emph{peeling property}
\cite{Sachs:1961zz,Sachs:1962wk,Newman:1961qr}, that outgoing
null hypersurfaces define sequences of $S^2$ spheres which are
conformal to unit spheres and a choice of coordinates that ensures
appropriate fall-off of the metric components in the
extraction frame.

Extraction of GWs at finite extraction radii $r_{\rm ex}$
is therefore affected by various potential errors. An attempt to estimate
the uncertainty arising from the use of finite $r_{\rm ex}$
consists in measuring the GW signal at different values of the
radius and analyzing its behaviour as the distance is increased.
Convergence of the signal as $1/r_{\rm ex} \rightarrow 0$
may then provide some estimate for the error incurred
and improved results may be obtained through extrapolation to
infinite $r_{\rm ex}$; see, e.g., \cite{Boyle:2009vi,Hinder:2013oqa}.
While such methods appear to work relatively well in practice (applying
balance arguments together with measurements of BH horizon masses
and the ADM mass or comparison with alternative extraction methods
provide useful checks), it is important
to bear in mind the possibility of systematic errors arising in the
extraction of GWs using this method.

In the following discussion we will assume
that the above requirements are met and describe a frequently used
recipe that leads from the metric components of a numerical simulation
to estimates of the energy and momenta contained in the gravitational
radiation.
The first step in the calculation of
$\Psi_4$ from the ADM metric is to construct the null tetrad.
An approximation to a quasi-Kinnersley tetrad is
given in terms of the unit timelike normal
vector $n^{\alpha}$ introduced in Section~\ref{sec:NRFormulations}, and a triad
$u^i$, $v^i$, $w^i$ of spatial vectors on each surface $\Sigma_t$ constructed
through Gram--Schmidt orthonormalization starting with
\begin{equation}
 u^i = [x,\,y,\,z]\,,~~~~~v^i=[xz,\,yz,\,-x^2-y^2]\,,~~~~~
 w^i = \epsilon^i{}_{mn} u^m v^n\,.
\end{equation}
Here $\epsilon^{imn}$ represents the three-dimensional Levi-Civita tensor on $\Sigma_t$
and $x,\,y,\,z$ are standard Cartesian coordinates. An orthonormal tetrad
is then obtained from
\begin{equation}
 k^{\alpha} = \frac{1}{\sqrt{2}}(n^\alpha + u^{\alpha})\,,
 ~~~~~
 \ell^{\alpha} = \frac{1}{\sqrt{2}}(n^{\alpha} - u^{\alpha})\,,
 ~~~~~
 m^{\alpha} = \frac{1}{\sqrt{2}}(v^{\alpha} + iw^{\alpha})\,,
 \label{eq:tetrad}
\end{equation}
where time components of the spatial triad vectors vanish by construction.

Then, the calculation of $\Psi_4$ from the ADM variables can be achieved either
by constructing the spacetime metric from the spatial metric, lapse and shift
vector and computing the spacetime Riemann or Weyl tensor through their
definitions (see the preamble on ``notation and conventions'').  Alternatively,
we can use the electric and magnetic parts of the Weyl tensor given by
\cite{Friedrich:1996hq} 
\begin{equation}
  E_{\alpha \beta} = \bot^{\mu}{}_{\alpha} \bot^{\nu}{}_{\beta}
        C_{\mu \rho \nu \sigma} n^{\rho} n^{\sigma}\,,~~~~~
  B_{\alpha \beta} = \bot^{\mu}{}_{\alpha} \bot^{\nu}{}_{\beta}
        \,{}^*C_{\mu \rho \nu \sigma} n^{\rho} n^{\sigma}\,,
\end{equation}
where the ${}^*$ denotes the Hodge dual. By using the Gauss--Codazzi
equations (\ref{eq:projections_Riemann}), one can express the
electric and magnetic parts in vacuum in terms of the ADM variables
according to%
\epubtkFootnote{The electric and magnetic part of the Weyl tensor may
  be interpreted as describing tidal effects and differential dragging
  of inertial frames, respectively, which has been employed to
  visualize spacetimes in terms of so-called ``Frame-Drag vortexes''
  and ``Tidal
  Tendexes''~\cite{Owen:2010fa,Nichols:2011pu,Nichols:2012jn,Zhang:2012jj}.}
\begin{equation}
  E_{ij} = \mathcal{R}_{ij} - \gamma^{mn}(K_{ij} K_{mn}
        - K_{im} K_{jn})\,,~~~~~
  B_{ij} = \gamma_{ik} \epsilon^{kmn}D_m K_{nj}\,.
  \label{eq:EBADM}
\end{equation}
In vacuum, the Weyl tensor is then given in terms of electric and magnetic parts
by Eq.~(3.10) in Ref.~\cite{Friedrich:1996hq}. Inserting this relation
together with (\ref{eq:tetrad}) and (\ref{eq:EBADM}) into the definition
(\ref{eq:allPsi4}) gives us the final expression for $\Psi_4$ in terms
of spatial variables
\begin{eqnarray}
  \Psi_4 &=& -\frac{1}{2} \left[
        E_{mn}(v^m v^n - w^m w^n) - B_{mn}(v^m w^n + w^m v^n) \right]
        \nonumber \\
     && + \frac{i}{2}
        \left[ E_{mn}(v^m w^n -w^mv^n) + B_{mn} ( w^m w^n + v^m v^n) \right]\,.
\end{eqnarray}
The GW signal is often presented in the form of multipolar components
$\psi_{\ell m}$ defined by projection of $\Psi_4$ onto spherical
harmonics of spin weight $-2$ ~\cite{Goldberg:1967sp}
\begin{equation}
  \Psi_4(t,\theta,\phi) = \sum_{l m} \psi_{l m}(t)
        Y^{(-2)}_{l m}(\theta, \phi)~\Leftrightarrow~
  \psi_{l m}(t) = \int \Psi_4(t,\theta,\phi)
        \overline{Y^{(-2)}_{l m}}(\theta,\phi) \, d\Omega_2\,,
  \label{eq:multipoles}
\end{equation}
where the bar denotes the complex conjugate. The $\psi_{lm}$
are often written in terms of amplitude and phase
\begin{equation}
  \psi_{lm} = A_{lm} e^{i\phi_{lm}}\,.
\end{equation}
The amount
of energy, linear and angular momentum carried by the GWs can be calculated
from $\Psi_4$ according to~\cite{Ruiz:2007yx}
\begin{eqnarray}
  \frac{dE}{dt} &=& \lim_{r\rightarrow \infty}
        \left[ \frac{r^2}{16\pi} \int_{\Omega_2} \left|
        \int_{-\infty}^t \Psi_4 d\tilde{t} \right|^2 \, d\Omega \right]\,,
  \\[5pt]
  \frac{dP_i}{dt} &=& -\lim_{r\rightarrow\infty}
        \left[ \frac{r^2}{16\pi} \int_{\Omega_2} \ell_i
        \left| \int_{-\infty}^t \Psi_4 d\tilde{t}\right|^2 \, d\Omega \right]\,,
  \\[5pt]
  \frac{dJ_i}{dt} &=& -\lim_{r\rightarrow\infty}
        \left\{ \frac{r^2}{16\pi} {\rm Re} \left[
        \int_{\Omega_2} \left( \hat{J}_i \int_{-\infty}^t \Psi_4 \, d\tilde{t}
        \right)
        \left( \int_{-\infty}^t \int_{-\infty}^{\hat{t}} \overline{\Psi}_4
        d\tilde{t} d\hat{t} \right) \, d\Omega \right] \right \}\,,
  \nonumber \\
\end{eqnarray}
where
\begin{eqnarray}
  \ell_i &=& [-\sin \theta\, \cos\phi,~-\sin \theta\,\sin \phi,~-\cos \theta]\,,
        \\[5pt]
  \hat{J}_x &=& -\sin \phi\,\partial_\theta - \cos \phi\,
        \left(\cot \theta\,\partial_\phi -\frac{2i}{\sin \theta} \right)\,,
  \\
  \hat{J}_y &=& \cos \phi\,\partial_\theta - \sin\phi\,
        \left( \cot \theta \,\partial_\phi-\frac{2i}{\sin \theta} \right)\,,
  \\
  \hat{J}_z &=& \partial_\phi\,.
\end{eqnarray}
In practice, one often starts the integration
at the start of the numerical simulation (or shortly thereafter to avoid
contamination from spurious GWs contained in the initial data) rather than
at $-\infty$.

We finally note that the GW strain commonly used in GW data analysis
is obtained from $\Psi_4$ by integrating twice in time
\begin{equation}
  h \equiv h_+ - ih_{\times} = \int_{-\infty}^t \left(
        \int_{-\infty}^{\tilde{t}}
        \Psi_4 d\hat{t} \right) \, d\tilde{t}\,.
\end{equation}
$h$ is often decomposed into multipoles in analogy to
Eq.~(\ref{eq:multipoles}).  As before, the practical
integration is often started at finite value rather than at $-\infty$.
It has been noted that this process of integrating
$\Psi_4$ twice in time is susceptible to large nonlinear drifts.
These are due to fundamental difficulties that arise in the integration
of finite-length, discretly sampled, noisy data streams which can be
cured or at least mitigated by performing the integration in
the Fourier instead of the time
domain~\cite{Reisswig:2010di,Hinder:2013oqa}.

\paragraph*{Perturbative wave extraction:}
The basis of this approach to extract GWs from numerical simulations in $D=4$ is
the Regge--Wheeler--Zerilli--Moncrief formalism developed for the study of
perturbations of spherically symmetric BHs. The assumption for applying this
formalism to numerically generated spacetimes is that at sufficiently large
distances from the GW sources, the spacetime is well approximated by a
spherically symmetric background (typically Schwarzschild or Minkowski
spacetime) plus non-spherical perturbations. These perturbations naturally
divide into odd and even multipoles which obey the
Regge--Wheeler~\cite{Regge:1957rw} (odd) and the
Zerilli~\cite{Zerilli:1971wd} (even) equations respectively (see
Section~\ref{classicpert}). Moncrief~\cite{Moncrief:1974am} developed
a gauge-invariant formulation for these perturbations in terms of a
master function which obeys a wave-type equation with a background
dependent scattering potential; for a review and applications of this formalism see for example~\cite{Nagar:2005ea,Sperhake:2005uf,Reisswig:2010cd}.

An extension of this formalism to higher-dimensional spacetimes
has been developed by Kodama \& Ishibashi~\cite{Kodama:2003jz}, and is discussed
in Section~\ref{pertdgt4}. This approach has been used to
develop wave extraction from NR simulations in $D>4$ with
$SO(D-2)$ symmetry~\cite{Witek:2010xi}. In particular, it has been applied to the extraction of GWs
from head-on collisions of BHs. As in our discussion of formulations of the
Einstein equations in higher dimensions in Section~\ref{sec:NRD}, it turns out
useful to introduce coordinates that are adapted to the rotational symmetry on a
$S^{D-2}$ sphere. Here, we choose spherical coordinates for this purpose which
we denote by $(t,\,r,\,\vartheta,\,\theta,\,\phi^a)$ where $a=4,\,\ldots,\,D-1$;
we use the same convention for indices as in Section~\ref{sec:NRD}.

We then assume that in the far-field region, the spacetime is perturbatively
close to a spherically symmetric BH background given in $D$ dimensions
by the Tangherlini~\cite{Tangherlini:1963bw} metric
\begin{equation}
  ds_{(0)}^2 = -A(r)^2dt^2 + A(r)^{-1}dr^2 + r^2
        \left[ d\vartheta^2 + \sin^2\vartheta\,(d\theta^2 + \sin^2 \theta\,
        d\Omega_{D-4}) \right]\,,
  \label{eq:KIBackground}
\end{equation}
where
\begin{equation}
  A(r) = 1 - \frac{R_{\rm S}^{D-3}}{r^{D-3}}\,,
\end{equation}
and the Schwarzschild radius $R_{\rm S}$ is related to the BH mass through Eq.~(\ref{eq:TangherliniRS}). For a spacetime with
$SO(D-3)$ isometry the perturbations away from the
background (\ref{eq:KIBackground}) are given by
\begin{equation}
  ds_{(1)}^2 = h_{\A \B} \, dx^{\A} \, dx^{\B} + h_{\A \vartheta}dx^{\A}
        d\vartheta + h_{\vartheta \vartheta} d\vartheta^2
        + h_{\theta \theta} d\Omega_{D-3}\,,
  \label{eq:KIPert1}
\end{equation}
where we introduce early upper case Latin indices $A,\,B,\,\ldots = 0,\,1$ and
$x^{\A}=(t,\,r)$. The class of axisymmetric spacetimes considered in
\cite{Witek:2010xi} obeys $SO(D-2)$ isometry which can be shown to imply that
$h_{\A \theta} = h_{\vartheta \theta} = 0$ and that the remaining components of
$h$ in Eq.~(\ref{eq:KIPert1}) only depend on the coordinates
$(t,\,r,\,\vartheta)$. As a consequence, only the perturbations which we have
called in Section~\ref{pertdgt4} ``scalar'' are non-vanishing, and are expanded in
tensor spherical harmonics; cf.~Section~II~C in Ref.~\cite{Witek:2010xi}.

As discussed in Section~\ref{pertdgt4}, the metric perturbations, decomposed in tensor harmonics,
can be combined in a gauge-invariant master function $\Phi_{\ell m}$.
From the master function, we can calculate the GW energy flux and the total radiated energy
as discussed in Section~\ref{pertdgt4}.

\subsubsection{Diagnostics in asymptotically AdS spacetimes}
The gauge/gravity duality, or AdS/CFT correspondence (see
Section~\ref{sec:gauge-gravity}), 
relates gravity in asymptotically AdS spacetimes
to conformal field theories on the boundary of this spacetime.
A key ingredient of the correspondence is the relation between
fields interacting gravitationally in the bulk spacetime and
expectation values of the field theory on the boundary. 
Here we restrict our attention to the extraction
of the expectation values of the energy-momentum
tensor $\langle T_{\I \J}\rangle$ of the field theory from the
fall-off behaviour of the AdS metric.

Through the AdS/CFT correspondence, the expectation values $\langle T_{\I \J} \rangle$ of the field theory
are given by the quasi-local Brown--York~\cite{Brown:1992br}
stress-energy tensor and thus are directly
related to the bulk metric. Following~\cite{deHaro:2000xn}, it
is convenient to consider the (asymptotically AdS) bulk metric in
Fefferman--Graham~\cite{Fefferman1985} coordinates
\begin{equation}
  ds^2 = g_{\mu \nu} \, dx^{\mu} \, dx^{\nu} =
       \frac{L^2}{r^2} \left[ dr^2 + \gamma_{\I \J} \, dx^{\I} \, dx^{\J} \right]\,,
  \label{eq:FeffermanGraham}
\end{equation}
where
\begin{equation}
  \gamma_{\I \J} = \gamma_{\I \J}(r,x^{\I}) =
      \gamma_{(0)\I \J} + r^2 \gamma_{(2)\I \J} + \ldots
      + r^d \gamma_{(d)\I \J} + h_{(d)\I \J} r^d \log r^2
      + \mathcal{O}(r^{d+1})\,.
\end{equation}
Here $d\equiv D-1$, the
$\gamma_{(a)\I \J}$ and $h_{(d)\I \J}$ are functions of the boundary coordinates
$x^i$, the logarithmic term only appears for even $d$ and powers of $r$ are
exclusively even up to order $d-1$.  As shown in Ref.~\cite{deHaro:2000xn}, the
vacuum expectation value of the CFT momentum tensor for $d=4$ is then obtained
from
\begin{eqnarray}
  \langle T_{\I \J} \rangle &=&
      \frac{4L^3}{16\pi} \left\{
      \gamma_{(4)\I \J} - \frac{1}{8}\gamma_{(0)\I \J} \left[
      \gamma_{(2)}^2- \gamma_{(0)}^{\K \M} \gamma_{(0)}^{\LL \N}
      \gamma_{(2)\K \LL} \gamma_{(2)\M \N} \right] \right.
      \nonumber \\
   && \left.
      - \frac{1}{2} \gamma_{(2)\I}{}^{\M}\gamma_{(2)\J \M}
      + \frac{1}{4} \gamma_{(2)\I \J} \gamma_{(2)}
      \right\} \,,
\end{eqnarray}
and $\gamma_{(2)\I \J}$ is determined in terms of $\gamma_{(0)\I \J}$.
The dynamical freedom of the CFT is thus encapsulated in the fourth-order term $\gamma_{(4)\I \J}$.
If $\gamma_{(0)\I \J}=\eta_{\I \J}$,
for $r \rightarrow 0$ the metric (\ref{eq:FeffermanGraham}) asymptotes
to the AdS metric in Poincar{\'e} coordinates (\ref{eq:AdSPoincare}).

The Brown--York stress tensor is also the starting point for
an alternative method to extract the $\langle T_{\I \J}\rangle$ that does
not rely on Fefferman-Graham coordinates. It is given by
\begin{equation}
  T^{\mu \nu} = \frac{2}{\sqrt{-\gamma}}
        \frac{\delta S_{\rm grav}}{\delta \gamma_{\mu \nu}}\,,
  \label{eq:BYtensor}
\end{equation}
where we have foliated the $D$-dimensional spacetime into \emph{timelike}
hypersurfaces $\Sigma_r$ in analogy to the foliation in terms of
\emph{spacelike} hypersurfaces $\Sigma_t$
in Section~\ref{sec:ADM}. The spacetime metric is given by
\begin{equation}
  ds^2 = \alpha^2 \, dr^2 + \gamma_{\I \J}(dx^{\I} + \beta^{\I} \, dr)
        (dx^{\J} + \beta^{\J}dr)\,.
\end{equation}
%
In analogy to the second fundamental form $K_{\alpha \beta}$ in
Section~\ref{sec:ADM}, we define the extrinsic curvature on $\Sigma_r$ by
\begin{equation}
  \Theta^{\mu \nu} \equiv -\frac{1}{2}(\nabla^{\mu} n^{\nu} + \nabla^{\nu}
        n^{\mu})\,,
\end{equation}
where $n^{\mu}$ denotes the outward pointing normal vector to $\Sigma_r$.
Reference~\cite{Balasubramanian:1999re} provides
a method to cure divergencies that appear in the Brown-York tensor
when the boundary is pushed to infinity by adding counterterms
to the action $S_{\rm grav}$. This work discusses asymptotically AdS
spacetimes of different dimensions. For AdS$_{5}$, the procedure
results in
\begin{equation}
  T^{\mu\nu} = \frac{1}{8\pi}\left[
        \Theta^{\mu \nu} - \Theta \gamma^{\mu \nu} - \frac{3}{L}
        \gamma^{\mu \nu} - \frac{L}{2} \mathcal{G}^{\mu \nu} \right]\,,
\end{equation}
where $\mathcal{G}_{\mu \nu}=\mathcal{R}_{\mu \nu}
-\mathcal{R}\gamma_{\mu \nu}/2$ is the Einstein tensor associated with
the induced metric $\gamma_{\mu \nu}$. Applied to the AdS$_5$ metric
in global coordinates, this expression gives a non-zero energy-momentum tensor $T^{\mu \nu}$ 
which, translated into
the expectation values $\langle T^{\mu \nu}\rangle$, can be interpreted
as the Casimir energy of a quantum field theory on the spacetime with topology
$\mathbb{R}\times S^3$~\cite{Balasubramanian:1999re}. This Casimir energy
is non-dynamical and in numerical applications to the AdS/CFT correspondence
may simply be subtracted from $T^{\mu \nu}$; see, e.g.,~\cite{Bantilan:2012vu}.

The role of additional (e.g., scalar) fields in the AdS/CFT dictionary is
discussed, for example, in Refs.~\cite{deHaro:2000xn,Skenderis:2002wp}.

\newpage
\section{Applications of Numerical Relativity}
\label{sec:NRapplications}

Numerical relativity was born out of efforts to solve the
two-body problem in GR, and aimed mainly at understanding
stellar collapse and GW emission from BH and NS binaries. There
is therefore a vast amount of important results and literature
on NR in astrophysical contexts. Because these results fall
outside the scope of this review, we refer the interested reader to
Refs.~\cite{Pretorius:2007nq,Ott:2008wt,Centrella:2010mx,
Sperhake:2011xk,Ajith:2012tt,Pfeiffer:2012pc,Hinder:2013oqa}
and to the relevant sections of Living Reviews%
\epubtkFootnote{\url{http://relativity.livingreviews.org/Articles/subject.html}}
for (much) more on this subject. Instead, we now focus on applications of
NR outside its traditional realm, most of which are relatively recent
new directions in the field.

\subsection{Critical collapse}
\label{sec:critical_collapse}
The nonlinear stability of Minkowski spacetime was established by Christodoulou and Klainerman, who showed
that arbitrarily ``small'' initial fluctations eventually disperse to infinity~\cite{1993gnsm.book.....C}.
On the other hand, large enough concentrations of matter are expected to collapse to BHs, therefore raising the question
of how the threshold for BH formation is approached.

Choptuik performed a thorough investigation of this issue, by evolving initial data for a minimally coupled massless scalar field~\cite{Choptuik:1992jv}. Let the initial data be described by a parameter $p$ which characterizes the initial scalar field wavepacket. 
For example, in Choptuik's analysis, the following family of initial
data for the scalar field $\Phi$ was considered,
\be
\phi=\phi_0 r^3\exp{\left(-\left[(r-r_0)/\delta\right]^q\right)}\,,\label{pulse_shape}
\ee
where $\Phi=\phi'$; therefore any of the quantities $\phi_0,r_0,\delta, q$ is a suitable parameter $p$.

\epubtkImage{}{%
  \begin{figure}[htbp]
    \centerline{\includegraphics[width=0.5\textwidth]{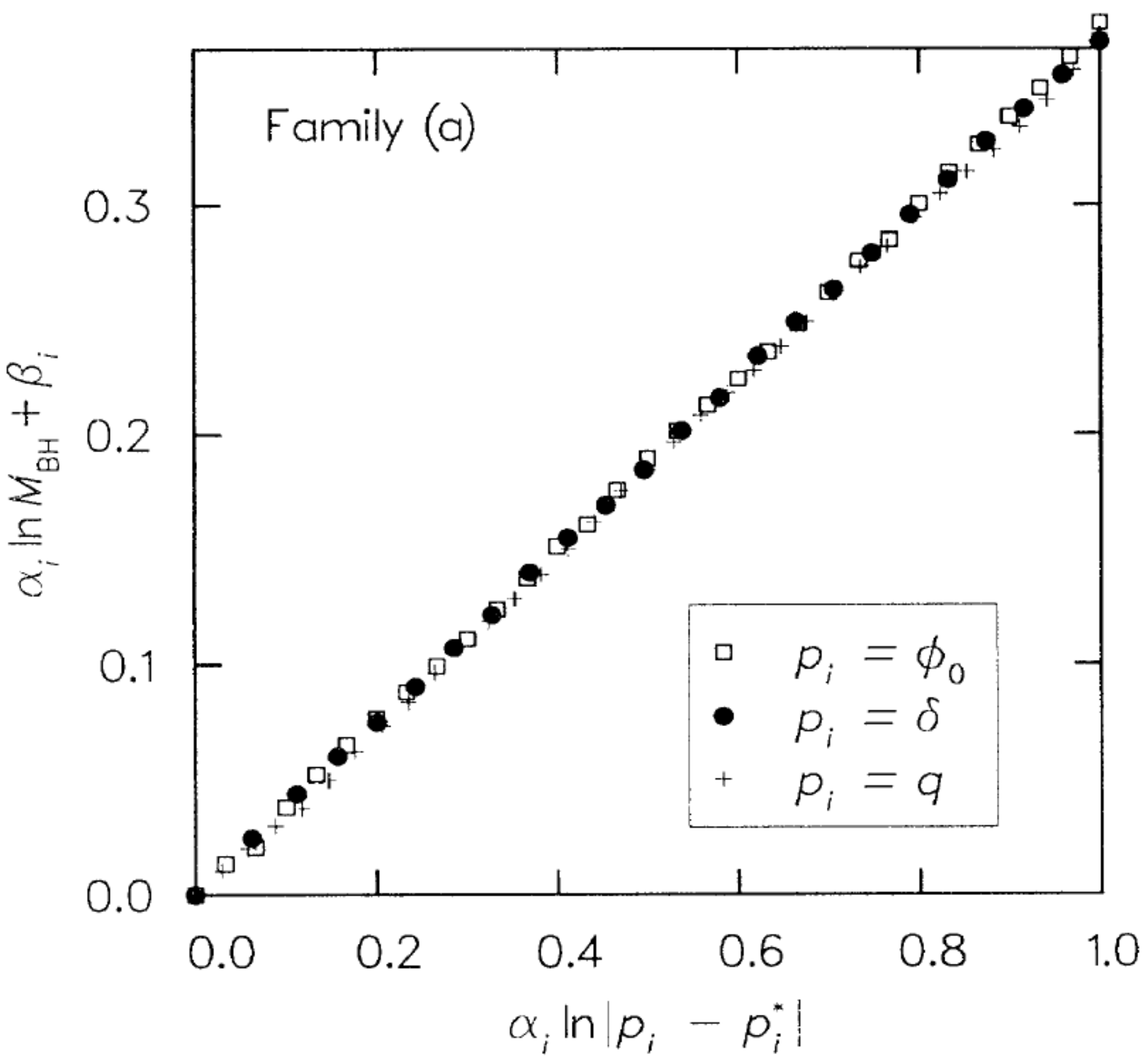}}
    \caption{Illustration of the conjectured mass-scaling relation
\eqref{eq:mass_scaling}. The data refer to three separate one-parameter variations
of the pulse shape \eqref{pulse_shape}.
The constants $\alpha_i$ and $\beta_i$ are chosen to normalize the ranges
of the abscissa and place the data point corresponding to the smallest BH in each family at the origin. From~\cite{Choptuik:1992jv}.}
\label{fig:chop}
\end{figure}}

The evolution of such initial data close to the threshold of BH formation is summarized in Figure~\ref{fig:chop}.
Fix all but one parameter, say the scalar field amplitude $p=\phi_0$. For large amplitudes $\phi_0$, a large BH is formed. As the amplitude of the initial data decreases, the mass of the formed BH decreases, until a critical threshold amplitude $\phi_{0*}$ is reached below which no BH forms and the initial data disperses away (consistently with the nonlinear stability of Minkowski). Near the threshold, BHs with arbitrarily small masses can be created, and the BH mass scales as
\be
M\propto (p-p_*)^{\gamma}\,,\label{eq:mass_scaling}
\ee
It was found that $\gamma\approx 0.37$ is a universal (critical) exponent which does not depend on the initial data,
or in other words it does not depend on which of the parameters $\phi_0,r_0,\delta, q$
is varied (but it may depend on the type of collapsing material).

The BH threshold in the space of initial data for GR shows both surprising
structure and surprising simplicity.  In particular, critical behavior was found
at the threshold of BH formation associated with universality, power-law scaling
of the BH mass, and discrete self-similarity, which bear resemblance to more
familiar statistical physics systems.  Critical phenomena also provide a route
to develop arbitrarily large curvatures visible from infinity (starting from
smooth initial data) and are therefore likely to be relevant for cosmic
censorship (see Section~\ref{sec:cosmic_censor_review}),
quantum gravity, astrophysics, and our general understanding of the dynamics of
GR.

Choptuik's original result was extended in many different directions, to encompass massive scalar fields~\cite{Brady:1997fj,Okawa:2013jba}, collapse in higher dimensions~\cite{Garfinkle:1999zy} or different gravitational theories~\cite{Deppe:2012wk}. Given the difficulty of
the problem, most of these studies have focused on 1+1 simulations; the first non spherically (but axially) symmetric simulations were performed in Ref.~\cite{Abrahams:1993wa}, whereas recently the first 3+1 simulations of the collapse of minimally coupled scalar fields
were reported~\cite{Healy:2013xia}. The attempt to extend these results to asymptotically AdS spacetimes would uncover a new surprising result, which
we discuss below in Section~\ref{sec_spacetime_stability}.
A full account of critical collapse along with the relevant references can be found in a Living Reviews article on the subject~\cite{Gundlach:2007gc}. 

\subsection{Cosmic censorship}
\label{sec:cosmic_censor_review}

As discussed in Section~\ref{sec_cc}, an idea behind cosmic censorship
is that classical GR is self-consistent for physical processes. That is,
despite the fact that GR predicts the formation of singularities, at which
geodesic incompleteness occurs and therefore failure of predictablity, such
singularities should be -- for physical processes%
\epubtkFootnote{Cosmic censorship does not apply to cosmological
  singularities, i.e., Big Bang or Big Crunch.}
-- causally disconnected from distant observers by virtue of horizon
cloaking. In a nutshell: a GR evolution does not lead, generically, to
a system GR cannot tackle. To test this idea, one must analyze strong
gravity dynamics, which has been done both using numerical evolutions
and analytical arguments. Here we shall focus on recent results based
on NR methods. The interested reader is referred to some historically
relevant numerical~\cite{Shapiro:1991zza,Goldwirth:1987nu} and
analytical~\cite{Christodoulou:1984mz,Roberts:1989sk} results, as well
as to reviews on the
subject~\cite{wald1999gravitational,Berger:2002st,Ringstrom:2010zz,Joshi:2012mk}
for further information.

The simplest (and most
physically viable) way to violate cosmic censorship would be through the
gravitational collapse of very rapidly rotating matter, possibly leading to a Kerr naked singularity with $a>M$. 
However, NR simulations of the collapse of a rotating NS to a BH
\cite{1981PThPh..65.1876N,Baiotti:2004wn,Giacomazzo:2011cv} have shown that when
the angular momentum of the collapsing matter is too large, part of the matter
bounces back, forming an unstable disk that dissipates the excess angular
momentum, and eventually collapses to a Kerr BH.
Simulations of the coalescence of rapidly rotating BHs
\cite{Washik:2008jr,Healy:2009ir} and NSs~\cite{Kastaun:2013mv} have shown that
the $a>M$ bound is preserved by these processes as well.
These simulations provide
strong evidence supporting the cosmic censorship conjecture. Let us remark that analytical computations and
NR simulations show that naked singularities can arise in the collapse of ideal
fluids~\cite{Joshi:2012mk} but these processes seem to require fine-tuned
initial conditions, such as in spherically symmetric collapse or in the critical
collapse~\cite{Gundlach:2002sx} discussed in Section~\ref{sec:critical_collapse}.

A claim of cosmic censorship violation in $D>4$ spacetime dimensions
was made in the context of the non-linear evolution of the
Gregory--Laflamme (see Section~\ref{sec_hi_dim}) instability for black
strings. In Ref.~\cite{Lehner:2010pn} long-term numerical simulations
were reported showing that the development of the instability leads to
a cascade of ever smaller spherical BHs connected by ever thinner
black string segments -- see Figure~\ref{lp_fig}, left (top, middle
and bottom) panel for a visualization of the (first, second and third)
generations of spherical BHs and string segments. Observe, from the
time scales presented in Figure~\ref{lp_fig}, that as viewed by an
asymptotic observer, each new generation develops more rapidly than
the previous one. The simulations therefore suggest that arbitrarily
thin strings, and thus arbitrarily large curvature at the horizon,
will be reached in \emph{finite} asymptotic time. If true, this system
is an example where a classical GR evolution is driving the system to
a configuration that GR cannot describe, a state of affairs that will
presumably occur when Planck scale curvatures are attained at the
horizon.
The relevance of this example 
%
for cosmic censorship, may, however, be questioned, based on
its higher dimensionality and the lack of asymptotic flatness: cosmic strings
with horizons require the spacetime dimension to be greater or equal to five and
the string is infinitely extended in one dimension. In addition, 
the simulations of~\cite{Lehner:2010pn} assume cylindrical symmetry, and
cylindrically symmetric matter configurations are
unstable~\cite{Cardoso:2006sj}; therefore, fine-tuning of initial conditions may
be required for the formation of a naked singularity.

\epubtkImage{}{%
  \begin{figure}[htbp]
    \centerline{
      \includegraphics[width=0.5\textwidth]{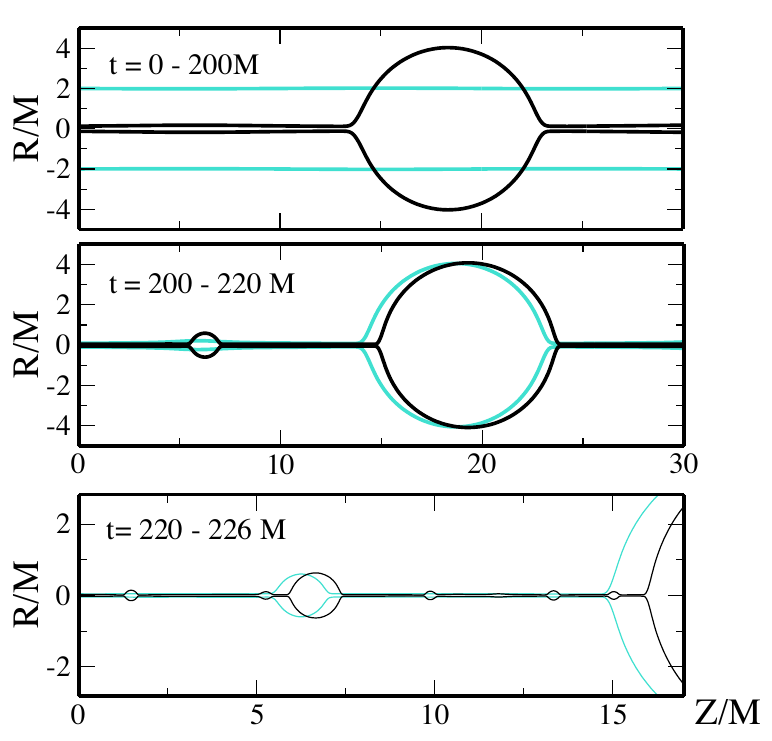}
      \includegraphics[width=0.5\textwidth]{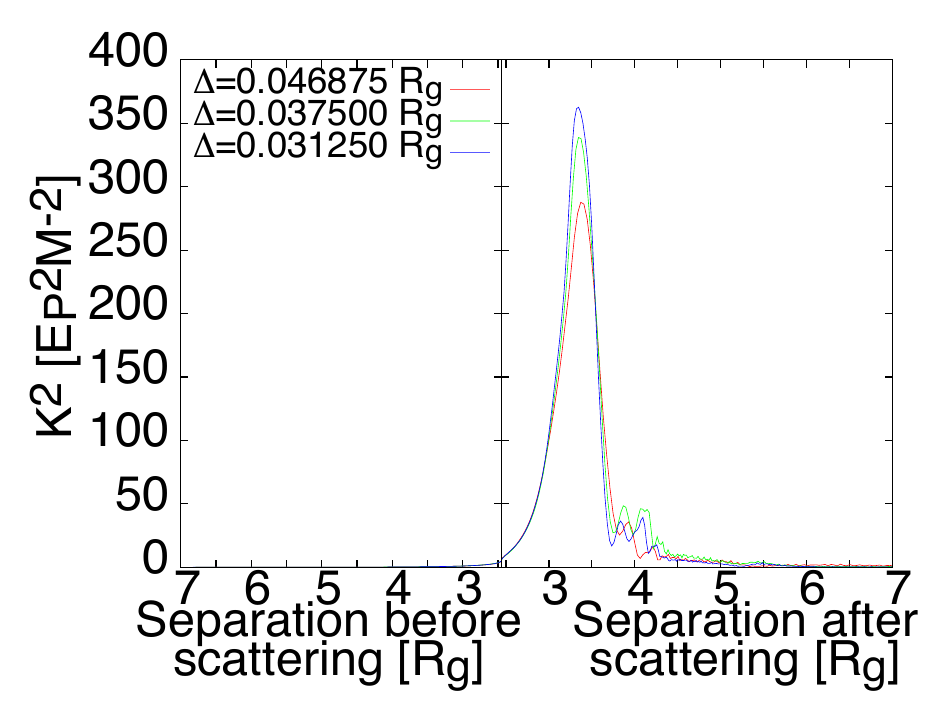}
}
\caption{\emph{Left panel:} Embedding diagram of the AH of the perturbed black string
  at different stages of the evolution.  The light (dark) lines denote the first
  (last) time from the evolution segment shown in the corresponding
  panel. From~\cite{Lehner:2010pn}. \emph{Right panel:} Dimensionless Kretschmann
  scalar $\mathcal{K}^2$ at the centre of mass of a binary BH system as a
  function of the (areal) coordinate separation between the two BHs in a $D=5$
  scattering, in units of $R_g=R_{\rm S}$.
    From~\cite{Okawa:2011fv}.}
    \label{lp_fig}
\end{figure}}

Another suggestion that Planckian scale curvature becomes visible in a classical
evolution in $D=5$ GR arises in the high-energy scattering of BHs. In
Ref.~\cite{Okawa:2011fv} NR simulations of the scattering of two non-spinning
boosted BHs with an impact parameter $b$ were reported.  For
sufficiently small initial velocities ($v\lesssim 0.6 c$) it is possible to find
the threshold impact parameter $b_{\rm scat}=b_{\rm scat}(v)$ such that the BHs merge into a (spinning) BH
for $b<b_{\rm scat}$ or scatter off to infinity for $b>b_{\rm scat}$. For high
velocities, however, only a lower bound on the impact parameter for scattering
$b_C=b_C(v)$ and an upper bound on the impact parameter for merger $b_B=b_B(v)$
could be found, since simulations with $b_B<b<b_C$ crashed before the final
outcome could be determined (cf. Figure~\ref{fig:d5_bscat} below). Moreover, an analysis of a scattering configuration
with $v=0.7$ and $b=b_C$, shows that very high curvature develops outside the individual BHs' AHs, shortly after they have
reached their minimum separation - see Figure~\ref{lp_fig} (right panel). The
timing for the creation of the high curvature region, i.e., that it occurs
\emph{after} the scattering, is in agreement with other simulations of high
energy collisions. For instance, in Refs.~\cite{Choptuik:2009ww,East:2012mb} BH
formation is seen to occur in the wake of the collision of non-BH objects, which
was interpreted as due to focusing effects~\cite{East:2012mb}. In the case of
Ref.~\cite{Okawa:2011fv}, however, there seems to be no (additional) BH
formation. Both the existence and significance of such a high curvature region,
seemingly uncovered by any horizon, remains mysterious and deserves further
investigation.

In contrast with the two higher-dimensional examples above, NR simulations that have tested the
cosmic censorship conjecture in different $D=4$ setups, found support for the
conjecture. We have already mentioned simulations of the gravitational collapse of rotating matter, and of the
coalescence of rotating BH and NS binaries. As we discuss below in Section~\ref{sec:HEcollisions},
the high-energy head-on collisions of BHs~\cite{Sperhake:2008ga}, boson stars
\cite{Choptuik:2009ww} or fluid particles~\cite{East:2012mb,Rezzolla:2013kt} in
$D=4$ result in BH formation but no naked singularities. A different check of the conjecture
involves asymptotically dS spacetimes~\cite{Zilhao:2012bb}. Here, the
cosmological horizon imposes an upper limit on the size of BHs. Thus one may ask
what is the outcome of the collision of two BHs with almost the maximum allowed
size. In Ref.~\cite{Zilhao:2012bb} the authors were able to perform the
evolution of two BHs, initially at rest with the cosmological expansion. They
observe that for all the (small) initial separations attempted, a cosmological
AH, as viewed by an observer at the center of mass of the binary BH system,
eventually forms in the evolution, and both BH AHs are outside the cosmological
one. In other words, the observer in the center of mass loses causal contact
with the two BHs which fly apart rather than merge. This suggests that the
background cosmological acceleration dominates over the gravitational attraction
between `large' BHs. It would be interesting to check if a violation of the
conjecture can be produced by introducing opposite charges to the BHs (to
increase their mutual attraction) or give them mutually directed initial boosts.

\subsection{Hoop conjecture}
\label{sec:hoop_review}
The hoop conjecture, first proposed by K.~Thorne in 1972
\cite{Thorne:1972}, states that when the mass $M$ of a system (in $D=4$
dimensions) gets compacted into a region whose circumference in every direction
has radius $R\lesssim R_{\rm s}=2M$, a horizon -- and thus a BH -- forms (for a
generalization in $D>4$, see~\cite{Ida:2002hg}).  This conjecture is important
in many contexts. In high-energy particle collisions, it implies that a classical BH forms
if the center-of-mass energy significantly exceeds the Planck energy.  This is the
key assumption behind the hypothesis -- in the TeV gravity scenario -- of BH
production in particle accelerators (see Section~\ref{sec:tev}). In the
trans-Planckian regime, the particles can be treated as classical objects. If
two such ``classical'' particles with equal rest mass $m_0$ and radius (corresponding to the de Broglie wavelength of the process) $R$ collide with a boost parameter $\gamma$, the mass-energy in the centre-of-mass frame is $M=2\gamma
m_0$. The threshold radius of Thorne's hoop is then $R_{\rm s}=4\gamma m_0$ and
the condition $R\lesssim R_{\rm s}=2M=4\gamma m_0$ translates into a bound on
the boost factor, $\gamma\gtrsim\gamma_{\rm h}\equiv R/(4m_0)$.

Even though the hoop conjecture seems plausible, finding a rigorous proof is not
an easy task. In the last decades, the conjecture has mainly been supported by
studies of the collision of two infinitely boosted point particles
\cite{DEath:1990de,Eardley:2002re,Yoshino:2002br,Yoshino:2002tx}, but it is
questionable that they give an accurate description of an actual particle
collision (see e.g. the discussion in Ref.~\cite{Choptuik:2009ww}). In recent years,
however, advances in NR have made it possible to model trans-Planckian
collisions of massive bodies and provided more solid evidence in favor of the
validity of the hoop conjecture.

The hoop conjecture has been first addressed in NR by Choptuik \& Pretorius
\cite{Choptuik:2009ww}, who studied head-on collisions of boson stars in four
dimensions (see Section~\ref{sec_elliptic}). The simulations show that the
threshold boost factor for BH formation is $\sim 1/3 \gamma_{\rm h}$ (where the
``hoop'' critical boost factor $\gamma_{\rm h}$ is defined above), well in
agreement with the hoop conjecture.
These results have been confirmed by NR simulations of fluid star collisions
\cite{East:2012mb}, showing that a BH forms when the boost factor is larger than
$\sim0.42\gamma_{\rm h}$.  Here, the fluid balls are modeled as two superposed
Tolman--Oppenheimer--Volkoff ``stars'' with a $\Gamma=2$ polytropic equation of
state.

\epubtkImage{}{%
\begin{figure}[htb]
 \centerline{\includegraphics[width=\textwidth]{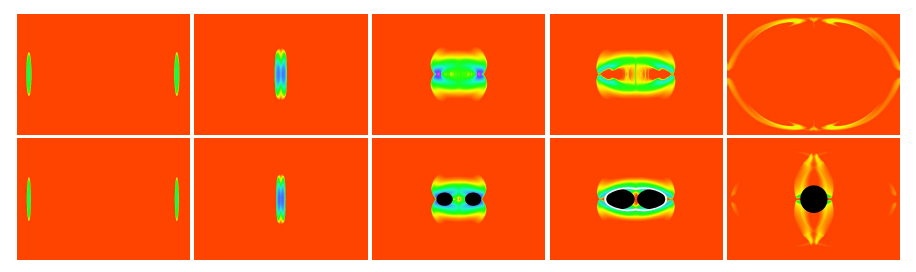}}
 \caption{Snapshots of the rest-mass density in the collision of
          fluid balls with boost factor $\gamma=8$ (upper panels)
          and $\gamma=10$ (lower panels) at the initial time,
          shortly after collision, at the time corresponding to
          the formation of separate horizons in the $\gamma=10$ case,
          and formation of a common horizon (for $\gamma=10$)
          and at late time in the dispersion ($\gamma=8$) or
          ringdown ($\gamma=10$) phase. Taken from~\cite{East:2012mb}.}
 \label{fig:east_collisions}
\end{figure}}

The simulations furthermore show that for boosts slightly above the threshold of
BH formation, there exists a brief period where two individual AHs
are present, possibly due to a strong focusing of the fluid elements of each
individual star caused by the other's gravitational field. These results are
illustrated in Figure~\ref{fig:east_collisions} which displays snapshots of a
collision for $\gamma = 8$ that does not result in horizon formation (upper
panels) and one at $\gamma=10$ that results in a BH (lower panels). We remark
that the similarity of the behaviour of boson stars and fluid stars provides
evidence supporting the ``matter does not matter'' hypothesis discussed below in
Section~\ref{sec:HEcollisions}.  A similar study of colliding fluid balls
\cite{Rezzolla:2013kt} has shown similar results. Therein it has been found that a BH forms when
the compactness of the star is $m_0/R\gtrsim 0.08\gamma^{-1.13}$, i.e., for
$\gamma^{1.13}\gtrsim R/(12m_0)=1/3\,\gamma_{\rm h}$. Type-I critical behaviour has also been identified, with BH formation for initial masses $m_0$ above a
critical value scaling as $\sim\gamma^{-1.0}$.

\subsection{Spacetime stability}
\label{sec_spacetime_stability}

Understanding the stability of stationary solutions to the Einstein field equations, or
generalisations thereof, is central to gauge their physical relevance. If the
corresponding spacetime configuration is to play a role in a given dynamical process,
it should be stable or, at the very least, its instabilities should have longer
time scales than those of that dynamical process. Following the evolution of unstable solutions, on the other
hand, may unveil smoking guns for establishing their transient
existence. NR provides a unique tool both for testing
non-linear stability and for following the non-linear development of unstable
solutions. We shall now review the latest developments in both these directions, but  before doing so let us make a remark. At the linear level, typical studies of space-time stability are in fact studies of \emph{mode stability}. A standard example is Whiting's study of the mode stability for Kerr BHs~\cite{Whiting:1988vc}. For BH spacetimes, however, mode stability does not guarantee \emph{linear stability}, cf. the discussion in~\cite{Dafermos:2010hd}. We refer the reader to this reference for further information on methods to analyse linear stability.

Even if a spacetime does not exhibit unstable modes in a linear analysis it may be unstable when fully non-linear
dynamics are taken into account. A remarkable illustration of
this possibility is the turbulent instability of the AdS spacetime
reported in Ref.~\cite{Bizon:2011gg}. These authors consider Einstein gravity with a negative cosmological constant
$\Lambda$ and minimally coupled to a massless real scalar field $\phi$ in $D=4$
spacetime dimensions.  The AdS metric is obviously a solution of the system
together with a constant scalar field. Linear scalar field perturbations around
this solution generate a spectrum of normal modes with real frequencies~\cite{Burgess:1984ti}:
$\omega_N L=2N+3+\ell$, where $N\in \mathbb{N}_0$ and $L,\ell$ are the AdS
length scale and total angular momentum harmonic index, respectively. The
existence of this discrete spectrum is quite intuitive from the global structure
of AdS: a time-like conformal boundary implies that AdS behaves like a
cavity. Moreover, the fact that the frequencies are real shows that the system
is stable against scalar field perturbations at linear level.

The latter conclusion dramatically changes when
going beyond linear analysis. Setting
up spherically symmetric Gaussian-type initial data of amplitude $\epsilon$, Bizo\'n and Rostworowski~\cite{Bizon:2011gg} made the following observations. For large
$\epsilon$ the wave packet collapses to form a BH, signalled by an AH at some radial coordinate. As $\epsilon$ is made smaller, the 
AH radius also decreases, reaching zero size at some (first) threshold
amplitude. This behaviour is completely analogous to that observed in
asymptotically flat spacetime by Choptuik~\cite{Choptuik:1992jv} and discussed in Section~\ref{sec:critical_collapse}; in fact, the
solutions obtained with this threshold amplitude asymptote -- far from the AdS
boundary -- to the self-similar solution obtained in the $\Lambda=0$ case. For
amplitudes slightly below the first threshold value, the wave packet
travels to the AdS boundary, where it is reflected, and collapses to form a BH
upon the second approach to the centre. By further decreasing the
amplitude, one finds a second threshold amplitude at which the size of the
AH formed in this second generation interaction decreases to
zero. This pattern seems to repeat itself indefinitely. In~\cite{Bizon:2011gg}
ten generations of collapse were reported, as shown in Figure~\ref{turbulence_fig}
(left panel). These results were confirmed and extended in subsequent
work~\cite{Buchel:2012uh}.  If indeed the pattern described in the previous
paragraph repeats itself indefinitely, a remarkable conclusion is that, no
matter how small the initial amplitude is, a BH will form in AdS after a time
scale $\mathcal{O}(\epsilon^{-2})$. A corollary is then that linear analysis
misses the essential physics of this problem; in other words, the evolution
always drives the system away from the linear regime.

The central property of AdS to obtain this instability is its global structure,
rather than its local geometry. This can be established by noting that a
qualitatively similar behaviour is obtained by considering precisely the same
dynamical system in Minkowski space enclosed in a
cavity~\cite{Maliborski:2012gx}, see Figure~\ref{turbulence_fig} (right
panel). Moreover, the mechanism behind the instability seems to rely on
non-linear interactions of the field that tend to shift its energy to higher
frequencies and hence smaller wavelenghts. This process stops in GR since the
theory has a natural cutoff: BH formation.

It has since been pointed out that collapse to BHs may not be the generic
outcome of evolutions in
AdS~\cite{Buchel:2013uba,Balasubramanian:2014cja,Dias:2012tq,Maliborski:2013jca,Maliborski:2014rma}. For example, in
Ref.~\cite{Buchel:2013uba} ``islands of stability'' were discovered for which
the initial data, chosen as a small perturbation of a boson star, remain in a
nonlinearly stable configuration. In Ref.~\cite{Okawa:2013jba}, the authors
raised the possibility that some of the features of the AdS instability could
also show up in asymptotically flat spacetimes in the presence of some
confinement mechanism. They observed that the evolution of minimally coupled,
massive scalar wavepackets in asymptotically flat spacetimes can also lead to
collapse after a very large number of ``bounces'' off the massive effective
potential barrier in a manner akin to that discovered in AdS. Similarly, in
some region of the parameter space the evolution drives the system towards
nonlinearly stable, asymptotically flat ``oscillatons''~\cite{Seidel:1991zh,Fodor:2013lza}. Nevertheless, for
sufficiently small initial amplitudes they observe a $t^{-3/2}$ decay of the
initial data, characteristic of massive fields, and showing that Minkowski is
nonlinearly stable.  The ``weakly turbulent'' instability discovered in AdS is
stimulating research on the topic of turbulence in GR. A full understanding of
the mechanism(s) will require further studies, including collapse in
non-spherically symmetric backgrounds, other forms of matter and boundary
conditions, etc.

\epubtkImage{}{%
\begin{figure}[htbp]
    \centerline{\includegraphics[width=1.0\textwidth]{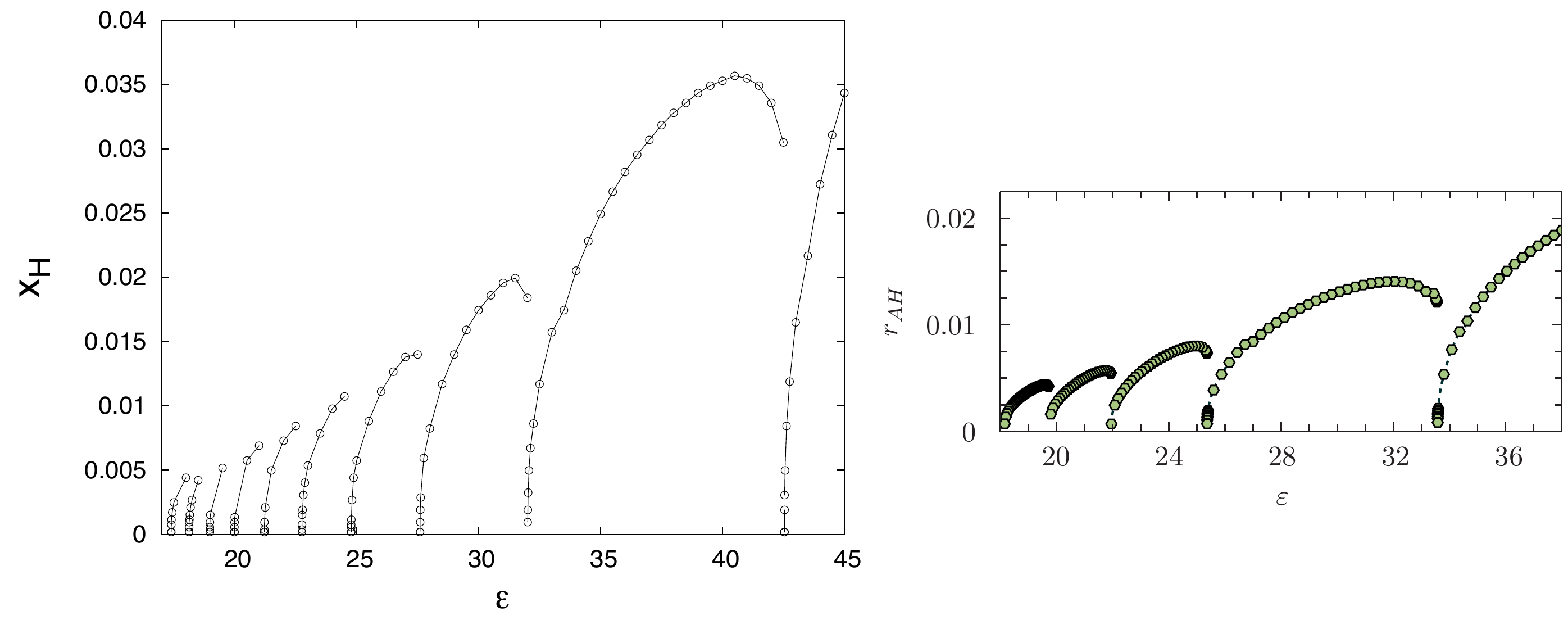}}
\caption{Instability against BH formation in AdS (left panel,
  from~\cite{Bizon:2011gg}) and Minkowski enclosed in a cavity (right panel,
  from~\cite{Maliborski:2012gx}). In both panels, the horizontal axis represents
  the amplitude of the initial (spherically symmetric) scalar field
  perturbation. The vertical axis represents the size of the BH
  formed. Perturbations with the largest plotted amplitude collapse to form a
  BH. As the amplitude of the perturbation is decreased so does the size of the
  BH, which tends to zero at a first threshold amplitude. Below this energy, no
  BH is formed in the first generation collapse and the scalar perturbation
  scatters towards the boundary. But since the spacetime behaves like a cavity,
  the scalar perturbation is reflected off the boundary and re-collapses,
  forming now a BH during the second generation collapse. At smaller amplitudes
  a second, third, etc, threshold amplitudes are found. The left (right)
  panel shows ten (five) generations of collapse. Near the threshold
  amplitudes, critical behavior is observed.}
\label{turbulence_fig}
\end{figure}}

\epubtkImage{}{%
\begin{figure}[htbp]
\centerline{
      \includegraphics[width=0.5\textwidth]{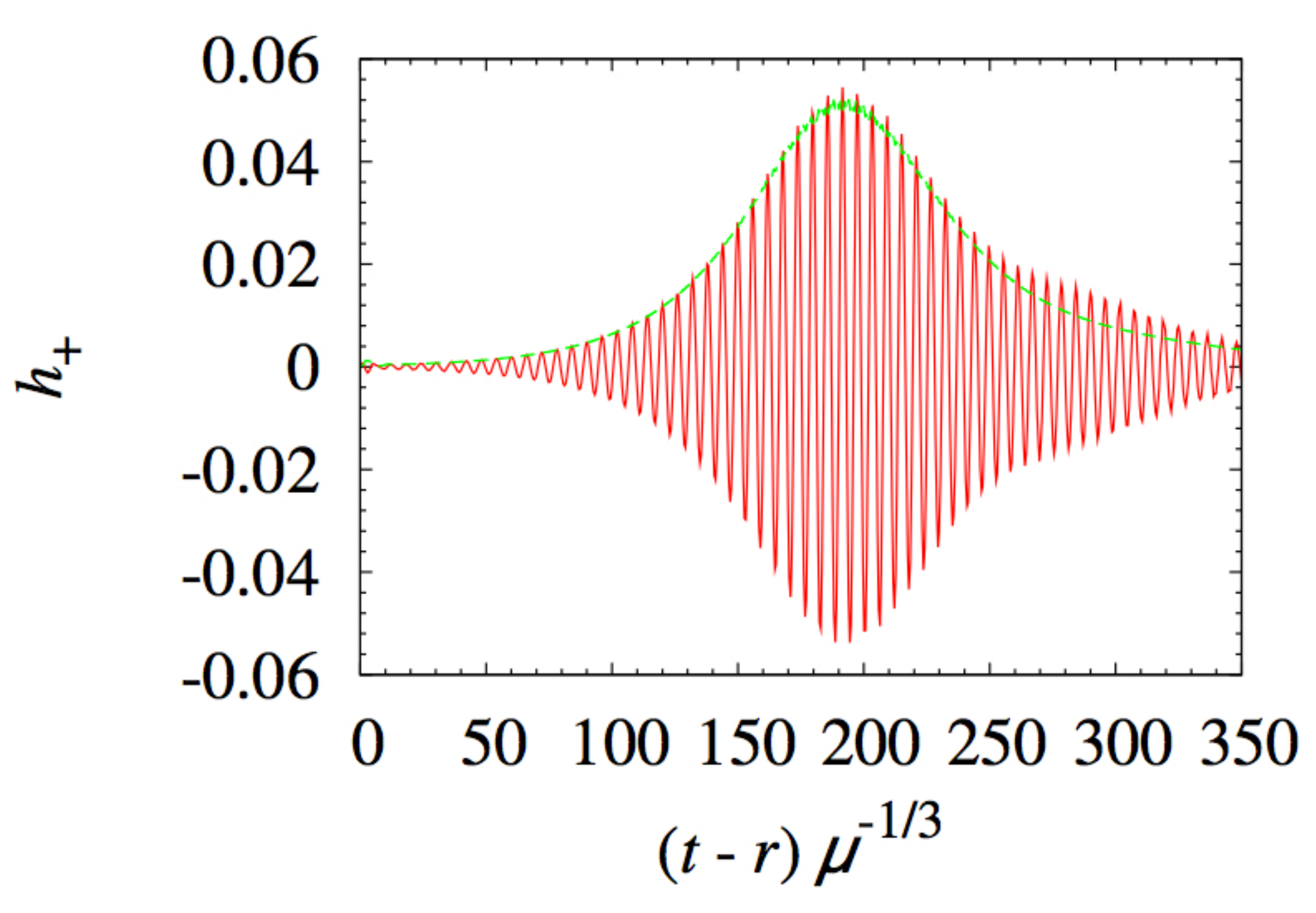}
      \includegraphics[width=0.5\textwidth]{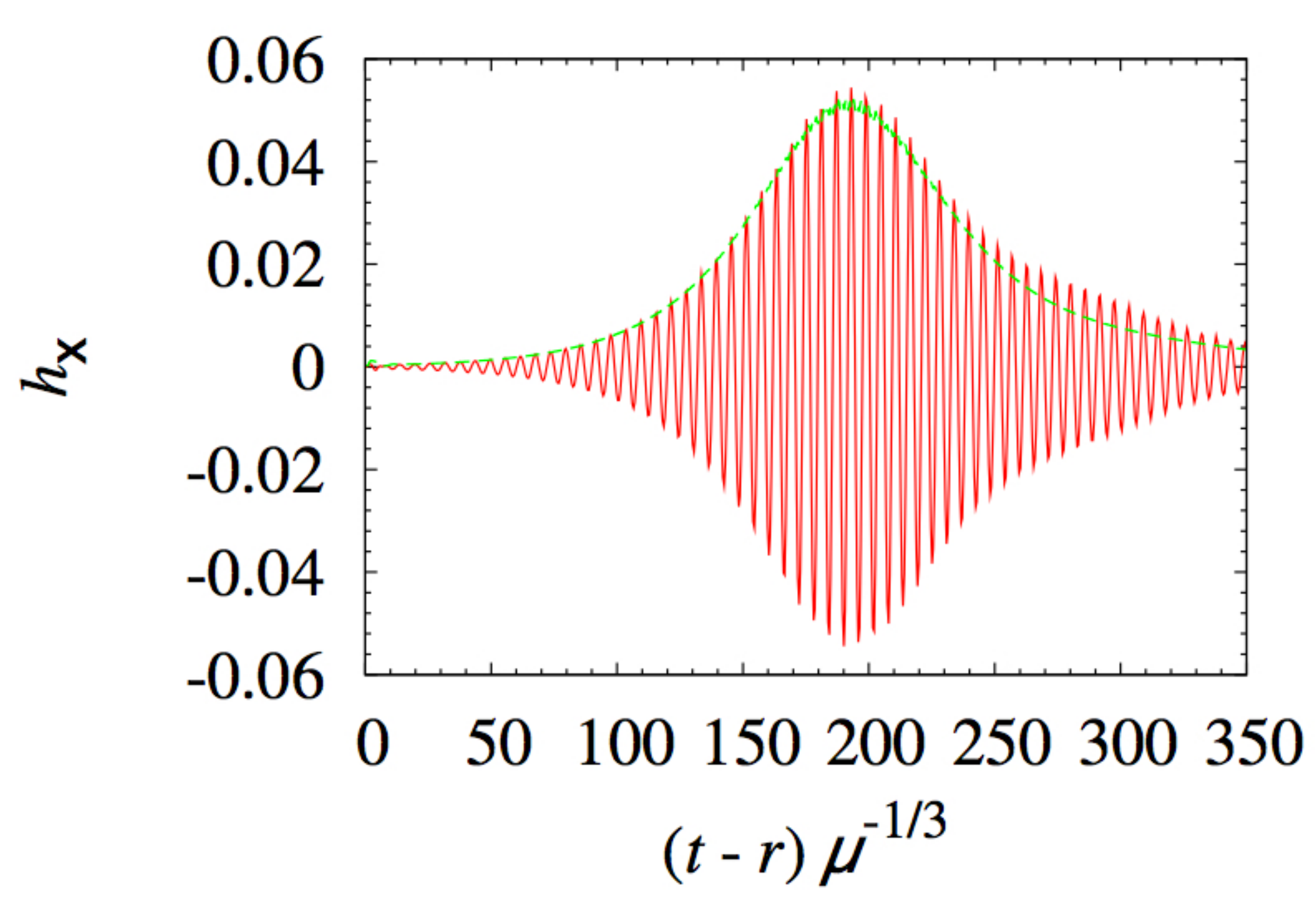}
}
    \caption{(a) and (b): $+$ and $\times$ modes of gravitational
      waveform (solid curve) from an unstable six-dimensional BH with
      $q = 0.801$ as a function of a retarded time defined by $t-r$
      where $r$ is the coordinate distance from the
      center. From~\cite{Shibata:2010wz}.}
\label{fig:shibata}
\end{figure}}

We now turn to solutions that display an instability at linear level, seen by a mode analysis, and to the
use of NR techniques to follow the development of such instabilities into the
non-linear regime.  One outstanding example is the Gregory--Laflamme instability
of black strings already described in Section~\ref{sec:cosmic_censor_review}. Such
black strings exist in higher dimensions, $D\ge 5$. It is expected that the
same instability mechanism afflicts other higher-dimensional BHs, even with a
topologically spherical horizon. A notable example are Myers--Perry BHs. In $D\ge
6$, the subset of these solutions with a single angular momentum parameter have
no analogue to the Kerr bound, i.e., a maximum angular momentum for a given
mass. As the angular momentum increases, they become \emph{ultra-spinning} BHs
and their horizon becomes increasingly flattened and hence resembling the
horizon of a black $p$-brane, which is subject to the
Gregory--Laflamme instability~\cite{Emparan:2003sy}. It was indeed shown in Ref.~\cite{Dias:2009iu,Dias:2010maa,Dias:2010eu}, by
using linear perturbation theory, that rapidly rotating Myers--Perry BHs for
$7\le D \le 9$ are unstable against axisymmetric perturbations. The non-linear
growth of this instability is unknown but an educated guess is that it may lead
to a deformation of the pancake like horizon towards multiple concentric
rings. 

A different argument -- of entropic nature -- for the instability of ultra-spinning BHs against \emph{non-axisymmetric} perturbations was given by Emparan and Myers~\cite{Emparan:2003sy}. Such type of instability has been tested in $6\le D\le
8$~\cite{Shibata:2010wz}, but also in $D=5$~\cite{Shibata:2009ad,Shibata:2010wz}
-- for which a slightly different argument for instability was given in~\cite{Emparan:2003sy}  -- by evolving a
Myers--Perry BH with a non-axisymmetric bar-mode deformation, using a NR code
adapted to higher dimensions.  In each case sufficiently rapidly rotating BHs
are found to be unstable against the bar-mode deformation. In terms of
a dimensionless spin parameter $q\equiv a/\mu^{1/(D-3)}$,
where $\mu,a$ are the standard mass and angular momentum parameters of the Myers--Perry solution, the
onset values for the instability were found to be: $D=5, \ q=0.87 ; D=6, \
q=0.74 ; D=7, \ q=0.73 ; D=8, \ \ q=0.77$. We remark that the corresponding
values found in~\cite{Dias:2009iu} for the Gregory--Laflamme instability in $7\le
D \le 9$ are always larger than unity. Thus, the instability triggered by
non-axisymmetric perturbations sets in for lower angular momenta than the
axisymmetric Gregory--Laflamme instability. Moreover, in~\cite{Shibata:2010wz},
long-term numerical evolutions have been performed to follow the non-linear
development of the instability. The central conclusion is that the unstable BHs
relax to stable configurations by radiating away the excess angular
momentum. These results have been confirmed for $D=6,7$ by a linear
analysis in Ref.~\cite{Dias:2014eua}; such linear analysis suggests
that for $D=5$, however, the single spinning Myers--Perry BH is
linearly stable.%
\epubtkFootnote{The $D=5$ analysis
  in~\cite{Shibata:2009ad,Shibata:2010wz} has been revised in yet
  unpublished work, obtaining better agreement with the linear results
  in~\cite{Dias:2014eua}, cf.\ Ref.~\cite{talkshibata}.}

Results for the $D=6$ ``gravitational waveforms'' $h_{+,\,\times}$ (see
Eqs.~(43)\,--\,(44) in Ref.~\cite{Shibata:2010wz} for a precise definition of these
quantities) for an initial dimensionless $q=0.801$
are shown in Figure~\ref{fig:shibata}. The early stage shows an exponential increase of the
amplitude, after which a saturation phase is reached, and where angular momentum
is being shed through GW emission.  After this stage,
exponential decay of the oscillations ensues. This bar-mode instability
discovered -- before linear perturbation analysis -- in
Refs.~\cite{Shibata:2009ad,Shibata:2010wz} has recently been seen
at linear level for BHs having all angular momentum parameters equal in Refs.~\cite{Hartnett:2013fba,Emparan:2014jca}.

Another spacetime instability seen at linear level
is the superradiant instability of rotating or charged BHs in the presence of
massive fields or certain boundary conditions. This instability will be
discussed in detail in the next section.

Concerning the non-linear stability of BH solutions, the only generic statement one can produce at the moment is that  hundreds of NR evolutions of binary Kerr or Schwarzschild BHs in vacuum, over the last decade, lend empirical support to the non-linear stability of these solutions. One must remark, however, on the limitations of testing instabilities with NR simulations. For instance, fully non-linear dynamical simulations cannot probe -- at least at present -- extremal Kerr BHs; they are also unable to find instabilities associated with very high harmonic indices $\ell,m$ (associated with very small scales), as well as instabilities that may grow very slowly. Concerning the first caveat, it was actually recently found by Aretakis that extremal RN and Kerr BHs are linearly unstable against scalar perturbations~\cite{Aretakis:2011ha,Aretakis:2011hc,Aretakis:2012ei}, an observation subsequently generalised to more general linear fields~\cite{Lucietti:2012sf} and to a non-linear analysis~\cite{Murata:2013daa,Aretakis:2013dpa}. This growth of generic initial data on extremal horizons seems to be a very specific property of extremal BHs, in particular related to the absence of a redshift effect~\cite{Murata:2013daa}, and there is no evidence a similar instability occurs for non-extremal solutions.

To conclude this section let us briefly address the stability of BH \emph{interiors} already discussed in Section~\ref{section_interior}. The picture suggested by Israel and Poisson~\cite{Poisson:1989zz, Poisson:1990eh} of mass inflation has been generically confirmed in a variety of toy models -- i.e., not Kerr -- by numerical evolutions~\cite{Burko:1997zy, Burko:1998jz, Burko:2002fv, Hansen:2005am, Avelino:2009vv, Hwang:2010im, Avelino:2011ee} and also analytical arguments~\cite{Dafermos:2003wr}.  Other numerical/analytical studies also suggest the same holds for the realistic Kerr case~\cite{Hamilton:2008zz, Hamilton:2009hu, Hamilton:2010hq, Luk:2013cqa}. As such, the current picture is that mass inflation will drive the curvature to Planckian values, near \emph{or} at the Cauchy horizon. The precise nature of the consequent singularity, that is, if it is space-like or light-like, is however still under debate (see, e.g., \cite{talkdafermos}).

\subsection{Superradiance and fundamental massive fields}
\label{sec:fundamental_fields}

There are several reasons to consider extensions of GR with
minimally, or non-minimally coupled massive scalar fields with mass parameter $\mu_S$.  As mentioned in
Section~\ref{sec:beyondastro}, ultra-light degrees of freedom appear in the
\emph{axiverse} scenario~\cite{Arvanitaki:2009fg,Arvanitaki:2010sy} and they play
an important role in cosmological models and also in dark matter models.
Equally important is the fact that massive scalar fields are a very simple proxy
for more complex, realistic matter fields, the understanding of which in full NR
might take many years to achieve.

At linearized level, the behavior of fundamental fields in the vicinities of
non-rotating BHs has been studied for decades, and the main features can
be summarized as follows:

\noindent(i) A prompt response at early times, whose features depend on the
initial conditions. This is the counterpart to light-cone propagation in flat
space.

\noindent(ii) An exponentially decaying ``ringdown'' phase at intermediate
times, where the BH is ringing in its characteristic QNMs.
Bosonic fields of mass $\mu_S \hbar$ introduce both
an extra scale in the problem and a potential barrier at distances $\sim 1/\mu_S$,
thus effectively trapping fluctuations. In this case, extra modes appear which
are quasi-bound states, i.e., extremely long-lived states effectively turning
the BH into a quasi-hairy
BH~\cite{Barranco:2012qs,Witek:2012tr,Dolan:2012yt,Rosa:2011my,Pani:2012vp,Brito:2013wya}.

\noindent (iii) At late times, the signal is dominated by a power-law fall-off,
known as ``late-time tail''
\cite{Price:1971fb,Leaver:1986gd,Ching:1995tj,Koyama:2001ee}. Tails are caused
by backscattering off spacetime curvature (and a potential barrier induced by
massive terms) and more generically by a failure of Huygens' principle. In other
words, radiation in curved spacetimes travels not only \emph{on}, but
\emph{inside} the entire light cone.

When the BH is rotating, a novel effect can be triggered:
{\emph{superradiance}}~\cite{zeldovich1,zeldovich2,Bekenstein:1973mi,Cardoso:2013krh}.
Superradiance consists of energy extraction from rotating BHs, and a transfer of
this energy to the interacting field~\cite{Cardoso:2013krh}.  For a
monochromatic wave of frequency $\omega$, the condition for superradiance
is~\cite{zeldovich1,zeldovich2,Bekenstein:1973mi,Cardoso:2013krh}
\begin{align}
\label{eq:MFSRcond}
\omega < m\Omega_H\,,
\end{align}
where $m$ is the azimuthal harmonic index and $\Omega_H$ is the angular velocity
of the BH horizon.  If, in addition, the field is massive, a ``BH bomb-type''
mechanism can
ensue~\cite{Damour:1976,Detweiler:1980uk,Zouros:1979iw,Cardoso:2004nk} leading
to an instability of the spacetime and the growth of a scalar condensate outside
the BH
horizon~\cite{Pani:2012bp,Yoshino:2012kn,Witek:2012tr,Dolan:2012yt,Okawa:2014nda,Cardoso:2013krh}.
The rich phenomenology of scenarios where fundamental fields couple to gravity
motivated recent work on the subject, where full nonlinear evolutions are
performed~\cite{Okawa:2014nda,East:2013mfa,Healy:2011ef,Berti:2013gfa}.  East
 et~al.~\cite{East:2013mfa} have performed nonlinear scattering experiments, solving the field
equations in the generalized harmonic formulation, and constructing initial data
representing a BH with dimensionless spin $a/M=0.99$, and an incoming
quadrupolar GW packet. Their results are summarized in
Figure~\ref{fig:ah_fig}, for three different wavepacket frequencies,
$M\omega=0.75,\,0.87,\,1$ (note that only the first is superradiant according to
condition \eqref{eq:MFSRcond}).  The wavepackets carry roughly $10\%$ of the
spacetime's total mass. These results confirm that low frequency radiation does
extract mass and spin from the BH (both the mass $M_{\rm BH}$ and spin $J_{\rm
  BH}$ of the BH decrease for the superradiant wavepacket with $M\omega=0.75$),
and that nonlinear results agree quantitatively with linear predictions for
small wavepacket amplitudes~\cite{Teukolsky:1974yv}. To summarize,
superradiance is confirmed at full nonlinear level, providing a rigorous
framework for the complex dynamics that are thought to arise for massive fields
around rotating BHs.

\epubtkImage{}{%
  \begin{figure}[htbp]
    \centerline{\includegraphics[width=0.7\textwidth]{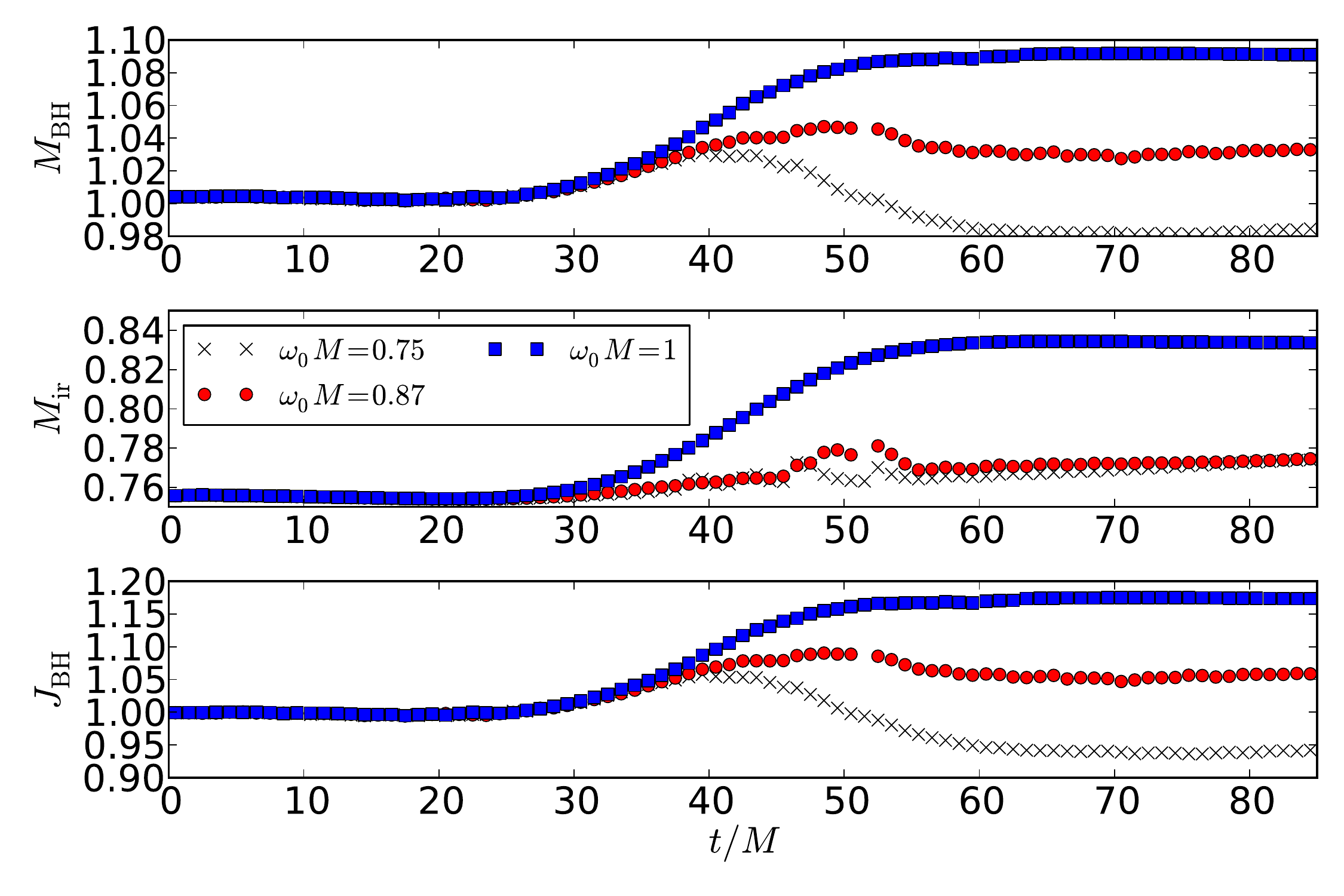}}
    \caption{Evolution of a highly spinning BH ($a/M=0.99$) during interaction with
  different frequency GW packets, each with initial mass $\approx 0.1M$.  Shown
  (in units where $M=1$) are the mass, irreducible mass, and angular momentum of
  the BH as inferred from AH properties.
  From~\cite{East:2013mfa}.}
\label{fig:ah_fig}
\end{figure}}

Self-interacting scalars can give rise to stable or very long-lived
configurations. For example, self-interacting complex scalar fields can form
\emph{boson stars} for which the scalar field has an oscillatory nature, but the
metric is
stationary~\cite{Jetzer:1991jr,Schunck:2003kk,Liebling:2012fv,Macedo:2013qea}.
Real-valued scalars can form oscillating solitons or ``oscillatons'', long-lived
configurations where both the scalar field and the metric are
time-dependent~\cite{Seidel:1991zh,Seidel:1993zk,Page:2003rd,Okawa:2013jba}.
Dynamical boson star configurations were studied by several
authors~\cite{Liebling:2012fv}, with focus on boson star collisions with
different velocities and impact
parameters~\cite{Palenzuela:2007dm,Palenzuela:2006wp,Choptuik:2009ww}.  These
are important for tests of the hoop and cosmic censorship conjectures, and were
reviewed briefly in Sections~\ref{sec:cosmic_censor_review} and
\ref{sec:hoop_review}. For a thorough discussion and overview of
results on dynamical boson stars we refer the reader to the Living
Reviews article by Liebling and Palenzuela~\cite{Liebling:2012fv}.

The first steps towards understanding the nonlinear interaction between massive
fields and BHs were taken by Okawa et~al.~\cite{Okawa:2014nda,Okawa:2014sxa}, who
found new ways to prescribe, and evolve, constraint-satisfying initial data,
analytically or semi-analytically, for minimally coupled self-interacting scalar
fields~\cite{Okawa:2014nda,Okawa:2013jba}. This construction was reviewed in
Section~\ref{sec:initdata}. In Ref.~\cite{Okawa:2014nda}, the authors used
this procedure to generate initial data and to evolve wavepackets of arbitrary
angular shape in the vicinity of rotating BHs. Their results are summarized in
Figure~\ref{fig:multipoles_massive}. Spherically symmetric initial data for
massless fields reproduce previous results in the
literature~\cite{Gundlach:1993tn}, and lead to power-law tails of integer
index.
The mass term adds an extra scale and a barrier at large distances,
resulting in characteristic late-time tails of massive fields.

Full nonlinear results from Ref.~\cite{Okawa:2014nda} are reproduced in
Figure~\ref{fig:multipoles_massive}, and agree with linearized predictions. Higher
multipoles ``feel'' the centrifugal barrier close to the light ring which,
together with the mass barrier at large distances, provides a confining
mechanism and gives rise to almost stationary configurations, shown in the right
panel of Figure~\ref{fig:multipoles_massive}. The beating patterns are a
consequence of the excitation of different overtones with similar ringing
frequency~\cite{Witek:2012tr,Okawa:2014nda}.  These ``scalar condensates'' are
extremely long-lived and can, under some circumstances, be considered as adding
hair to the BH. They are not however really stationary: the changing quadrupole
moment of the ``scalar cloud'' triggers the simultaneous release of
gravitational radiation~\cite{Okawa:2014nda,Yoshino:2013ofa,Okawa:2014sxa}. In fact,
gravitational radiation is one of the most important effects not captured by
linearized calculations.  These nontrivial results extend to higher multipoles,
which display an even more complex behavior~\cite{Okawa:2014nda,Okawa:2014sxa}.

Although only a first step towards understanding the physics of fundamental
fields in strong-field gravity, these results are encouraging. We expect 
that with more robust codes and longer simulations one will be able to fully
explore the field, in particular, the following features.

\epubtkImage{}{%
  \begin{figure}[htbp]
    \centerline{
      \includegraphics[width=0.5\textwidth]{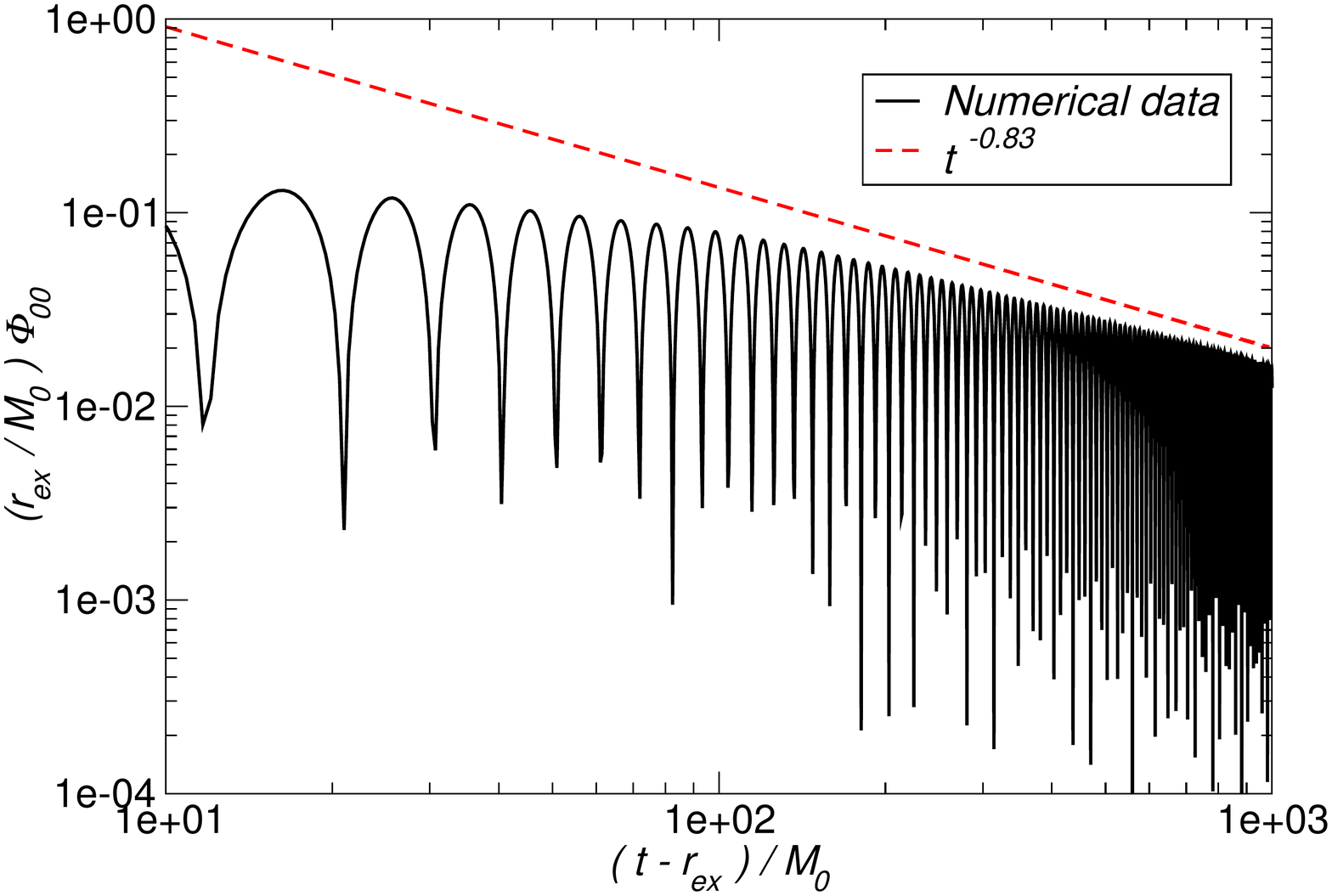}
      \includegraphics[width=0.5\textwidth]{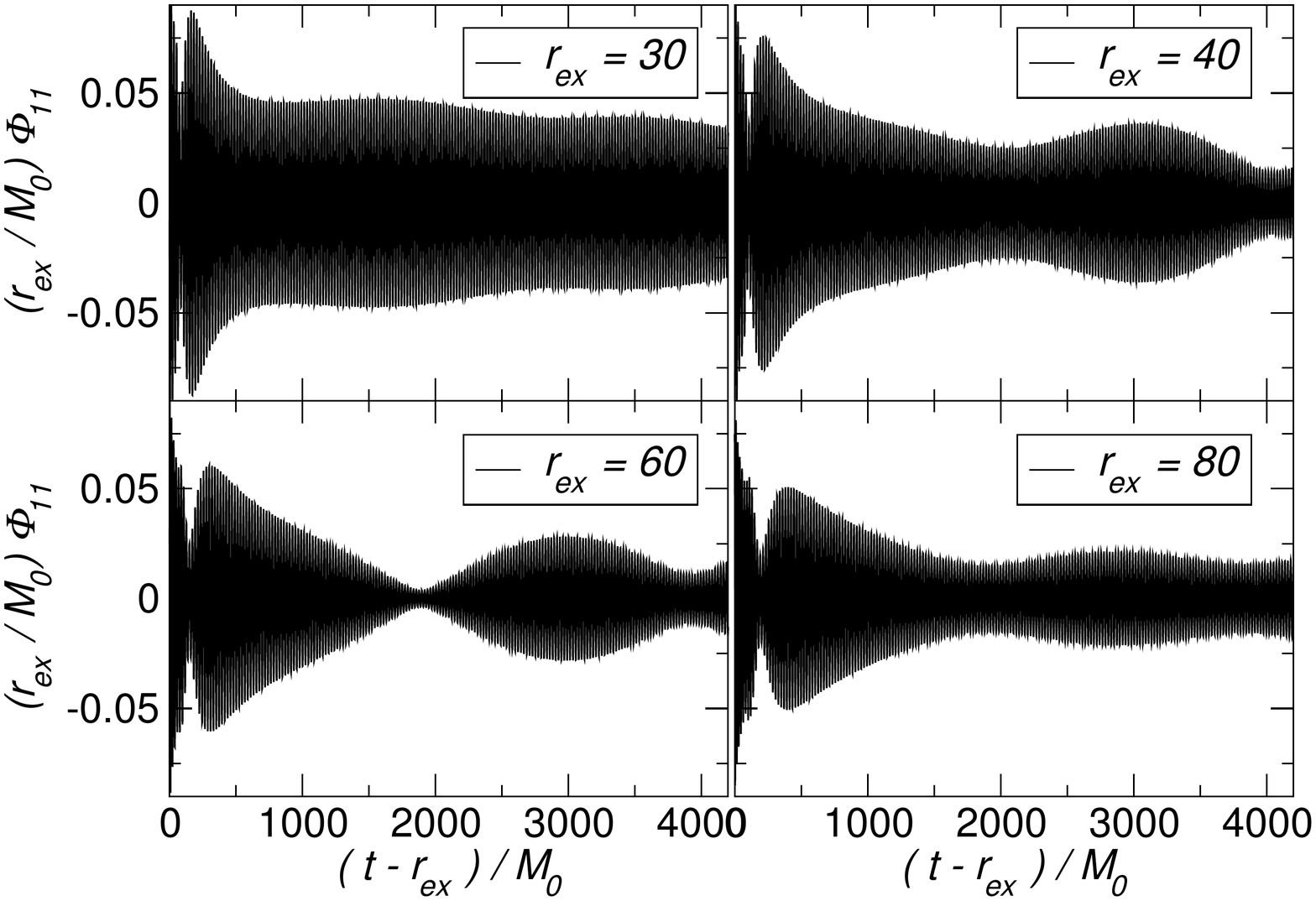}
    }
    \caption{Massive scalar field (nonlinear) evolution of the spacetime of an initially non-rotating BH, with $M\mu=0.29$. 
    \emph{Left panel:} Evolution of a spherically symmetric $l=m=0$ scalar waveform, 
measured at $r_{\rm{ex}}=40M$, with $M$ the initial BH mass. In addition to the numerical 
data (black solid curve) we show a fit to the late-time tail 
 (red dashed curve) with $t^{-0.83}$, in excellent agreement with linearized analysis.
 \emph{Right panel:} The dipole signal resulting from the evolution of an $l=m=1$
 massive scalar field around a non-rotating BH.  The waveforms, extracted at
 different radii $r_{\rm{ex}}$ exhibit pronounced beating patterns caused by
 interference of different overtones.  The critical feature is however, that
 these are extremely long-lived configurations.  From
 Ref.~\cite{Okawa:2014nda}.}
    \label{fig:multipoles_massive}
\end{figure}}

\begin{description}
\item[$\bullet$]{Superradiant instability and its saturation.}  The timescales
  probed in current nonlinear simulations are still not sufficient to
  unequivocally observe superradiance with test scalar fields.  The main reason
  for this is the feebleness of such instabilities: for scalar fields they have,
  at best, an instability timescale of order $10^7 M$ for carefully tuned scalar
  field mass. However, current long-term simulations are able to extract
  GWs induced by the scalar cloud~\cite{Okawa:2014nda}.

  The biggest challenge ahead is to perform simulations which are accurate
  enough and last long enough to observe the scalar-instability growth
  \emph{and} its subsequent saturation by GW emission. This will allow
  GW templates for this mechanism to finally be released.

  Due to their simplicity, scalar fields are a natural candidate to carry on
  this program, but they are not the only one.  Massive vector fields, which
  are known to have amplification factors one order of magnitude larger, give
  rise to stronger superradiant instabilities, and might also be a good
  candidate to finally observe superradiant instabilities at the nonlinear
  level. We note that the development of the superradiant instability may, in some special cases, lead to a truly
  asymptotically flat, hairy BH solution of the type recently discussed in~\cite{Herdeiro:2014goa}.

\item[$\bullet$]{Turbulence of massive fields in strong gravity.} Linearized
  results indicate that the development of superradiant instabilities leaves
  behind a scalar cloud with scalar particles of frequency $\omega\sim m\Omega$,
  in a nearly stationary state.  This system may therefore be prone to
  turbulent effects, where nonlinear terms may play an important role.  One
  intriguing aspect of these setups is the possibility of having gravitational
  turbulence or collapse on sufficiently large timescales. Such effects were
  recently observed in ``closed'' systems where scalar fields are forced to
  interact gravitationally for long
  times~\cite{Bizon:2011gg,Maliborski:2012gx,Buchel:2013uba}. It is
  plausible that quasi-bound states are also prone to such effects, but in
  asymptotically flat spacetimes.

\item[$\bullet$]{Floating orbits.} Our discussion until now has focused on
  minimally coupled fundamental fields. If couplings to matter exist, new
  effects are possible: a small object (for example, a star), orbiting a
  rotating, supermassive BH might be able to extract energy and angular momentum
  from the BH and convert it to gravitational radiation.  For this to
  happen, the object would effectively stall at a superradiant orbit, with a
  Newtonian frequency $\Omega=2\mu_S$ for the dominant quadrupolar emission. These
  are called \emph{floating orbits}, and were verified at linearized
  level~\cite{Cardoso:2011xi,Cardoso:2013krh}.%
\epubtkFootnote{Floating orbits would manifest themselves in
  observations by depleting the inner part of accretion disks of stars
  and matter, and modifying the emitted gravitational waveform.}
Nonlinear evolutions of systems on floating
  orbits are extremely challenging on account of all the different extreme
  scales involved.

\item[$\bullet$]{Superradiant instabilities in AdS.} The mechanism
  behind superradiant instabilities relies on amplification close to the horizon
  and reflection by a barrier at large distances. Asymptotically AdS
  spacetimes provide an infinite-height barrier, ideal for the instability to
  develop~\cite{Cardoso:2004nk,Cardoso:2004hs,Cardoso:2006wa,Uchikata:2009zz,Cardoso:2013pza}.

Because in these backgrounds there is no dissipation at infinity, it is both
possible and likely that new, non-symmetric final states arise as a consequence
of the superradiant instability~\cite{Cardoso:2006wa,Cardoso:2009bv,Li:2013pra,Dias:2011at}.
Following the instability growth and its final state remains a challenge for
NR in asymptotically AdS spacetimes.

\item[$\bullet$]{Superradiant instabilities of charged BHs.}
  Superradiant amplification of \emph{charged} bosonic fields can occur in the
  background of charged BHs, in quite a similar fashion to the rotating case
  above, as long as the frequency of the impinging wave $\omega$ obeys
\begin{align}
\label{eq:MFSRcond2}
\omega < q\Phi_H\,,
\end{align}
where $q$ is the charge of the field and $\Phi_H$ is the electric potential on
the BH horizon. In this case both charge and Coulomb energy are extracted from
the BH in a way compatible with the first and second law of BH thermodynamics
\cite{Bekenstein:1973mi}. In order to have a recurrent scattering, and hence, an
instability, it is not enough, however, to add a mass term to the field~\cite{Furuhashi:2004jk,Hod:2013nn,Hod:2013eea,Degollado:2013eqa,Sampaio:2014swa}; but an
instability occurs either by imposing a mirror like boundary condition at some
distance from the BH (i.e., a boxed BH) or by considering an asymptotically AdS
spacetime. In Refs.~\cite{Herdeiro:2013pia,Degollado:2013bha} it has been
established, through both a frequency and a time domain analysis, that the time
scales for the development of the instability for boxed BHs can be made much
smaller than for rotating BHs (in fact, arbitrarily small
\cite{Hod:2013fvl}). Together with the fact that even
$s$-waves, i.e., $\ell=0$ modes, 
can trigger the instability in charged BHs, makes the numerical study of the
non-linear development of this type of superradiant instabilities particularly
promising. One should be aware, however, that there may be qualitative
differences in both the development and end-point of superradiant instabilities
in different setups. For instance, for AdS and boxed BHs, the end-point is
likely a hairy BH, such as those constructed in Ref.~\cite{Dias:2011at} (for rotating BHs), since the scalar field cannot be dissipated anywhere. This
applies to both charged and rotating backgrounds. By contrast, this does not
apply to asymptotically flat spacetimes, wherein rotating (but not charged)
superradiance instability may occur. It is then an open question if the system approaches a hairy BH -- of the type constructed in~\cite{Herdeiro:2014goa} -- or if the field is completely radiated/absorbed by the BH. Concerning the development of the
instability, an important difference between the charged and the rotating cases
may arise from the fact that a similar role, in Eqs.~\eqref{eq:MFSRcond} and
\eqref{eq:MFSRcond2}, is played by the field's azimuthal quantum number $m$ and
the field charge $q$; but whereas the former may change in a non-linear
evolution, the latter is conserved~\cite{Cardoso:2013pza}.

\end{description}

\subsection{High-energy collisions}
\label{sec:HEcollisions}

Applications of NR to collisions of BHs or
compact matter sources near the speed of light are largely
motivated by probing GR in its most violent regime and by
the modelling of BH formation in TeV gravity scenarios. The most important
questions that arise in these contexts can be summarized as follows.
\begin{itemize}
\item Does cosmic censorship still apply under the extreme conditions of
  collisions near the speed of light? As has already been discussed in
  Section~\ref{sec:cosmic_censor_review}, numerical simulations of these collisions in four dimensions
  have so far identified horizon formation in agreement with the censorship
  conjecture. The results of higher-dimensional simulations are still not fully understood, cf. Section~\ref{sec:cosmic_censor_review}.
\item Do NR simulations of high-energy particle collisions provide evidence
  supporting the validity of the hoop conjecture?  As discussed in
  Section~\ref{sec:hoop_review}, NR results have so far confirmed the hoop
  conjecture.
\item In collisions near the speed of light, the energy mostly consists of the
  kinetic energy of the colliding particles such that their internal structure
  should be negligible for the collision dynamics. Furthermore, the
  gravitational field of a particle moving at the speed of light is
  non-vanishing only near the particle's worldline~\cite{Aichelburg:1970dh},
  suggesting that the gravitational interaction in high-energy collisions should be dominant
  at the instant of collision and engulfed inside the horizon that forms.  This
  conjecture has sometimes been summarized by the statement that ``matter does
  not matter''~\cite{Choptuik:2009ww}, and is related to the hoop conjecture
  discussed above.  Do NR simulations of generic high-energy collisions of compact objects support this
  argument in the classical regime, i.e., does the modelling of the colliding
  objects as point particles (and, in particular, as BHs) provide an accurate
  description of the dynamics?
\item Assuming that the previous question is answered in the affirmative, what
  is the scattering threshold for BH formation? This corresponds to determining
  the threshold impact parameter $b_{\rm scat}$ that separates collisions
  resulting in the formation of a single BH ($b<b_{\rm scat}$) from scattering
  encounters ($b>b_{\rm scat}$), as a function of the number of spacetime
  dimensions $D$ and the collision velocity $v$ in the
  center-of-mass frame or boost parameter $\gamma =
  1/\sqrt{1-v^2}$.
\item How much energy and momentum is lost in the form of GWs
  during the collision? By conversion of energy and momentum, the GW
  emission determines the mass and spin of the BH (if formed) as a function
  of the spacetime dimension $D$, scattering parameter $b$, and boost factor
  $\gamma$ of the collision. Collisions near the speed of light are also
  intriguing events to probe the extremes of GR; in particular what is the
  maximum radiation that can be extracted from any collision and does the
  luminosity approach Dyson's limit $dE/dt \lesssim 1$~\cite{Dyson}? (See discussion in Section~\ref{sec_most_lum} about this limit.)
\end{itemize}
These issues are presently rather well understood through NR simulations in
$D=4$ spacetime dimensions but remain largely unanswered for the important cases
$D\ge 5$.

The relevance of the internal structure of the colliding bodies has been studied
in Ref.~\cite{Sperhake:2012me}, comparing the GW emission and scattering
threshold in high-energy collisions of rotating and non-rotating BHs in
$D=4$. The BH spins of the rotating configurations are either aligned or
anti-aligned with the orbital angular momentum corresponding to the so-called
\emph{hang-up} and \emph{anti-hang-up} cases which were found to have particularly
strong effects on the dynamics in quasi-circular BH binary inspirals
\cite{Campanelli:2006uy}.  In high-energy collisions, however, this
(anti-)hang-up effect disappears; the GW emission as well as the scattering
threshold are essentially independent of the BH spin at large collision
velocities (cf.~Figure~\ref{fig:v2d4_bscat_emax2} which will be discussed in more
detail further below). These findings suggest that ultra-relativistic collisions
are indeed well modelled by colliding point-particles or BHs in GR.  In the
center-of-mass frame, and assuming that the two particles have equal mass, the
collisions are characterized by three parameters. (i) The number $D$ of
spacetime dimensions, (ii) the Lorentz factor $\gamma$ or, equivalently, the
collision velocity $v$, and (iii) the impact parameter $b=L/P$, where $L$ and
$P$ are the initial orbital angular momentum and the linear momentum of either
BH in the center-of-mass frame.

The simplest set of configurations consists of head-on collisions with $b=0$ in
$D=4$ dimensions and was analysed in Ref.~\cite{Sperhake:2008ga} varying the
boost parameter in the range $1.07 \le \gamma \le 3$. In agreement with the
cosmic censorship conjecture, these collisions always result in the formation of
a single BH that settles into a stationary configuration through quasi-normal
ringdown. The total energy radiated in the form of GWs is well modelled by the
following functional form predicted by Smarr's~\cite{Smarr:1977fy}
zero-frequency limit (see Section~\ref{sec:ZFLsection})
\begin{equation}
  \frac{E}{M} = E_{\infty} \left(
        \frac{1+2\gamma^2}{2\gamma^2}
        + \frac{(1-4\gamma^2)\,\log(\gamma + \sqrt{\gamma^2-1})}
          {2\gamma^3 \sqrt{\gamma^2 - 1}}\right) \,,
  \label{eq:ho_EZFL}
\end{equation}
where $E_{\infty}$ is a free parameter that corresponds to the fraction of
energy radiated in the limit $\gamma \rightarrow \infty$. Fitting
the numerical results with Eq.~(\ref{eq:ho_EZFL}) yields $E_{\infty}=14\pm 3~\%$
which is about half of Penrose's upper limit~\cite{Penrose1974,Eardley:2002re}.
Observe the good agreement with the second order result in
Eq.~(\ref{eq:shock}) as discussed in Section~\ref{sec:shock_waves}.

\epubtkImage{}{%
  \begin{figure}
    \centerline{\includegraphics[height=200pt,clip=true]{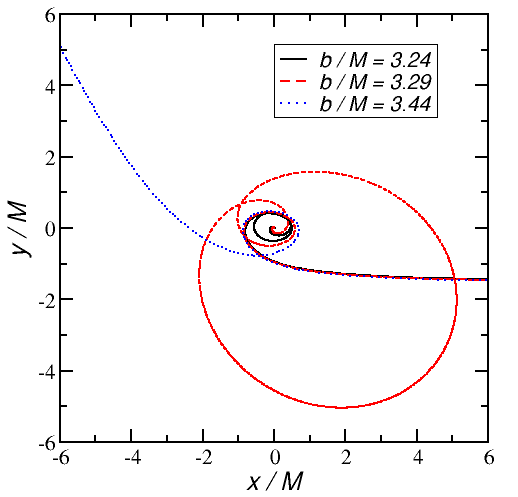}}
      \caption{BH trajectories in grazing collisions for $\gamma=1.520$
           and three values of the impact parameter corresponding to
           the regime of prompt merger (solid, black curve), of
           delayed merger (dashed, red curve), and scattering
           (dotted, blue curve). Note that for each case, the trajectory
           of one BH is shown only; the other BH's location is given
           by symmetry across the origin.}
         \label{fig:traj_EGW1}
\end{figure}}

\epubtkImage{}{%
  \begin{figure}
    \centerline{
      \includegraphics[height=170pt,clip=true]{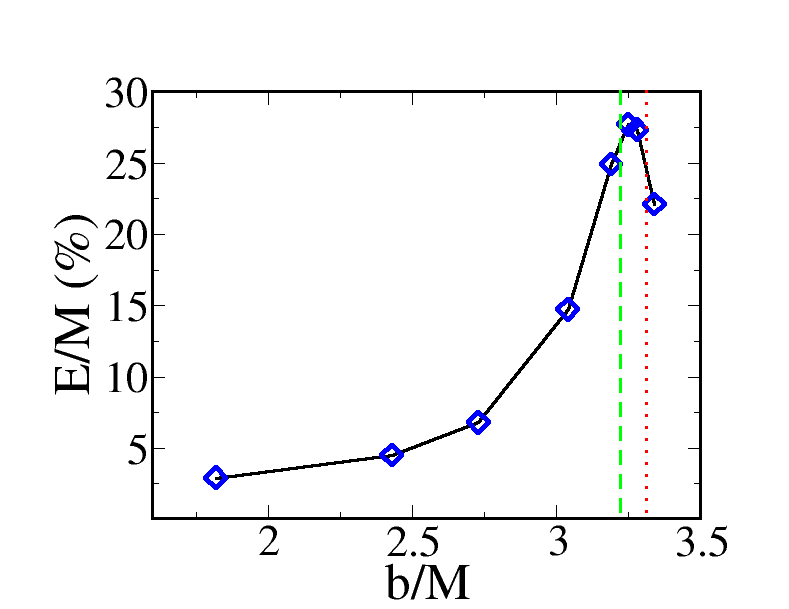}
      \includegraphics[height=170pt,clip=true]{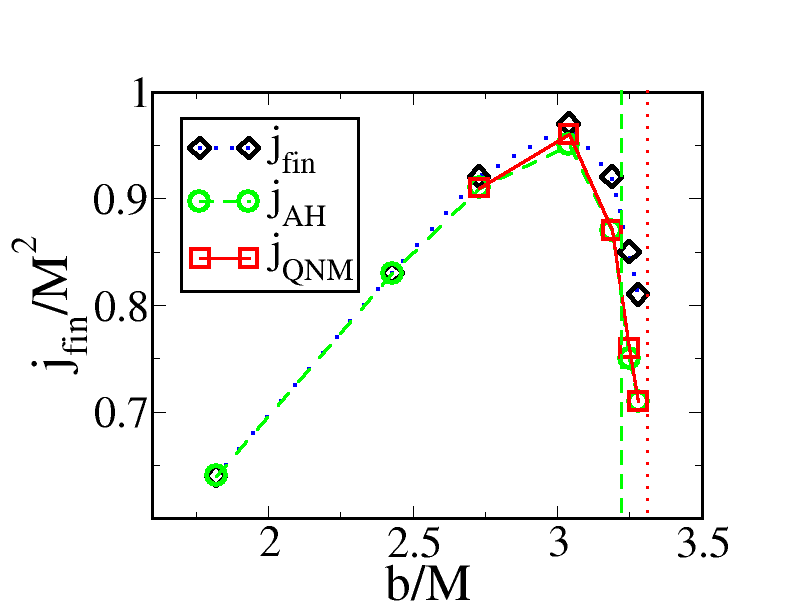}
    }
    \caption{Total energy radiated in GWs (left panel) and final dimensionless
           spin of the merged BH (right panel) as a function of impact parameter $b$
           for the same grazing collisions with $\gamma=1.520$.
           The vertical dashed (green) and dash-dotted (red) lines mark
           $b^*$ and $b_{\rm scat}$, respectively. From Ref.~\cite{Sperhake:2009jz}.}
          \label{fig:traj_EGW2}
\end{figure}}

Grazing collisions in four dimensions represent two-parameter studies, where the
boost factor $\gamma$ and the impact parameter $b$ are varied, and have been
investigated in Refs.~\cite{Shibata:2008rq,Sperhake:2009jz}.  At fixed Lorentz
boost, such grazing collisions exhibit three distinct regimes as the impact
parameter is increased from the head-on limit $b=0$: (i) prompt mergers, (ii)
delayed mergers, and (iii) the scattering regime where no common horizon
forms. These regimes are marked by two special values of the impact parameter
$b$, the scattering threshold $b_{\rm scat}$ that we have already mentioned
above and the \emph{threshold of immediate merger} $b^*$. This threshold has been
identified in numerical BH simulations by Pretorius \& Khurana
\cite{Pretorius:2007jn} as marking the onset of a regime where the two
BHs \emph{whirl} around each other prior to merging or scattering off
for a number of orbits proportional to $\log|b-b^*|$;
see also~\cite{Healy:2009zm,Gold:2009hr}. This \emph{zoom-whirl-like}
behaviour has also been identified in high-energy grazing collisions in
\cite{Sperhake:2009jz}.  The three different regimes are illustrated in Figure~\ref{fig:traj_EGW1} which shows the BH trajectories for $\gamma =
1.520$ and $b/M=3.24$, $3.29$ and $3.45$. For this boost factor, the thresholds
are given by $b^*/M\approx 3.25$ and $b_{\rm scat}/M\approx 3.35$.  For impact
parameters close to the threshold values $b^*$ and $b_{\rm scat}$, grazing
collisions can generate enormous amounts of GWs. This is shown in the left
panel of Figure~\ref{fig:traj_EGW2}. Starting from $E/M \approx 2.2~\%$ in the
head-on limit $b=0$, the radiated energy increases by more than an order of
magnitude to $\gtrsim 25~\%$ for $b^* < b < b_{\rm scat}$.  These simulations
can also result in BHs spinning close to the extremal Kerr limit as is shown in
the right panel of the figure which plots the dimensionless final spin as a
function of $b$.
By fitting their numerical results, Shibata et~al.~\cite{Shibata:2008rq}
have found an empirical relation for the scattering threshold given by
\begin{equation}
  \frac{b_{\rm scat}}{M} = \frac{2.50 \pm 0.05}{v}\,.
  \label{eq:shibata_bscat}
\end{equation}
For such a value of the impact parameter, they observe that the dimensionless
final spin of the merged BH is given by $J_{\rm fin} = (0.6 \pm 0.1) J$, where
$J/M^2 = 1.25 \pm 0.03$ is the initial angular momentum of the system.

\epubtkImage{}{%
  \begin{figure}
    \centerline{
      \includegraphics[height=182pt,clip=true]{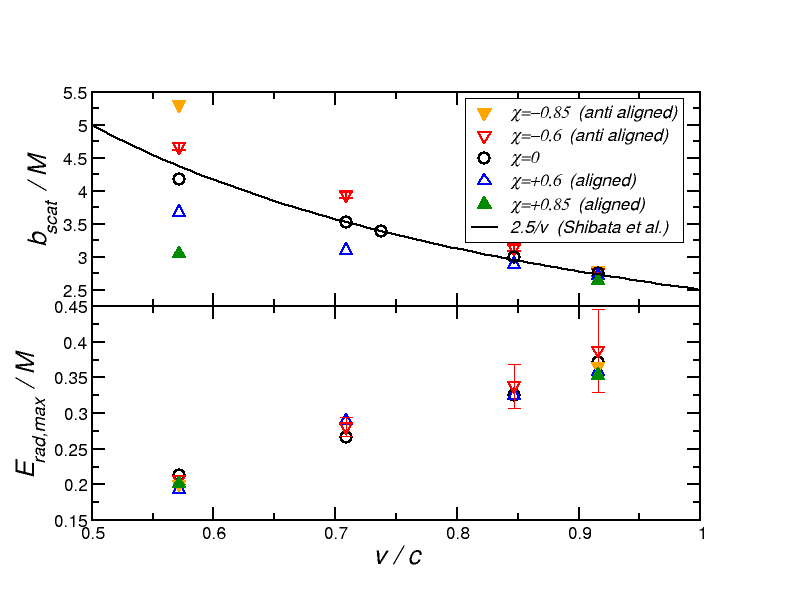}
      \includegraphics[height=182pt,clip=true]{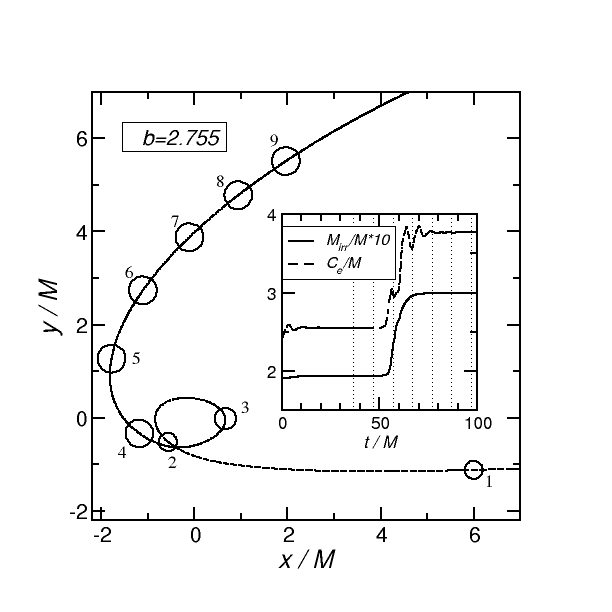}
    }
    \caption{\emph{Left panels:} Scattering threshold (upper panel) and maximum
  radiated energy (lower panel) as a function of $v$.
  Colored ``triangle'' symbols pointing up and down
  refer to the aligned and antialigned cases,
  respectively. Black ``circle'' symbols represent the thresholds for
  the nonspinning configurations. Right panel: Trajectory of one BH
  for a delayed merger configuration with anti-aligned spins $j=0.65$.
  The circles represent the BH location at equidistant intervals
  $\Delta t=10~M$ corresponding to the vertical lines in the inset
  that shows the equatorial circumference of the BH's AH
  as a function of time.}
  \label{fig:v2d4_bscat_emax2}
\end{figure}}

Grazing collisions of spinning and non-spinning BHs have been compared in
Ref.~\cite{Sperhake:2012me}. The initial configurations for these simulations
have been chosen with $\gamma$ factors up to 2.49 and equal spin for both BHs
of dimensionless magnitude $\chi =0.85$ and 0.65 aligned or anti-aligned with
the orbital angular momentum $L$. This set has been complemented with collisions
of non-spinning BHs covering the same range in $\gamma$. The scattering
threshold and the energy radiated in GWs in these simulations are shown in the
left panel of Figure~\ref{fig:v2d4_bscat_emax2}.  As expected from the hang-up
effect, aligned (anti-aligned) spins result in a smaller (larger) value of the
scattering threshold $b_{\rm scat}$ at low collision speeds.  At velocities
above $\sim 80\%$ of the speed of light, however, this effect is washed out
and, in agreement with the matter-does-not-matter conjecture mentioned above,
the collision dynamics are barely affected by the BH spins. Furthermore, the
scattering threshold determined for non-spinning BHs agrees very well with the
formula~(\ref{eq:shibata_bscat}).  As demonstrated in the bottom panel of the
figure, the radiated energy is barely affected by the BH spin even in the mildly
relativistic regime.  The simulations also suggest an upper limit of the
fraction of kinetic energy that can be converted into GWs.  Extrapolation of the
data points in Figure~\ref{fig:v2d4_bscat_emax2} to $v=1$ predicts that at most
about half of the total energy can be dissipated in GWs in any four dimensional
collision.  The other half, instead, ends up as
rest mass inside the common horizon formed in merging configurations or is
absorbed by the individual BHs during the close encounter in scattering
processes (the result of this extrapolation is consistent with the calculation
in Ref.~\cite{Gundlach:2012aj}). This is illustrated in the right panel of
Figure~\ref{fig:v2d4_bscat_emax2}, where the trajectory of one BH in a delayed
merger configuration with anti-aligned spins is shown. The circles, with radius
proportional to the horizon mass, represent the BH location at intervals $\Delta
t=10~M$.  During the close encounter, (i) the BH grows in size due to absorption
of gravitational energy and (ii) slows down considerably.

Collisions of BHs with electric charge have been simulated by Zilh{\~a}o et~al.~\cite{Zilhao:2012gp,Zilhao:2013nda}.  For the special case of BHs with
equal charge-to-mass ratio $Q/M$ and initially at rest, constraint-satisfying
initial data are available in closed analytic form. The electromagnetic wave
signal generated in these head-on collisions reveals three regimes similar to
the pattern known for the GW signal, (i) an infall phase prior to formation of a
common horizon, (ii) the non-linear merger phase where the wave emission reaches
its maximum and (iii) the quasi-normal ringdown.  As the charge-to-mass ratio is
increased towards $Q/M \lesssim 1$, the emitted GW energy decreases by about 3
orders of magnitude while the electromagnetic wave energy reaches a maximum at
$Q/M\approx 0.6$, and
drops towards $0$ in both the uncharged and the extreme limit.
This behaviour of the radiated energies is expected because of the decelerating
effect of the repulsive electric force between equally charged BHs. For
opposite electric charges, on the other hand, the larger collision velocity
results in an increased amount of GWs and electromagnetic
radiation~\cite{Zilhao:2013nda}.

An extended study of BH collisions using various analytic approximation
techniques including geodesic calculations and the ZFL has been
presented in Berti et~al.~\cite{Berti:2010ce}; see also
\cite{Berti:2010gx} for a first exploration in higher dimensions. Weak
scattering of BHs in $D=4$, which means large scattering parameters $b/M\sim
10$, and for velocities $v\approx 0.2$, has been studied by Damour et
al.~\cite{Damour:2014afa} using NR as well as PN and EOB
calculations. Whereas PN calculations start deviating significantly
from the NR results for $b/M \lesssim 10$, the NR calibrated EOB model
yields good agreement in the scattering angle throughout the weak scattering regime.

BH collisions in $D\ge 5$ spacetime dimensions are not as well understood as
their four-dimensional counterparts. This is largely a consequence of the fact
that NR in higher dimensions is not yet that robust and suffers more strongly from numerical
instabilities. Such complications in the higher-dimensional numerics do not
appear to cause similar problems in the construction of constraint satisfying
initial data. The spectral elliptic solver originally developed by Ansorg et~al.~\cite{Ansorg:2004ds} for $D=4$ has been successfully generalized to
higher $D$ in Ref.~\cite{Zilhao:2011yc} and provides
solutions with comparable accuracy as in $D=4$.

\epubtkImage{}{%
  \begin{figure}
    \centerline{\includegraphics[height=170pt,clip=true]{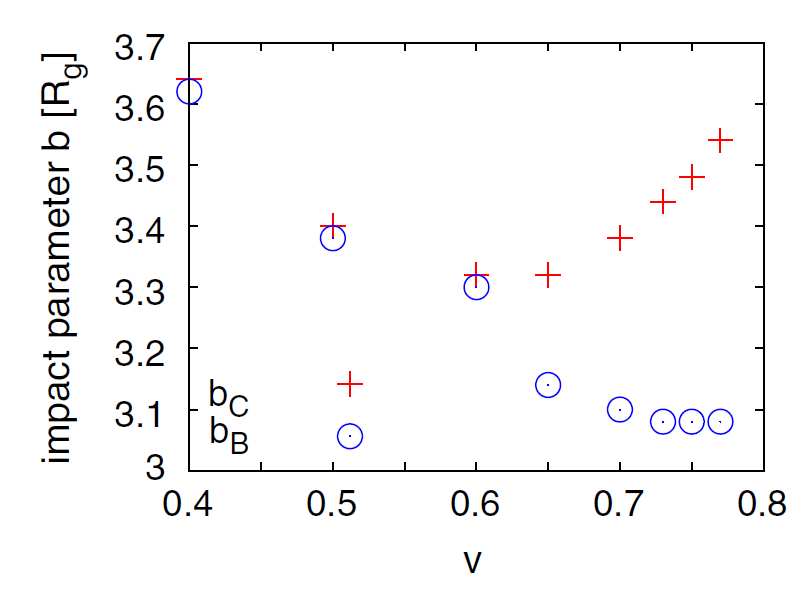}}
    \caption{The (red) plus and (blue) circle symbols mark scattering and merging
BH configurations, respectively, in the $b-v$ plane of impact parameter and
collision speed, for $D=5$ spacetime dimensions.}
\label{fig:d5_bscat}
\end{figure}}

A systematic exploration of the scattering threshold in $D=5$ dimensions has
been performed in Ref.~\cite{Okawa:2011fv}. By superposing
non-rotating, boosted single BH initial data, they have evolved grazing
collisions up to $v\lesssim 0.8$. Their results are summarized in
Figure~\ref{fig:d5_bscat}, where scattering (merging) BH collisions are marked by
``plus'' and ``circle'' symbols in the plane spanned by the collision velocity
$v$ and the impact parameter $b$.  The simulations show a decrease of the
scattering threshold at increasing velocity up to $v\approx 0.6$, similar to the
$D=4$ case in the upper left panel of Figure~\ref{fig:v2d4_bscat_emax2}. At larger
$v$, the threshold cannot yet be determined because simulations with near
critical impact parameter become numerically unstable. By monitoring the
Kretschmann scalar at the point of symmetry between the two BHs, a large curvature regime was furthermore identified in~\cite{Okawa:2011fv}, as discussed in Section~\ref{sec:cosmic_censor_review}.

\epubtkImage{}{%
  \begin{figure}
    \centerline{\includegraphics[height=220pt,clip=true]{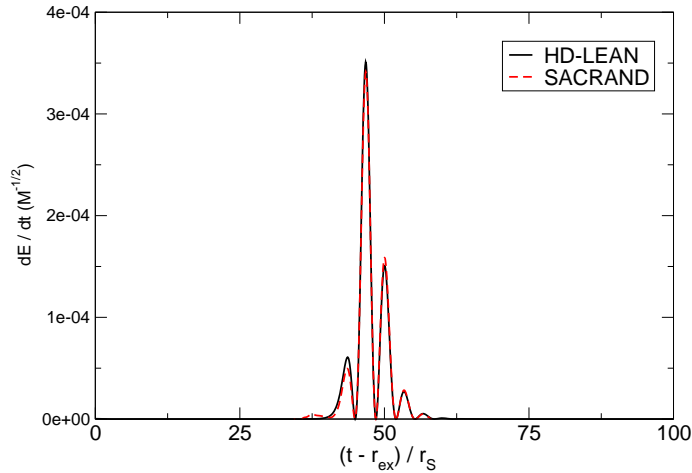}}
    \caption{Energy fluxes for head-on collisions of two BHs in $D=5$
      spacetime dimensions, obtained with two different codes,
      \textsc{HD-Lean}~\cite{Zilhao:2010sr,Witek:2010xi} (solid black
      line) and \textsc{SacraND}~\cite{Yoshino:2009xp,Okawa:2011fv}
      (red dashed line). The BHs start off at an initial coordinate
      separation $d/R_{\rm S}=6.47$. Taken from
      Ref.~\cite{Witek:2014mha}.}
    \label{fig:D5energyflux_comparison}
\end{figure}}

The GW emission in BH mergers in $D>4$ has
so far only been studied for collisions starting from rest. In Refs.~\cite{Witek:2010xi,Witek:2010az,Witek:2014mha} the wave signal was extracted using the
Kodama-Ishibashi formalism discussed in Section~\ref{sec:GWextraction}; the GW emission contains about 0.09\%, 0.08\% of the center-of-mass
energy for equal-mass binaries in $D=5,6$ respectively (note that in $D=4$ only 0.055\% of the
center-of-mass-eanergy goes into GWs), and decreases with the mass ratio. The
dependency of the radiated energy and momentum on the mass ratio is well
modelled by point particle calculations~\cite{Berti:2010gx}. A comparison of the predicted GW
emission in higher-dimensional collisions using two different numerical codes
with different formulations of the Einstein equations, namely those discussed in
Sections~\ref{sec:isometry} and \ref{sec:Cartoon}, has been presented in
Ref.~\cite{Witek:2014mha}.  The predictions from the two
codes using respectively the Kodama--Ishibashi formalism (cf.~Section~\ref{sec:GWextraction}) and
a direct extraction through the metric components (cf.~Section~IV B 1 in
\cite{Shibata:2010wz}) agree within numerical uncertainties~\cite{Witek:2014mha}. This result, illustrated in Figure~\ref{fig:D5energyflux_comparison}, represents an important validation of both the numerical evolution techniques and the
diagnostics of the simulations, along with the first estimate of the emitted energy for head-on collisions in $D=6$.

The main challenges for future numerical work in the field of high-energy
collisions are rather evident. For applications in the analysis of experimental
data in the context of TeV gravity scenarios (cf. Section~\ref{sec:tev}), it will be vital to generalize the results obtained in four dimensions to
$D\ge5$. Furthermore, it is currently not known whether the impact of electric
charge on the collision dynamics becomes negligible at high velocities, as
suggested by the matter-does-not-matter conjecture, and as is the case for the BH spin.

\subsection{Alternative theories}

As discussed in Section~\ref{sec:matter}, one of the most straightforward \emph{extensions} of Einstein's theory is obtained by the addition of minimally
coupled scalar fields. When the scalar couples to the Ricci scalar however, one
gets a modification of Einstein gravity, called scalar-tensor theory. In
vacuum, scalar-tensor theories are described by the generic action in
Eq.~(\ref{SJordan}), where $R$ is the Ricci scalar associated to the metric
$g_{\mu\nu}$, and $F(\phi), Z(\phi)$ and $U(\phi)$ are arbitrary functions (see
e.g.~\cite{Berti:2013gfa,Yunes:2011aa} and references therein). The matter
fields minimally coupled to $g_{\mu\nu}$ are collectively denoted by
$\Psi_m$. This form of the action corresponds to the choice of the so-called
``Jordan frame'', where the matter fields $\Psi_m$ obey the equivalence principle. For
$F=\phi, Z=\omega_{\rm BD}/\phi,\,U=0$, the action~(\ref{SJordan}) reduces to
the standard Brans--Dicke theory.

The equations of motion derived from the action~(\ref{SJordan}) are second-order
and the theory admits a well-posed initial-value problem~\cite{Salgado:2008xh}.
These facts turn scalar-theories into an attractive alternative to Einstein's
equations, embodying at least some of the physics one expects from an ultimate
theory of gravity, and have been a major driving force behind the efforts
to understand scalar-tensor theories from a NR point of
view~\cite{Shibata:1994qd,Scheel:1994yr,Scheel:1994yn,Novak:1997hw,Healy:2011ef,Berti:2013gfa,Barausse:2012da,Salgado:2008xh}.
In fact, scalar-tensor theories remain the only alternative theory to date where
full nonlinear dynamical evolutions of BH spacetimes have been performed.

Scalar-tensor theories can be recast in such a way as to be formally equivalent,
in vacuum, to GR with a \emph{minimally coupled} scalar field, i.e., to the theory
described previously in Section~\ref{sec:matter}. This greatly reduces the
amount of work necessary to extend NR to these setups. The explicit
transformations that recast the previous action in the ``Einstein frame''
are~\cite{Damour:1996ke}
\begin{eqnarray}
g^E_{\mu\nu}&=&F(\phi)g_{\mu\nu}\,,\qquad V=\frac{U}{16\pi F(\phi)^2}\,,
\label{JEtransf1}
\\
\varphi(\phi)&=&\frac{1}{\sqrt{4\pi G}}\int d\phi\,\left[\frac{3}{4}
\frac{F'(\phi)^2}{F(\phi)^2}+\frac{4\pi G Z(\phi)}{F(\phi)}\right]^{1/2}\,.
\label{JEtransf2}
\end{eqnarray}
The Einstein-frame action is then
\begin{equation}
S=\int {\rm d}^{4}x \sqrt{-g}\left( \frac{R^{E}}{16\pi G} -\frac{1}{2} g^{E,\,\mu\nu}\partial_{\mu}\varphi^{\ast}{}\partial_{\nu}\varphi
- V(\varphi)\right)+S_{\rm matter}(\Psi_m;\,F^{-1}g^{E}_{\mu\nu})\,,
\end{equation}
which is the action for a minimally coupled field (\ref{eq:action}) enlarged to
allow for a generic self-interaction potential (which could include the mass
term). The label $E$ denotes quantities constructed from
the Einstein-frame metric
$g_{\mu \nu}^{E}$. In the Einstein frame the scalar field is minimally coupled
to gravity, but any matter field $\Psi_m$ is coupled to the metric
$F^{-1}g^{E}_{\mu\nu}$.

In vacuum, this action leads to the following equations of motion:
\beq
&&R^{E}_{\mu\nu} - \frac{1}{2}g^{E}_{\mu\nu} R^{E} - 8\pi T_{\mu\nu}= 0\,,\label{eq:EoMEinstein1}
\\
&& \nabla^{\mu}\nabla_{\mu} \varphi - V^{'}(\varphi)=0\,,\label{eq:EoMEinstein2}
\eeq
where the energy-momentum tensor of the scalar field is determined by
\be
\label{eq:TmnScalar}
T_{\mu\nu}=-\frac{1}{2}g^{E}_{\mu\nu}\left( \partial_{\lambda}\varphi^{\ast}{}\partial^{\lambda}\varphi  \right)
-g^{E}_{\mu\nu} V(\varphi) +\frac{1}{2}\left(\partial_{\mu}\varphi^{\ast}{}\partial_{\nu}\varphi + \partial_{\mu}\varphi\partial_{\nu}\varphi^{\ast}{}\right)\,.
\ee

In summary, the study of scalar-tensor theories of gravity can be directly
translated, in vacuum, to the study of minimally coupled scalar fields.
For a trivial potential $V={\rm const}$, the equations of motion in the Einstein frame,
(\ref{eq:EoMEinstein1})\,--\,(\ref{eq:EoMEinstein2}) admit GR (with $\varphi=
\mathrm{const}$) as a solution. Because stationary BH spacetimes in GR are stable, i.e., any scalar
fluctuations die away rather quickly, the dynamical evolution of vacuum BHs is
expected to be the same as in GR.  This conclusion relies on hand-waving
stability arguments, but was verified to be true to first PN order by Will and
Zaglauer~\cite{Will:1989sk}, at 2.5 PN order by Mirshekari and
Will~\cite{Mirshekari:2013vb} and to all orders in the point particle limit in
Ref.~\cite{Yunes:2011aa}.

Thus, at least one of the following three ingredients are necessary to generate
interesting dynamics in scalar-tensor theories:
\begin{description}

\item[$\bullet$]{Nontrivial potential $V$ and initial conditions.} Healy et~al. studied an equal-mass BH binary in an inflation-inspired potential $V=\lambda\left(\varphi^2-\varphi_0^2\right)^2/8$ with nontrivial initial
  conditions on the scalar given by
  $\varphi=\varphi_0\tanh{\left(r-r_0\right)/\sigma}$~\cite{Healy:2011ef}.
  This setup is expected to cause deviations in the dynamics of the inspiralling
  binary, because the binary is now accreting scalar field energy.
  The larger the
  initial amplitude of the field, the larger those deviations are expected to
  be. This is summarized in Figure~\ref{fig:inspiral_healy}, where the BH
  positions are shown as a function of time for varying initial scalar
  amplitude.

\epubtkImage{}{%
  \begin{figure}[htbp]
    \centerline{
     \includegraphics[width=0.5\textwidth]{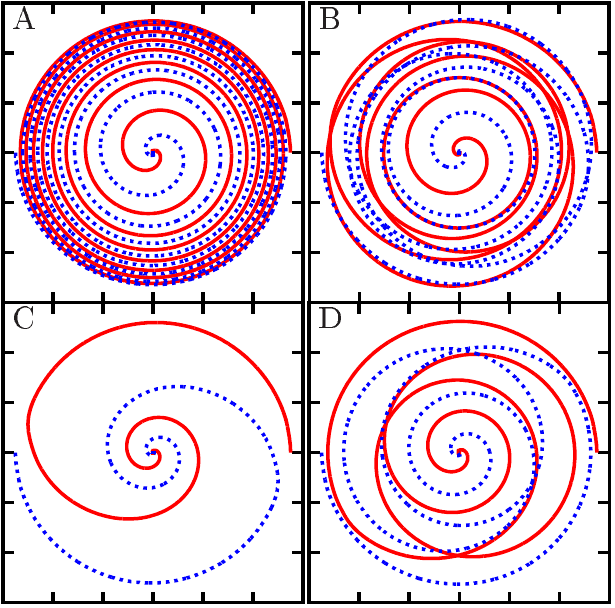}
    }
    \caption{Trajectories of BHs immersed in a scalar field bubble of different
      amplitudes.  The BH binary consists of initially non-spinning, equal-mass
      BHs in quasi-circular orbit, initially separated by $11M$, where $M$
      is the mass of the binary system.
      The scalar field bubble surrounding the binary has a
      radius $r_0=120M$ and thickness $\sigma=8M$. Panels $A,B,C$ correspond to
      $\varphi_0=0 ({\rm GR}),1/80,1/40$ 
      and a zero potential amplitude $\lambda$.
      Panel $D$ corresponds to $\varphi_0=1/80, 4\pi \lambda=10^3M^2$. From
      Ref.~\cite{Healy:2011ef}.}
    \label{fig:inspiral_healy}
\end{figure}}

\item[$\bullet$]{Nontrivial boundary conditions.} As discussed, GR is
  recovered for constant scalar fields.  For nontrivial
  time-dependent boundary conditions or background scalar fields, however,
  nontrivial
  results show up. These boundary conditions could mimic cosmological
  scenarios or dark matter profiles in
  galaxies~\cite{Horbatsch:2011ye,Berti:2013gfa}. Reference~\cite{Berti:2013gfa}
  modelled a BH binary evolving nonlinearly in a constant-gradient
  scalar field.  The scalar-field gradient induces scalar charge on the BHs, and
  the accelerated motion of each BH in the binary generates scalar radiation at
  large distances, as summarized in Figure~\ref{fig:inspiral_gradient}.

\epubtkImage{}{%
  \begin{figure}[htbp]
    \centerline{
     \includegraphics[width=0.5\textwidth]{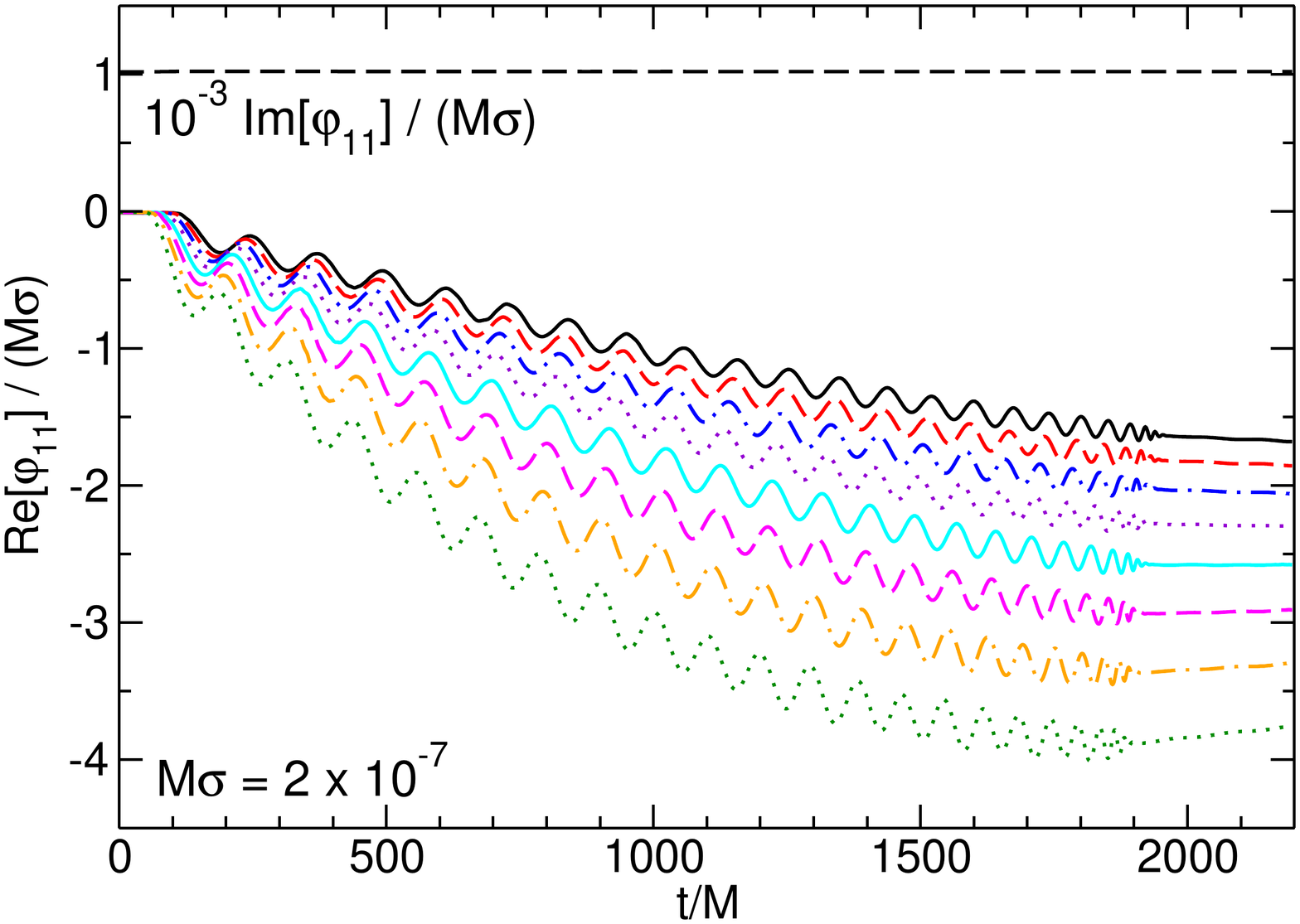}
     \includegraphics[width=0.5\textwidth]{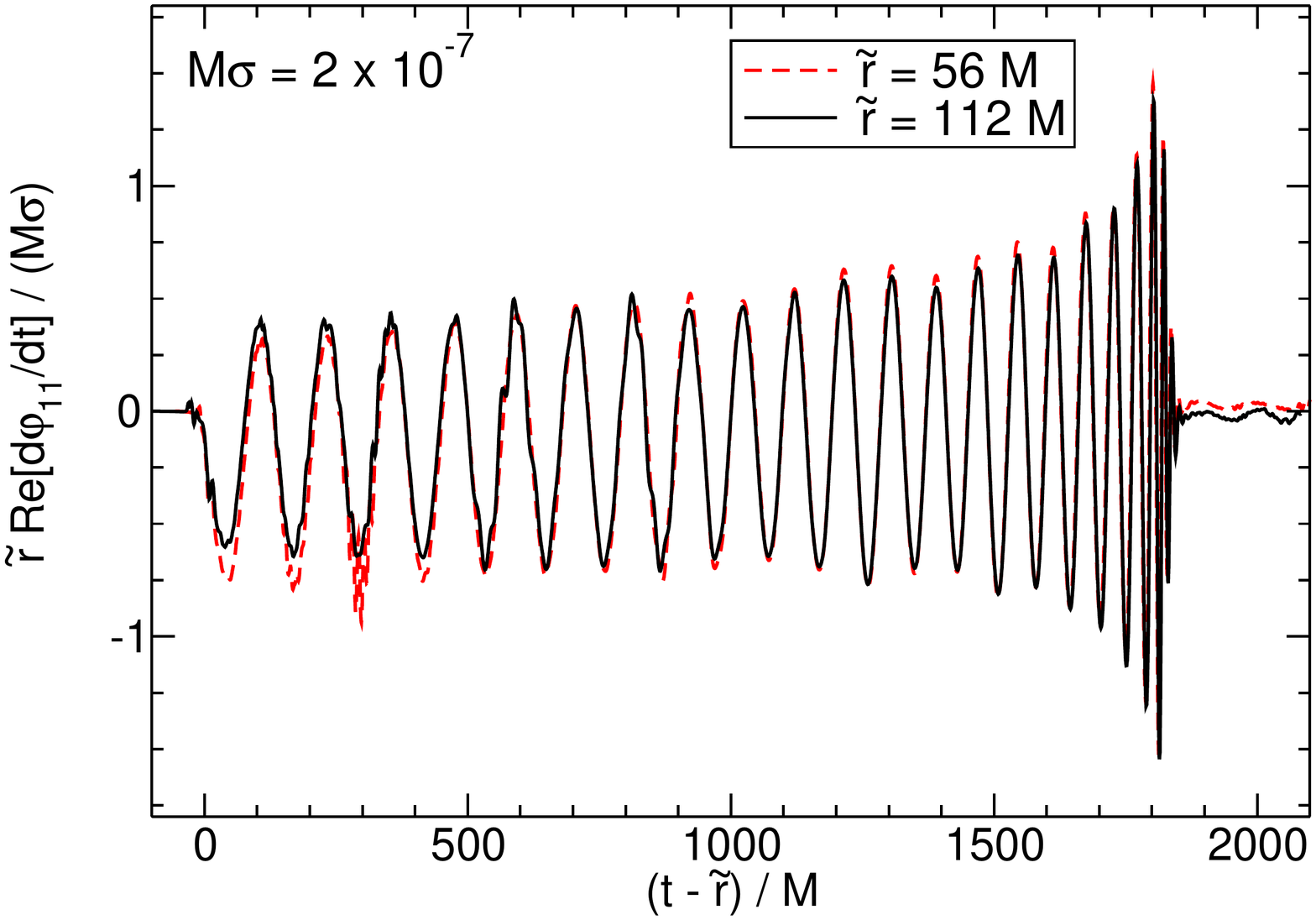}
    }
    \caption{Numerical results for a BH binary inspiralling in a scalar field
      gradient $M\sigma=10^{-7}$. \emph{Left panel:} dependence of the various components of
      the scalar radiation ${\rm Re}(\varphi_{11})/(M\sigma)$ on the extraction
      radius (top to bottom: 112$M$ to 56$M$ in equidistant steps).  The dashed
      line corresponds instead to $10^{-3}{\rm Im}(\varphi_{11})/(M\sigma)$ at
      the largest extraction radius. This is the dominant mode and corresponds
      to the fixed-gradient boundary condition, along the $z$-direction, at
      large distances. \emph{Right panel:} time-derivative of the scalar field at the largest
      and smallest extraction radii, rescaled by radius and shifted in
      time. Notice how the waveforms show a clean and typical merger
      pattern, and that they overlap showing that the field scales to good
      approximation as $1/\tilde{r}$. From Ref.~\cite{Berti:2013gfa}.}
    \label{fig:inspiral_gradient}
\end{figure}}

The scalar-signal at large distances, shown in the right panel of
Figure~\ref{fig:inspiral_gradient}, mimics the inspiral, merger and ringdown
stages in the GW signal of an inspiralling BH binary.

\item[$\bullet$]{Matter.} When matter is present, new effects (due to the
  coupling of matter to the effective metric $F^{-1}g^{E}_{\mu\nu}$) can
  dominate the dynamics and wave emission. For example,
  it has been shown that, for $\beta\equiv\partial_\varphi^2(\ln F(\varphi))\lesssim-4$,
  NSs can ``spontaneously scalarize,'' i.e., for
  sufficiently large compactnesses the GR solution is unstable. The stable
  branch has a nonzero expectation value for the scalar field  ~\cite{Damour:1992we}.

\epubtkImage{}{%
  \begin{figure}[htbp]
    \centerline{
     \includegraphics[width=0.5\textwidth]{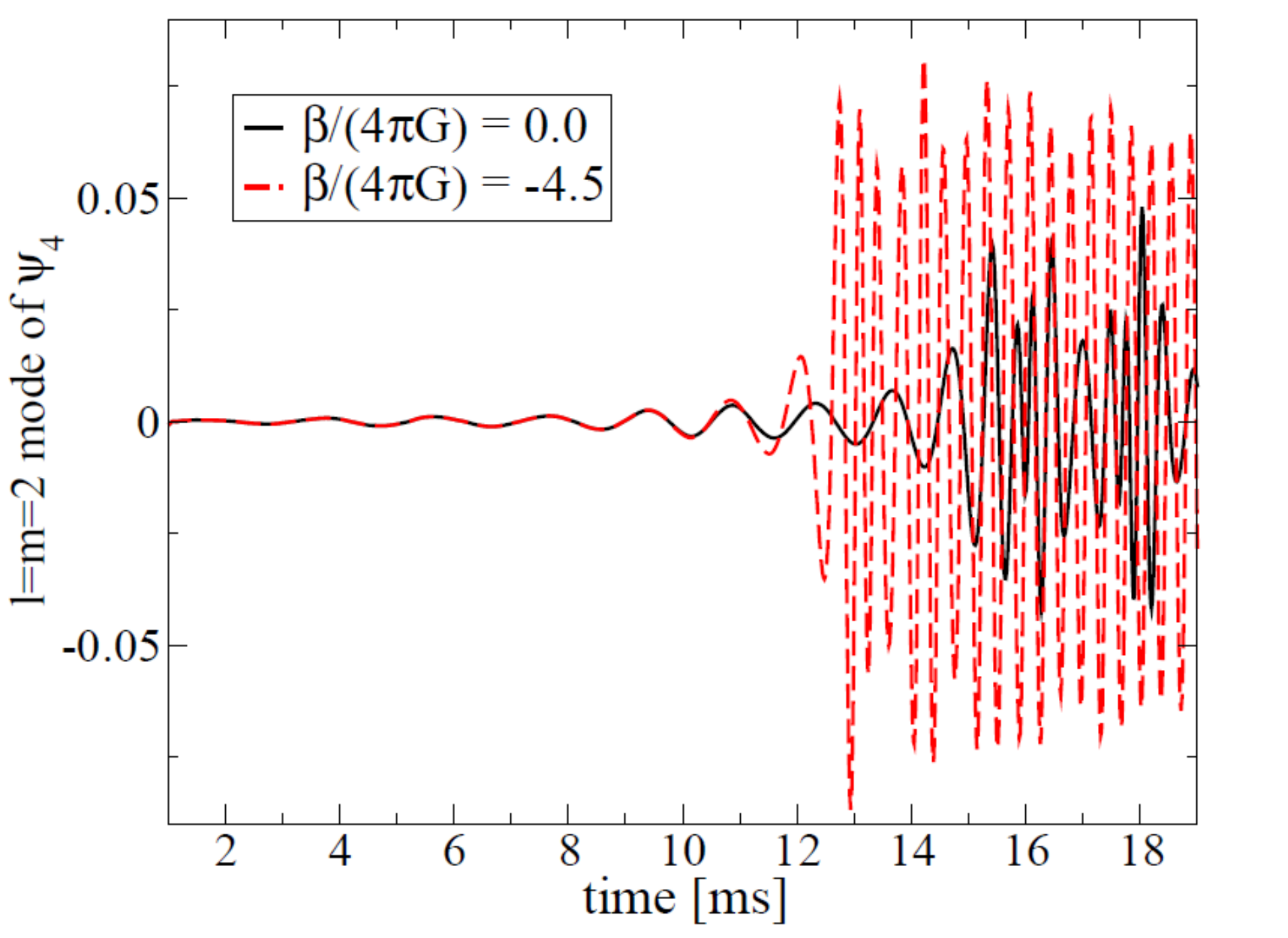}
   }
   \caption{The dominant quadrupolar component of the gravitational $\psi_4$
     scalar for an equal-mass, non-spinning NS binary with individual baryon masses of $1.625
     M_{\odot}$.  The solid (black) curve refers to GR, and the
     dashed (red) curve to a scalar-tensor theory with $\beta/(4\pi)=-4.5,\,\varphi_0=10^{-5}$. From   Ref.~\cite{Barausse:2012da}.}
    \label{fig:spontaneous_scalarization}
\end{figure}}

Scalarized matter offers a rich new phenomenology. For example, the dynamics and
GW emission of scalarized NSs can be appreciably
different (for given coupling function $F^{-1}(\phi)$) from the corresponding GR
quantities, as shown by Palenzuela et~al.~\cite{Barausse:2012da,Palenzuela:2013hsa} and summarized in
Figure~\ref{fig:spontaneous_scalarization}. Strong-field gravity can
even \emph{induce} dynamical scalarization of otherwise GR stars
during inspiral, offering new ways to constrain such
theories~\cite{Barausse:2012da,Palenzuela:2013hsa}.

\end{description}

The application of NR methods to the understanding of alternative theories of
gravity and tests of GR is still in its infancy.  Among various possible
directions, we point out the following.
\begin{description}

\item[$\bullet$]{Understanding the well posedness} of some 
  theory, in particular those having some motivation from fundamental physics,
  as for example Einstein-Dilaton--Gauss--Bonnet and Dynamical Chern--Simons gravity~\cite{Pani:2009wy,Alexander:2009tp}. A study on the well posedness of the latter has recently been presented in Ref.~\cite{Delsate:2014hba}.

\item[$\bullet$]{Building initial data} describing interesting setups for such
  theories. Unless the theory admits particularly simple analytic solutions, it
  is likely that initial data construction will also have to be done
  numerically. Apart from noteworthy exceptions, such as Gauss-Bonnet gravity in
  higher dimensions~\cite{Yoshino:2011qp}, initial data have hardly been
  considered in the literature.
 
\end{description}

Once well-posedness is established and initial data are constructed,
NR evolutions will help us understanding how these theories behave in
the non-linear regime.

\subsection{Holography}
\label{sec_holography7}

Holography provides a fascinating new source of problems for NR. As such, in recent years, a number of numerical frameworks have been explored in asymptotically AdS spacetimes, as to face the various pressing questions raised
within the holographic correspondence, cf.\ Section~\ref{sec:gauge-gravity}. At
the moment of writing, no general purpose code has been reported, comparable to
existing codes in asymptotically flat spacetime, which can evolve, say, BH
binaries with essentially arbitrary masses, spins and momenta. Progress has
occured in specific directions to address specific issues.  We shall now review
some of these developments emphasizing the gravity side of the problems.

\epubtkImage{}{%
\begin{figure}[htbp]
\centerline{
      \includegraphics[width=0.5\textwidth]{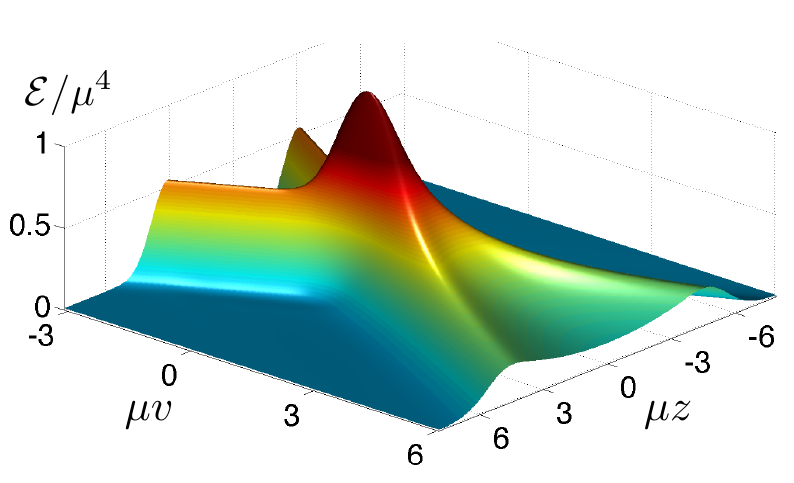}
      \includegraphics[width=0.5\textwidth]{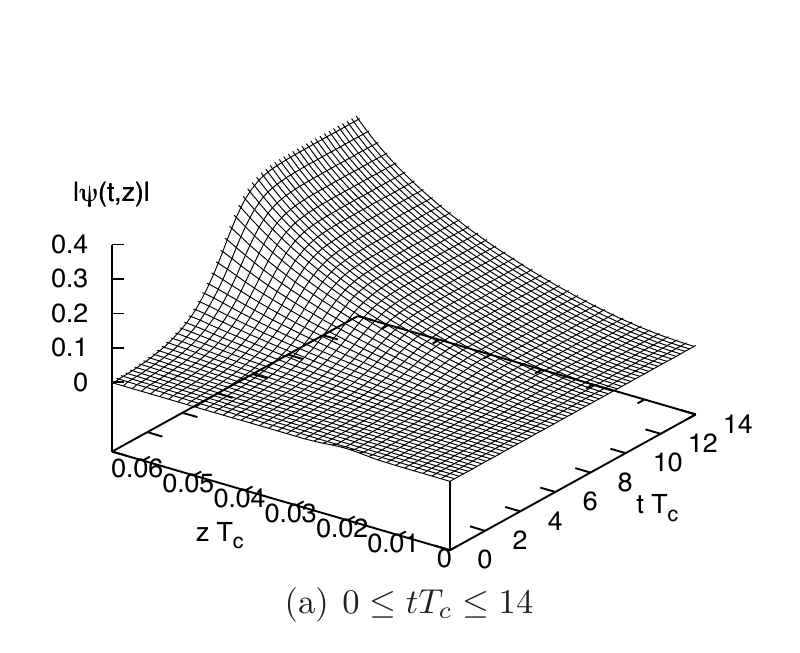}
}
\caption{\emph{Left panel:} Collision of two shock waves in ${\rm AdS}_5$. The energy
  density $\mathcal{E}/\mu^4$ is represented as a function of
an (advanced) time
  coordinate $v$ and a longitudinal coordinate $z$. $\mu$ defines the amplitude
  of the waves. From~\cite{Chesler:2010bi}. \emph{Right panel:} Evolution of the
  scalar field in an unstable RN-AdS BH. $z$ is a radial coordinate and the
  AdS boundary is at $z=0$. Due to the instability of the BH, the scalar
  density grows exponentially for $0<tT_c\lesssim 6$. Then, the scalar density
  approaches some static function. From~\cite{Murata:2010dx}.}
\label{chesler}
\end{figure}}

As mentioned in Section~\ref{sec:gauge-gravity} an important problem in the
physics of heavy ion collisions is to understand the ``early thermalization
problem''. In Ref.~\cite{Chesler:2010bi}, the gauge/gravity duality was used to
address this issue. On the gravity side, the problem at hand was to study a
head-on collision of two shock waves in asymptotically ${\rm AdS}_5$ spacetime.  The numerical scheme was to perfom a null
(characteristic) evolution. By choosing a specific metric ansatz, it was
possible to unveil in the non-linear Einstein equations a nested linear
structure: the equations can be integrated as linear ordinary differential
equations if an appropriate sequence is chosen. The AdS boundary condition was
implemented by an adequate radial expansion near the boundary and the initial
data consisted of two well-separated planar shocks, with finite thickness and
energy density, moving toward each other. In this setup an AH is always found
(even before the collision) and excision was performed by restricting the
computational domain to start at this horizon. The evolution of the two shock
waves is displayed in Figure~\ref{chesler} (left panel). By following the
evolution and using the gauge/gravity dictionary, the authors reported that the
total time required for apparent thermalization was 0.35 fm/c. This is within
the same order of magnitude as the thermalization scale obtained from
accelerator data, already discussed in Section~\ref{sec:gauge-gravity}. A discussion of numerical
approaches using null evolutions applied to asymptotically AdS problems can be
found in~\cite{Chesler:2013lia}. Other recent applications of shock wave
collisions in ${\rm AdS}_5$ to describe phenomenological properties of heavy ion
collisions can be found in
Refs.~\cite{Wu:2011yd,Romatschke:2013re,vanderSchee:2013pia}.

Time-plus-space decompositions have also been initiated, both based on a
generalized harmonic evolution scheme~\cite{Bantilan:2012vu} and in an ADM
formulation~\cite{Heller:2012je}. In particular the latter formulation seems
very suited for extracting relevant physical quantities for holography, such as
the boundary time for the thermalization process discussed in
Section~\ref{sec:gauge-gravity}.

Evolutions of BHs deformed by a scalar field in ${\rm AdS}_5$ have been
presented in Ref.~\cite{Bantilan:2012vu}.  The evolution leads the system
to oscillate in a (expected) superposition of quasi-normal modes, some
of which are nonlinearly driven. On the boundary, the dual CFT stress
tensor behaves like that of a thermalized ${\cal N}=4$ super-Yang--Mills
fluid, with an equation of state consistent with conformal invariance
and transport coefficients that match holographic calculations \emph{at
all times}. Similar conclusions were
reached in Ref.~\cite{Heller:2012km}, where the numerical scheme of
Ref.~\cite{Chesler:2010bi,Chesler:2013lia} was used to study the isotropization of a
homogeneous, strongly coupled, non-Abelian plasma by means of its gravity dual,
comparing the time evolution of a large number of initially anisotropic
states. They find that the linear approximation seems to work well even for
initial states with large anisotropies. This unreasonable effectiveness of
linearized predictions hints at something more fundamental at work, perhaps a
washing out of nonlinearities close to the horizon.  Such effects were observed
before in asymptotically flat spacetimes, for example the already mentioned
agreement between ZFL (see Section~\ref{sec:ZFLsection}) or close limit
approximation predictions (see Section~\ref{sec:st_perturbation}) and full
nonlinear results.

Also of interest for accelerator physics, and the subject of intense work in
recent years, are holographic descriptions of jet-quenching, i.e., the loss of
energy of partons as they cross strongly coupled plasmas produced in heavy ion
collisions~\cite{Aad:2010bu,Chatrchyan:2011sx,Chatrchyan:2012nia}. Numerical
work using schemes similar to that of
Refs.~\cite{Chesler:2010bi,Chesler:2013lia} have been used to evolve dual
geometries describing the quenches~\cite{Buchel:2012gw,Buchel:2013lla,Chesler:2014jva}; see also~\cite{Figueras:2012rb} for numerical \emph{stationary} solutions in this context.

Another development within the gauge/gravity duality that gained much
 attention, also discussed in Section~\ref{sec:gauge-gravity}, is related to
 condensed matter physics. In asymptotically AdS spacetimes, a simple theory,
 say, with a scalar field minimally coupled to the Maxwell field and to gravity
 admits RN-AdS as a solution. Below a critical temperature,
 however, this solution is unstable against perturbations of the scalar field,
 which develops a tachyonic mode. Since the theory admits another set of charged
 BH solutions, which have scalar hair, it was suggested that the
 development of the instability of the RN-AdS BHs leads the system to a hairy
 solution. From the dual field theory viewpoint, this corresponds to a phase
 transition between a normal and a superconducting phase. A numerical simulation
 showing that indeed the spacetime evolution of the unstable RN-AdS BHs leads to
 a hairy BH was reported in~\cite{Murata:2010dx}. Therein, the authors performed
 a numerical evolution of a planar RN-AdS BH perturbed by the scalar field and
 using Eddington-Finkelstein coordinates. A particular numerical scheme was
 developed, adapted to this problem. The development of the scalar field density
 is shown in Figure~\ref{chesler}. The initial exponential growth of the scalar
 field is eventually replaced by an approach to a fixed value, corresponding to
 the value of the scalar condensate on the hairy BH.

 Finally, the gauge/gravity duality itself may provide insight into
 turbulence. Turbulent flows of CFTs are dual to dynamical BH solutions in
 asymptotically AdS spacetimes. Thus, urgent questions begging for answers
 include how and when do turbulent BHs arise, and what is the (gravitational)
 origin of Kolmogorov scaling observed in turbulent fluid flows. These problems are
 just now starting to be addressed~\cite{Carrasco:2012nf,Adams:2012pj,Adams:2013vsa}.

\subsection{Applications in cosmological settings}

Some initial applications of NR methods addressing specific issues in cosmology
have been reviewed in the Living Reviews article by
Anninos~\cite{Anninos:1998bka}, ranging from the Big Bang singularity dynamics
to the interactions of GWs and the large-scale structure of the Universe. The
first of these problems -- the understanding of cosmological singularities --
actually motivated the earlier applications of NR to cosmological settings,
\emph{cf.} the Living Reviews article~\cite{Berger:2002st}.  The set of
homogeneous but \emph{anisotropic} universes was classified by Bianchi in 1898
into nine different types (corresponding to different independent groups of
isometries for the 3-dimensional space).  Belinskii, Khalatnikov and Lifshitz
(BKL) proposed that the singularity of a generic inhomogeneous cosmology is a
``chaotic'' spacelike curvature singularity, and that it would behave
asymptotically like a Bianchi IX or VIII homogeneous cosmological model.  This
is called BKL dynamics or mixmaster universe.  The accuracy of the BKL dynamics
has been investigated using numerical evolutions in
Refs.~\cite{Berger:1993fn,1973AnPhy..79..558M,Rugh:1990aq}, and the BKL
sensitivity to initial conditions in various references (see for instance
Ref.~\cite{Berger:1996gk}).  For further details we refer the reader to
Ref.~\cite{Berger:2002st}.

More recently, NR methods have been applied to the study of bouncing cosmologies, by studying the evolution of adiabatic perturbations in a nonsingular bounce~\cite{Xue:2013bva}. The results of Ref.~\cite{Xue:2013bva} show that the bounce is disrupted in regions of the Universe with significant inhomogeneity and anisotropy over the background energy density, but is achieved in regions that are relatively homogeneous and isotropic. Sufficiently small perturbations, consistent with observational constraints, can pass through the nonsingular bounce with negligible alteration from nonlinearity. 

In parallel, studies of ``bubble universes'', in which our Universe is one of many nucleating and growing inside an
ever-expanding false vacuum, have also been made with NR tools. In particular, Refs.~\cite{Wainwright:2013lea,Wainwright:2014pta} investigated
the collisions between bubbles, by computing the cosmological observables arising from bubble collisions
directly from the Lagrangian of a single scalar field.

\epubtkImage{}{%
\begin{figure}[htbp]
\centerline{
      \includegraphics[width=0.45\textwidth]{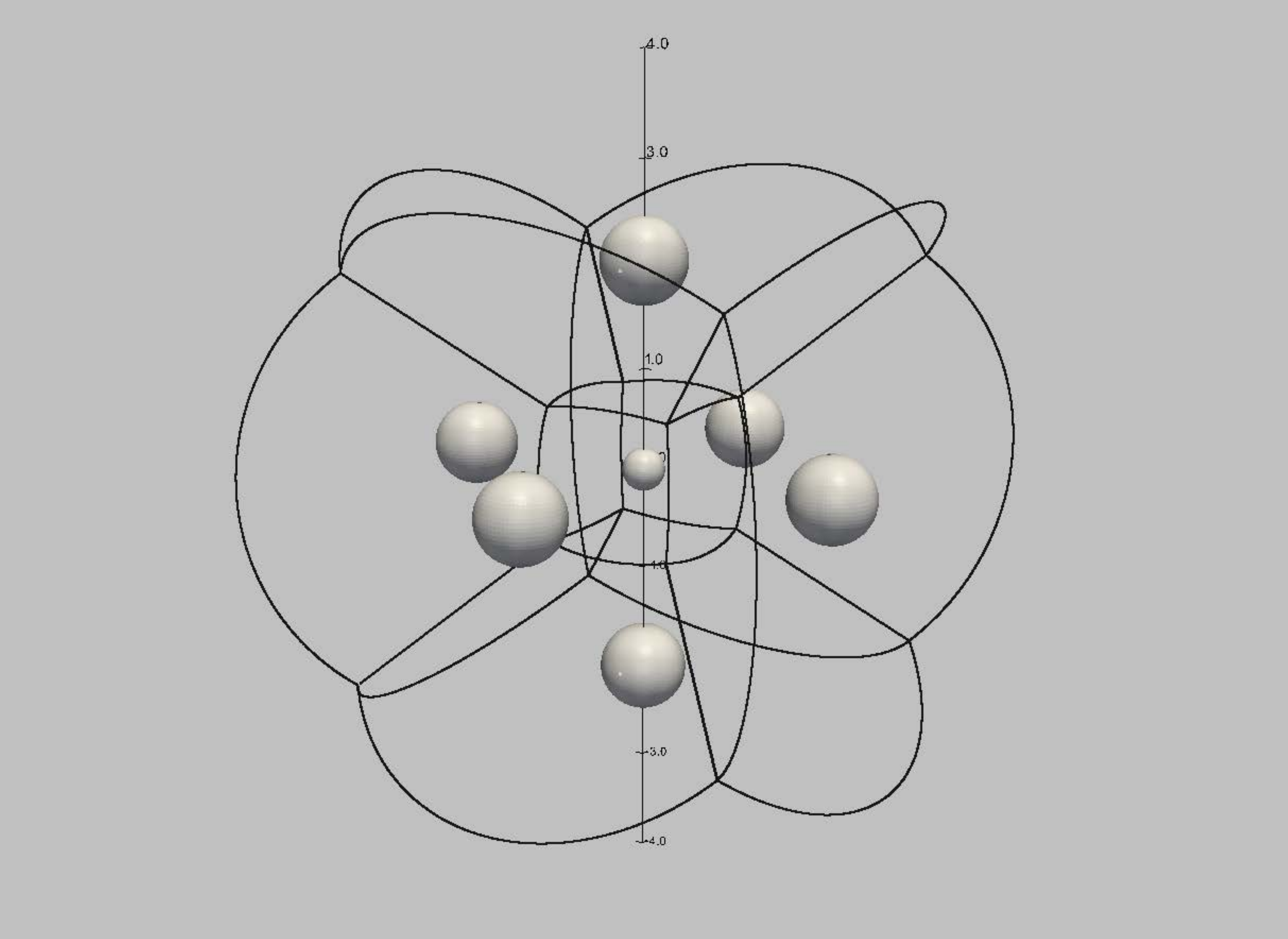} \hskip 1cm
      \includegraphics[width=0.45\textwidth]{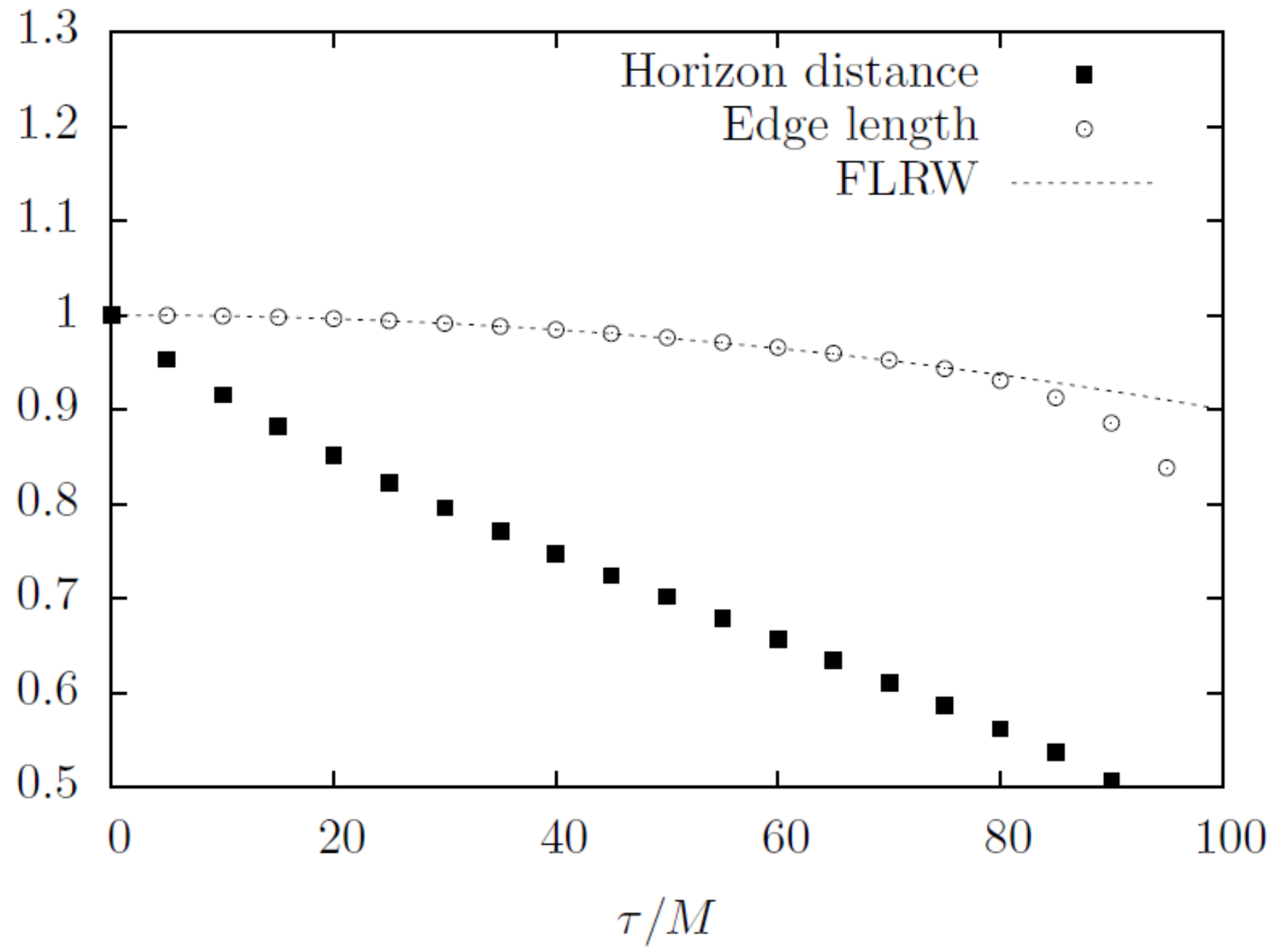}
}
    \caption{\emph{Left:} Elementary cells for the 8-BH configuration,
projected to $\RealNum^3$. The marginal surface corresponding to the BH at infinity encompasses the whole configuration. 
Note that the 8 cubical lattice cells are isometric after the conformal rescaling.
\emph{Right:} Several measures of scaling in the eight-BH universe, as functions of
proper time $\tau$, plotted against a possible identification of the corresponding FLRW model
(see Ref.~\cite{Bentivegna:2012ei} for details). All the quantities have been renormalized to their respective 
values at $\tau=0$. From~\cite{Bentivegna:2012ei}.}
\label{cosmo}
\end{figure}}

Applications of NR in more standard cosmological settings are still in their infancy, but
remarkable progress has been achieved. One of these concerns the
impact of cosmic inhomogeneities on the value of the cosmological constant and
the acceleration of the Universe. In other words, how good are models of
homogeneous and isotropic universes, -- the paradigmatic
Friedmann--Lema{\^{\i}}tre--Robertson--Walker (FLRW) geometry -- when we know that our
Universe has structure and is inhomogeneous?

Studies of this (long-standing, see for instance Ref.~\cite{RevModPhys.29.432})
question within NR have considered the evolution of BH lattices (the BHs
mimicking strong, self-gravitating inhomogeneities)
\cite{Bentivegna:2012ei,Yoo:2013yea}. In Ref.~\cite{Bentivegna:2012ei} the
authors explicitly constructed and evolved a three-dimensional, fully
relativistic, eight-BH lattice with the topology of $S^3$. The puncture
locations in that work projected down to $R^3$ are shown in the left panel of
Figure~\ref{cosmo} (one of the punctures is projected out to infinity, see
Ref.~\cite{Bentivegna:2012ei} for further details). The evolution of this 8-BH configuration is summarized in
  the right panel of Figure~\ref{cosmo}, showing the (minimal) proper distance
  between neighbouring surfaces and the proper length of each cell's edges.
  These quantities are then compared against a reference FLRW closed model with
  spatial slices of spherical topology. The comparison procedure is not
  straightforward, but adopting the procedure of Ref.~\cite{Bentivegna:2012ei}
  it yields good agreement.

The effects of local inhomogeneities have been investigated in Ref.~\cite{Yoo:2012jz} using different initial data, describing
an expanding inhomogeneous universe model composed of regularly aligned BHs of identical mass.
The evolution of these initial data also indicates that local inhomogeneities do not significantly affect the global expansion law of the universe, despite the fact that the inhomogeneities themselves are extremely nonlinear~\cite{Yoo:2012jz,Yoo:2013yea}.
Similar conclusions were reached in Ref.~\cite{2011PhRvD..83b3524Z}, where the ADM formalism is used to develop a practical scheme to calculate a proposed domain averaging effect in an inhomogeneous cosmology within the context of numerical large-scale structure simulations. This study finds that in
the weak-field, slow-motion limit, the proposed effect implies a small correction to the global expansion rate of the
Universe. In this limit, their simulations are always dominated by the expanding underdense regions, hence the
correction to the energy density is negative and the effective pressure is positive. The effects of strong gravity in more general scenarios are yet to be understood~\cite{2011PhRvD..83b3524Z}. For an earlier NR code developed to address inhomogeneous cosmologies see Ref.~\cite{Hern:1999dq}.

More complex NR codes aimed at understanding cosmological evolutions are currently being developed. 
NR simulations of large scale dynamical processes in the early Universe have recently been reported~\cite{Garrison:2012ex}. These take into account interactions of dark matter, scalar perturbations, GWs, magnetic fields and turbulent plasma.
Finally, Ref.~\cite{Zilhao:2012bb} considers the effect of (extreme) cosmological expansion on the head-on collision and merger of two BHs,
by modelling the collision of BHs in asymptotically dS spacetimes.

\newpage
\section{Conclusions}

\textit{``Somewhere, something incredible is waiting to be known.''}%
\epubtkFootnote{This quote, usually attributed to Carl Sagan,
was published in \textit{Seeking Other Worlds} in Newsweek magazine (April 15, 1977), a tribute to Carl Sagan by D.~Gelman, S.~Begley, D.~Gram and E.~Clark.}
\\

\noindent
Einstein's theory of general relativity celebrates its 100th anniversary in 2015
as perhaps the most elegant and successful attempt by humankind to capture the
laws of physics.  

Until recently, this theory has been studied mostly in the weak-field regime,
where it passed all experimental and observational tests with flying
colors. Studies in the strong-field regime, in contrast, largely concerned the
mathematical structure of the theory but made few and indirect connections with
observation and experiment.  Then, a few years ago, a phase transition in the
field of strong gravity occurred: on one hand, new experimental efforts
are promising to test gravity for the first time in the strong
field regime; on the other hand, a new tool -- numerical relativity -- has made
key breakthroughs opening up the regime of strong-field gravity phenomena for
accurate modelling. Driven by these advances, gravitation in the strong-field
regime has proven to have remarkable connections to other branches of physics.

With the rise of numerical relativity as a major tool to model and study
physical processes involving strong gravity, decade-old problems -- brushed
aside for their complexity -- are now tackled with the use of personal or
high-performance computers.  Together with analytic methods, old and new, the
new numerical tools are pushing forward one of the greatest human endeavours:
understanding the Universe.

\newpage
%
\section*{Acknowledgements}
\label{sec:acknowledgements}
This work benefited greatly from discussions over the years with many colleagues.
In particular, we thank E.~Berti, R.~Brito, J.~C.~Degollado,
R.~Emparan, P.~Figueras, D.~Hilditch, P.~Laguna, L.~Lehner, D.~Mateos,
A.~Nerozzi, H.~Okawa, P.~Pani, F.~Pretorius, E.~Radu, H.~Reall,
H.~R\'unarsson, M.~Sampaio, J.~Santos, M.~Shibata, C.~F.~Sopuerta, A.~Sousa,
H.~Witek and M.~Zilh\~ao for very useful comments and suggestions.
This work was partially funded by the NRHEP 295189 FP7-PEOPLE-2011-IRSES and PTDC/FIS/116625/2010 grants and the CIDMA strategic project UID/MAT/04106/2013.
V.C. acknowledges partial financial support provided under the European
Union's FP7 ERC Starting Grant
``The dynamics of black holes: testing the limits of Einstein's theory'' 
grant agreement no.~DyBHo--256667. 
L.G. acknowledges partial financial support from NewCompStar (COST Action MP1304).
C.H. is funded by the FCT-IF programme.
U.S. acknowledges support from
FP7-PEOPLE-2011-CIG Grant No. 293412 CBHEO,
STFC GR Roller Grant No. ST/L000636/1,
NSF XSEDE Grant No. PHY-090003 and support by the Cosmos Shared Memory system
at DAMTP, University of Cambridge, operated on behalf of the STFC
DiRAC facility and funded by BIS National E-infrastructure capital grant
ST/J005673/1 and STFC Grant Nos.~ST/H008586/1, ST/K00333X/1.
This research was supported in part by Perimeter Institute for Theoretical
Physics. Research at Perimeter Institute is supported by the Government of
Canada through Industry Canada and by the Province of Ontario through the Ministry of
Economic Development. 

\newpage


\bibliography{BeyondAstro_ref_v3}

\end{document}